\documentclass[10pt,journal,compsoc]{IEEEtran}
\usepackage{ifpdf}

\ifCLASSOPTIONcompsoc
  \usepackage[nocompress]{cite}
\else
  \usepackage{cite}
\fi
\ifCLASSINFOpdf
  \usepackage[pdftex]{graphicx}
  \DeclareGraphicsExtensions{.pdf,.jpeg,.jpg,.png}
\else
  \usepackage[dvips]{graphicx}
  \DeclareGraphicsExtensions{.eps}
\fi
\ifpdf
\else
	\usepackage{psfrag}
\fi

\usepackage{amsmath}
\usepackage{array}

\ifCLASSOPTIONcompsoc
  \usepackage[caption=false,font=footnotesize,labelfont=sf,textfont=sf]{subfig}
\else
  \usepackage[caption=false,font=footnotesize]{subfig}
\fi
\usepackage{dblfloatfix}
\usepackage{url}

\usepackage{threeparttable}

\newcommand{\arXiv}[2]{#1}

\newcommand{\TCBB}[2]{#1}		
\TCBB{
\newcommand{\SkipTCBBToMerge}[1]{#1}	
\newcommand{\SkipNonTCBBToMerge}[1]{}	
}{
\newcommand{\SkipTCBBToMerge}[1]{}	
\newcommand{\SkipNonTCBBToMerge}[1]{#1}	
}

\ifdefined\SUPPLEMENT
\else
\newcommand{\SUPPLEMENT}[1]{}
\fi

\newcommand{\SkipSupplToMerge}[1]{}
\SUPPLEMENT{
\renewcommand{\SkipSupplToMerge}[1]{#1}
}

\ifdefined\TEXT
\else
\newcommand{\TEXT}[1]{#1}
\fi

\newcommand{\SkipTextToMerge}[1]{}
\TEXT{
\renewcommand{\SkipTextToMerge}[1]{#1}
}

\TCBB{
\usepackage{ifthen}
\usepackage{xifthen}
\usepackage{ifthenx}

\ifdefined\Skip
\else
\newcommand{\Skip}[1]{}
\fi

\usepackage{color}
\usepackage{xcolor}
\usepackage{times}
\usepackage{txfonts}
\usepackage{bm}

\usepackage[hidelinks]{hyperref}

\ifdefined\TCBB
\else
\newcommand{\TCBB}[2]{#2}
\fi
\ifdefined\arXiv
\else
\newcommand{\arXiv}[2]{#2}
\fi
\ifdefined\BMC
\else
\newcommand{\BMC}[2]{#2}
\fi

\ifdefined\Skip
\else
\newcommand{\Skip}[1]{}                 
\fi

\newcommand{\script}[1]{{\mbox{\scriptsize #1}}}

\newcommand{\CITE}[1]{\cite{#1}}
\newcommand{\VEC}[1]{\text{\boldmath{$#1$}}}
\newcommand{\VECS}[1]{\text{\boldmath{$#1$}}}
\newcommand{\Eqno}[1]{{#1}}
\newcommand{\Eq}[1]{Eq. \Eqno{#1}}
\newcommand{\Eqs}[1]{Eqs. {#1}}

\newcommand{\Fig}[1]{Fig. {#1}}
\newcommand{\Figs}[1]{Figs. {#1}}
\newcommand{\SFig}[1]{Fig. {#1}}
\newcommand{\SFigs}[1]{Figs. {#1}}
\newcommand{\Table}[1]{Table {#1}}
\newcommand{\Tables}[1]{Tables {#1}}

\newcommand{\SkipFigure}[1]{}
\newcommand{\FigureInText}[1]{}

\newcommand{\FigureInLegends}[1]{}

\newcommand{\TableInText}[1]{}
\newcommand{\TableLegends}[1]{}

\newcommand{\EQ}{eq}
\newcommand{\TBL}{tbl}
\newcommand{\FIG}{fig}
\newcommand{\SECTION}[1]{\section{#1}}
\newcommand{\SUBSECTION}[1]{\subsection{#1}}
\newcommand{\SUBSUBSECTION}[1]{\subsubsection{#1}}
\newcommand{\PARAGRAPH}[1]{\paragraph{#1}}

\newcommand{\TEXTBF}[1]{#1}

\ifdefined\EQ
\renewcommand{\EQ}{eq}
\else
\newcommand{\EQ}{eq}
\fi
\ifdefined\TBL
\renewcommand{\TBL}{tbl}
\else
\newcommand{\TBL}{tbl}
\fi
\ifdefined\FIG
\renewcommand{\FIG}{fig}
\else
\newcommand{\FIG}{fig}
\fi

\newcommand{\Detail}[1]{}

\ifdefined\text
\else
\newcommand{\text}[1]{\textrm{#1}}
\fi

\renewcommand{\VEC}[1]{\text{\boldmath{$#1$}}} 
\renewcommand{\VECS}[1]{\text{\boldmath{$#1$}}}

\newcommand{\RED}[1]{\textcolor{red}{#1}}
\newcommand{\BLACK}[1]{\textcolor{black}{#1}}

\renewcommand{\SFig}[1]{\Fig{#1}}
\renewcommand{\SFigs}[1]{\Figs{#1}}

\newcommand{\SupplementaryBibliography}[1]{#1}

\newcommand{\TextMaterial}[1]{#1}
\TextMaterial{
\renewcommand{\SupplementaryBibliography}[1]{}
}

\renewcommand{\FigureInText}[1]{#1}
\renewcommand{\TableInText}[1]{#1}

\newcommand{\SkipFigsInText}[1]{}
\FigureInText{
\renewcommand{\SkipFigsInText}[1]{#1}
}
\newcommand{\SkipTablesInText}[1]{}
\TableInText{
\renewcommand{\SkipTablesInText}[1]{#1}
}

}{

}

\newcommand{\NAG}[2]{#1}	
\newcommand{\FiguresWithoutCaption}[1]{}
\newcommand{\withSup}[2]{#1}
\FiguresWithoutCaption{
\renewcommand{\withSup}[2]{#2}
}

\begin{document}
%
\title{
Boltzmann Machine Learning and Regularization Methods
for Inferring
Evolutionary Fields and
Couplings
from a Multiple Sequence Alignment
}
%
%
%
%

\author{Sanzo~Miyazawa
\IEEEcompsocitemizethanks{\IEEEcompsocthanksitem	
Email: sanzo.miyazawa@gmail.com
}%
\thanks{Manuscript received 20 Sept. 2019; revised 8 Apr. 2020; accepted 2 May 2020;
DOI: 10.1109/TCBB.2020.2993232 ; The corrections made afterward are indicated in red.}
}

%
%

\markboth{IEEE/ACM Trans. Comput. Biol. Bioinform., 2020}%
{Sanzo Miyazawa: Boltzmann machine learning and regularization methods for an inverse Potts problem}
%



\IEEEtitleabstractindextext{%
\begin{abstract}
The inverse Potts problem to infer
a Boltzmann distribution for homologous protein sequences
from their single-site and pairwise amino acid frequencies
recently attracts a great deal of attention
in the studies of protein structure and evolution.
We study
regularization and learning methods
and how to tune regularization parameters
to correctly infer interactions
in Boltzmann machine learning.	
Using $L_2$ regularization for fields, group $L_1$ for couplings
is shown to be very effective
for sparse couplings in comparison with $L_2$ and 
$L_1$.
Two regularization parameters 
are tuned to yield equal values for both
the sample and ensemble averages of evolutionary energy.
Both averages
smoothly change and converge, but
their learning profiles are very different between 
learning methods.
The Adam method is modified
to make stepsize proportional to the 
gradient
for sparse couplings\RED{.}
It is shown
by first inferring interactions from protein sequences and then from Monte Carlo samples
that the fields and couplings can be well recovered, but
that recovering the pairwise 
correlations
in the resolution of a total energy is harder for the natural proteins than
for the protein-like sequences.
Selective temperature for folding/structural constrains in protein evolution is also estimated.

\end{abstract}

\begin{IEEEkeywords}
group $L_1$,
inverse Potts problem,
learning method,
optimum regularization, 
sparse interactions,
selective temperature.
\end{IEEEkeywords}}

\maketitle

\IEEEdisplaynontitleabstractindextext

%
\IEEEpeerreviewmaketitle

\IEEEraisesectionheading{\section{Introduction}\label{sec:introduction}}


\ifdefined\NAG

\withSup{
\newcommand{\SFigPFvModAdamvsAdamhJPiaCijab}{\SFig{\ref{sfig: PF00595_ModAdam_vs_Adam_hJ_PiaCijab}}}
\newcommand{\SFigPFvmaxJijcomparison}{\SFig{\ref{sfig: PF00595_maxJij_comparison}}}

\newcommand{\SFigPFvModAdamvsNAGhJPiaCijab}{\SFig{\ref{sfig: PF00595_ModAdam_vs_NAG_hJ_PiaCijab}}}

\newcommand{\SFigPFvModAdamvsRPROPLRhJPiaCijab}{\SFig{\ref{sfig: PF00595_ModAdam_vs_RPROP-LR_hJ_PiaCijab}}}
\newcommand{\SFigPFvmaxJijvsrpd}{\SFig{\ref{sfig: PF00595_maxJij_vs_rpd}}}
\newcommand{\SFigPFvhJcomparison}{\SFig{\ref{sfig: PF00595_hJ_comparison}}}
\newcommand{\SFigPFihJcomparison}{\SFig{\ref{sfig: PF00153_hJ_comparison}}}
\newcommand{\SFigsPFvPFihJcomparison}{\SFigs{\ref{sfig: PF00595_hJ_comparison} and \ref{sfig: PF00153_hJ_comparison}}}

\newcommand{\SFigsPFvPFiKL}{\SFig{\ref{sfig: learning_process_KL}}}

\newcommand{\SFigsPFvPFiPiaCijab}{\SFigs{\ref{fig: PF00595_PiaCijab} and \ref{fig: PF00153_PiaCijab}}}

\newcommand{\SFigPFvMCPiaCijab}{\SFig{\ref{sfig: PF00595MC_PiaCijab}}}
\newcommand{\SFigPFiMCPiaCijab}{\SFig{\ref{sfig: PF00153MC_PiaCijab}}}
\newcommand{\SFigsPFvMCPFiMCPiaCijab}{\SFigs{\ref{sfig: PF00595MC_PiaCijab} and \ref{sfig: PF00153MC_PiaCijab}}}

}{

\TCBB{

\newcommand{\SFigPFvModAdamvsAdamhJPiaCijab}{\SFig{S1}}
\newcommand{\SFigPFvmaxJijcomparison}{\SFig{S2}}

\newcommand{\SFigPFvModAdamvsNAGhJPiaCijab}{\SFig{S3}}

\newcommand{\SFigPFvModAdamvsRPROPLRhJPiaCijab}{\SFig{S4}}
\newcommand{\SFigPFvmaxJijvsrpd}{\SFig{S5}}

\newcommand{\SFigPFvhJcomparison}{\SFig{S6}}
\newcommand{\SFigPFihJcomparison}{\SFig{S7}}

\newcommand{\SFigsPFvPFihJcomparison}{\SFigs{S6 and S7}}

\newcommand{\SFigsPFvPFiKL}{\SFigs{S8}}

\newcommand{\SFigsPFvPFiPiaCijab}{\SFigs{S9 and S10 }}

\newcommand{\SFigPFvMCPiaCijab}{\SFig{S11}}
\newcommand{\SFigPFiMCPiaCijab}{\SFig{S12}}

\newcommand{\SFigsPFvMCPFiMCPiaCijab}{\SFigs{S11 and S12}}

}{

\newcommand{\SFigPFvModAdamvsAdamhJPiaCijab}{\SFig{S1}}
\newcommand{\SFigPFvmaxJijcomparison}{\SFig{S2}}

\newcommand{\SFigPFvModAdamvsNAGhJPiaCijab}{\SFig{S3}}

\newcommand{\SFigPFvModAdamvsRPROPLRhJPiaCijab}{\SFig{S4}}
\newcommand{\SFigPFvmaxJijvsrpd}{\SFig{S5}}

\newcommand{\SFigPFvhJcomparison}{\SFig{S6}}
\newcommand{\SFigPFihJcomparison}{\SFig{S7}}

\newcommand{\SFigsPFvPFihJcomparison}{\SFigs{S6 and S7}}

\newcommand{\SFigsPFvPFiKL}{\SFigs{S8}}

\newcommand{\SFigsPFvPFiPiaCijab}{\SFigs{S9 and S10 }}

\newcommand{\SFigPFvMCPiaCijab}{\SFig{S11}}
\newcommand{\SFigPFiMCPiaCijab}{\SFig{S12}}

\newcommand{\SFigsPFvMCPFiMCPiaCijab}{\SFigs{S11 and S12}}

}

}

\withSup {

\newcommand{\SecPotts}{methods \ref{section: Potts}}
\newcommand{\SecSampleAve}{methods \ref{section: sample_ave}}
\newcommand{\SecEnsembleAve}{methods \ref{section: ensemble_ave}}
\newcommand{\SecSampleAndEnsembleAve}{methods \ref{section: sample_ave} and \ref{section: ensemble_ave}}
\newcommand{\SecEvolution}{methods \ref{section: evolution}}
\newcommand{\SecTsTgTm}{methods \ref{section: Ts_Tg_Tm}}
\newcommand{\SecEvolutionAndTsTgTm}{methods \ref{section: evolution} and \ref{section: Ts_Tg_Tm}}
\newcommand{\SecBML}{methods \ref{section: BML}}
\newcommand{\SecReg}{methods \ref{section: Regularization}}
\newcommand{\SecParamUpdates}{methods \ref{section: Parameter_updates}}
\newcommand{\SecIterations}{methods \ref{section: Iterations}}
\newcommand{\SecGauge}{methods \ref{sec: Ising_gauge_for_comparison}}

\newcommand{\EQelasticNet}{\ref{\EQ: elastic_net}}
\newcommand{\EQgroupL}{\ref{\EQ: group_L1}}
\newcommand{\EQdefSampleAvepsi}{\ref{\EQ: def_sample_ave_of_psi}}
\newcommand{\EQsampleAvepsi}{\ref{\EQ: sample_ave_of_psi}}
\newcommand{\EQensAvepsi}{\ref{\EQ: ensemble_ave_of_psi}}

\newcommand{\EQIsingGauge}{\ref{\EQ: Ising gauge}}

\newcommand{\EQModAdam}{\ref{\EQ: ModAdam}}
\newcommand{\EQgradh}{\ref{\EQ: gradients-h}}
\newcommand{\EQgradJ}{\ref{\EQ: gradients-J}}

\newcommand{\EQKLA}{\ref{\EQ: KL1}}
\newcommand{\EQKLB}{\ref{\EQ: KL2}}

\newcommand{\EQdefGammaij}{\ref{\EQ: def_gamma_ij}}

\newcommand{\EQTs}{\ref{\EQ: equilibrium_distr_of_seq_of_dG}}

\newcommand{\EQequilDistrOfSeqOfdG}{\ref{\EQ: equilibrium_distr_of_seq_of_dG}}

}{

}

\TCBB{

\withSup {
}{
\newcommand{\SecPotts}{methods S.1.1}

\newcommand{\SecSampleAve}{methods S.1.2}
\newcommand{\SecEnsembleAve}{methods S.1.3}
\newcommand{\SecSampleAndEnsembleAve}{methods S.1.2 and S.1.3}
\newcommand{\SecEvolution}{methods S.1.4}
\newcommand{\SecTsTgTm}{methods S.1.5}
\newcommand{\SecEvolutionAndTsTgTm}{methods S.1.4 and S.1.5}
\newcommand{\SecBML}{methods S.1.6}
\newcommand{\SecReg}{methods S.1.7}
\newcommand{\SecParamUpdates}{methods S.1.8}
\newcommand{\SecIterations}{methods S.1.8.3}
\newcommand{\SecGauge}{methods S.1.9}

\newcommand{\EQelasticNet}{S46}
\newcommand{\EQgroupL}{S54}
\newcommand{\EQdefSampleAvepsi}{S5}
\newcommand{\EQsampleAvepsi}{S6}

\newcommand{\EQensAvepsi}{S9}

\newcommand{\EQIsingGauge}{S72}

\newcommand{\EQModAdam}{S60}
\newcommand{\EQgradh}{S40}
\newcommand{\EQgradJ}{S41}

\newcommand{\EQKLA}{S71}
\newcommand{\EQKLB}{S70}

\newcommand{\EQdefGammaij}{S57}

\newcommand{\EQTs}{S12}

\newcommand{\EQequilDistrOfSeqOfdG}{S12}

}

}{

\withSup {
}{
}

}
\else
\fi

\renewcommand{\TEXT}[1]{#1}
\renewcommand{\SUPPLEMENT}[1]{}
\renewcommand{\SkipSupplToMerge}[1]{}

\TEXT{

\renewcommand{\TEXT}[1]{#1}
\renewcommand{\SUPPLEMENT}[1]{}

\TCBB{
\renewcommand{\EQ}{seq}
}{
}

\TCBB{
}{
\SECTION{Introduction}
}

\TCBB{
\IEEEPARstart{T}{he}
}{
The
}
maximum entropy model,
$P(\VECS{\sigma}) \propto \exp(- \psi_N(\VECS{\sigma}))$, where
$\psi_N \equiv - \, [ \, \sum_i \, \{ \, h_i(\sigma_i) + \sum_{j > i} J_{ij}(\sigma_i, \sigma_j)\, \}\, ]$,
sequence $\VECS{\sigma} \equiv (\ldots, \sigma_i, \ldots)$,
and $\sigma_i \in \{\text{amino acids, deletion}\}$,
for the distribution of homologous proteins in sequence space
recently attracts a great deal of attention in particular
due to its capacity to accurately predict residue-residue contacts 
in a 3D protein structure and complex\CITE{MPLBMSZOHW:11,MCSHPZS:11,SMWHO:12,CASP11:16,OPVHPKKKB:17,M:17b,M:18}.
Because 
genome-wide analyses require computationally fast methods,
approximate methods such as the mean field approximation\CITE{LGJ:02,LGJ:12,MPLBMSZOHW:11,MCSHPZS:11} and
pseudo-likelihood maximization methods\CITE{ELLWA:13,EHA:14} have been employed
for the inverse Potts problem that is to infer 
fields ($h_i(a_k)$) and couplings ($J_{ij}(a_k, a_l)$) 
from single-site frequencies,
$P_i(a_k)$ where $a_k \in \{\text{amino acids, deletion}\}$, 
and pairwise frequencies, $P_{ij}(a_k,a_l)$,
observed in a multiple sequence alignment (MSA); see \SecPotts \ of the supplementary material.
The performance of contact prediction by the mean-field or pseudo-likelihood maximization method
is sufficiently good\CITE{CASP11:16}, but it has been reported\CITE{BLCC:16,CFFMW:17,FBW:18} that these methods
can recover the structure of the interaction network but typically
not the correct strength of interactions.
The estimates of the fields and couplings
are also employed, however,
to discuss protein evolution\CITE{MSCOW:14,M:17}, particularly 
to analyze coevolution between residue substitutions,
and to discuss protein folding\CITE{JGSCM:16,BCCJM:16}.
Unlike contact predictions,
accurate estimations of the fields and couplings
are required in these studies; for instance,
quantitative analyses of the effects of amino acid substitutions on protein stability, which 
are also discussed in this manuscript.
One of generative methods that can 
better recover sequence statistics
\CITE{BLCC:16,CFFMW:17,FBW:18}
is a Boltzmann machine learning (BML) \CITE{HS:83,AHS:85,H:07,WWSHH:09},
in which pairwise marginal distributions are estimated by 
Markov chain Monte Carlo (MCMC) samplings with the Metropolis-Hastings algorithm\CITE{MRRTT:53,H:70}
or Gibbs sampler\CITE{GG:84},
and the fields and couplings are iteratively inferred 
by maximizing log-likelihood, equivalently by minimizing cross entropy; see \SecBML.

The number of parameters, fields and couplings, to be optimized in the inverse Potts problem
is very large in comparison with learning data.
To prevent over-fitting, regularization terms are often employed as part of the objective function.
Including the regularization terms in the cross entropy also fixes
the gauge for the evolutionary potential $\psi_N(\VECS{\sigma})$, which
is gauge-invariant, that is, 
invariant under a certain transformation of fields and
couplings; see \SecGauge.
An appropriate regularization model and hyper-parameters, 
however, must be employed 
to correctly infer fields and couplings.
Also, a learning method must be one
leading to reasonable values for fields and couplings.
However, problems 
are:
1) Natural proteins with known fields and couplings are not available to
optimize hyper-parameters and to choose a better regularization model and
gradient-descent method.
2) In addition, in the Boltzmann machine (BM)
the learning process fluctuates, but
the cross entropy/likelihood cannot be used to pick up the best set of parameters,
because it can hardly be evaluated, 
even though its partial derivatives can be easily evaluated.

Then, what characteristics are required for the fields and couplings
in protein sequences?
1) Couplings ($J_{ij}$) 
should
be sparse and 
their strength is expected to negatively correlate with the distance between residues,
because
strong
residue-residue correlations/coevolutions are expected 
for 
closely-located, interacting residue pairs in a 3D protein structure and complex
\CITE{LGLS:99,LGJ:02,LGJ:12,RLMYR:05,SPSLAGL:08,BN:08,WWSHH:09,HRLR:09,BN:10,MPLBMSZOHW:11,MCSHPZS:11,AVBMN:88,GSSV:94,SKS:94,PT:97,PTG:99,AWFTD:00,FOVC:01,FA:04,FYB:04,DPJG:05,MGDW:05,FT:06,DP:07,DG:07,DWG:08,PLFP:08,D:12,G-K:12},
although weak coevolutions may occur to select less-attractive residue pairs
for distantly-located residue pairs\CITE{JGSCM:16}.
2) 
The sample mean of 
$\psi_N(\VECS{\sigma}) (\equiv - [ \sum_i \{ h_i(\sigma_i) + \sum_{j > i} J_{ij}(\sigma_i, \sigma_j) \} ] )$
over homologous sequences $\VECS{\sigma}(=\VECS{\sigma}_N)$ 
is equal to the ensemble average over the Boltzmann distribution, 
which may be evaluated by approximating
the distribution of $\psi_N(\VECS{\sigma})$ of random sequences
as a Gaussian distribution,
$\bar{\psi} - \delta\psi^2$, where 
the $\bar{\psi}$ and $\delta\psi^2$ are the mean and variance 
of $\psi_N(\VECS{\sigma})$
expected for the random sequences
with the same amino acid composition
\CITE{SG:93a,SG:93b,RS:94,PGT:97,M:17};
see \SecSampleAve\ and \SecEnsembleAve.
In contact prediction, hyper-parameters have been optimized 
by maximizing the precision of contact prediction.
However, the second requirement for fields and couplings above should be also
satisfied by any method, if the evolutionary energies of natural sequences can be approximated to be at equilibrium
in the Boltzmann distribution.

In the Boltzmann machine, 
statistical errors cannot be avoided in the estimation of
the partial derivatives of cross entropy/log-likelihood, because
they are evaluated on the basis of the pairwise marginal distributions
estimated by 
MCMC samplings.
As a result, 
even though the learning rate is sufficiently small,
the cross entropy/log-likelihood are not expected to be 
monotonically improved but
to fluctuate in the minimization/maximization process.
In the present case, in which the first-order methods based on
gradients are employed, the objective function will further fluctuate.
Here, the average ($D_2^{KL}$) of the Kullback-Leibler divergences for pairwise marginal distributions
over all residue pairs is monitored as an approximate measure of fitting to the target distribution.
Although $D_2^{KL}$ significantly fluctuates,
the sample and ensemble averages of evolutionary energy
along a learning process smoothly change and converge, but
their profiles are very different between the learning methods,
indicating which method is better than the others.

It is well-known that $L_1$ regularization is better for a sparse-parameter system
than $L_2$.  In the present system, the couplings $J$ must be sparse in terms of 
residue pair $(i,j)$. 
Hence, $L_2$ for the fields and group $L_1$ for the couplings (L2-GL1) are employed, and
it is shown that the L2-GL1 model makes the estimate of the couplings more reasonable than 
the other models, $L_2$ for fields and $L_1$ for couplings (L2-L1) and $L_2$ for both (L2-L2);
in the present work, the L1 for couplings means the elastic net including
a small contribution of $L_2$ 
in addition to the $L_1$ regularization,
because it is known\CITE{ZhHt:05} that the regularization of pure $L_1$ can occasionally
produce non-unique solutions; see \SecReg.

Secondly, we show that 
it is preferable for the stepsize of parameter updates to be proportional to 
the partial derivative of the objective function,
on 
estimating the dependencies of couplings $J_{ij}$ on the distance between residues $i$ and $j$.
Various stochastic gradient-descent methods to minimize loss functions
have been invented for machine learning;
the momentum method\CITE{RHW:86},
and
Nesterov's Accelerated Momentum (NAG) (Fast proximal gradient method)\CITE{N:04}
that manipulate the learning rate equally for all parameters,
and AdaGrad\CITE{DHS:11}, AdaDelta\CITE{Z:12}, 
RPROP\CITE{RB:93,R:94},
RMSprop\CITE{TH:12}, and Adam\CITE{KB:14}
that employ adaptive learning rates for each parameter.
The per-parameter adaptive learning rate methods, particularly
Adam method, are ones that are often used in neural networks.
They are stable and fast methods for stochastic gradient-descent.
In the RPROP, a stepsize does not depend on the partial derivative but only on
the temporal behavior of its sign.
In the other per-parameter adaptive learning rate methods,
a stepsize is proportional to the partial derivative
but each partial derivative is normalized
in such a way that
stepsizes for all parameters
are essentially a similar order.
This characteristic of stepsizes
appears to be inappropriate for
the present case in which 
couplings are expected to be very sparse
and to correlate with residue-residue distance.
For the present Potts problem, the RPROP method 
was modified\CITE{BLCC:16,FBW:18}
in such a way that a stepsize is proportional to the partial derivative with
the proportional constant determined by the RPROP method;
we call this modified RPROP method RPROP-LR;
the RPROP-LR stands for resilient propagation learning rate.
\RED{Also}, we invent and employ the modified Adam (ModAdam) method, in which
the stepsize of parameter updates is proportional
to the partial derivative, and
the proportional constant is not per-parameter but
adaptive; see \SecParamUpdates.
\ifdefined\NAG
Couplings inferred by the Adam, NAG, and RPROP-LR in the L2-L2 regularization model 
\else
Couplings inferred by the Adam and RPROP-LR in the L2-L2 regularization model 
\fi
are compared with those by the ModAdam to show 
that the stepsize of parameter updates must be proportional to the partial derivative
in order to 
better
estimate the dependencies of couplings $J_{ij}$ on the distance 
between residues $i$ and $j$.

Thirdly, we discuss 
how to tune regularization parameters.
In the present model,
hyper-parameters that directly affect the estimates 
of fields and couplings
are two proportional
constants, $\lambda_1$ and $\lambda_2$, for their regularization terms.
These hyper-parameters are tuned in such a way that
the sample mean of 
$\psi_N(\VECS{\sigma}_N)$
over homologous sequences $\VECS{\sigma}_N$ 
is equal to 
$\bar{\psi} - \delta\psi^2$, where 
the $\bar{\psi}$ and $\delta\psi^2$ are the mean and variance 
of $\psi_N(\VECS{\sigma})$ 
for the random sequences
with the same amino acid composition.

By the L2-GL1 model and the ModAdam method, 
single-site frequencies and pairwise correlations
can be well recovered.  
Also,  
by first estimating fields and couplings from protein sequences and then again from 
MCMC samples
obtained by the Boltzmann machine learning,
we show that fields and couplings in protein-like sequences can be well recovered, too.
However, the distribution of evolutionary energies over natural proteins
is shifted towards lower energies from that of 
MCMC samples,
indicating that
recovering the pairwise amino acid frequencies
in the resolution of a total energy is harder for the natural proteins than
for the protein-like sequences.

Lastly, 
based on the present estimates of fields and couplings,
the constancy of the standard deviation of evolutionary energy changes 
due to single nucleotide nonsynonymous substitutions over protein families,
which was found\CITE{M:17} by the mean field method, 
is confirmed.
Then, selective temperature, 
which quantifies how strong the folding/structural constraints are in the evolution of a protein family
\CITE{DS:01,MSCOW:14,M:17},
is estimated 
based on the evolutionary energy changes due to single amino acid substitutions\CITE{M:17};
see \SecEvolutionAndTsTgTm\ for details.

Methods are described in detail in the supplementary file.


\SECTION{Results}

The multiple sequence alignments (MSA) of 
the Pfam\CITE{EMBELPQRSSSHPPTF:19}, protein families PF00153\CITE{FBW:18} and PF00595\CITE{M:17},
which include at least one member whose atomic coordinates are available,
are only employed here, because of intensive computation,
in order to demonstrate 
what regularization model and
what type of gradient-descent method are preferable,
how to tune regularization parameters,
and  how well the fields and couplings of protein-like sequences can be reproduced.

The protein family PF00595 (PDZ domain),
which is a common structural domain of 80-90 amino-acids found in diverse signaling proteins,
is chosen because
experimental data of 31 folding free energy changes 
($\Delta\Delta G_{ND}$)
due to various types of single amino acid changes at many sites is available, which are required to
estimate selective temperatures defined in \Eq{\EQTs}.
On the other hand, PF00153 (Mitochondrial substrate/solute carrier) 
that consists of more family members than PF00595 has been chosen 
to examine the effects of alignment depth on the recoverability of single-site and
pairwise amino acid frequencies and the estimation of fields and couplings.

The length of the proteins,
the number of sequences ($N$) in the MSAs
and their effective number ($N_{\text{eff}}$) 
are listed in \Table{\ref{tbl: MSAs}}.
The number of sequences ($M$) that do not contain deletions and their effective number ($M_{\text{eff}}$) 
are also listed in this table as well as
the PDB IDs of the protein structures employed to calculate
contacting residue pairs.

\TableInText{
\ifdefined\ThreeDigitsInTable
\else
\TCBB{
\newcommand{\ThreeDigitsInTable}[2]{#1}
}{
\newcommand{\ThreeDigitsInTable}[2]{#2}
}
\fi

\begin{table}[!h]
\begin{threeparttable}[b]
\caption{
\TCBB{
\TEXTBF{
Protein Families Employed.
}
}{
\TEXTBF{
Protein families employed.
}
}
}
\label{tbl: MSAs}
\vspace*{1em}
\begin{tabular}{lccrl}
\hline
Pfam ID
	& $N$ / $N_{\text{eff}}$ $^{a}$ & $M$ $^{b}$ / $M_{\text{eff}}$ $^{a}$ & $L$ $^c$	
	& PDB ID
	\\
\hline
PF00595$^\dagger$ 	& 13814 / 4748.8 & 1255 / 340.0 & 81
	& 1GM1-A:16-96
	\\
PF00153	 		& 54582 / 19473.9 & 255 / 139.8 & 97
	& 2LCK-A:112-208
	\\
\hline
\end{tabular}
\begin{tablenotes}
\item [$^\dagger$] Identical sequences are removed.

\item [$^a$] The effective number of sequences, $\sum_{\VECS{\sigma}_N} w_{\VECS{\sigma}_N}$,
where the sample weight $w_{\VECS{\sigma}_N}$ for a natural sequence $\VECS{\sigma}_N$ is
equal to the inverse of the number of sequences
that are less than 20\% different from a given sequence.

\item [$^b$] The number of unique sequences that include no deletion 
for PF00595 and no more than 2 for PF00153.

\item [$^c$] The number of residues in a sequence.
\end{tablenotes}
\end{threeparttable}
\end{table}
}
\TCBB{
\SUBSECTION{A Markov Chain Monte Carlo (MCMC) Method}
}{
\SUBSECTION{A Markov chain Monte Carlo (MCMC) method}
}

Multiple Markov chains from different initial configurations with the same potential
are generated by the Metropolis-Hastings algorithm\CITE{MRRTT:53,H:70} in parallel computation and
Markov chain Monte Carlo (MCMC) samplings are done in each chain to estimate
pairwise marginal distributions after 
the fluctuation of $\psi_N(\VECS{\sigma})$ passes a statistical test for an equilibrium condition. 
The Metropolis-Hastings algorithm was employed
due to less computation time rather than the Gibbs sampler\CITE{GG:84}.
Then, the partial derivatives of the cross entropy including
the regularization terms are evaluated according to \Eqs{\EQgradh\ and \EQgradJ},
and it is iteratively minimized by a gradient-descent method.
The estimates of the partial derivatives 
on the basis of the marginal distributions estimated by the 
MCMC
samplings strongly depend on the number of samples employed. 
The more samples are employed, the more precisely they can be estimated,
although computational loads also increase. 
In the present case, the target frequencies are estimated
from the MSA consisting of a limited number of sequences, limiting
the accuracy of fields and couplings inferred.
Employing too many samples to estimate marginal probabilities
would cause over-fitting.
Here, we employ samples whose number is equal to the effective number of sequences in 
the MSA in order to estimate the pairwise marginal distributions; $N_{\text{MC}} \simeq N_{\text{eff}}$.

Because the first-order methods based on gradients
are used for minimization, and also the estimates of the gradients include 
statistical errors due to 
MCMC
samplings, the cross entropy including the regularization terms 
fluctuates during the process of the minimization.
However, an optimum set of fields and couplings cannot be chosen 
on the basis of the cross entropy,
because the cross entropy itself can hardly be evaluated 
although its partial derivatives can be.  Here, 
the average of the Kullback-Leibler divergences of the pairwise marginal distributions
over all site pairs
is employed as an approximate measure of fitting to the target distribution; see
\Eq{\EQKLB}.

\FigureInText{

\begin{figure*}[!h]
\centerline{
\includegraphics[width=44mm,angle=0]{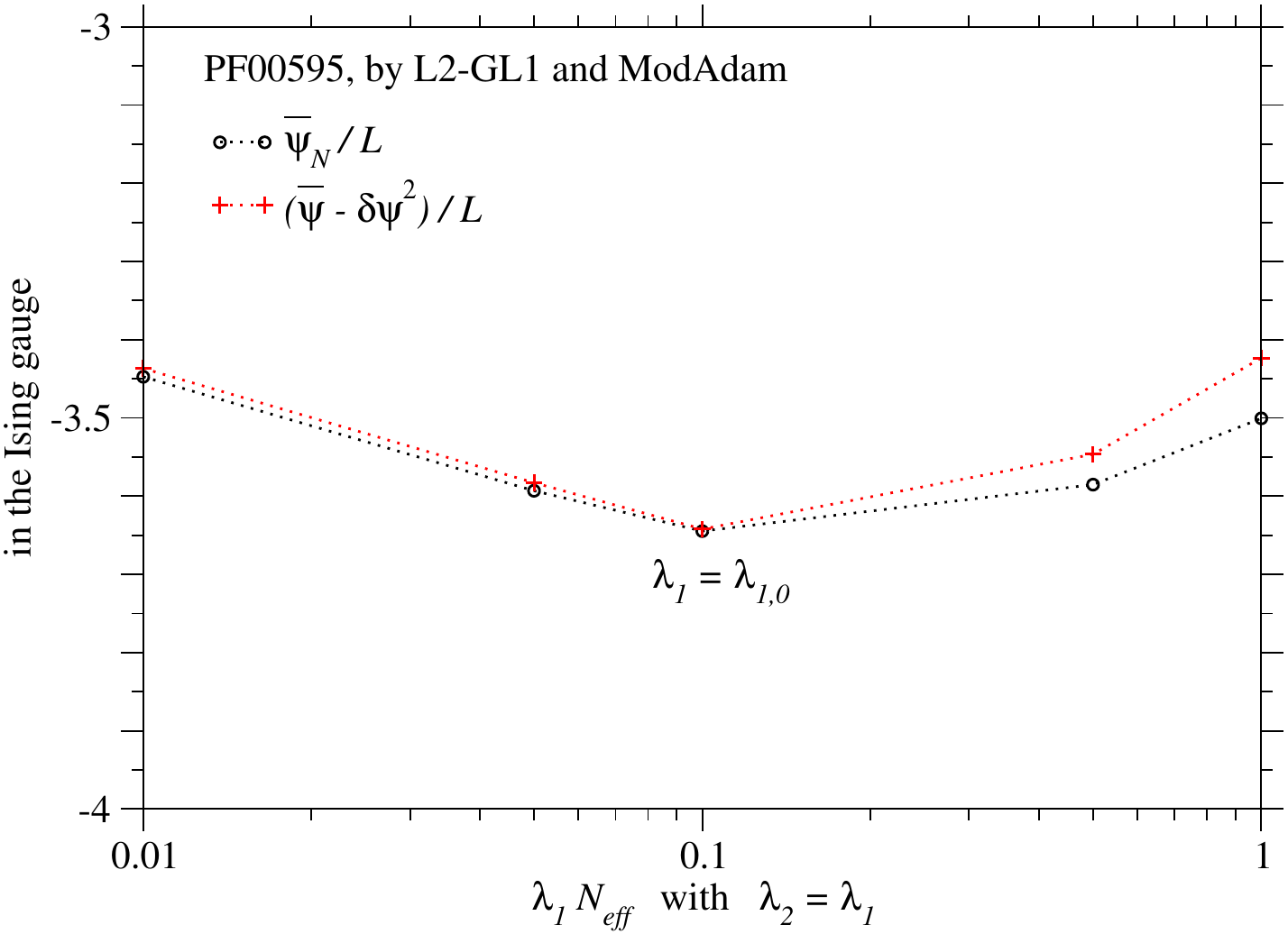}
\includegraphics[width=43mm,angle=0]{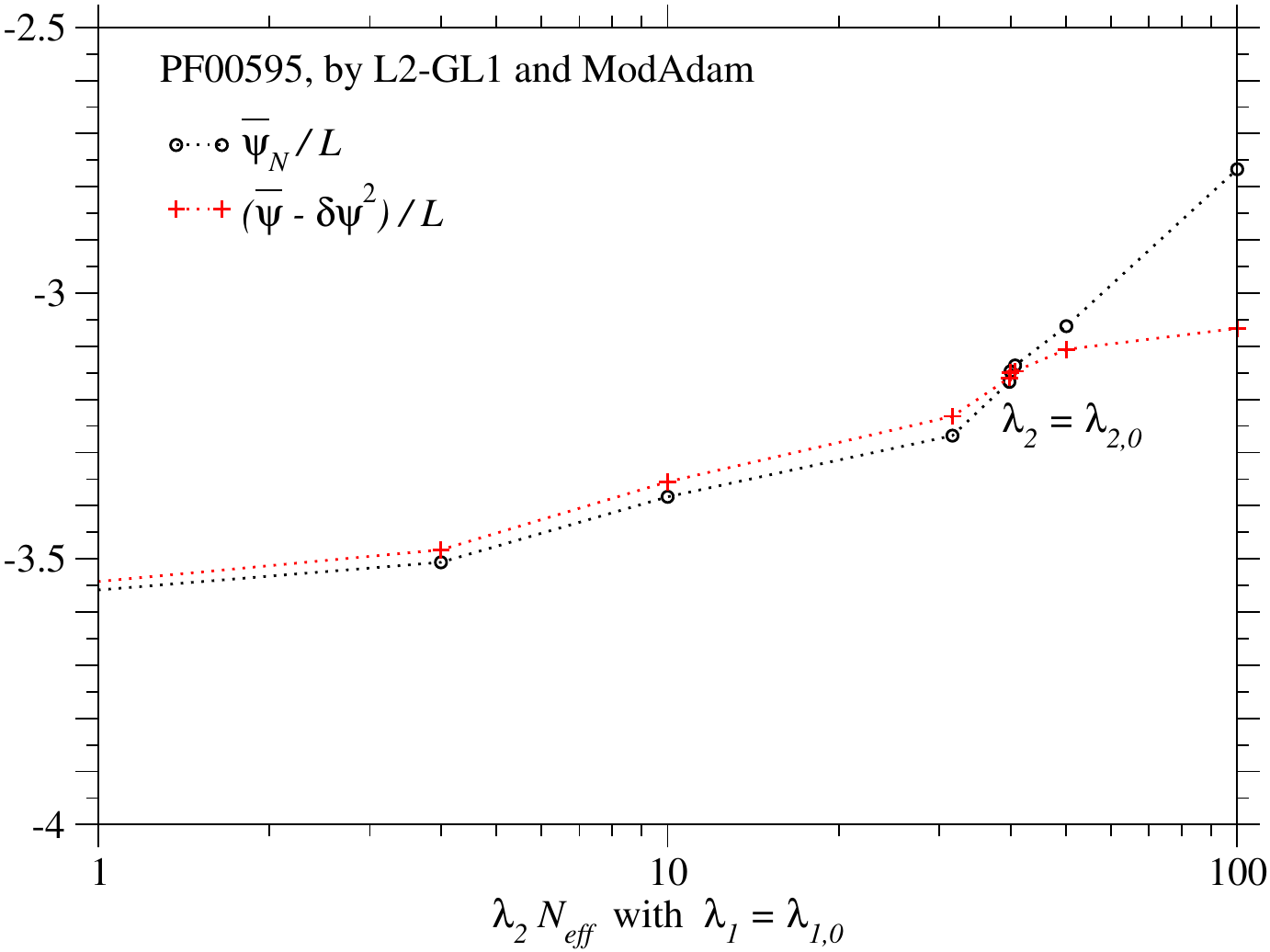}
\includegraphics[width=44mm,angle=0]{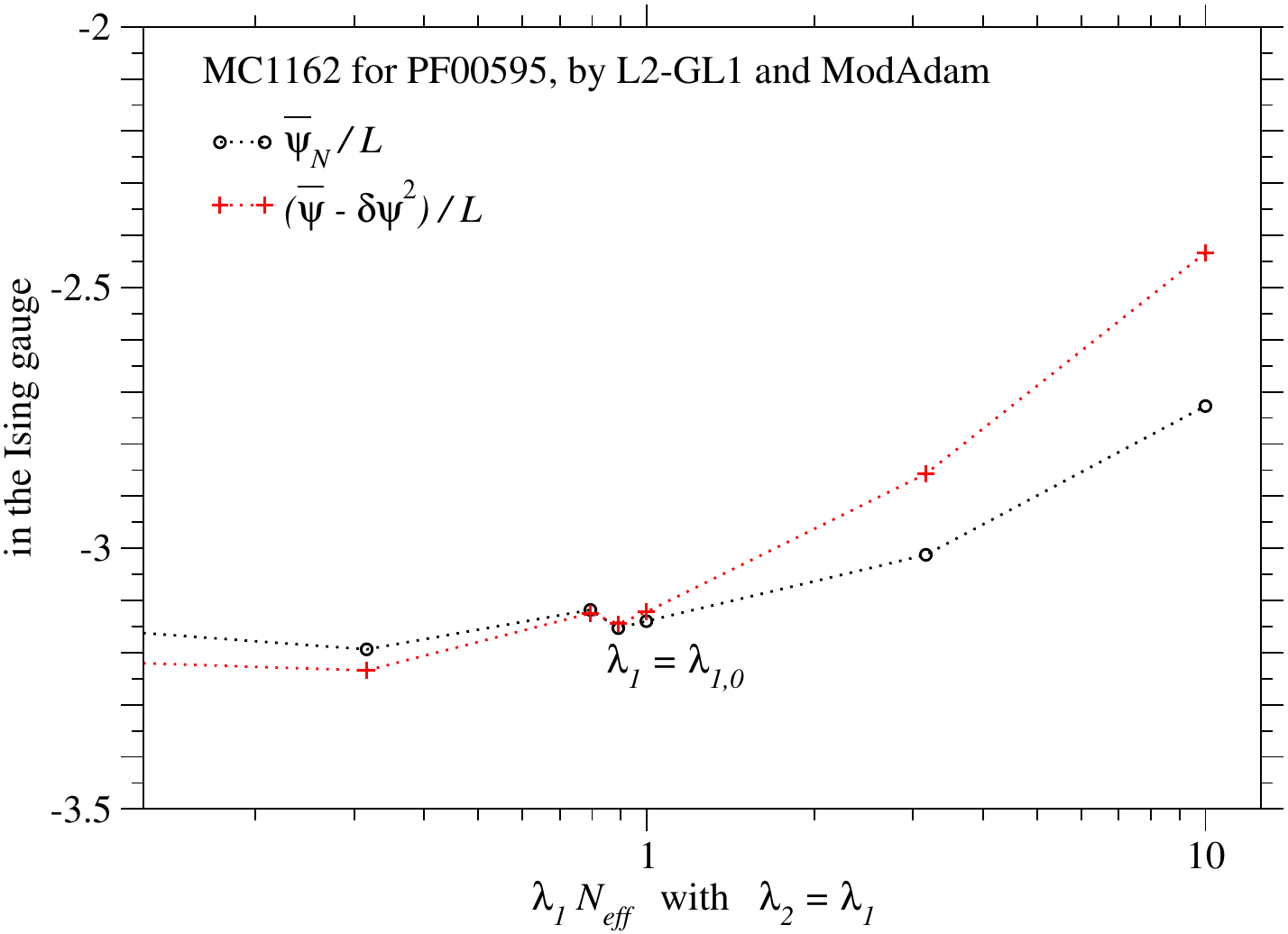}
\includegraphics[width=43mm,angle=0]{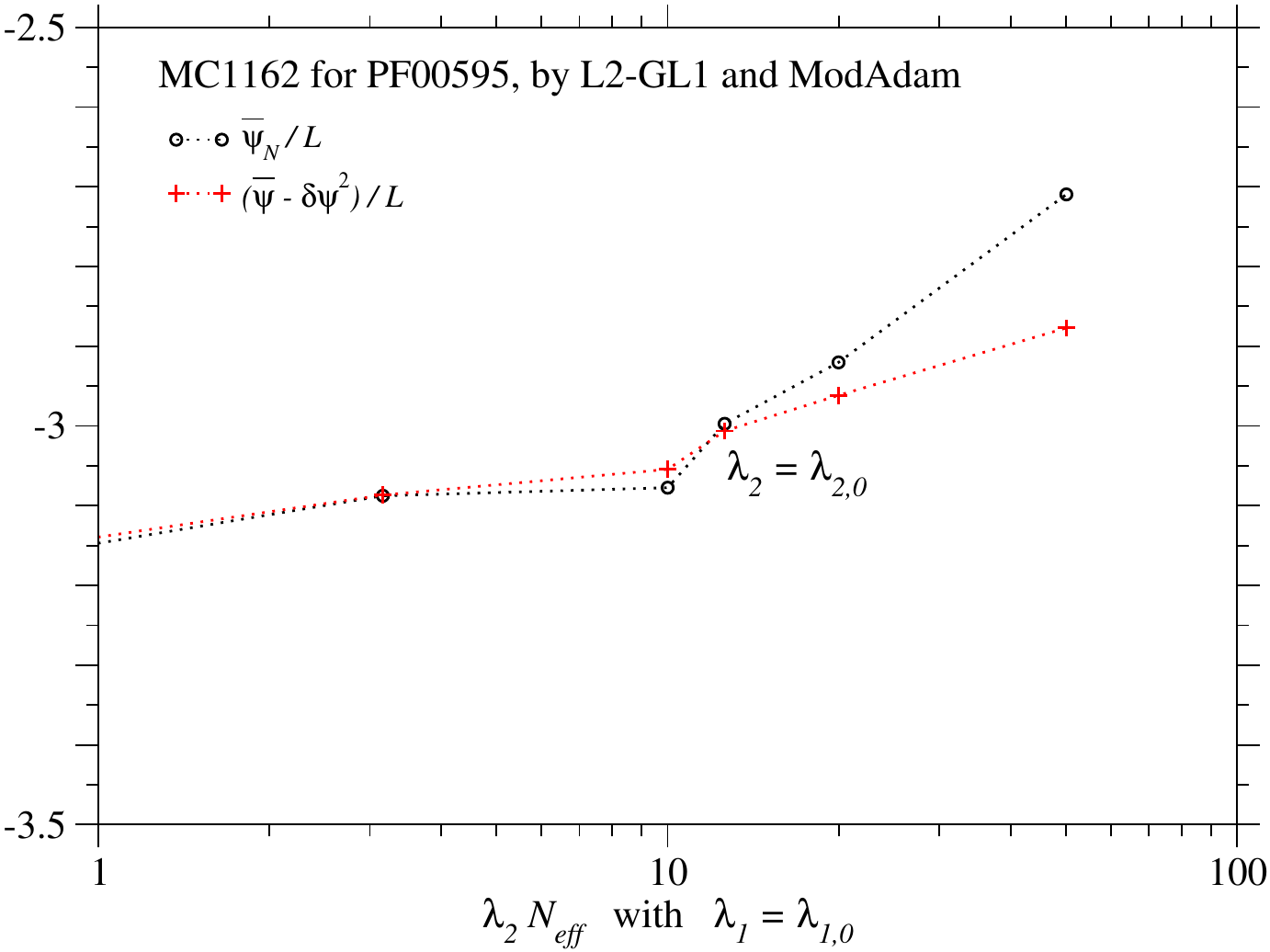}
}
\centerline{
\includegraphics[width=44mm,angle=0]{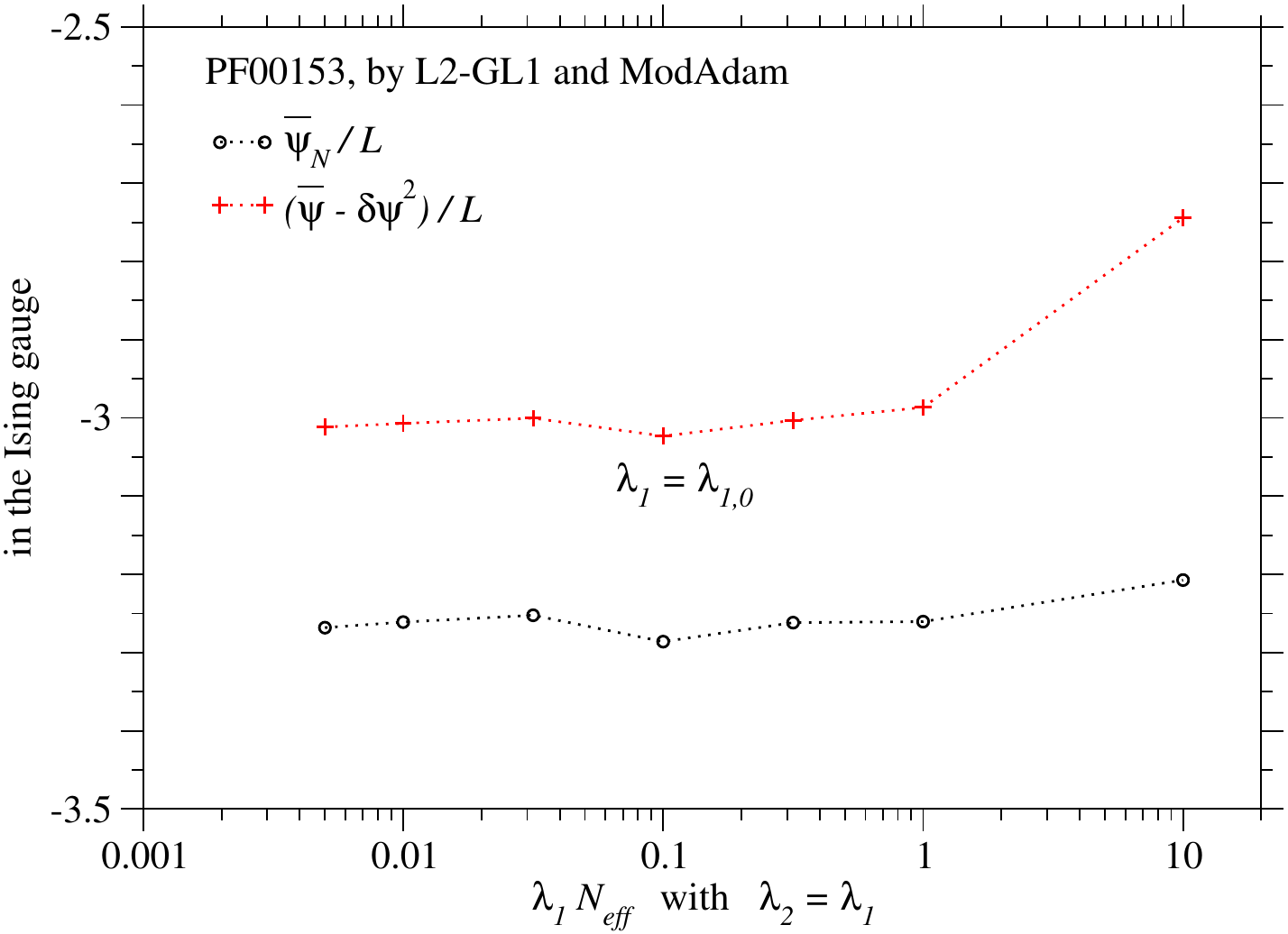}
\includegraphics[width=43mm,angle=0]{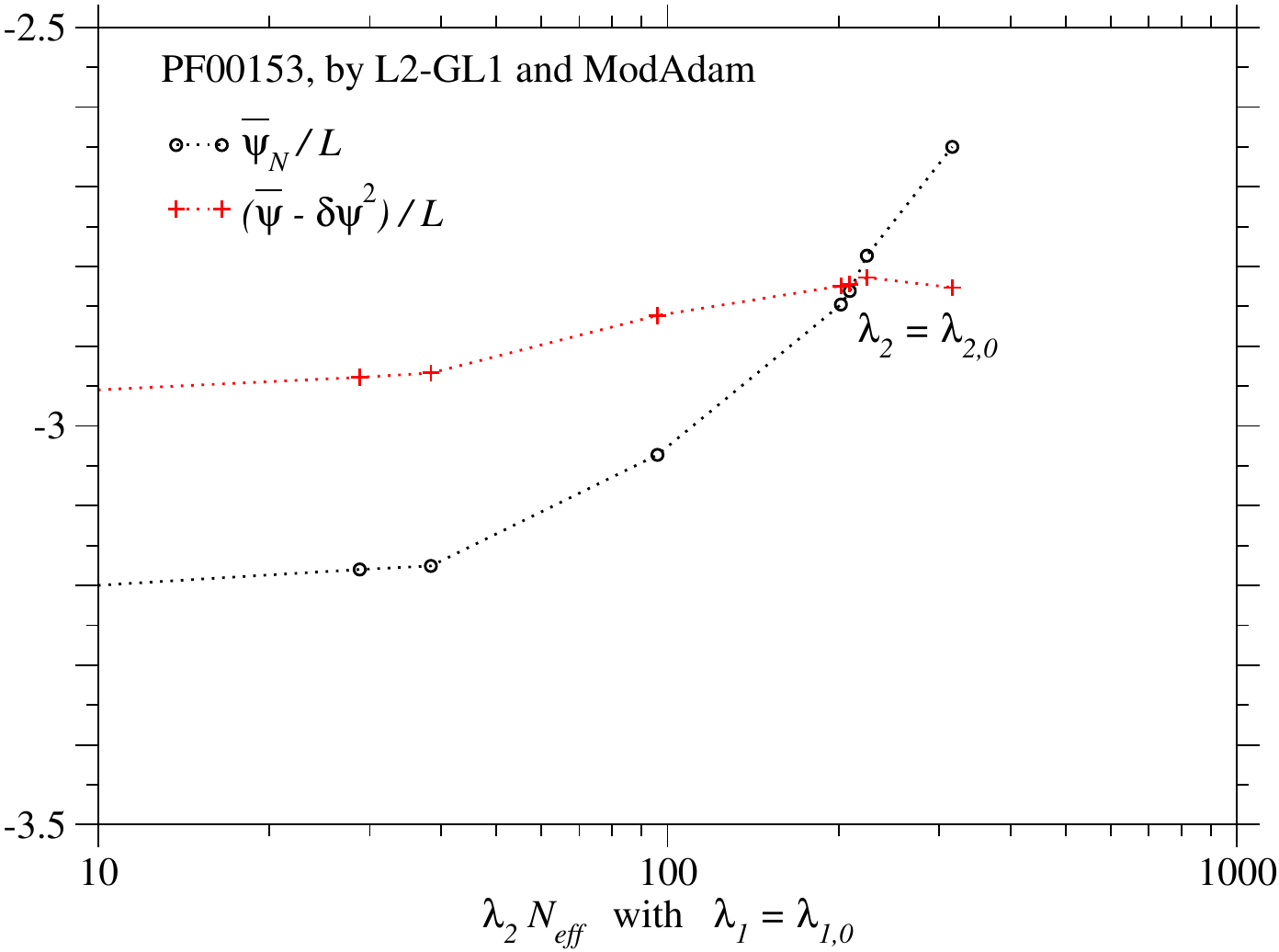}
\includegraphics[width=44mm,angle=0]{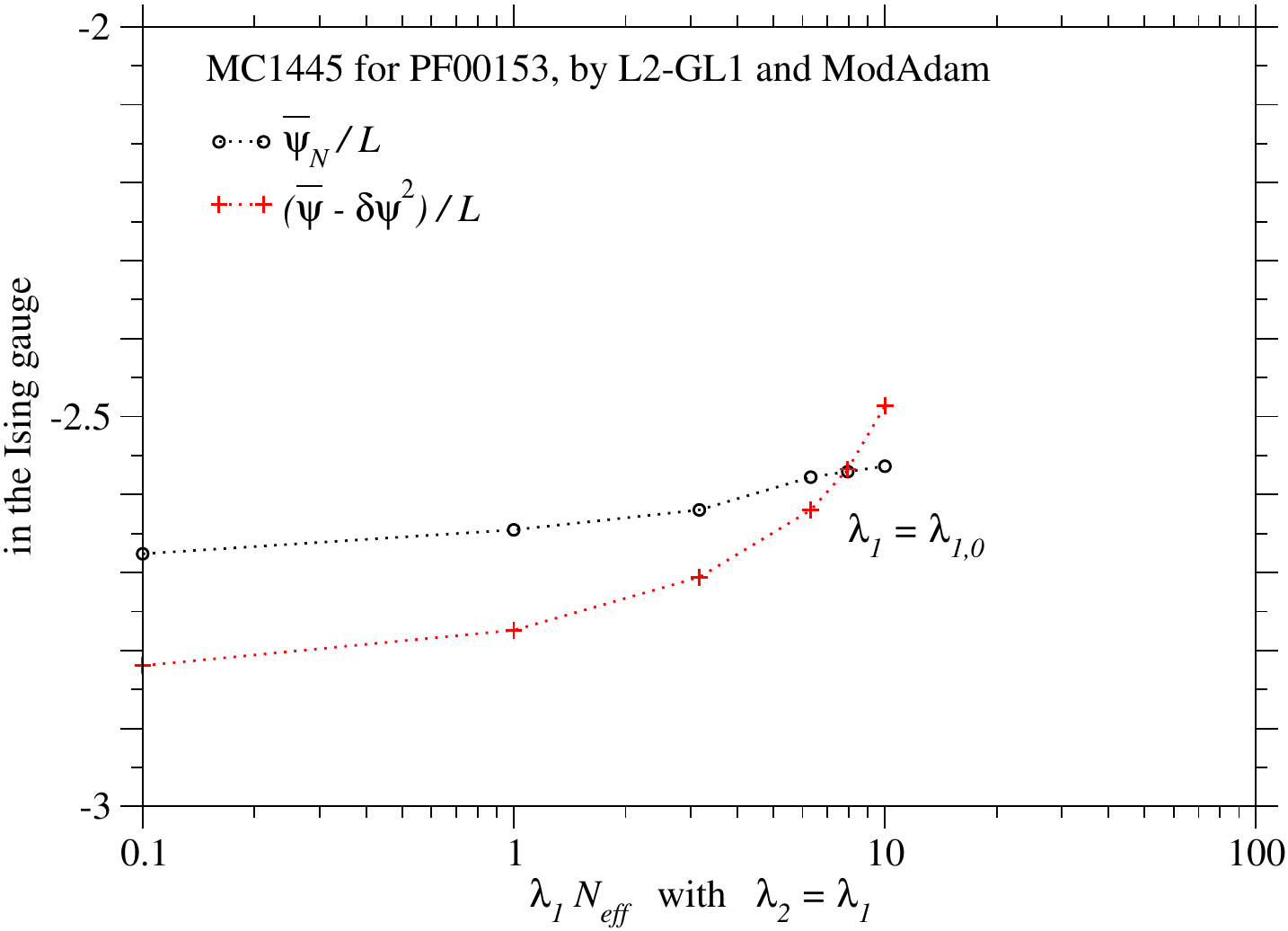}
\includegraphics[width=43mm,angle=0]{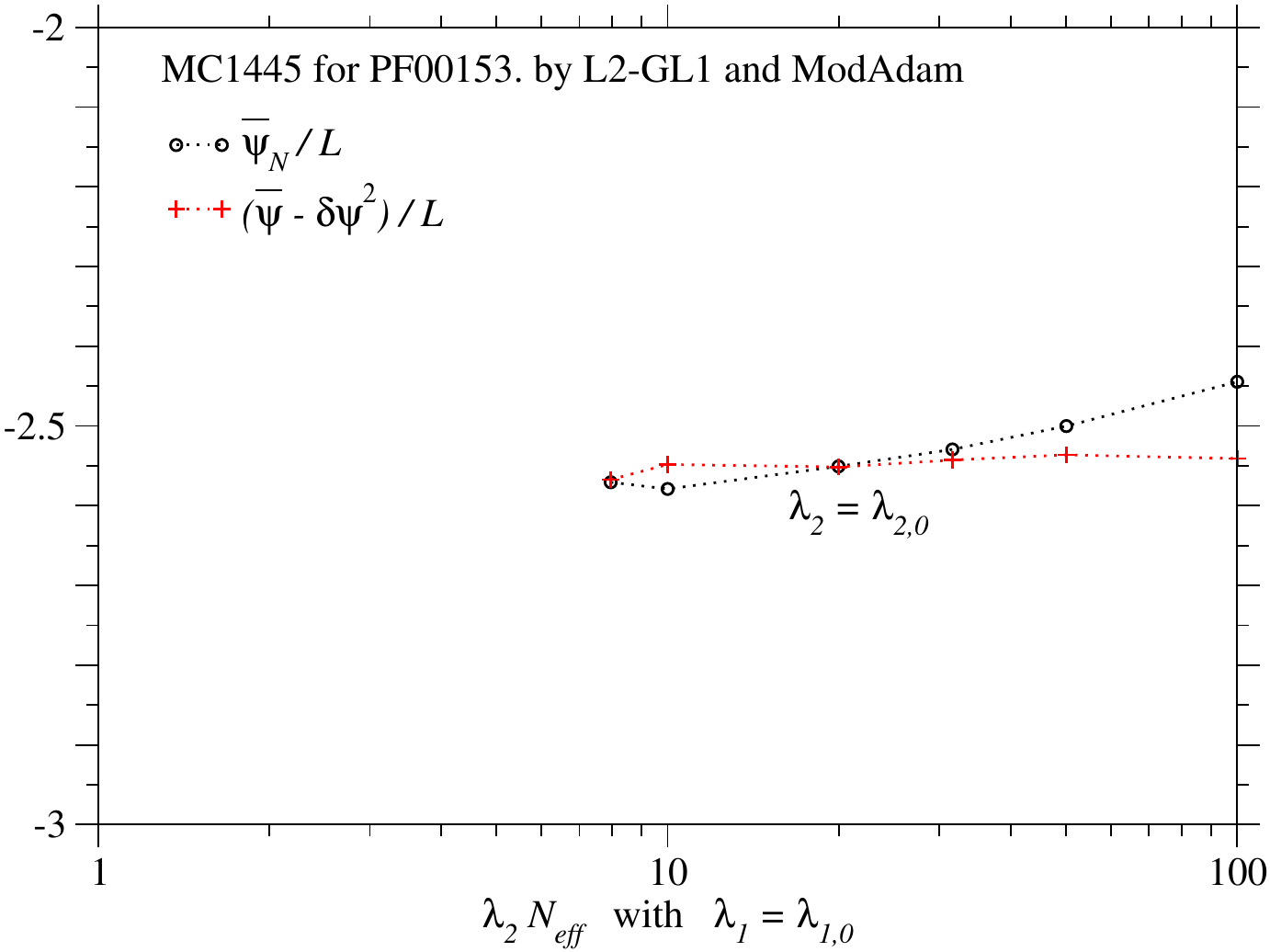}
}
\caption
{
\TEXTBF{A schematic representation of how to tune regularization parameters, $\lambda_1$ for fields and $\lambda_2$ for couplings. }
The average evolutionary energy per residue $\overline{\psi_N}/L$ of natural sequences 
and the ensemble average of evolutionary energy per residue $(\bar{\psi} - \delta\psi^2)/L$
in the Ising gauge are plotted by circle and plus markes, respectively, against one of the regularization parameters, 
$\lambda_1 N_{eff}$ with $\lambda_2=\lambda_1$ 
or $\lambda_2 N_{eff}$ with $\lambda_1=\lambda_{1,0}$;
$\lambda_{1,0}$ and $\lambda_{2,0}$ are the optimum values of $\lambda_{1}$ and $\lambda_{2}$,
respectively.
The regularization model L2-GL1 and the modified Adam method are employed;
see \Tables{\ref{tbl: PF00595_parameters} and \ref{tbl: PF00153_parameters}} 
for the values of $\lambda_{1,0}$ and $\lambda_{2,0}$.
The upper and lower rows correspond to the figures for PF00595 and PF00153, respectively. 
The left four figures are for natural sequences of PF00595 and PF00153,
and the right four figures are for 
the MCMC
samples, MC1162
and MC1445 that are obtained in the Boltzmann machine learnings for 
PF00595 and PF00153, respectively.
\label{fig: tuning}
}
\end{figure*}

}

\TCBB{
\SUBSECTION{How to Tune Regularization Parameters, $\lambda_1$ and $\lambda_2$}
}{
\SUBSECTION{How to tune regularization parameters, $\lambda_1$ and $\lambda_2$}
}

\label{section: How_to_tune}

Hyper-parameters that directly affect the estimated values of parameters are
two regularization parameters, $\lambda_1$ for fields $\phi_i(a_k)$ and 
$\lambda_2$ for couplings $\phi_{ij}(a_k, a_l)$; see \Eqs{\EQelasticNet\ and \EQgroupL}.

These parameters are tuned for inferred fields and couplings to satisfy
the condition that
the sample mean $\overline{\psi_N(\VECS{\sigma}_N)}$ of evolutionary energies
over homologous sequences $\VECS{\sigma}_N$
is equal to 
the ensemble average 
$\langle \psi_N(\VECS{\sigma}) \rangle_{\VECS{\sigma}}$
of evolutionary energy in the Boltzmann distribution;
according to the random energy model for protein folding\CITE{SG:93a,SG:93b,RS:94,PGT:97},
the ensemble average is evaluated to be 
$\langle \psi_N(\VECS{\sigma}) \rangle_{\VECS{\sigma}} = \bar{\psi} - \delta\psi^2$
by approximating
the distribution of the evolutionary energies of random sequences
as a Gaussian distribution,
where its mean ($\bar{\psi}$) and variance ($\delta\psi^2$) 
are evaluated to be equal to those of $\psi_N(\VECS{\sigma})$
expected for
random sequences whose amino acid composition is equal to
the average amino acid composition of sequences in the MSA
\CITE{M:17};
see \Eqs{\EQsampleAvepsi\ and \EQensAvepsi}.

Increasing $\lambda_1$ and $\lambda_2$ enforces 
the values of $\phi_{i}(a_k)$ and $\phi_{ij}(a_k, a_l)$ toward zero, respectively.
First, reduce the regularization terms by 
decreasing $\lambda_1$ with $\lambda_2 = \lambda_1$.  
The more it is reduced, the better single-site and pairwise probabilities become recovered.
If $\lambda_1$ is reduced beyond a certain value, they will be over-fitted.
In \Fig{\ref{fig: tuning}},  $\overline{\psi_N(\VECS{\sigma}_N)}$ and
$\langle \psi_N(\VECS{\sigma}) \rangle_{\VECS{\sigma}}$ are 
plotted against $\lambda_1$ with $\lambda_2 = \lambda_1$.
Energies are expressed in the Ising gauge for comparison; 
see 
\Eq{\EQIsingGauge}.
The ensemble average  of evolutionary energy in the Gaussian approximation 
as well as the average evolutionary energy of homologous sequences in the MSA would decrease
to favor those sequences as the regularization terms are reduced, 
At the certain value of $\lambda_1$, where the over-fitting begins to occur,
$\langle \psi_N(\VECS{\sigma}) \rangle_{\VECS{\sigma}}$ and $\overline{\psi_N(\VECS{\sigma}_N)}$
almost take their minima. 
This value $\lambda_{1,0}$ is the optimum value for $\lambda_1$.
In the case of PF00153 that consists of more sequences than PF00595, 
$\overline{\psi_N(\VECS{\sigma}_N)}$ is more negative than $\langle \psi_N(\VECS{\sigma}) \rangle_{\VECS{\sigma}}$,
indicating that the natural sequences are over-favored.
In the case of PF00595, they are almost equal to each other at the $\lambda_1 = \lambda_2 = \lambda_{1,0}$.

Then, increase $\lambda_2$ fixing $\lambda_1$ at $\lambda_1 = \lambda_{1,0}$.
Increasing $\lambda_2$  increases sparsity in couplings, 
reducing coupling interactions including noises.
As a result, the average evolutionary energy of homologous sequences in the MSA
as well as the ensemble average of evolutionary energy in the Gaussian approximation
would increase, and the former would become higher than the latter, because
reducing couplings makes the target sequences unfavorable.
A crossing point $\lambda_{2,0}$ of 
$\langle \psi_N(\VECS{\sigma}) \rangle_{\VECS{\sigma}}$ and $\overline{\psi_N(\VECS{\sigma}_N)}$
is the optimum value for $\lambda_2$, where the over-fitting disappears.
The values of the regularization parameters determined in this way 
are listed in \Table{\ref{tbl: PF00595_parameters}} for PF00595 and \Table{\ref{tbl: PF00153_parameters}}
for PF00153.  These tables include those for the L2-GL1, L2-L2, and L2-L1 models.

In the case of employing the Monte Carlo samples obtained in Boltzmann machines as protein-like sequences,
as the $\lambda_1$ is decreased below a certain value $\lambda_{1,0}$ 
with $\lambda_2 = \lambda_1$,
$\langle \psi_N(\VECS{\sigma}) \rangle_{\VECS{\sigma}}$ 
becomes more negative than $\overline{\psi_N(\VECS{\sigma})}$,
before both take their minima,
because 
the average evolutionary energy of the Monte Carlo samples 
is much higher than that of the natural sequences as discussed later;
this $\lambda_{1,0}$ is taken as the optimum value for $\lambda_{1}$.

\TableInText{
\begin{table*}[!h]
\begin{threeparttable}[b]
\caption{
\TCBB{
\TEXTBF{
Regularization Parameters and Characteristics of Boltzmann Machines$^a$ with the ModAdam Method for PF00595.
}
}{
\TEXTBF{
Regularization parameters and characteristics of Boltzmann machines$^a$ with the ModAdam method for PF00595.
}
}
}
\label{tbl: PF00595_parameters}
\ifdefined\ThreeDigitsInTable
\begin{tabular}{llrrllllrccc}
\hline
\\
\multicolumn{2}{c}{MSA \hspace*{2em} regularizers}
			& \multicolumn{1}{c}{$\lambda_1$} & \multicolumn{1}{c}{$\lambda_2$} &  \multicolumn{1}{c}{\#Iter $^b$} & \multicolumn{1}{c}{$D^{KL}_1$} & \multicolumn{1}{c}{$D^{KL}_2$} 
		& \multicolumn{4}{c}{$\delta\psi^2 / L$ $^c$ \hspace*{0em} $(\bar{\psi} - \delta\psi^2) / L$ $^{d}$ \hspace*{0em} $\overline{\psi_N} / L$ $^e$ \hspace*{1em} $\overline{\psi_{\text{MC}}} / L$ $^f$}
		& Precision $^g$
	\\
\hline
\hline
PF00595
	& L2-GL1 & $0.100/N_{\text{eff}}$ & $=\lambda_1$ & 1250
		& 0.00506	
		& 0.0709	
		& 3.23	
			& \hspace*{1em} $-3.64$  	
			& $-3.64$	
			& $-3.29$	
			& 0.565		
			\\
\textbf{PF00595}$^h$
	& L2-GL1 & $0.100/N_{\text{eff}}$ & $40.0/N_{\text{eff}}$ & 1162$^\dagger$
		& 0.00369	
		& 0.0759	
		& 2.75	
			& $-3.15$ 	
			& $-3.15$	
			& $-2.79$ 
			& 0.588	
			\\
	&	&	&	&
		&
		&
		&
			&
			&
			& ($-2.81$$^i$)	
			&
			\\
\hline
MC1162$^\ddagger$
	& L2-GL1 & $0.100/N_{\text{eff}}$ & $40.0/N_{\text{eff}}$ & 1151
		& 0.00283	
		& 0.0689	
		& 2.61	
			& $-2.98$	
			& $-2.80$ 
			& $-2.63$	
			& 0.500	
			\\
	&	&	&	&
		&
		&
		&
			&
			& ($-2.82$$^j$)	
			&
			&
			\\
MC1162$^\ddagger$
	& L2-GL1 & $0.891/N_{\text{eff}}$ & $=\lambda_1$ & 1280
		& 0.00296	
		& 0.0621	
		& 2.76	
			& $-3.14$	
			& $-3.15$	
			& $-2.93$	
			& 0.457	
			\\
\textbf{MC1162}$^\ddagger$$^h$
	& L2-GL1 & $0.891/N_{\text{eff}}$ & $12.6/N_{\text{eff}}$ & 1183
		& 0.00275	
		& 0.0646	
		& 2.63	
			& $-3.00$	
			& $-3.00$ 
			& $-2.79$	
			& 0.483	
			\\
	&	&	&	&
		&
		&
		&
			&
			& \hspace*{0em} ($-2.93$$^j$)	
			&
			&
			\\
\hline
\hline
PF00595
	& L2-L1 $^k$ & $0.100/N_{\text{eff}}$ & $=\lambda_1$ & 1201
		& 0.00674 	
		& 0.0747	
		& 3.19	
			& $-3.60$   	
			& $-3.61$	
			& $-3.31$	
			& 0.563	
			\\
\textbf{PF00595}$^h$
	& L2-L1 $^k$ & $0.100/N_{\text{eff}}$ & $0.316/N_{\text{eff}}$ & 1007 
		& 0.00497	
		& 0.0736	
		& 3.08	
			& $-3.48$   	
			& $-3.49$	
			& $-3.13$	
			& 0.560	
			\\
\hline
\hline
PF00595
	& L2-L2 & $0.100/N_{\text{eff}}$ & $=\lambda_1$ & 1047 
		& 0.00580 	
		& 0.0737	
		& 3.13	
			& $-3.54$   	
			& $-3.55$	
			& $-3.27$	
			& 0.557	
			\\
\textbf{PF00595}$^h$
	& L2-L2 & $0.100/N_{\text{eff}}$ & $25.1/N_{\text{eff}}$ & 1119 
		& 0.00387	
		& 0.0725	
		& 2.99	
			& $-3.39$  	
			& $-3.39$	
			& $-3.04$	
			& 0.551	
			\\
\hline
\end{tabular}
\else
\fi
\begin{tablenotes}
\item [$^a$]
Boltzmann machine learning is carried out
with more iterations than 1200 by the modified Adam gradient-descent method 
with $\rho_m = 0.9$, $\rho_v = 0.999$, and an initial learning rate, $\kappa_0=0.01$;
see \SecParamUpdates.
The number of 
MCMC
samples in the Metropolis-Hastings method is equal to 
the effective number of sequences; $N_{\text{MC}} \simeq N_{\text{eff}}$.
\item [$^b$] The iteration number corresponding to $\min D^{KL}_2$ over the iteration numbers larger than 1000. 
\item [$^c$] The variance per residue of the evolutionary energies of random sequences.
\item [$^d$] $\langle \psi_N(\VECS{\sigma}) \rangle / L$: the ensemble average of evolutionary energy per residue in the Boltzmann distribution
by the Gaussian approximation
for the distribution of the evolutionary energies of random sequences with the mean $\bar{\psi}$ and the variance $\delta \psi^2$;
the Ising gauge is employed.
\item [$^e$] The sample average of evolutionary energies per residue over the sequences with no deletion in the MSA; 
the Ising gauge is employed.
\item [$^f$] The average of evolutionary energies per residue over 
the MCMC
samples with no deletion;
the Ising gauge is employed.
\item [$^g$] Precision of contact prediction; the number of predicted contacts is 352,
which is equal to the total number of closely located residue pairs
within $8$ \AA\  between side-chain centers in the 3D protein structure.
The corrected Frobenius norm of couplings
is employed for the contact score\CITE{ELLWA:13,EHA:14}.
\item [$^h$] The optimum set of regularization parameters, which is indicated by bold fonts.
\item [$^i$] The average of evolutionary energies per residue over the 
MCMC
samples with no deletion, 3317 of the total 100000 samples.
\item [$^j$] The sample average of evolutionary energies per residue over the natural proteins with no deletion.
\item [$^k$] The L1 means the elastic net with $\theta_2 = 0.9$ in \Eq{\EQelasticNet}.
\item [$^\ddagger$] 
MCMC
samples corresponding to $\dagger$.
\end{tablenotes}
\end{threeparttable}
\end{table*}
}

\TableInText{
\begin{table*}[!h]
\begin{threeparttable}[b]
\caption{
\TCBB{
\TEXTBF{
Regularization Parameters and Characteristics of Boltzmann Machines$^a$ with the ModAdam Method for PF00153.
}
}{
\TEXTBF{
Regularization parameters and characteristics of Boltzmann machines$^a$ with the ModAdam method for PF00153.
}
}
}
\label{tbl: PF00153_parameters}
\ifdefined\ThreeDigitsInTable
\begin{tabular}{llrrllllrccc}
\hline
\\
\multicolumn{2}{c}{MSA \hspace*{2em} regularizers}
			& \multicolumn{1}{c}{$\lambda_1$} & \multicolumn{1}{c}{$\lambda_2$} &  \multicolumn{1}{c}{\#Iter $^b$} & \multicolumn{1}{c}{$D^{KL}_1$} & \multicolumn{1}{c}{$D^{KL}_2$} 
		& \multicolumn{4}{c}{$\delta\psi^2 / L$ $^c$ \hspace*{0em} $(\bar{\psi} - \delta\psi^2) / L$ $^{d}$ \hspace*{0em} $\overline{\psi_N} / L$ $^e$ \hspace*{1em} $\overline{\psi_{\text{MC}}} / L$ $^f$}
		& Precision $^g$
	\\
\hline
\hline
PF00153
	& L2-GL1 & $0.100/N_{\text{eff}}$ & $=\lambda_1 $ & 1084
		& 0.00342	
		& 0.0264 	
		& 2.71	
			& \hspace*{1em} $-3.02$  	
			& $-3.29$	
			& $-3.04$	
			& 0.596	
			\\
\textbf{PF00153}$^h$
	& L2-GL1 & $0.100/N_{\text{eff}}$ & $209/N_{\text{eff}}$ & 1445$^\dagger$
		& 0.00112	
		& 0.0318 	
		& 2.50	
			& $-2.82$  	
			& $-2.83$	
			& $-2.54$ 
			& 0.630	
			\\
	&	&	&	&
		&
		&
		&
			&
			&
			& ($-2.51$$^i$)	
			&
			\\
\hline
MC1445$^\ddagger$
	& L2-GL1 & $0.100/N_{\text{eff}}$ & $209/N_{\text{eff}}$ & 1390 
		& 0.00151	
		& 0.0323	
		& 2.48	
			& $-2.82$   	
			& $-2.54$ 
			& $-2.52$	
			& 0.630	
			\\
	&	&	&	&
		&
		&
		&
			&
			& ($-2.83$$^j$)	
			&
			&
			\\
MC1445$^\ddagger$
	& L2-GL1 & $7.94/N_{\text{eff}}$ & $=\lambda_1$ & 1181 
		& 0.000975 	
		& 0.0160	
		& 2.25	
			& $-2.57$   	
			& $-2.57$	
			& $-2.47$	
			& 0.551	
			\\
\textbf{MC1445}$^\ddagger$$^h$
	& L2-GL1 & $7.94/N_{\text{eff}}$ & $20.0/N_{\text{eff}}$ & 1197
		& 0.000985	
		& 0.0162	
		& 2.24	
			& $-2.55$   	
			& $-2.55$ 
			& $-2.43$	
			& 0.557	
			\\
	&	&	&	&
		&
		&
		&
			&
			& ($-2.64$$^j$)	
			&
			&
			\\
\hline
\hline
PF00153
	& L2-L1 $^k$ & $0.100/N_{\text{eff}}$ & $=\lambda_1 $ & 1149
		& 0.00313 	
		& 0.0265	
		& 2.73	
			& $-3.05$   	
			& $-3.32$	
			& $-3.09$	
			& 0.599	
			\\
\textbf{PF00153}$^h$
	& L2-L1 $^k$ & $0.100/N_{\text{eff}}$ & $25.1/N_{\text{eff}}$ & 1208
		& 0.00165	
		& 0.0318	
		& 2.57	
			& $-2.91$   	
			& $-2.91$	
			& \RED{$-2.66$}	
			& 0.557	
			\\
\hline
\hline
PF00153
	& L2-L2 & $0.100/N_{\text{eff}}$ & $=\lambda_1 $ & 1223
		& 0.00329 	
		& 0.0264	
		& 2.76	
			& $-3.08$   	
			& $-3.35$	
			& $-3.10$	
			& 0.605	
			\\
\textbf{PF00153}$^h$
	& L2-L2 & $0.100/N_{\text{eff}}$ & $398/N_{\text{eff}}$ & 1066 
		& 0.00119	
		& 0.0336	
		&  2.55	
			& $-2.87$   	
			& $-2.86$	
			& $-2.52$	
			& 0.569	
			\\
\hline
\end{tabular}
\else
\fi
\begin{tablenotes}
\item [$^a$]
Boltzmann machine learning is carried out
with more iterations than 1200 by the modified Adam gradient-descent method
with $\rho_m = 0.9$, $\rho_v = 0.999$, and an initial learning rate, $\kappa_0=0.01$;
see \SecParamUpdates.
The number of 
MCMC
samples in the Metropolis-Hastings method is equal to
the effective number of sequences; $N_{\text{MC}} \simeq N_{\text{eff}}$.
\item [$^b$] The iteration number corresponding to $\min D^{KL}_2$ over the iteration numbers larger than 1000.
\item [$^c$] The variance per residue of the evolutionary energies of random sequences
\item [$^d$] $\langle \psi_N(\VECS{\sigma}) \rangle / L$: the ensemble average of evolutionary energy per residue in the Boltzmann distribution 
by the Gaussian approximation
for the distribution of the evolutionary energies of random sequences with the mean $\bar{\psi}$ and the variance $\delta \psi^2$;
the Ising gauge is employed.
\item [$^e$] The sample average of evolutionary energies per residue over the sequences with no more than 2 deletions for PF00153 and
with no more than 3 for the 
MCMC
samples;
the Ising gauge is employed.
\item [$^f$] The average of evolutionary energies per residue over 
the
MCMC
samples with no more than 3 deletions;
the Ising gauge is employed.
\item [$^g$] Precision of contact prediction; the number of predicted contacts is 332,
which is equal to the total number of closely located residue pairs
within $8$ \AA\  between side-chain centers in the 3D protein structure.
The corrected Frobenius norm of couplings 
is employed for the contact score\CITE{ELLWA:13,EHA:14}.
\item [$^h$] The optimum set of regularization parameters, which is indicated by bold fonts.
\item [$^i$] The average of evolutionary energies per residue over the 
MCMC
samples with no deletion, 207 of the total 100000 samples.
\item [$^j$] The sample average of evolutionary energies per residue over the natural proteins with no more than 2 deletions.
\item [$^k$] The L1 means the elastic net with $\theta_2 = 0.9$ in \Eq{\EQelasticNet}.
\item [$^\ddagger$] 
MCMC
samples corresponding to $\dagger$.
\end{tablenotes}
\end{threeparttable}
\end{table*}
}

\TableInText{
\ifdefined\NAG
\begin{table*}[!ht]
\begin{threeparttable}[b]
\caption{
\TCBB{
\TEXTBF{
Comparison of the Learning Methods$^a$ for Gradient Descent on PF00595.
}
}{
\TEXTBF{
Comparison of the learning methods$^a$ for gradient descent on PF00595.
}
}
}
\label{tbl: PF00595_learning_methods}
\ifdefined\ThreeDigitsInTable
\begin{tabular}{llrrllllrrcc}
\hline
Learning \\
\multicolumn{2}{c}{method \hspace*{1em} regularizers}
			& \multicolumn{1}{c}{$\lambda_1$} & \multicolumn{1}{c}{$\lambda_2$} &  \multicolumn{1}{c}{\#Iter $^b$} & \multicolumn{1}{c}{$D^{KL}_1$} & \multicolumn{1}{c}{$D^{KL}_2$} 
		& \multicolumn{4}{c}{$\delta\psi^2 / L$ $^c$ \hspace*{0em} $(\bar{\psi} - \delta\psi^2) / L$ $^{d}$ \hspace*{0em} $\overline{\psi_N} / L$ $^e$ \hspace*{1em} $\overline{\psi_{\text{MC}}} / L$ $^f$}
		& Precision $^g$
	\\
\hline
ModAdam $^h$	
		& L2-L2
		& $0.100/N_{\text{eff}}$ & $25.1/N_{\text{eff}}$ & 1119 
		& 0.00387	
		& 0.0725	
		& 2.99	
			& \hspace*{1em} $-3.39$ 	
			& \hspace*{1em} $-3.39$	
			& $-3.04$	
			& 0.551	
			\\
\multicolumn{2}{l}{\hspace*{1em}(second run) $^i$}	
		&	&	
		& 2018 
		& 0.00372	
		& 0.0696	
		& 3.12	
			& $-3.53$  	
			& $-3.52$	
			& $-3.16$	
			& 0.568	
			\\
Adam $^h$	
		& L2-L2
		& $0.100/N_{\text{eff}}$ & $25.1/N_{\text{eff}}$ & 1012
		& 0.00320 	
		& 0.0681	
		& 3.35	
			& $-3.73$   	
			& $-3.59$	
			& $-3.23$	
			& 0.563	
			\\
NAG $^h$	
		& L2-L2
		& $0.100/N_{\text{eff}}$ & $25.1/N_{\text{eff}}$ & 1110 
		& 0.00381	
		& 0.0724	
		& 2.94	
			& \hspace*{1em} $-3.34$ 	
			& \hspace*{1em} $-3.34$	
			& $-3.01$	
			& 0.557	
			\\
\multicolumn{2}{l}{\hspace*{1em}\hspace*{5em}$^i$}	
		&	&	
		& 2095
		& 0.00361	
		& 0.0690	
		& 3.08	
			& $-3.48$  	
			& $-3.48$	
			& $-3.12$	
			& 0.565	
			\\
RPROP-LR $^j$  & L2-L2
		& $0.100/N_{\text{eff}}$ & $25.1/N_{\text{eff}}$ & 1052
		& 0.00391 	
		& 0.0766	
		& 2.97	
			& $-3.36$   	
			& $-3.28$	
			& $-2.95$	
			& 0.560	
			\\
\hline
\end{tabular}

\else
\fi
\begin{tablenotes}

\item [$^a$] Boltzmann machine learning is carried out
with more iterations than 1200 by each gradient-descent method;
see \SecParamUpdates.
The number of 
MCMC
samples in the Metropolis-Hastings method is equal to
the effective number of sequences; $N_{\text{MC}} \simeq N_{\text{eff}}$.
\item [$^b$] The iteration number corresponding to $\min D^{KL}_2$ over the iteration numbers larger than 1000. 
\item [$^c$] The variance per residue of evolutionary energies of random sequences
\item [$^d$] $\langle \psi_N(\VECS{\sigma}) \rangle / L$: the ensemble average of evolutionary energy per residue in the Boltzmann distribution 
by the Gaussian approximation 
for the distribution of the evolutionary energies of random sequences with the mean $\bar{\psi}$ and the variance $\delta \psi^2$;
the Ising gauge is employed.
\item [$^e$] The average of evolutionary energies per residue over the homologous sequences with no deletion in the MSA;
the Ising gauge is employed.
\item [$^f$] The average of evolutionary energies per residue over 
the MCMC
samples with no deletion;
the Ising gauge is employed.
\item [$^g$] Precision of contact prediction; the number of predicted contacts is 352, 
which is equal to the total number of closely located residue pairs
within $8$ \AA\ between side-chain centers in the 3D protein structure.
The corrected Frobenius norm of couplings 
is employed for the contact score\CITE{ELLWA:13,EHA:14}.
\item [$^h$] The initial learning rates and other hyper-parameters are $(\kappa_0 = 0.01, \rho_m=0.9, \rho_v=0.999)$ 
for the ModAdam, $(\kappa_0 = 0.001, \rho_m=0.9, \rho_v=0.999)$ for the Adam, and
$(\kappa_0 = 0.1, \rho_m=0.95)$ for the NAG; see \SecParamUpdates.
\item [$^i$] The iteration number corresponding to $\min D^{KL}_2$ over the iteration numbers
larger than 2000; more than 2100 iterations are carried out.
\item [$^j$] The RPROP learning rate method\CITE{BLCC:16}: the learning rates are limited between $10^{-5}$ and $10$.
\end{tablenotes}
\end{threeparttable}
\end{table*}
\else
\fi
}

\TCBB{
\SUBSECTION{Dependences of Inferred Parameters on the Gradient-Descent Methods}
}{
\SUBSECTION{Dependences of inferred parameters on the gradient-descent methods}
}

\FigureInText{
\ifdefined\NAG

\begin{figure*}[hbt]
\centerline{
\includegraphics[width=60mm,angle=0]{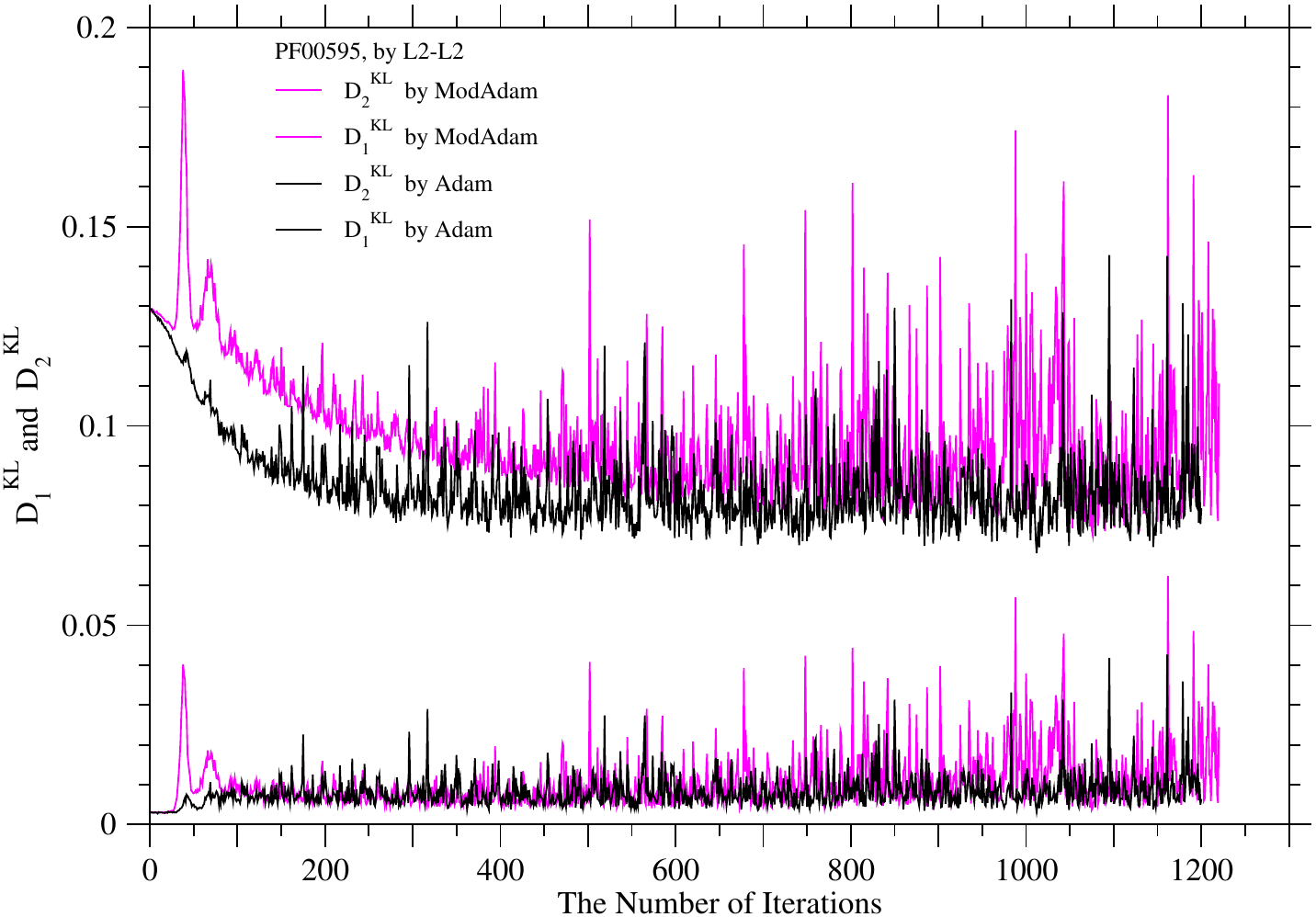}
\includegraphics[width=60mm,angle=0]{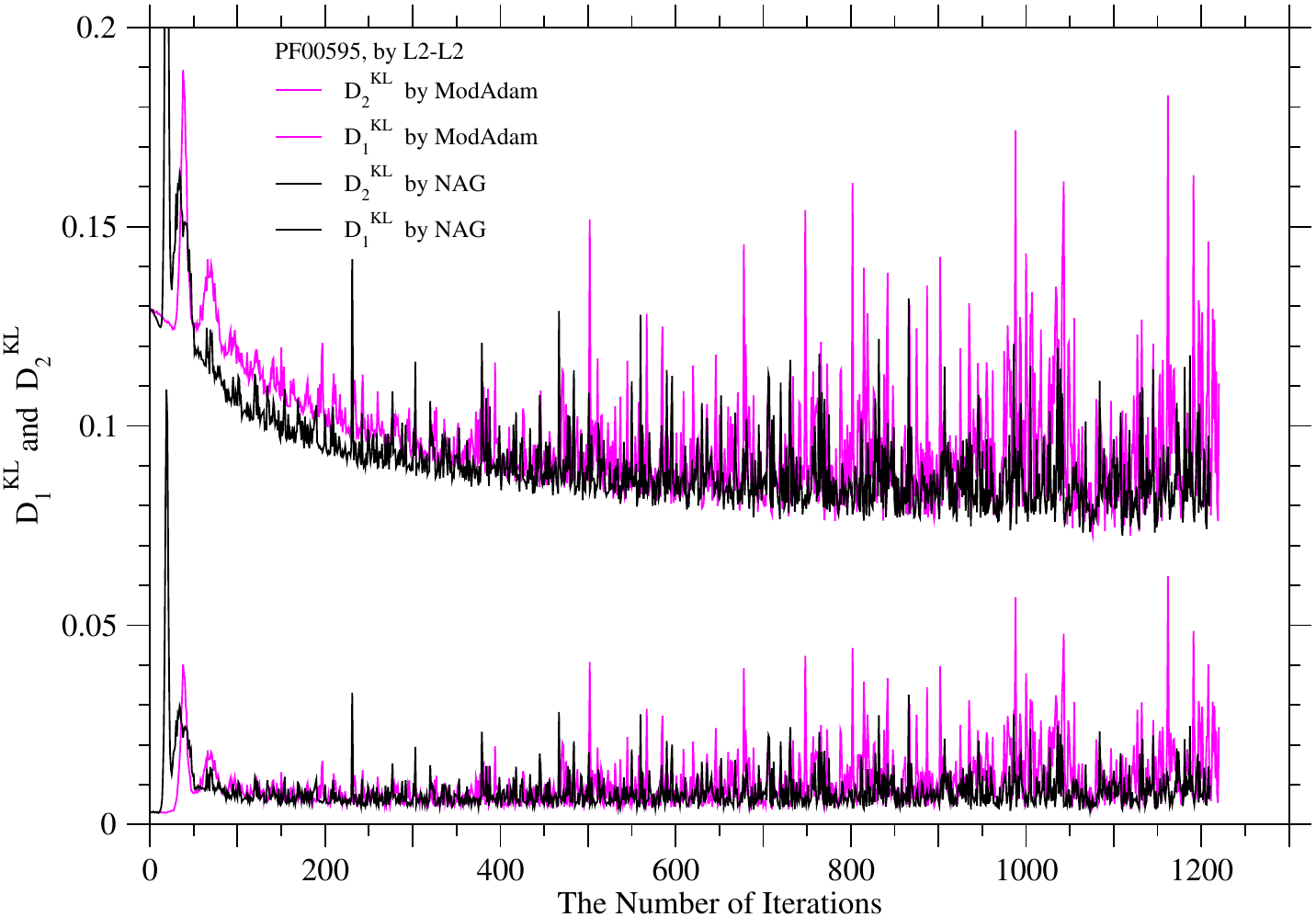}
\includegraphics[width=60mm,angle=0]{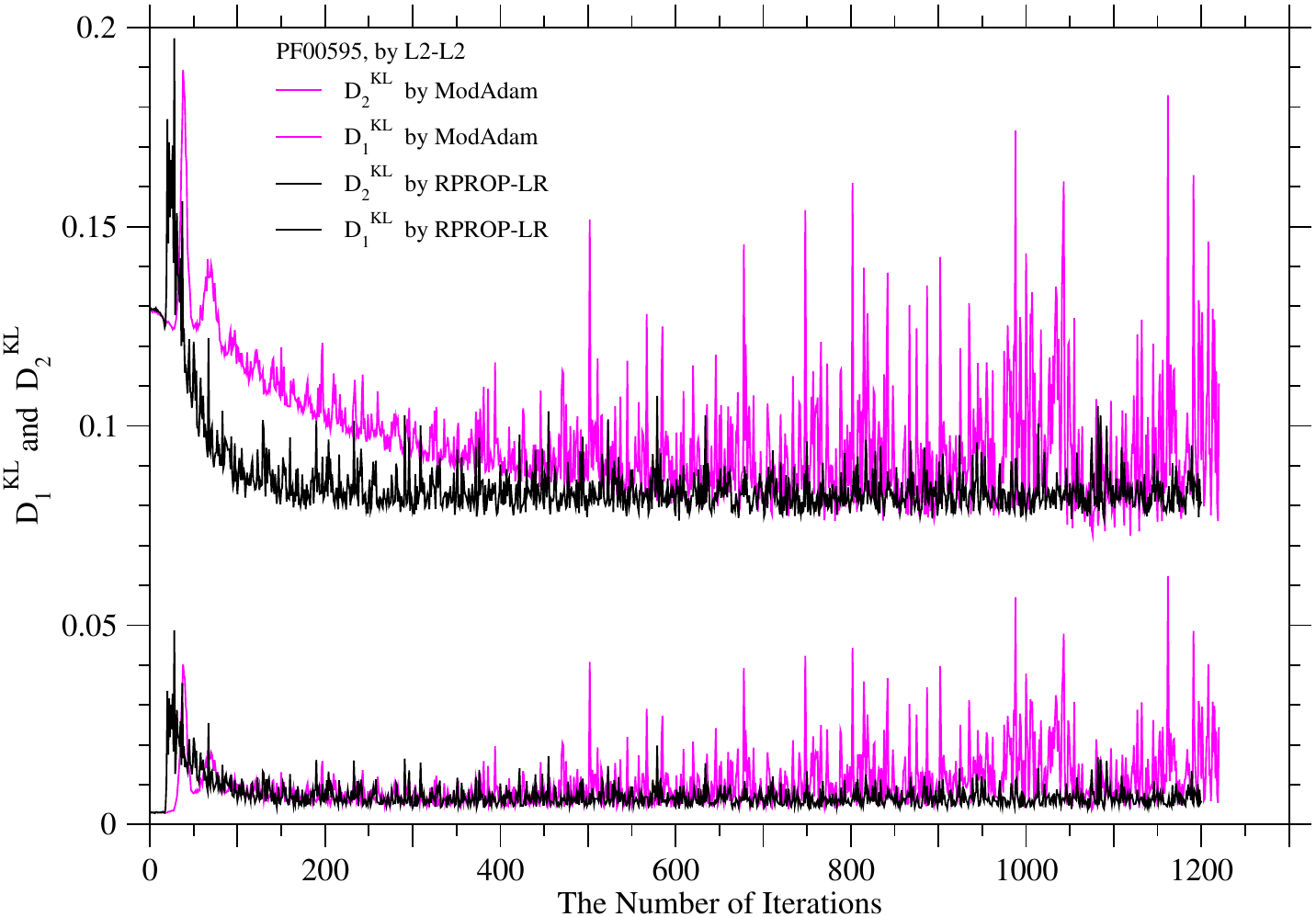}
}
\caption
{
\noindent
\TEXTBF{Learning processes by the ModAdam, Adam, NAG, and RPROP-LR gradient-descent methods for PF00595.}
The averages of Kullback-Leibler divergences,  
$D^{2}_{\text{KL}}$ for pairwise marginal distributions
and $D^{1}_{\text{KL}}$ for single-site marginal distributions, are
drawn against iteration number in the learning processes.
$D^{2}_{\text{KL}}$ and $D^{1}_{\text{KL}}$ 
for the Adam, NAG, and RPROP-LR
are indicated by the upper and lower black lines
in the left, middle, and right figures, respectively,
and those for the ModAdam are shown in pink in all figures 
for comparison.
The L2-L2 regularization model is employed.
The values of hyper-parameters
are listed in \Table{\ref{tbl: PF00595_learning_methods}} as well as others.
\label{fig: PF00595_learning_process_KL_by_ModAdam_and_Adam}
\label{fig: PF00595_learning_process_KL_by_ModAdam_and_RPROP-LR}
\label{fig: PF00595_learning_process_KL}
}
\end{figure*}
\else
\fi
}

First, the dependences of inferred parameters on the gradient-descent methods,
Adam\CITE{KB:14}, 
\ifdefined\NAG
NAG\CITE{N:04},
\fi
RPROP-LR\CITE{BLCC:16,FBW:18}, and ModAdam, 
are examined on PF00595; 
see \SecParamUpdates.
\RED{Here} the L2-L2 regularization is employed
instead of the L2-GL1 regularization.
The same regularization parameters, which have been tuned
for the L2-L2 regularization
as described in the preceding section,
are employed for all gradient-descent methods.
Although
all the methods attain similar precisions in contact prediction
as shown in \Table{\ref{tbl: PF00595_learning_methods}},
the inferred couplings and 
the profile of evolutionary energy along the learning process are very different 
among the gradient-descent methods.  

In \Fig{\ref{fig: PF00595_learning_process_KL_by_ModAdam_and_RPROP-LR}},
the averages of Kullback-Leibler divergences,
$D_{2}^{KL}$ for pairwise marginal distributions 
and $D_{1}^{KL}$ for single site marginal distributions,
which are defined by \Eqs{\EQKLB\ and \EQKLA}, 
are plotted against the iteration number of learning 
in pink for the ModAdam, in black for 
\ifdefined\NAG
the Adam,
NAG,
\else
the Adam
\fi
and RPROP-LR.
The fluctuations of $D_{2}^{KL}$ and $D_{1}^{KL}$ 
at large iteration numbers primarily originate in the statistical error
of marginal frequencies estimated by the 
MCMC
sampling.
In the ModAdam method, $D_{2}^{KL}$ converge more slowly
and its fluctuations seem to be larger than in the other methods.
However, 
a more important thing is 
the reasonable inference of fields and couplings.

Let us see 
how the single-site frequencies and pairwise correlations are recovered in each method,
and how the inferred fields and couplings differ among the gradient-descent methods. 
The fields and couplings 
as well as the marginal single-site frequencies and pairwise correlations
are compared between the ModAdam and Adam methods in 
\Fig{\ref{fig: PF00595_ModAdam_vs_Adam_hJ}} and 
\SFigPFvModAdamvsAdamhJPiaCijab.
It should be noticed here that fields and couplings are
expressed in the Ising gauge for comparison.
Although the single-site marginal probabilities in both models almost coincide with each other,
there are some fields that are significantly more positive in the Adam than in the ModAdam.
On the other hand,
strong couplings
are significantly underestimated by the Adam in the comparison with the ModAdam.
Consistently,
the pairwise correlations are slightly under-reproduced in the region of 
strong correlations.
As a result, 
the negative correlation between couplings and residue-residue distance is
better
detected by the ModAdam than by the Adam method
as shown in \Fig{\ref{fig: PF00595_maxJij_comparison}}, in which
$J_{ij}(a_k,a_l)$, where $(a_k,a_l) = \text{argmax}_{a_k,a_l \neq \text{deletion}} | J_{ij}(a_k,a_l) |$ in the Ising gauge, are plotted
against the distance between $i$th and $j$th residues;
all residue pairs with $| J_{ij}(a_k,a_l) | >= 0.73$ in either method are in contact within 8 \AA ,  
but there are only 6 such pairs in the Adam but 16 pairs in the ModAdam;
only amino acid pairs
are taken into account in the \text{argmax}, because
deletions within gaps in Pfam alignments tend to have large positive correlations.
This tendency is very clear even at the small number of iterations;
see \SFigPFvmaxJijcomparison.
Even if gradient-descent methods attain to a similar solution via different intermediate states,
it will be desirable to attain 
with a limited number of iterations 
to approximate solutions that satisfy characteristics required for the solution.
\ifdefined\NAG
Here it is noteworthy that the fields and couplings inferred by the NAG almost coincide
with those by the ModAdam; see \Fig{\ref{fig: PF00595_ModAdam_vs_NAG_hJ}} and \SFigPFvModAdamvsNAGhJPiaCijab.
\fi
Hence, 
these results indicate
that
the stepsize of parameter updates must be proportional to the
partial derivative in order to correctly estimate the dependencies
of couplings $J_{ij}$ on the distance between residues $i$ and $j$.
The parameters inferred by the Adam likely include as much information of contact ranks as those
by the ModAdam, because as shown in \Table{\ref{tbl: PF00595_learning_methods}} 
the precision of contact prediction by the Adam is 
as good as that by the ModAdam.
It should be noticed, however, that the present purpose is
to correctly infer not only the ranks but 
also couplings and fields.
In the Adam method, 
a stepsize for each parameter update is proportional to 
the partial derivative that is normalized for each parameter at each step,
so that stepsizes are essentially in the same order for all parameters at each iteration,
making the Adam suitable to similarly dense interaction systems as well as
the $L_2$ regularization.

\TCBB{
\FigureInText{
\ifdefined\NAG

\begin{figure*}[hbt]
\centerline{
\includegraphics[width=60mm,angle=0]{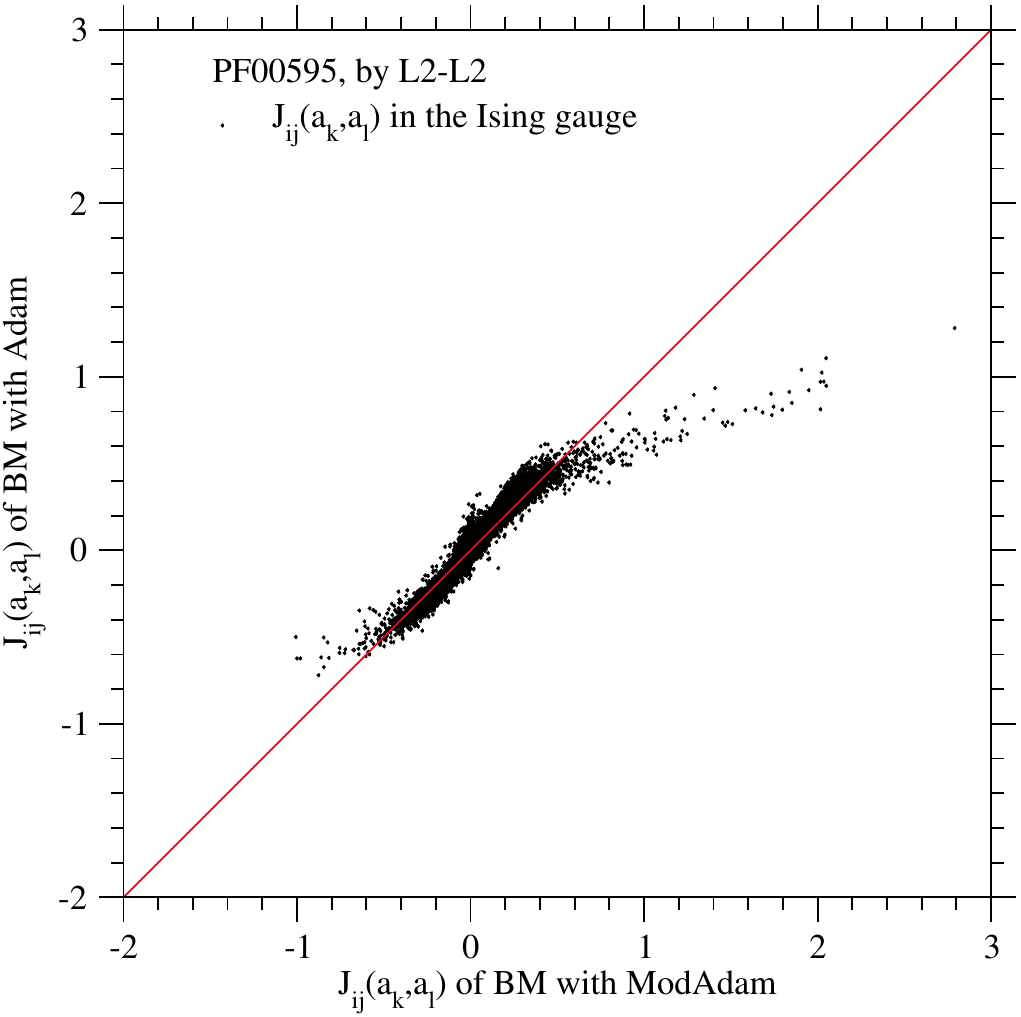}
\includegraphics[width=60mm,angle=0]{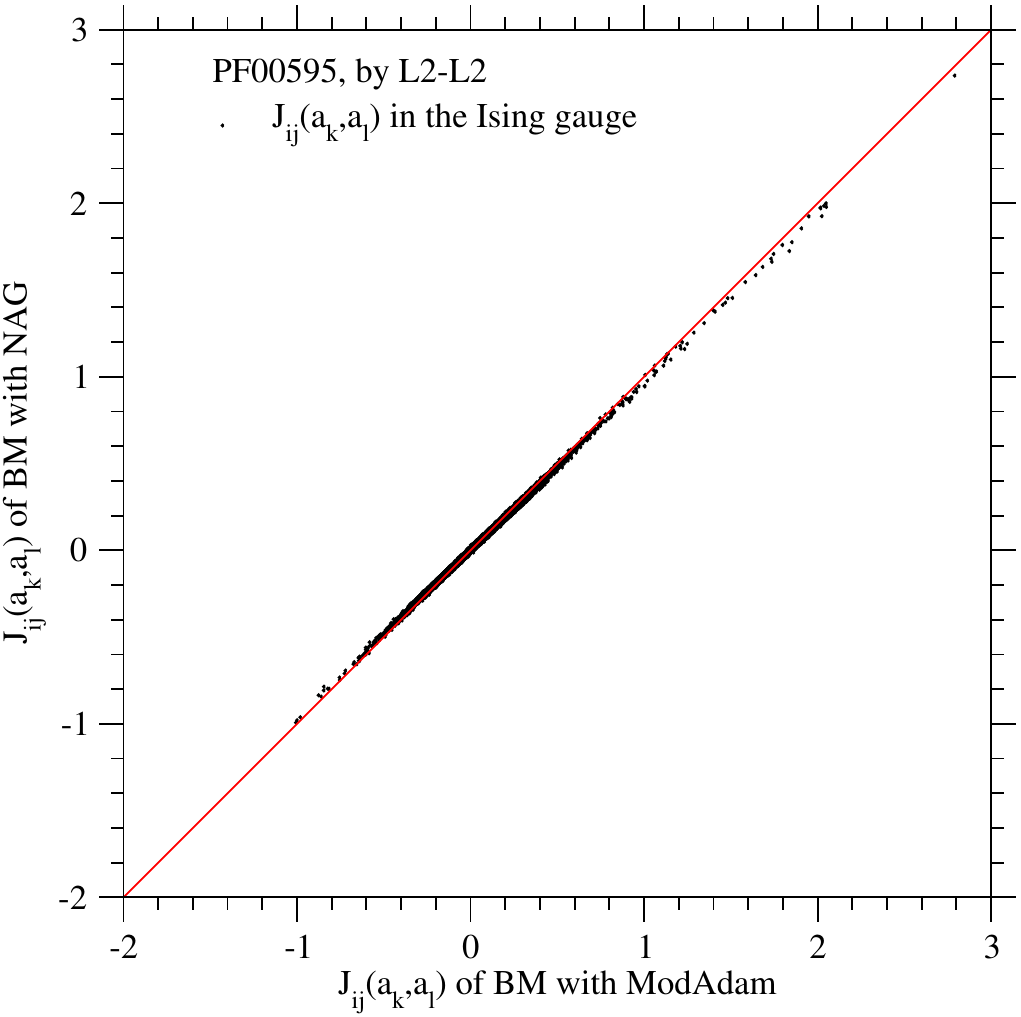}
\includegraphics[width=60mm,angle=0]{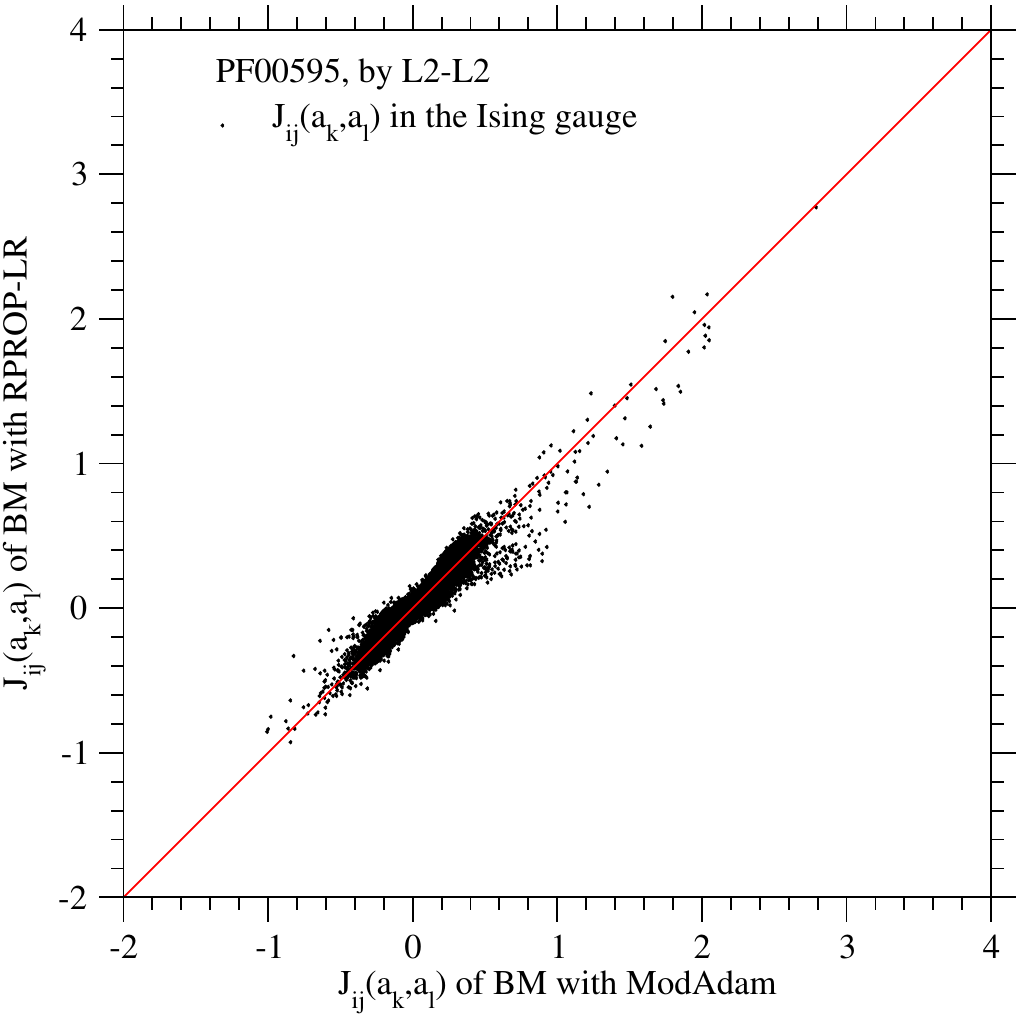}
}
\caption
{
\noindent
\TEXTBF{Comparisons of the inferred 
couplings $J_{ij}(a_k,a_l)$ 
in the Ising gauge
between the ModAdam and the other gradient-descent methods, Adam, NAG, and RPROP-LR, for PF00595.}
The abscissas correspond to the 
couplings 
inferred by the modified Adam, and 
the ordinates correspond to those by
the Adam, NAG, and RPROP-LR  
in order from the left to the right.
The regularization model L2-L2 is employed for all methods.
The solid lines show the equal values between the ordinate and abscissa.
The values of hyper-parameters are listed in \Table{\ref{tbl: PF00595_parameters}}.
The overlapped points of $J_{ij}(a_k,a_l)$ in the units 0.001 are removed.
\label{fig: PF00595_ModAdam_vs_Adam_hJ}
\label{fig: PF00595_ModAdam_vs_RPROP-LR_hJ}
\label{fig: PF00595_ModAdam_vs_NAG_hJ}
\label{fig: PF00595_ModAdam_vs_others_hJ}
\label{fig: PF00595_ModAdam_vs_others_J}
}
\end{figure*}
\else
\fi
}

}{
}

\FigureInText{

\begin{figure}[hbt]
\centerline{
\includegraphics[width=43mm,angle=0]{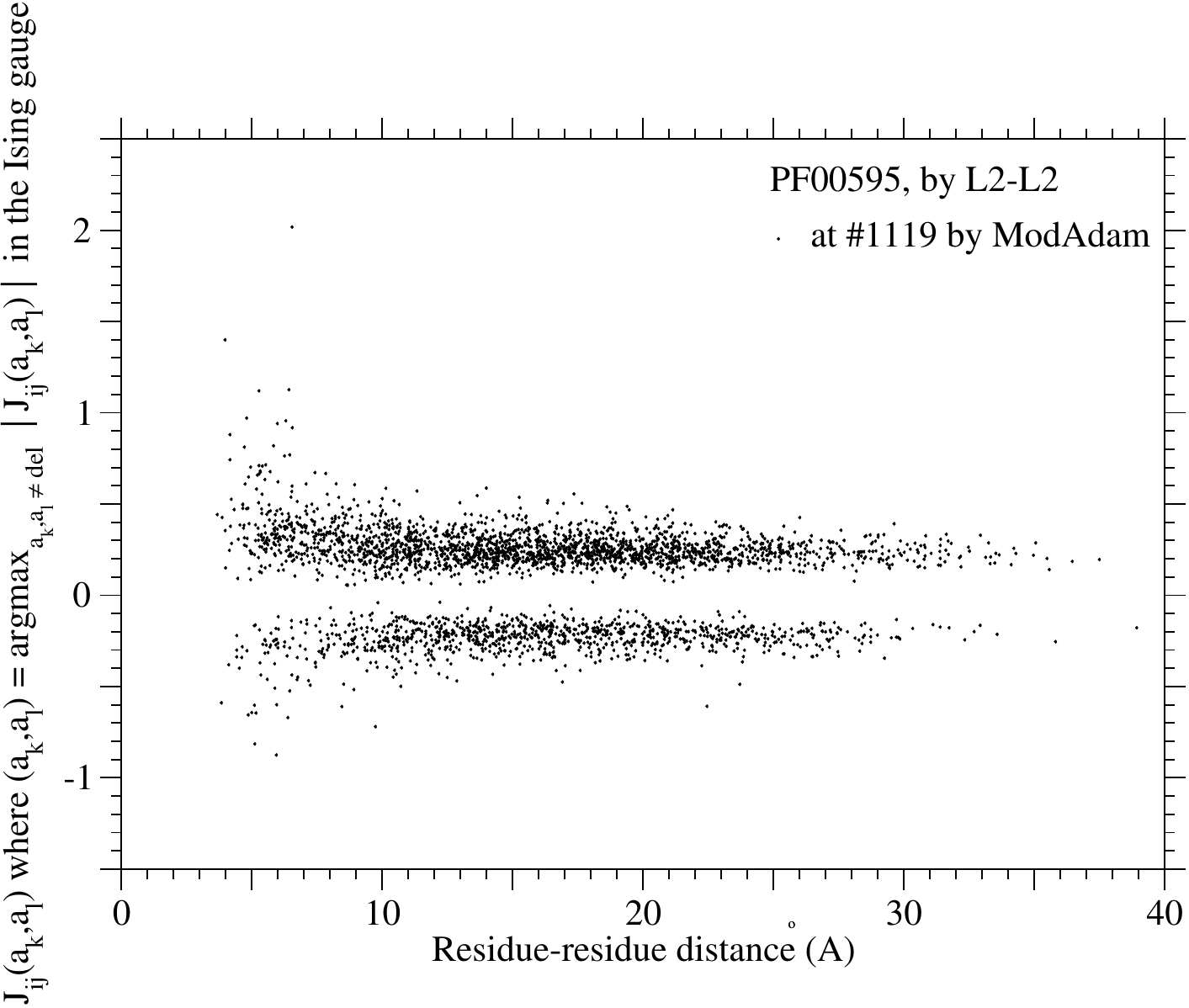}
\includegraphics[width=43mm,angle=0]{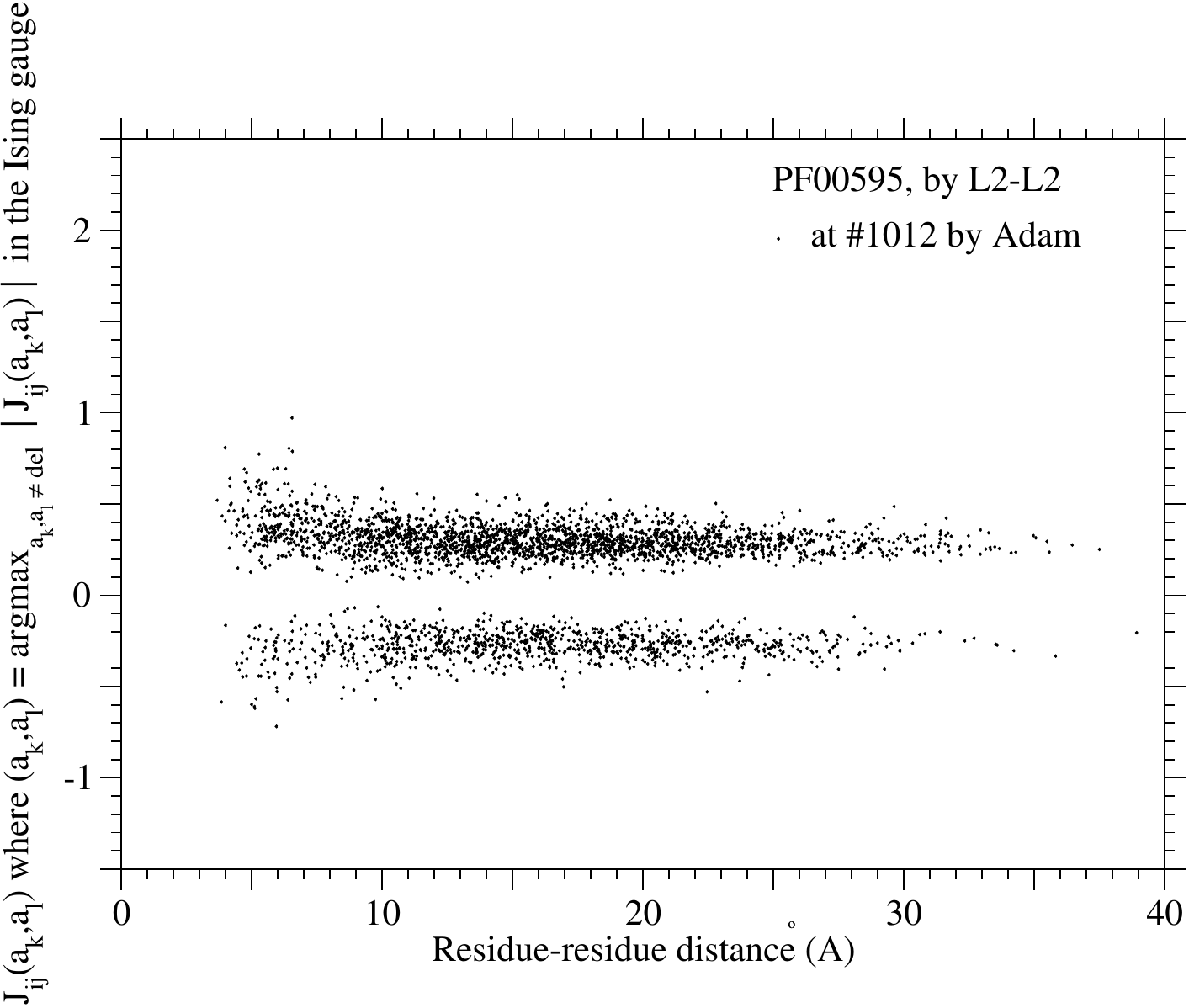}
}
\caption
{
\noindent
\TEXTBF{Differences in the learning of coupling parameters, $J_{ij}(a_k,a_l)$,
between the ModAdam and Adam gradient-descent methods for PF00595.}
All $J_{ij}(a_k,a_l)$ where $(a_k,a_l) = \text{argmax}_{a_k,a_l \neq \text{deletion}} | J_{ij}(a_k,a_l) |$ 
in the Ising gauge
are plotted
against the distance between $i$th and $j$th residues.
The ModAdam and Adam methods are employed for the left and right figures,
respectively.
The regularization model L2-L2 is employed for both methods.
The learning processes by both methods are shown in
\Fig{\ref{fig: PF00595_learning_process_KL_by_ModAdam_and_Adam}}.
Please notice that 
more strong couplings tend to be inferred for closely located residues pairs by 
the ModAdam method than by the Adam method.
The values of hyper-parameters are listed in \Table{\ref{tbl: PF00595_parameters}}.
\label{fig: PF00595_maxJij_comparison}
}
\end{figure}
}

\TCBB{
}{
}

RPROP, in which the stepsize does not depend on the value of the partial derivative but on
its sign, was modified\CITE{BLCC:16} for inverse Potts problems to be proportional to the partial derivative.
This modified RPROP is called here RPROP-LR.
In \Fig{\ref{fig: PF00595_ModAdam_vs_RPROP-LR_hJ} } and 
\SFigPFvModAdamvsRPROPLRhJPiaCijab,
the fields and couplings inferred by the RPROP-LR method 
are shown
in comparison with those by the ModAdam
as well as the marginal single-site probabilities and pairwise correlations.
The RPROP-LR method is, like Adam, a per-parameter adaptive learning rate method but
unlike Adam it does not normalize each partial derivative of parameter.
As a result, the RPROP-LR infers couplings similarly to the ModAdam. 
From these figures, it is hard to judge which is better, ModAdam or RPROP-LR.
However,
The sample and ensemble averages of evolutionary energy
along the learning process provide more useful information 
with respect to the characteristics of each gradient-descent method.

In \Fig{\ref{fig: average_energy_comparison}},
the average evolutionary energy per residue ($\overline{\psi_{N}}/L$) of natural sequences and
the ensemble average of evolutionary energy per residue ($(\bar{\psi} - \delta\psi^2)/L$) 
are plotted against the iteration number for each method.
These profiles are well reproducible by another run of Boltzmann machine learning.
In the figure for the ModAdam method, 
both the results of the first and second runs are shown by dots and solid lines, respectively. 
They indistinguishably overlap to each other.
The profile of evolutionary energy along the learning process is very different among
the gradient-descent methods.  
The 
average of evolutionary energies over natural proteins, $\overline{\psi_N}$,
as well as $D_{2}^{KL}$
more quickly converges in the RPROP-LR and Adam than in the ModAdam. 
However, 
a more important feature is that 
the sample average of evolutionary energies over natural sequences, $\overline{\psi_N(\VECS{\sigma_N})}$, is
higher than $(\bar{\psi} - \delta\psi^2)$ 
with the interaction parameters inferred by the Adam and RPROP-LR methods 
under the regularization parameters under which they are equal to each other in the ModAdam method.
It should be recalled that
$(\bar{\psi} - \delta\psi^2)/L$ approximates
the ensemble average $\langle \psi_N(\VECS{\sigma}) \rangle_{\VECS{\sigma}}$ of evolutionary energy in the Boltzmann distribution;
see \Eq{\EQensAvepsi}.
The fact that $\overline{\psi_N(\VECS{\sigma_N})} > (\bar{\psi} - \delta\psi^2)$ indicates that
the fields and couplings inferred by the Adam and the RPROP-LR are less favorable to the natural proteins 
than those by the ModAdam under the same condition.
In other words, 
the ModAdam method is better to infer more reasonable interaction parameters 
for protein sequences than the Adam and the RPROP-LR. 
In addition, $D_2^{KL}$ of the 
MCMC
samples obtained by the RPROP-LR is higher 
than that of the ModAdam as shown in \Table{\ref{tbl: PF00595_learning_methods}},
indicating that the recoverability of pairwise frequencies is less in the RPROP-LR than in the ModAdam.

\ifdefined\NAG
  The NAG method 
is not a per-parameter learning rate method and
employs a stepsize that is proportional to the partial derivative in common with the ModAdam.
As a result, 
the fields and couplings inferred by the NAG almost coincide with those by the ModAdam.
However, the sample and ensemble averages of evolutionary energy
converge a little higher values in the NAG than in the ModAdam, indicating
that the fields and couplings are slightly more optimized by the ModAdam.
\fi

The Adam method as well as $L_2$ regularization is not appropriate to 
the present Potts problem, because
residue-residue interactions in proteins are very sparse.
On the other hand,
the RPROP-LR appears to be inferior to the ModAdam 
\ifdefined\NAG
and NAG 
\fi
with respect to the quality of inferred interactions,
although it quickly converges and infers couplings similar to those by ModAdam\RED{.}
\RED{Thus}, the ModAdam method is 
\ifdefined\NAG
\else
invented and 
\fi
employed in the following;
see \SecParamUpdates. 

\FigureInText{

\begin{figure*}[hbt]
\centerline{
\includegraphics[width=44mm,angle=0]{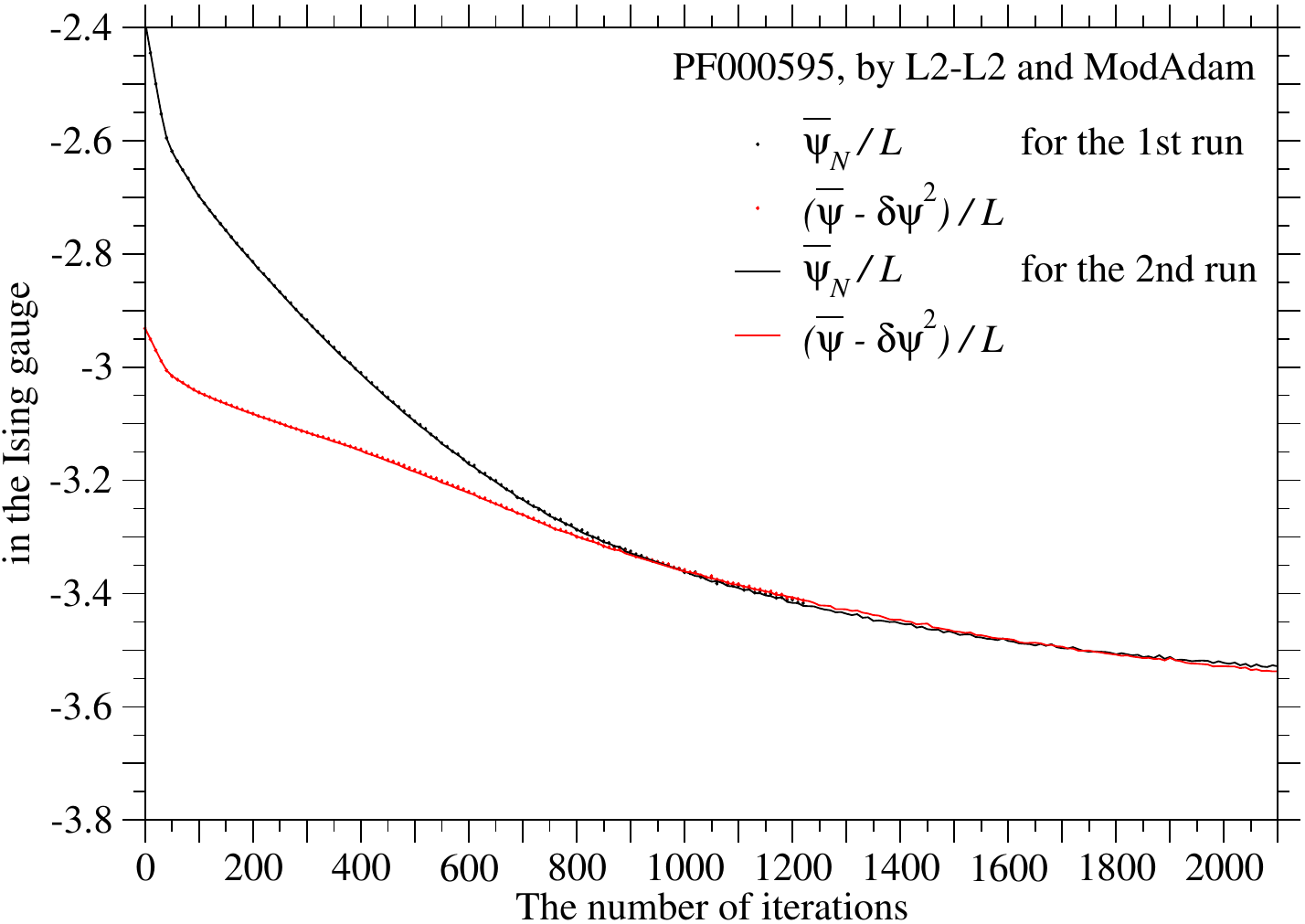}
\includegraphics[width=44mm,angle=0]{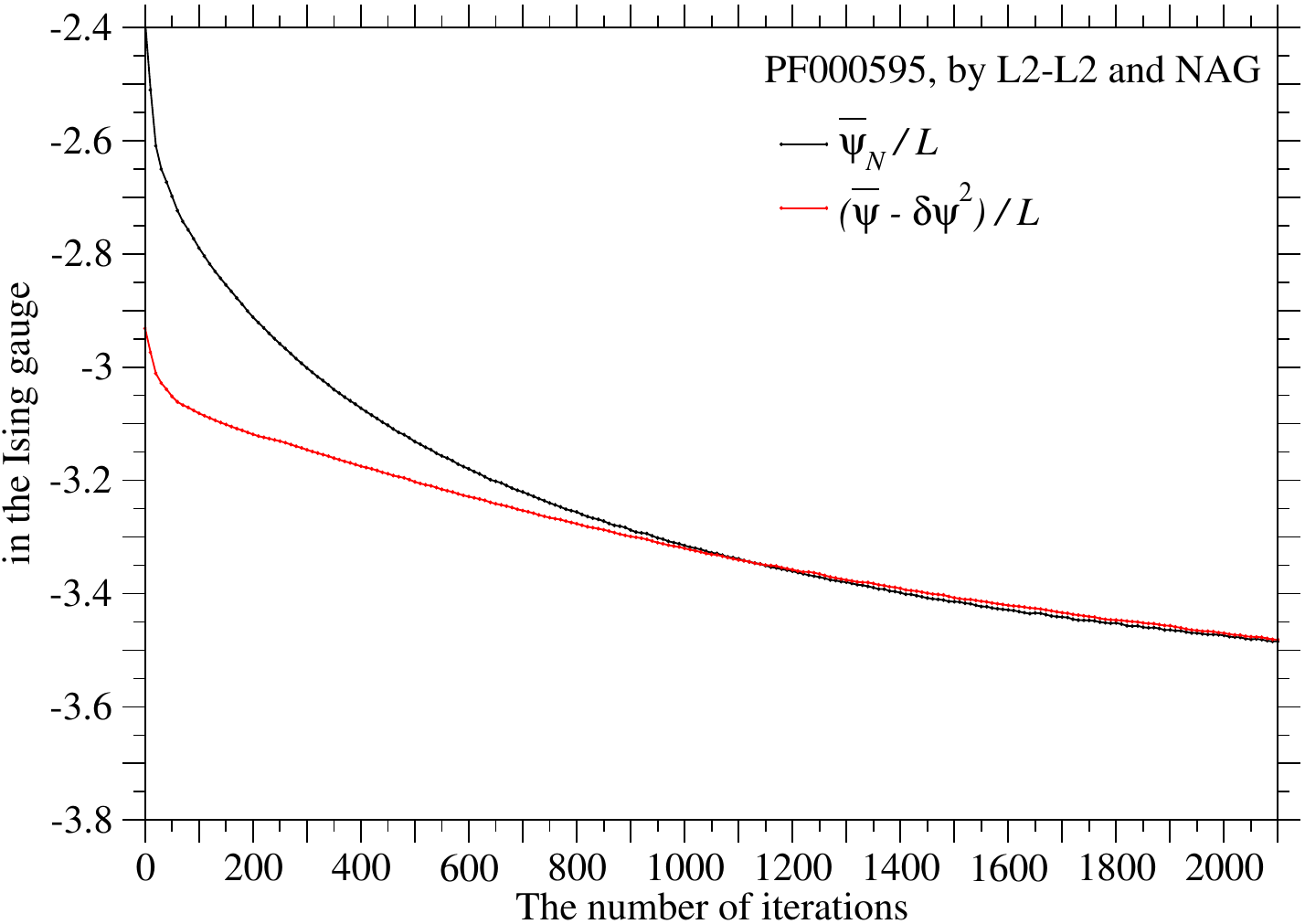}
\includegraphics[width=44mm,angle=0]{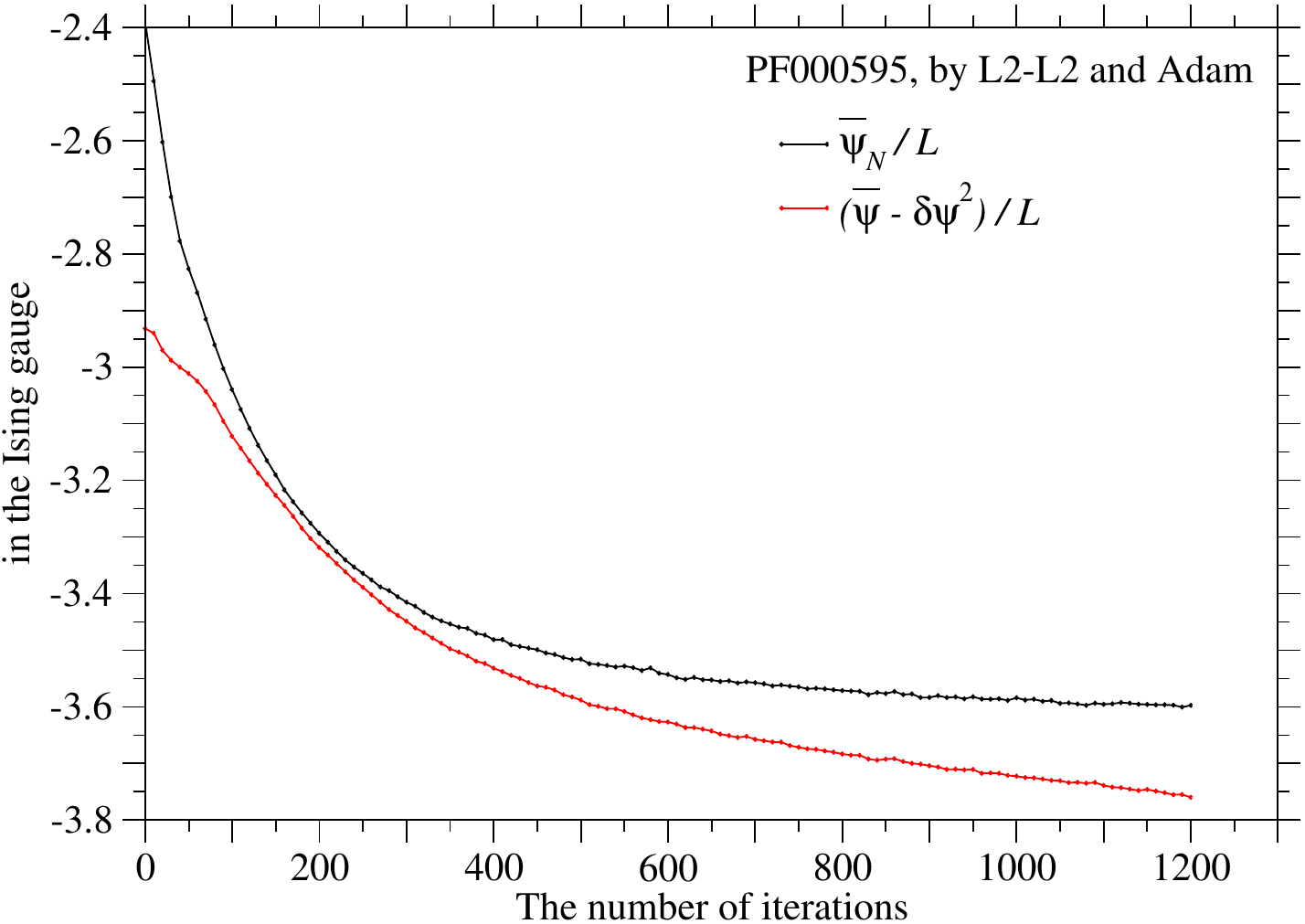}
\includegraphics[width=44mm,angle=0]{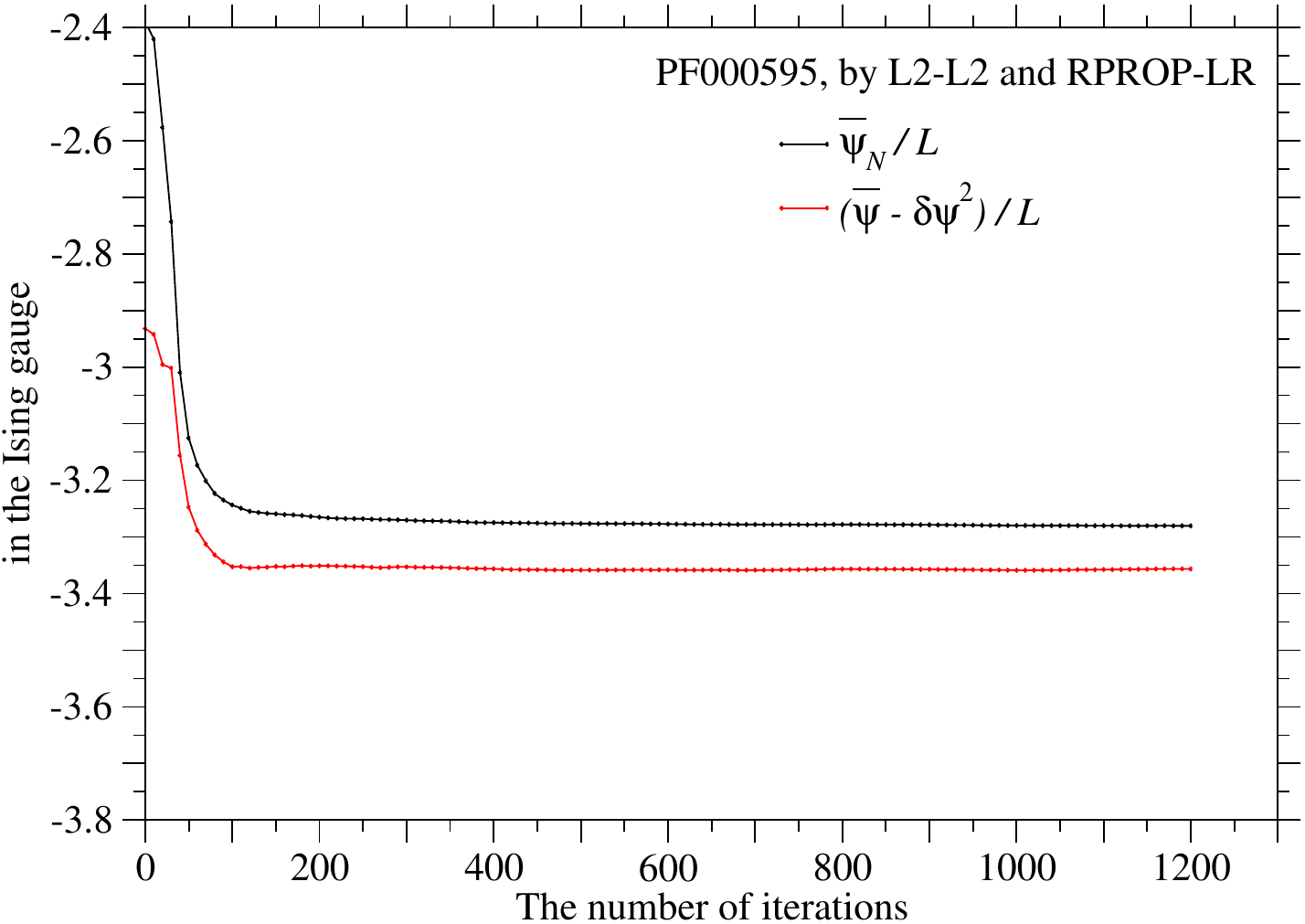}
}
\caption
{
\noindent
\TEXTBF{The profile of the average evolutionary energies along the learning process in the L2-L2 model 
by each gradient-descent method for PF00595.}
The average evolutionary energy per residue $\overline{\psi_N}/L$ of natural sequences
and 
the ensemble average of evolutionary energy per residue in the Gaussian approximation $(\bar{\psi} - \delta\psi^2)/L$
in the Ising gauge
are plotted every 10 iterations against iteration number in the learning by each of the ModAdam,
NAG, Adam, and RPROP-LR in the order of the left to the right;
the sample and ensemble averages are indicated by the upper and lower lines, respectively.
The L2-L2 regularization model is employed.
For the ModAdam in the leftmost figure, those for the first run of 1220 iterations
and for the second run, which is conditioned to run by more than 2000 iterations,
are plotted by dots and solid lines, 
respectively;
they indistinguishably overlap.
\label{fig: average_energy_comparison}
}
\end{figure*}
}

\TCBB{
\SUBSECTION{Dependences of Inferred Parameters on the Regularization Models:
the Effects of the Group $L_1$ Regularization} 	
}{
\SUBSECTION{Dependences of inferred parameters on the regularization models:
the effects of the group $L_1$ regularization} 	
}

\FigureInText{

\begin{figure}[hbt]
\centerline{
\includegraphics[width=43mm,angle=0]{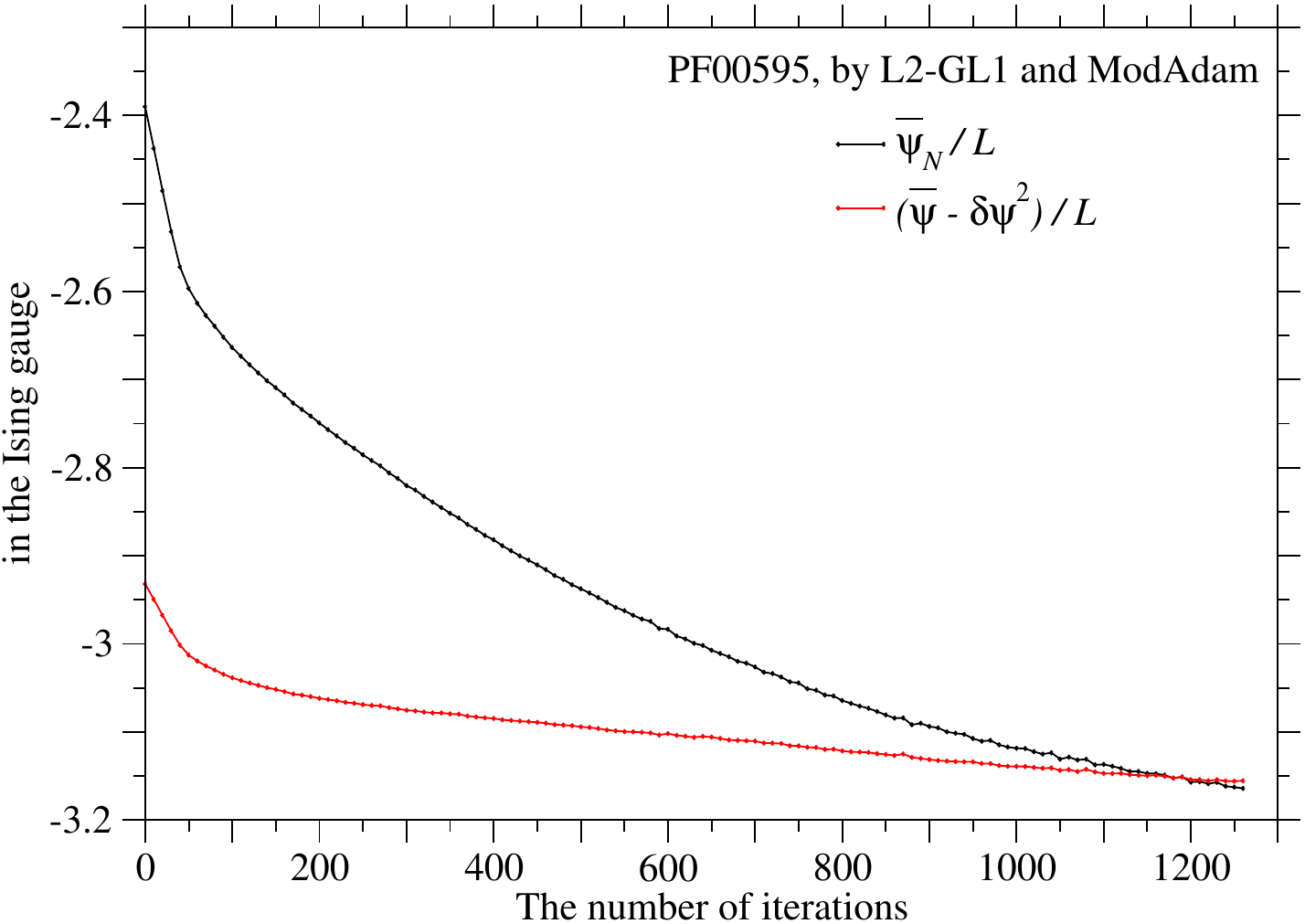}
\includegraphics[width=43mm,angle=0]{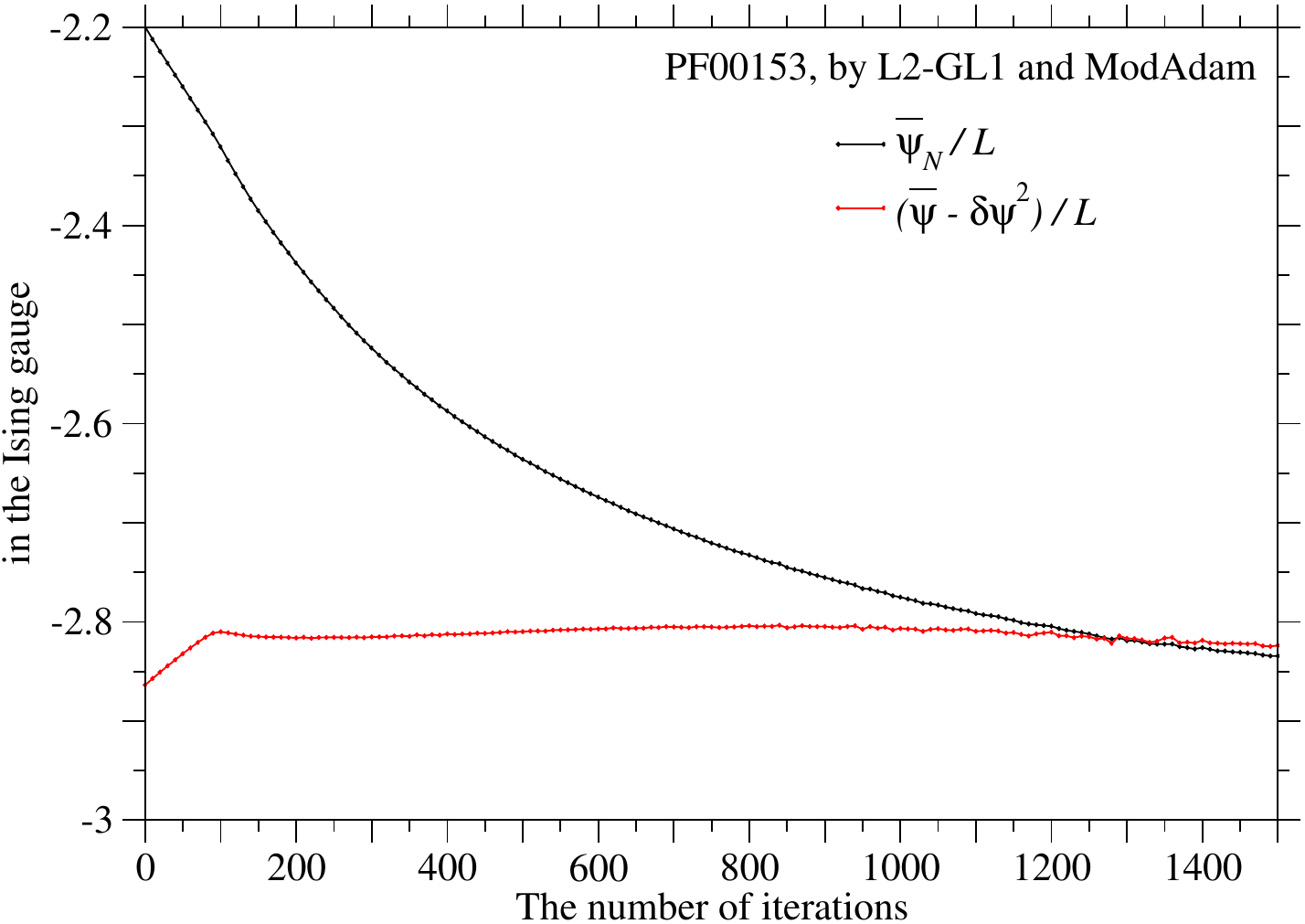}
}
\caption
{
\noindent
\TEXTBF{The profiles of the average evolutionary energies along the leaning process in the L2-GL1 model by the ModAdam
for PF00595 and PF00153.}
The average evolutionary energy per residue $\overline{\psi_N}/L$ of natural sequences
and the ensemble average of evolutionary energy per residue in the Gaussian approximation $(\bar{\psi} - \delta\psi^2)/L$
in the Ising gauge 
are plotted every 10 iterations against iteration number in the learning by the ModAdam;
the sample and ensemble averages are indicated by the upper and lower lines, respectively.
The left and right figures are for PF00595 and PF00153, respectively. 
The L2-GL1 regularization model is employed.
The values of the regularization parameters are listed in 
\Tables{\ref{tbl: PF00595_parameters} and \ref{tbl: PF00153_parameters}}.
\label{fig: average_energy}
\label{fig: aveH}
}
\end{figure}
}

\FigureInText{
\begin{figure*}[hbt]
\centerline{
}
\vspace*{1em}
\centerline{
\includegraphics[width=60mm,angle=0]{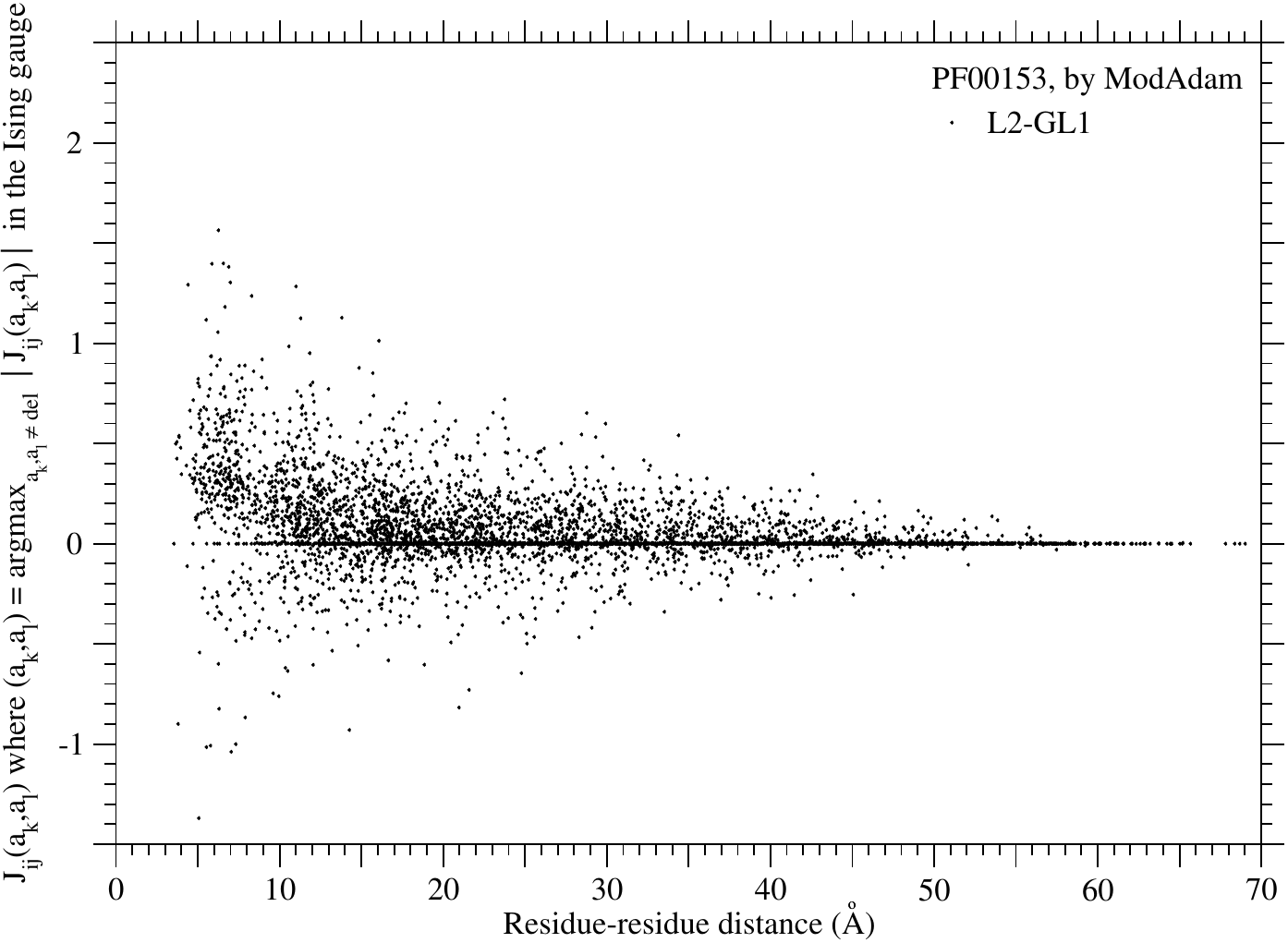}
\includegraphics[width=60mm,angle=0]{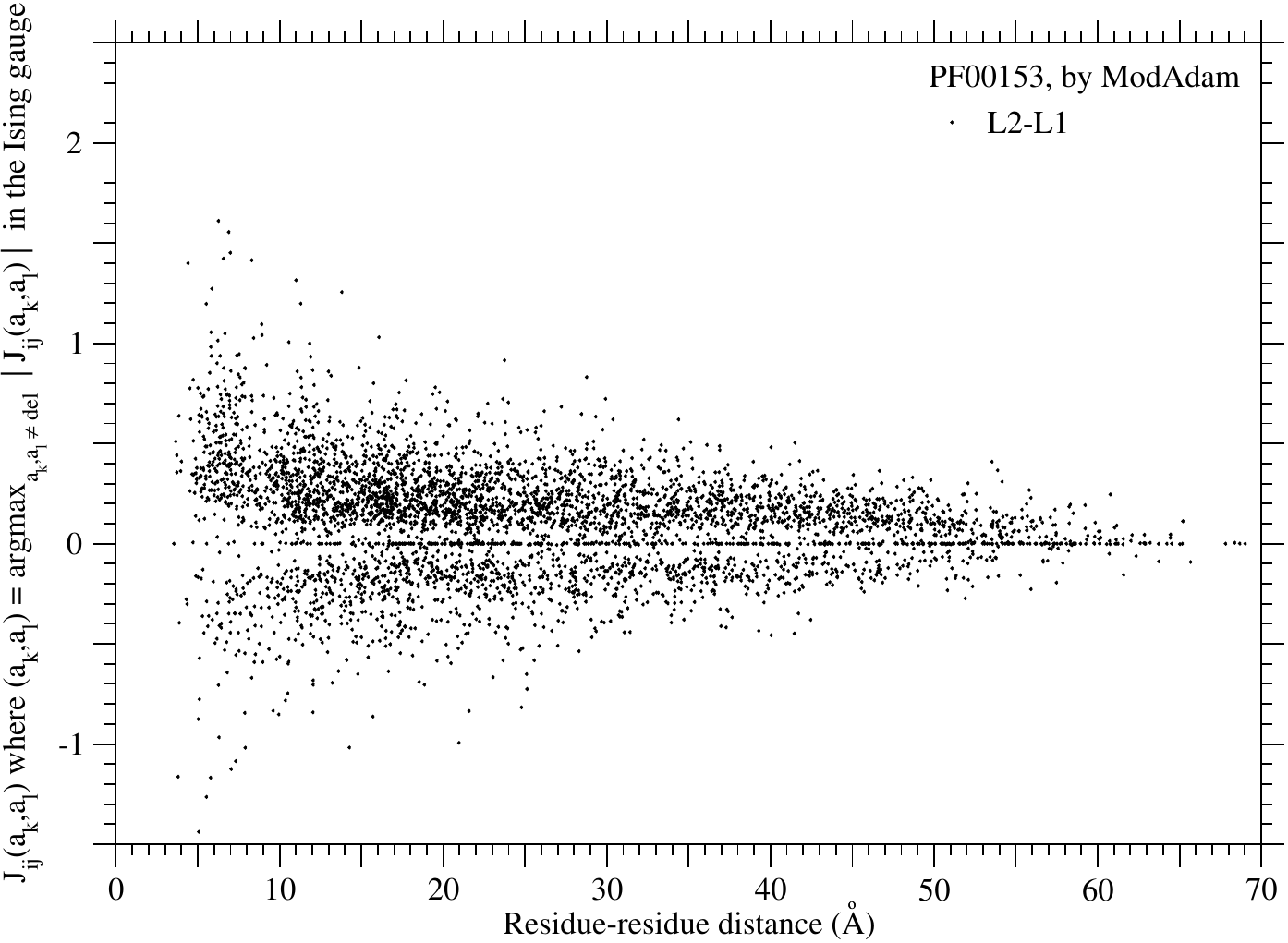}
\includegraphics[width=60mm,angle=0]{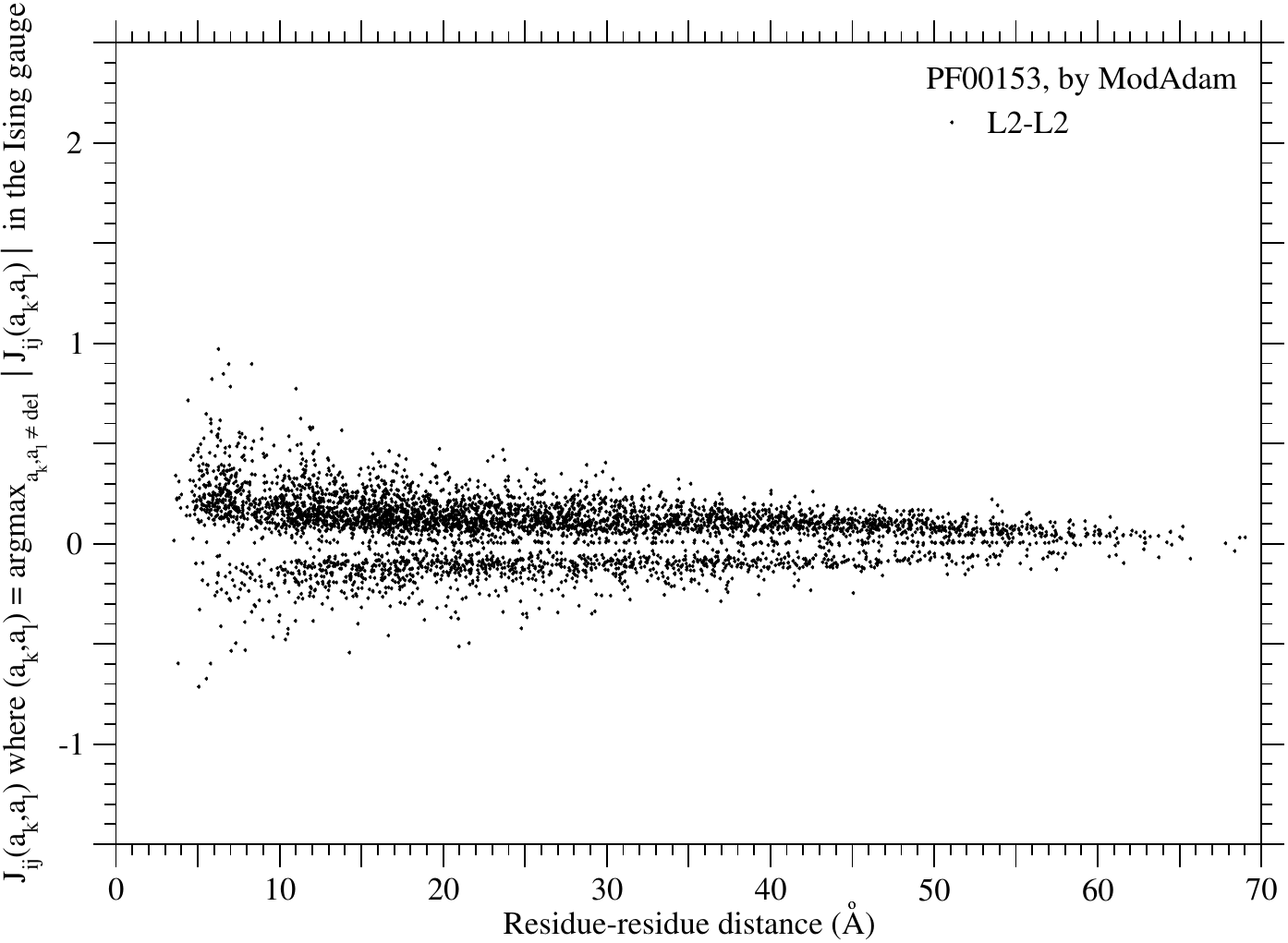}
}
\caption
{
\noindent
\TEXTBF{Differences of inferred couplings $J_{ij}(a_k,a_l)$ among the regularization models for PF00153.}
All $J_{ij}(a_k,b_l)$ where $(a_k,a_l) = \text{argmax}_{a_k,a_l \neq \text{deletion}} | J_{ij}(a_k,a_l) |$
in the Ising gauge are plotted 
against the distance between $i$th and $j$th residues.
The regularization models L2-GL1, L2-L1, and L2-L2 are
employed for the left, middle, and right figures, respectively.
The protein family PF00153 is employed.
The values of regularization parameters are listed in \Table{\ref{tbl: PF00153_parameters}}.
\label{fig: PF00153_maxJij_vs_rpd}
}
\end{figure*}
}

The profiles of the average evolutionary energies of PF00595 and PF00153
in the learning processes with the L2-GL1 regularization 
are shown in \Fig{\ref{fig: aveH}}; the values of
the regularization parameters are listed in \Tables{\ref{tbl: PF00595_parameters} 
and \ref{tbl: PF00153_parameters}}.
It is clear that
both $\overline{\psi_N}$ and $\bar{\psi} - \delta\psi^2$ almost converge.
In \Fig{\ref{fig: PF00153_maxJij_vs_rpd}} and 
\SFigPFvmaxJijvsrpd,
the 
$J_{ij}(a_k,a_l)$, where $(a_k,a_l) = \text{argmax}_{a_k,a_l \neq \text{deletion}} | J_{ij}(a_k,a_l) |$ in the Ising gauge, are plotted
against the distance between $i$th and $j$th residues
for each regularization model.
The negative correlation of coupling interactions on residue-residue distance is clearly
shown in all models.
On the other hand, there is not an energy gap at the value zero
for $J_{ij}(a_k,a_l)$ in the L2-GL1 but in the other models, 
indicating that the group-L1 regularization is effective
to yield sparsity in the couplings.
It should be noticed here that
in the present work the L1 for couplings means the elastic net 
including
a small contribution of $L_2$
in addition to the $L_1$ regularization
to avoid non-unique solutions;
see \Eq{\EQelasticNet}.

The direct comparisons of the inferred fields and couplings
between the L2-GL1 and the other regularization models
are shown in \Fig{\ref{fig: PF00595_hJ_comparison}} for PF00595 
and \Fig{\ref{fig: PF00153_hJ_comparison}} for PF00153;
also see \SFigsPFvPFihJcomparison\ 
for marginal single-site probabilities and pairwise correlations.
There is no significant difference in the inferred fields
between the regularization models.
On the other hand,
weak couplings are differently inferred
between the L2-GL1 and the other models, and 
the differences of their estimates clearly show the typical characteristics of the group-L1 model that
the coupling interactions are estimated sparsely in the L2-GL1 model.
Coupling interactions except for nearly non-interacting residue pairs 
are inferred to be weaker in the L2-L2 than L2-GL1, but
in the L2-L1 similarly to the L2-GL1.
This tendency is more clearly shown for PF00153 consisting of more sequences 
than for PF00595.

In \Fig{\ref{fig: Prec_8A}},
the precisions of contact prediction based on contact score, 
which is defined as 
the corrected Frobenius norm of couplings\CITE{ELLWA:13,EHA:14},
are compared among the regularization models;
residues whose side chain centers are within $8$ \AA\  in a three-dimensional
protein structure are defined here to be in contact, including
neighboring residue pairs along a sequence.
The L2-GL1 model performs better in a whole range of contact ranks than the L2-L1 and L2-L2 models.
Consistently,
the better performance of the L2-GL1 model is more clear for
PF00153 that consists of more effective sequences than PF00595.
The better performance of the L2-GL1 than the others is, however, less than 0.04 in precision for PF00595 and 0.08 for PF00153.

\FigureInText{

\begin{figure}[hbt]
\centerline{
\includegraphics[width=43mm,angle=0]{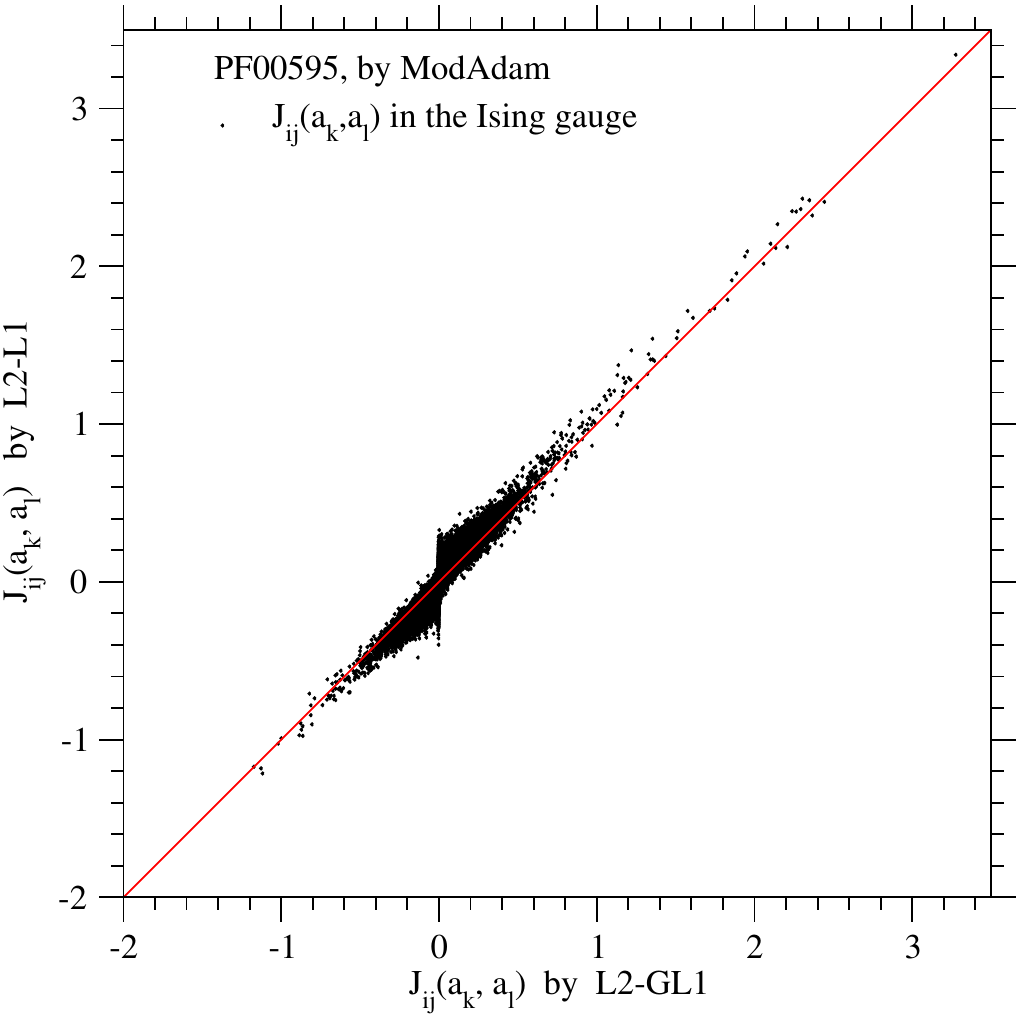}
\includegraphics[width=43mm,angle=0]{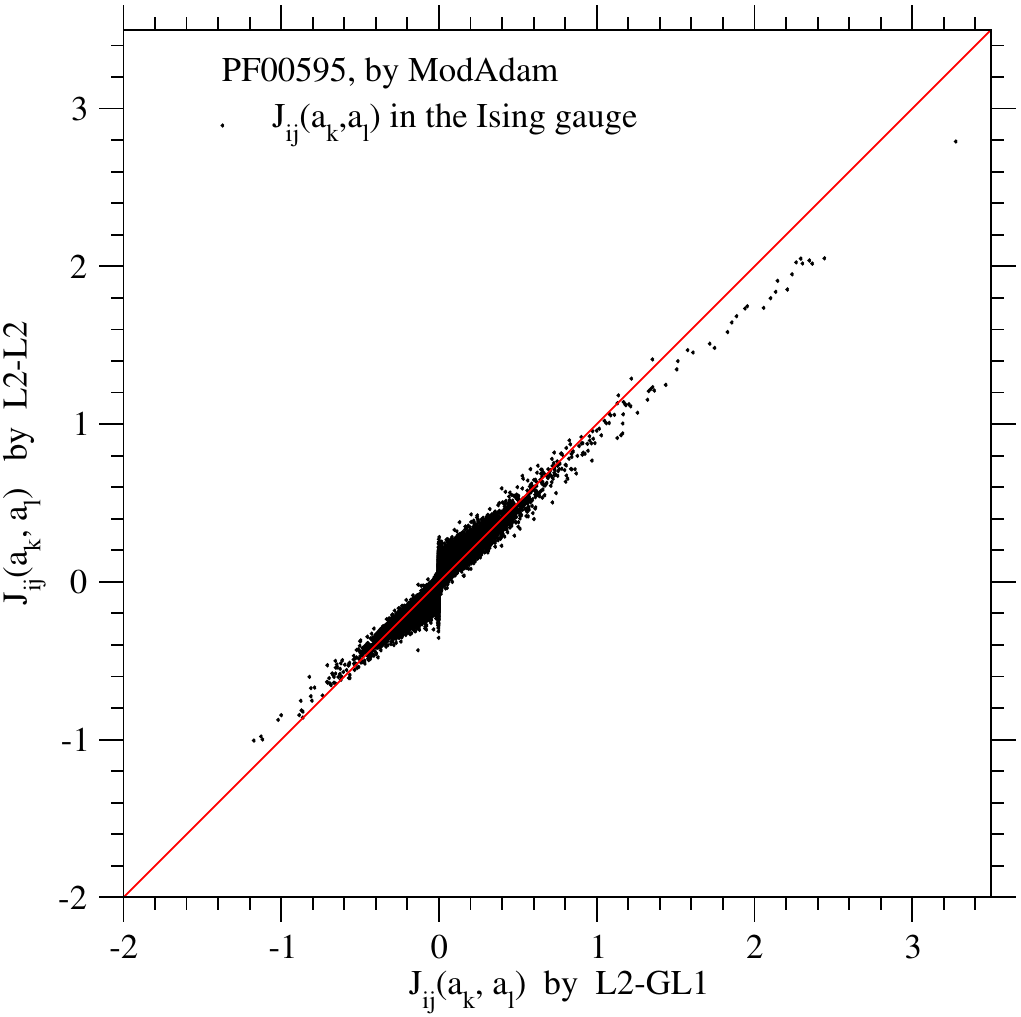}
}
\caption
{
\noindent
\TEXTBF{Comparisons of 
inferred couplings $J_{ij}(a_k,a_l)$ 
in the Ising gauge
between the regularization models for PF00595.}
Both abscissa correspond to 
the
couplings 
inferred by the L2-GL1.
The ordinates in the left and right figures 
correspond to 
the
couplings 
inferred by the L2-L1 and L2-L2 models,
respectively.
The values of regularization parameters are listed in \Table{\ref{tbl: PF00595_parameters}}.
The solid lines show the equal values between the ordinate and abscissa.
The overlapped points of $J_{ij}(a_k,a_l)$ in the units 0.001 are removed.
\label{fig: PF00595_hJ_comparison}
}
\end{figure}
}

\FigureInText{

\begin{figure}[hbt]
\centerline{
\includegraphics[width=43mm,angle=0]{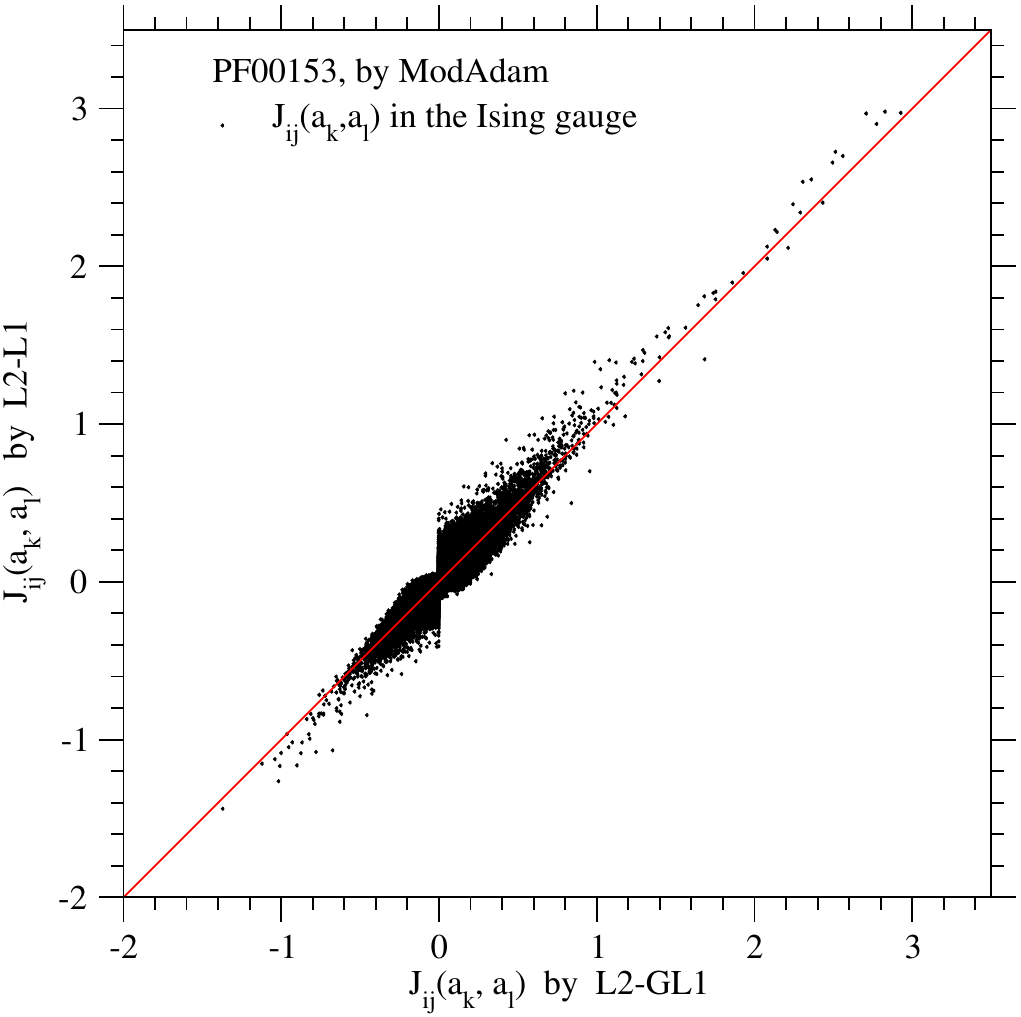}
\includegraphics[width=43mm,angle=0]{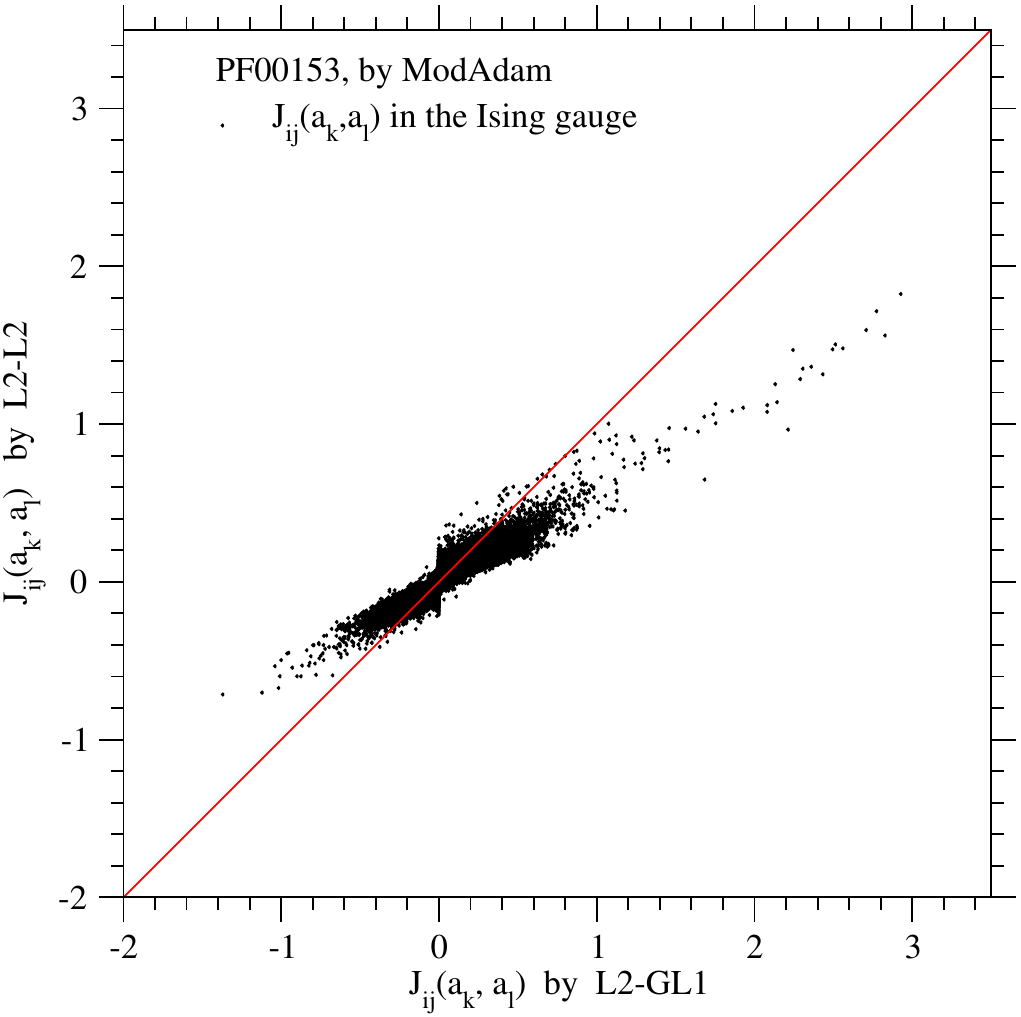}
}
\caption
{
\noindent
\TEXTBF{Comparisons of 
inferred couplings $J_{ij}(a_k,a_l)$ 
in the Ising gauge
between the regularization models for PF00153.}
Both abscissa correspond to 
the
couplings inferred by the L2-GL1.
The ordinates in the left and right figures 
correspond to 
the
couplings inferred by the L2-L1 and L2-L2 models,
respectively.
The values of regularization parameters are listed in \Table{\ref{tbl: PF00153_parameters}}.
The solid lines show the equal values between the ordinate and abscissa.
The overlapped points of $J_{ij}(a_k,a_l)$ in the units 0.001 are removed.
\label{fig: PF00153_hJ_comparison}
}
\end{figure}
}

\FigureInText{

\begin{figure}[hbt]
\centerline{
\includegraphics[width=43mm,angle=0]{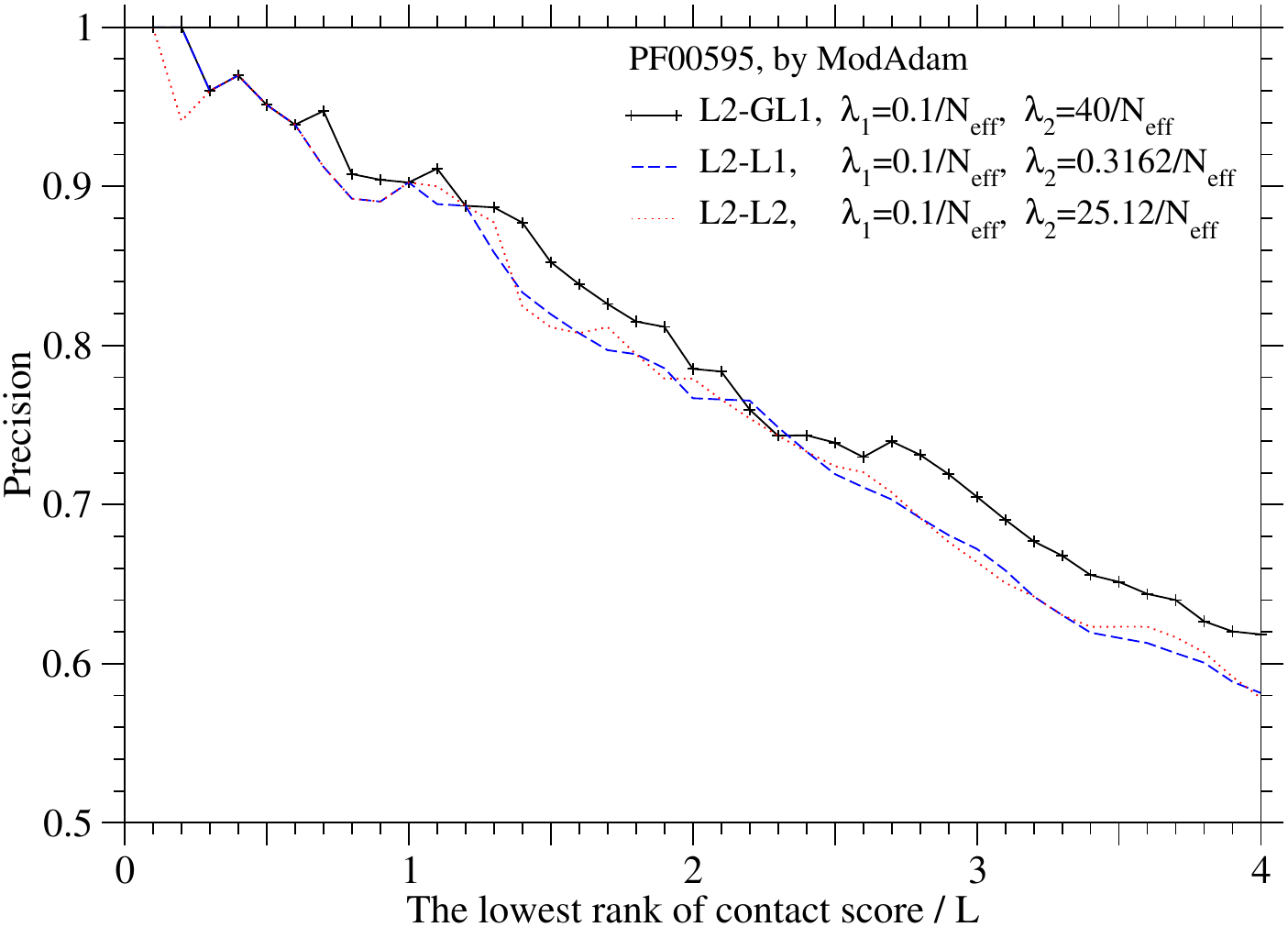}
\includegraphics[width=43mm,angle=0]{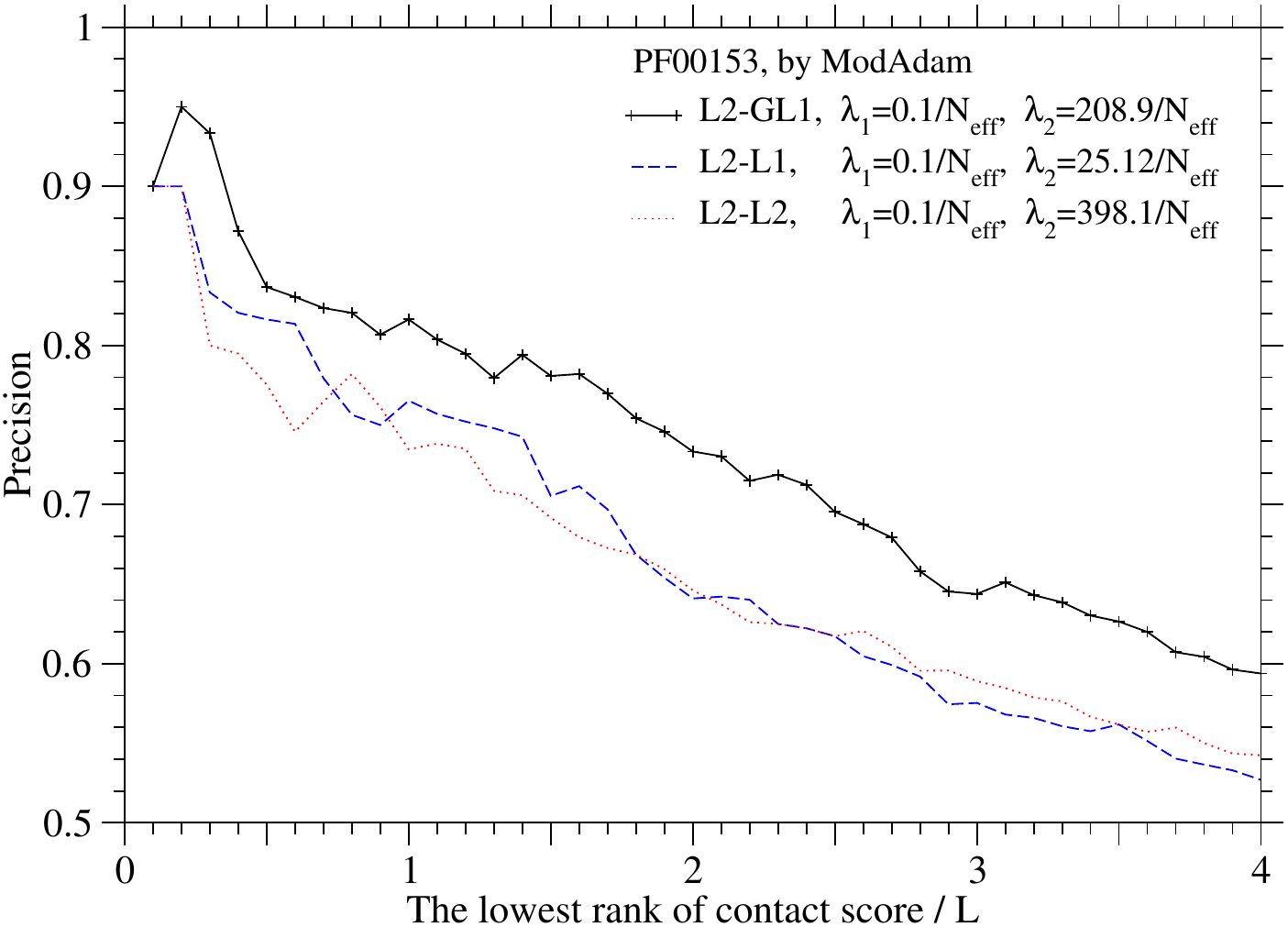}
}
\caption
{
\noindent
\TEXTBF{Precision of each regularization model in the contact predictions for PF00595 and PF00153.}
Precisions of contact predictions
are compared between 
the regularization models.
The ordinate of each figure corresponds to
the precision of contact prediction, in which 
residue pairs are predicted as contacts in the decreasing order of contact score,
and the number of predicted contacts is indicated 
as (the lowest rank of contact score) $/ L$ by the abscissa.
Residues whose side chain centers are within $8$ \AA\  in the 3D protein structure
are defined to be in contact; 
neighboring residue pairs along the sequence are included.
The left and right figures are for the protein families PF00595 and PF00153,
respectively.
The solid, broken, and dotted lines correspond to the regularization models,
L2-GL1, L2-L1, and L2-L2, respectively.
The corrected Frobenius norm of couplings
is employed for the contact score\CITE{ELLWA:13,EHA:14}.
\label{fig: Prec_8A}
}
\end{figure}
}

\TCBB{
\SUBSECTION{Recoverabilities of Single-Site Frequencies and Pairwise Correlations in the L2-GL1 Regularization Model}
}{
\SUBSECTION{Recoverabilities of single-site frequencies and pairwise correlations in the L2-GL1 regularization model}
}

\TCBB{
\FigureInText{

\begin{figure}[hbt]
\centerline{
\includegraphics[width=43mm,angle=0]{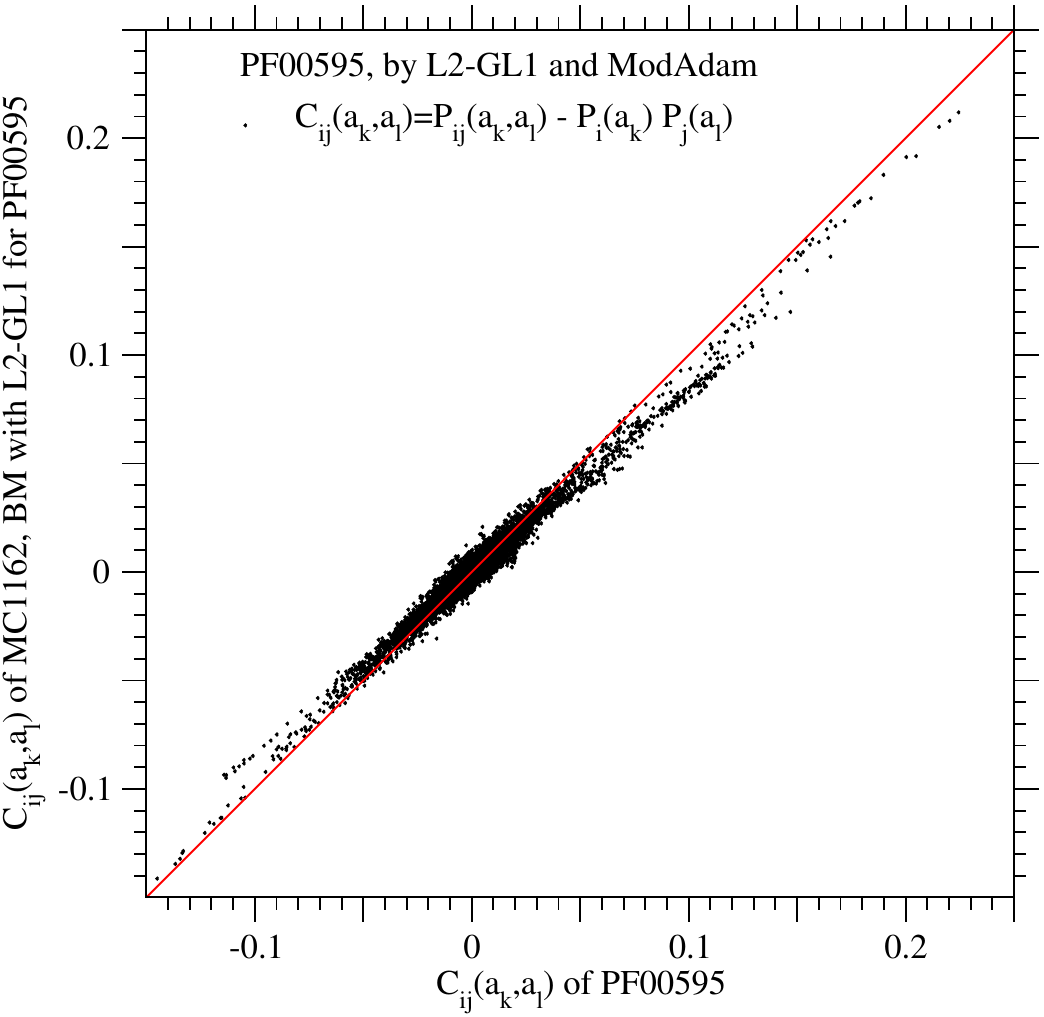}
\includegraphics[width=43mm,angle=0]{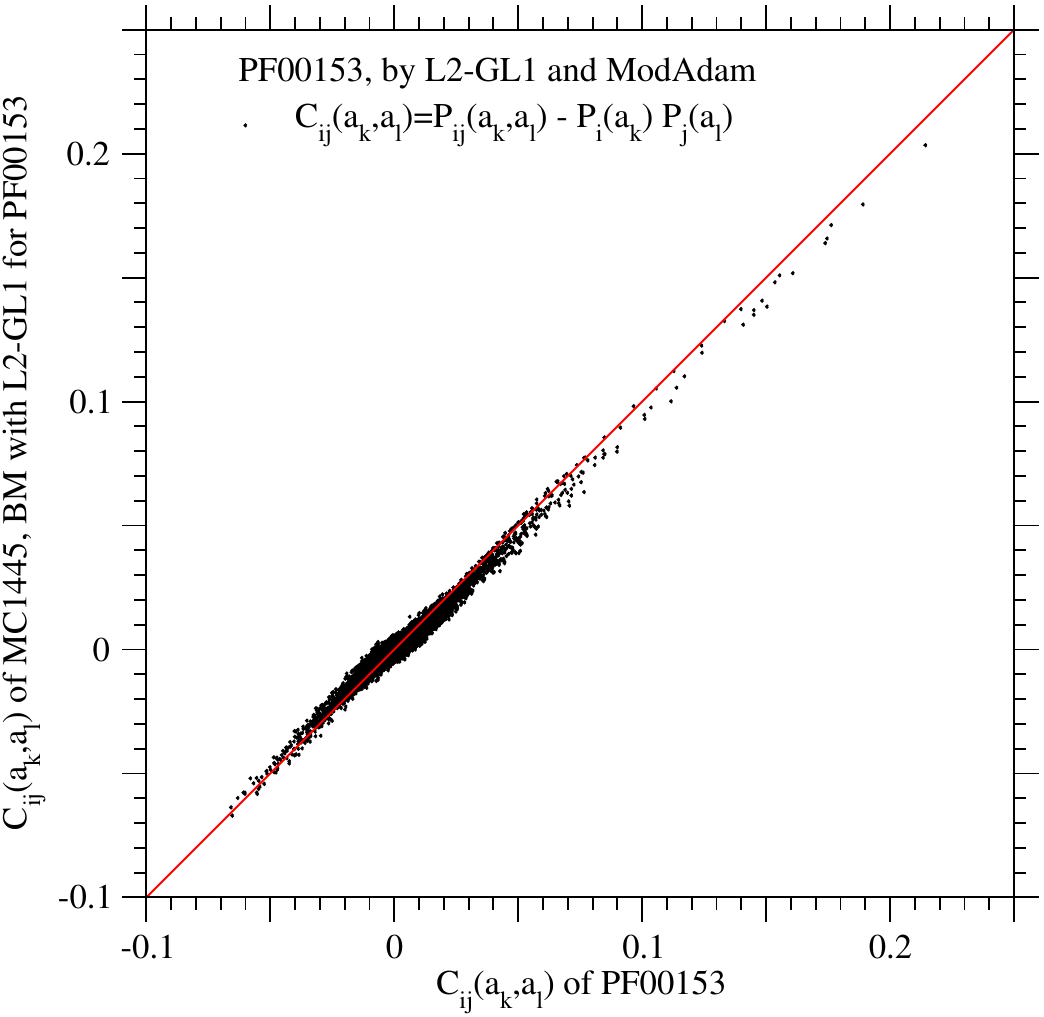}
}
\caption
{
\noindent
\TEXTBF{Recoverabilities of the 
pairwise correlations
of PF00595 and PF00153
by the Boltzmann machine learning 
with the L2-GL1 model and the ModAdam method.}
The left and right figures are for 
PF00595 and PF00153,
respectively;
$D_2^{KL} = 0.0759$ for PF00595 and $D_2^{KL} = 0.0318$ for PF00153.  
The solid lines show the equal values between the ordinate and abscissa.
The overlapped points of $C_{ij}(a_k,a_l)$ in the units 0.0001 are removed.
See \Tables{\ref{tbl: PF00595_parameters} and \ref{tbl: PF00153_parameters}} for the regularization parameters employed.
\label{fig: PF00595_PF00153_PiaCijab}
\label{fig: PF00595_PF00153_PiaPijab}
}
\end{figure}
}
}{
}

Recoverabilities of single-site frequencies and pairwise correlations in the L2-GL1 model
are shown 
\TCBB{
in \Fig{\ref{fig: PF00595_PF00153_PiaCijab}} 
and in \SFigsPFvPFiPiaCijab,
}{
in \Fig{\ref{fig: PF00595_PiaPijab}} for PF00595 and
in \Fig{\ref{fig: PF00153_PiaPijab}} for PF00153,
}
and are also indicated by
the values of $D_1^{KL}$ and $D_2^{KL}$ listed 
in \Tables{\ref{tbl: PF00595_parameters} and \ref{tbl: PF00153_parameters} }.
Although single-site and pairwise amino acid probabilities are well recovered in both protein families,
they are better recovered as expected in PF00153 consisting of more effective number of sequences than PF00595;
Both $D_1^{KL} = 0.00112$ and $D_2^{KL} = 0.0318$ for PF00153
are less than half of $D_1^{KL} = 0.00369$ and $D_2^{KL} = 0.0759$ for PF00595, 
even though $\lambda_2$ is larger for PF00153 than PF00595.
It should be noticed, however, that
the correlations, $C_{ij}(a_k,a_l) \equiv P_{ij}(a_k,a_l) - P_{i}(a_k)P_{j}(a_l)$,
are under-reproduced with this set of the regularization parameters 
for strongly correlated site pairs in both proteins.

\TCBB{
\SUBSECTION{Reproducibilities of Fields and Couplings in the L2-GL1 Model}
}{
\SUBSECTION{Reproducibilities of fields and couplings in the L2-GL1 model}
}

\TCBB{
\FigureInText{

\begin{figure}[hbt]
\centerline{
\includegraphics[width=43mm,angle=0]{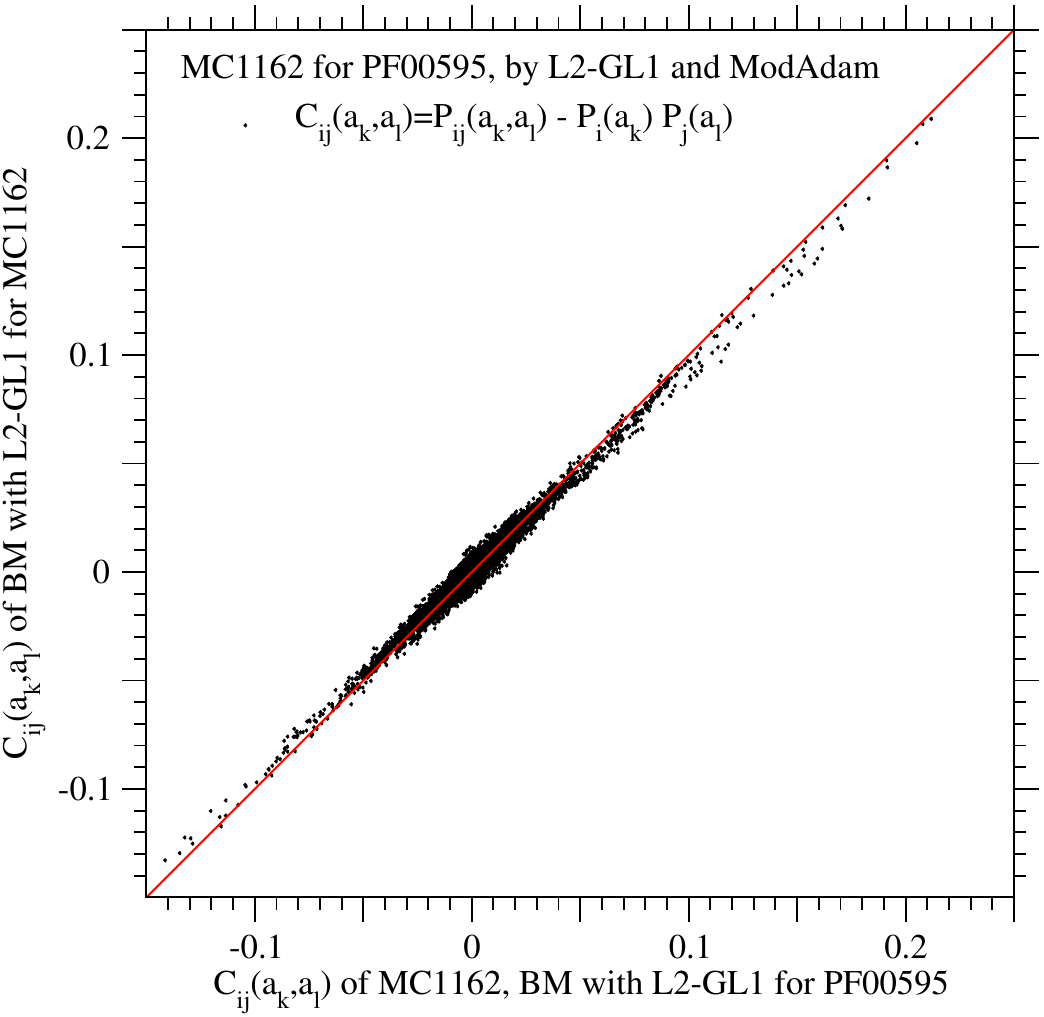}
\includegraphics[width=43mm,angle=0]{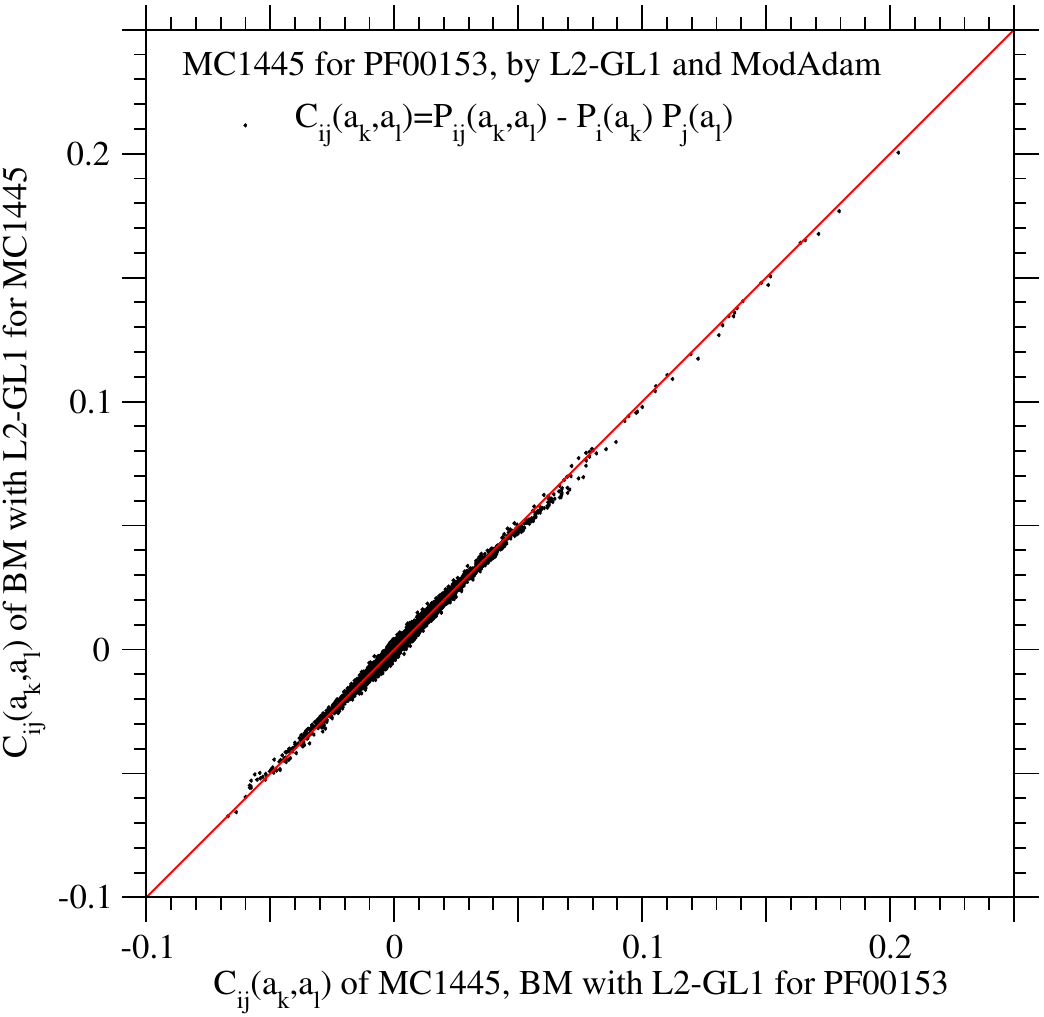}
}
\caption
{
\noindent
\TEXTBF{Recoverabilities of 
the pairwise correlations
by the Boltzmann machine learning
with the L2-GL1 model and the ModAdam method
for the protein-like sequences,
the MCMC
samples that are obtained by
the same Boltzmann machine
for PF00595 and PF00153.
}
The 
MCMC
samples obtained by the Boltzmann machine learning
with the L2-GL1 model and the ModAdam method 
for 
PF00595 and PF00153
are employed as protein-like sequences for which the Boltzmann machine learning 
with the same model and method is executed again in order to examine
how precisely the marginals of protein-like sequences
can be recovered.
The marginals recovered by the Boltzmann machine learning 
for the 
MCMC
samples
are compared to those of the 
MCMC
samples. 
The left and right figures are for 
the single-site probabilities and pairwise correlations,
respectively.
The solid lines show the equal values between the ordinate and abscissa.
The overlapped points of $C_{ij}(a_k,a_l)$ in the units 0.0001 are removed.
See \Tables{\ref{tbl: PF00595_parameters} and \ref{tbl: PF00153_parameters} } for the regularization parameters employed.
\label{fig: PF00595MC_PF00153MC_Cijab}
}
\end{figure}
}
}{
}

\FigureInText{

\begin{figure}[hbt]
\centerline{
\includegraphics[width=43mm,angle=0]{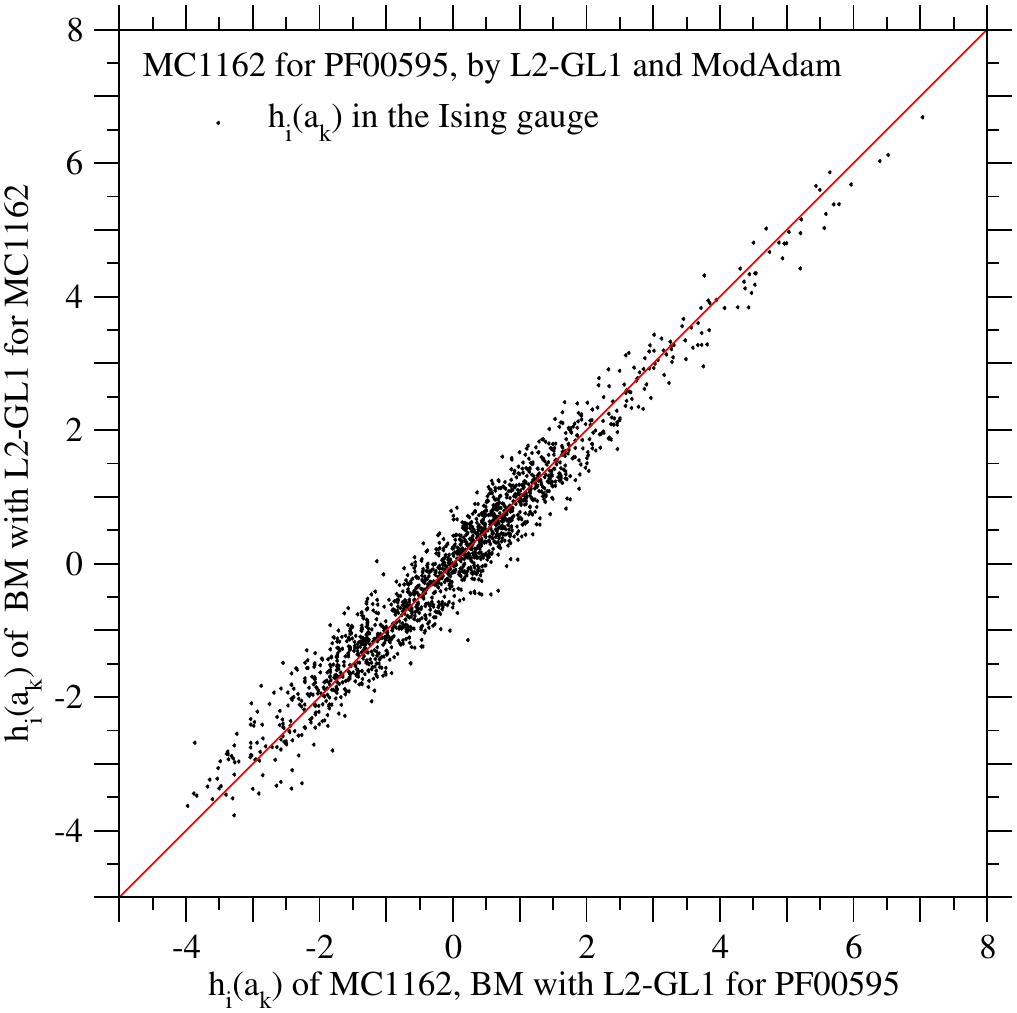}
\includegraphics[width=43mm,angle=0]{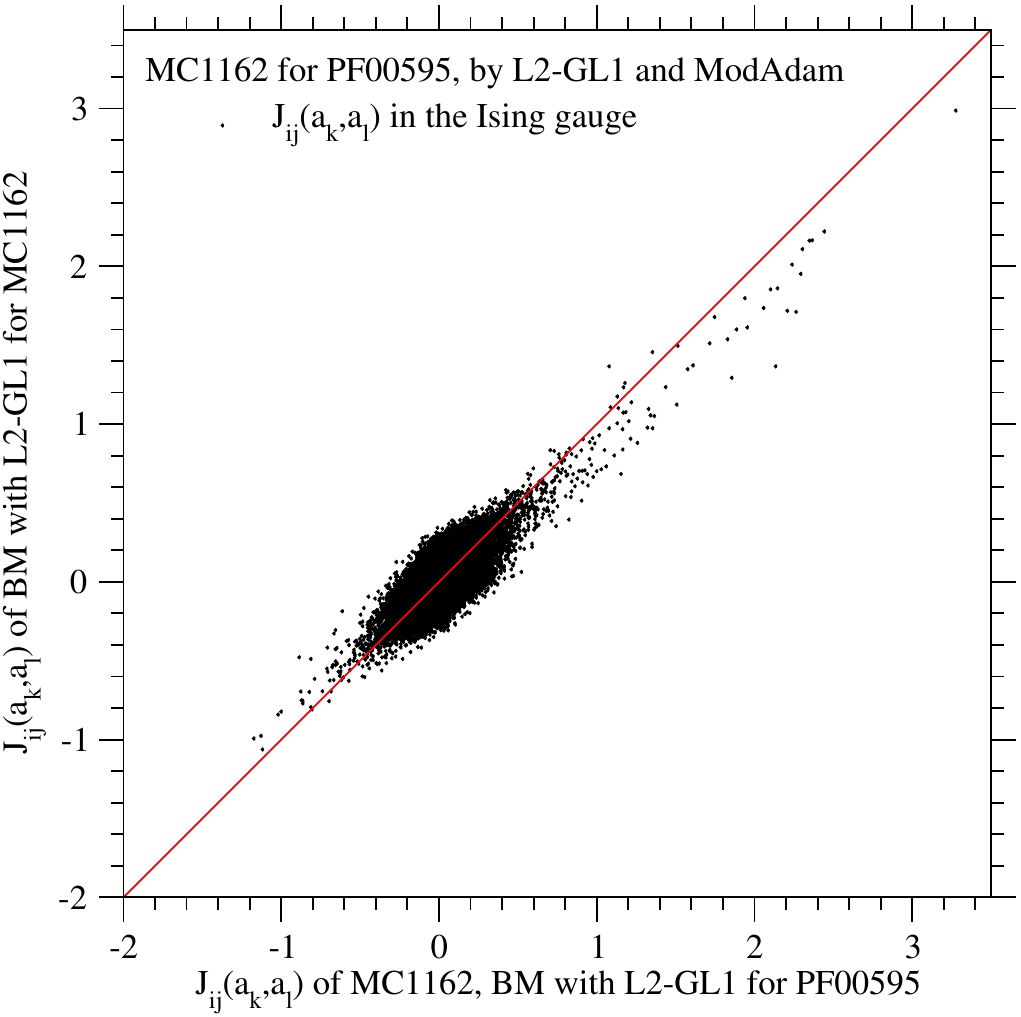}
}
\caption
{
\noindent
\TEXTBF{
Reproducibility of the fields and couplings
in the Ising gauge
by the Boltzmann machine learning
with the L2-GL1 model and the ModAdam method 
for the protein-like sequences,
the MCMC
samples that are obtained by
the same Boltzmann machine for PF00595.}
The 
MCMC
samples obtained by the Boltzmann machine learning
with the L2-GL1 model and the ModAdam method for PF00595
are employed as protein-like sequences
for which the Boltzmann machine learning
with the same model and method is executed again in order to examine 
how well the fields and couplings in the protein-like sequences 
can be reproduced.
The fields and couplings inferred by the Boltzmann machine learning for the 
MCMC
samples are plotted against the actual values of their interactions
in the left and right figures, respectively.
The solid lines show the equal values between the ordinate and abscissa.
The overlapped points of $J_{ij}(a_k,a_l)$ in the units 0.001 are removed.
See \Table{\ref{tbl: PF00595_parameters}} for the regularization parameters employed.
\label{fig: PF00595MC_hJ}
}
\end{figure}
}

Now let us consider how well the Boltzmann machine can infer fields and couplings from protein sequence data.
Reproducibilities of fields and couplings have been examined 
for artificial interactions on a lattice protein and others\CITE{BLCC:16}.
Here, 
MCMC 
samples that are generated 
with the fields and couplings inferred by the Boltzmann machine
for the protein families
are employed as protein-like sequences for which
the Boltzmann machine learning with the same
regularization model and gradient-descent method 
is executed again in order to examine how well the Boltzmann machine
can reproduce the fields and couplings in protein-like sequences.

MC1162 listed in \Table{\ref{tbl: PF00595_parameters}} and 
MC1445 in \Table{\ref{tbl: PF00153_parameters}} are
MCMC
samples generated with the fields and couplings inferred for PF00595
and for PF00153, respectively.
First,
the regularization parameters optimized in the first stage
have been employed for the Boltzmann machine in the second stage.
Ideally, the condition, $(\bar{\psi} - \delta\psi^2) \sim \overline{\psi_N}$,
should be satisfied with this set of regularization parameters.
In the interactions inferred in the first stage,
however, 
the average evolutionary energies of 
MCMC
samples
are higher 
than those of the natural protein families.
As a result, 
$\langle \psi_N(\VECS{\sigma}) \rangle_{\VECS{\sigma}} < \overline{\psi_N}$ 
is obtained 
with the same regularization parameters in the second stage, 

The sample averages of the following evolutionary energies
take similar values in both set of interactions inferred 
in the first and second stages except $\overline{\psi_{\text{PF00595}}}$;
$\overline{\psi_{\text{PF00595}}} = -3.15 < -2.82$, 
$\overline{\psi_{\text{MC1162}}} = -2.79 \sim -2.80$, 
$\overline{\psi_{\text{PF00153}}} = -2.83 \sim -2.83$, 
and $\overline{\psi_{\text{MC1445}}} = -2.54 \sim -2.54$
in the interactions inferred in the first and second stages,
respectively, and 
$\langle \psi_N(\VECS{\sigma}) \rangle_{\VECS{\sigma}} \sim \overline{\psi_{\text{PF00595}}} \leq \overline{\psi_{\text{MC1162}}}$
and
$\langle \psi_N(\VECS{\sigma}) \rangle_{\VECS{\sigma}} \sim \overline{\psi_{\text{PF00153}}} \leq \overline{\psi_{\text{MC1445}}}$
are satisfied in both the interactions.
This fact indicates that 
the interactions inferred in the second stage also favor the natural protein families,
and in this aspect those inferred in the first stage are well recovered in the second stages, 
although the optimum condition, 
$\langle \psi_N(\VECS{\sigma}) \rangle_{\VECS{\sigma}} \sim \overline{\psi_N}$, 
is not satisfied.
However, let us tune the regularization parameters without a priori knowledge 
according to the procedure described in the preceding section.
Regularization parameters optimized in the second stage are listed in
\Table{\ref{tbl: PF00595_parameters} and \ref{tbl: PF00153_parameters}}.

The single-site frequencies and pairwise correlations recovered in the second stage are compared with those in the first stage and also with
the natural proteins 
\TCBB{
in \Fig{\ref{fig: PF00595MC_PF00153MC_Cijab}}
and in \SFigsPFvMCPFiMCPiaCijab.
}{
in \Figs{\ref{fig: PF00595MC_PiaCijab} and \ref{fig: PF00153MC_PiaCijab}}
for PF00595 and PF00153, respectively; see also
\SFigsPFvMCPFiMCPiaCijab. 
}
The single-site frequencies and pairwise correlations are extremely well recovered 
in the second stage for PF00153 
consisting of more sequences than PF00595.
Pairwise correlations are slightly under-reproduced for strongly correlated site pairs 
of PF00595 as well as in the first stage.
The smaller values of $\lambda_2$ in the second stage than in the first stage cause 
single-site frequencies and pairwise correlations to be better recovered
and also smaller $D_1^{KL}$ and $D_2^{KL}$ in the second stage than in the first stage.

In \Figs{\ref{fig: PF00595MC_hJ} and \ref{fig: PF00153MC_hJ}},
the fields and couplings inferred in the second stage are compared with true ones, that is, 
those inferred in the first stage.
Both are well recovered, although strong couplings are always underestimated in both proteins.
Large errors may be included in the estimates for weak couplings of $|J_{ij}(a_k,a_l)| < -0.6$.
Noise being included in the couplings is expected to increase as
$\lambda_2$ decreases in the second stage.

\TCBB{
}{
}

\FigureInText{

\begin{figure}[hbt]
\centerline{
\includegraphics[width=43mm,angle=0]{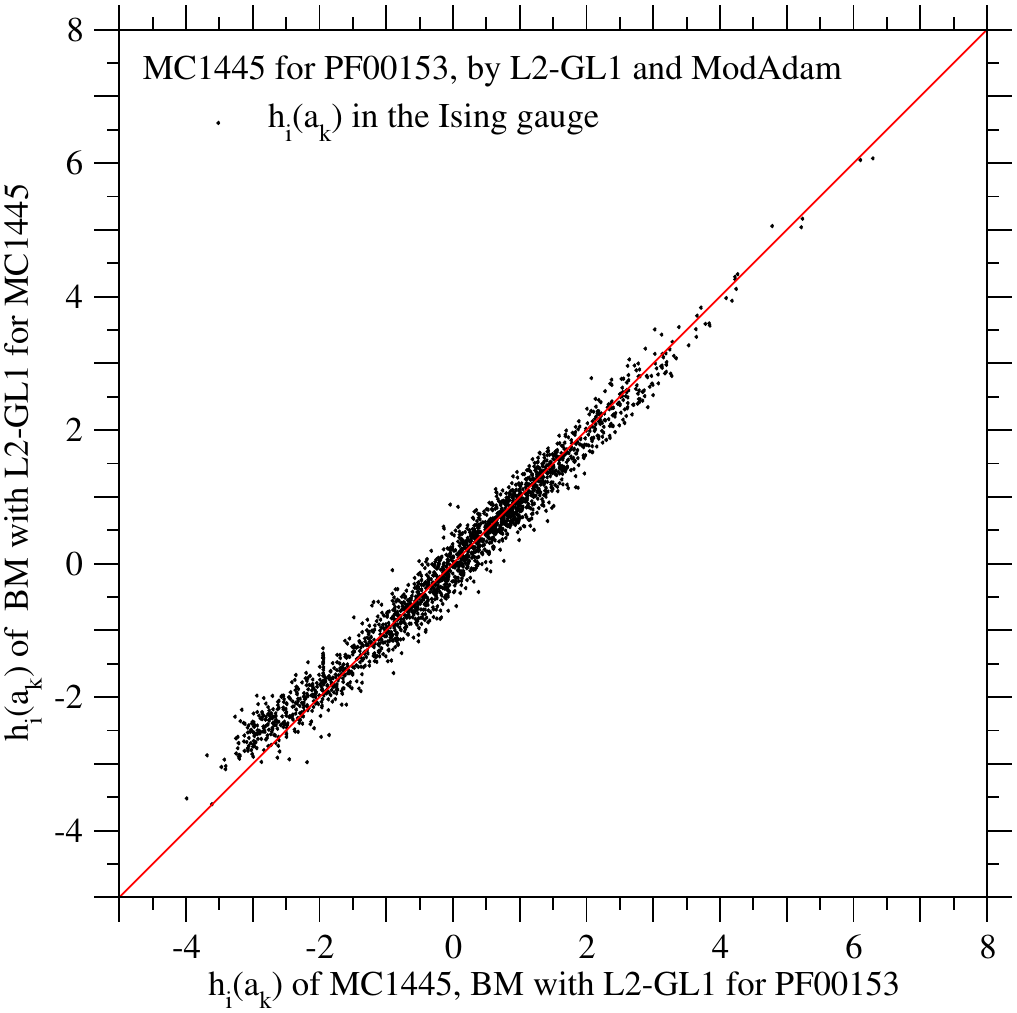}
\includegraphics[width=43mm,angle=0]{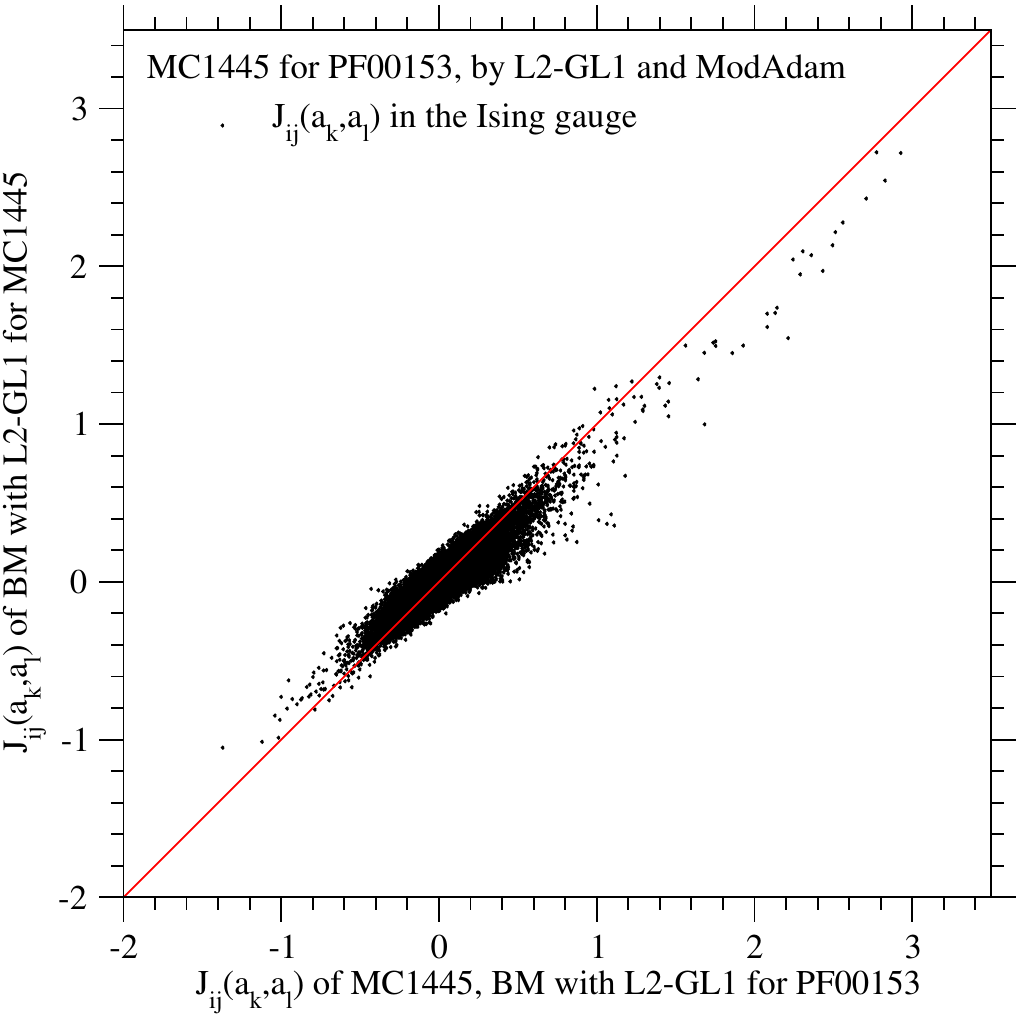}
}
\caption
{
\noindent
\TEXTBF{
Reproducibility of the fields and couplings
in the Ising gauge
by the Boltzmann machine learning
with the L2-GL1 model and the ModAdam method
for the protein-like sequences, 
the MCMC
samples that are obtained by
the same Boltzmann machine for PF00153.
}
The 
MCMC
samples obtained by the Boltzmann machine learning
with the L2-GL1 model and the ModAdam method for PF00153
are employed as protein-like sequences
for which the Boltzmann machine learning
with the same model and method is executed again in order to examine
how well the fields and couplings in the protein-like sequences
can be reproduced.
The fields and couplings inferred by the Boltzmann machine learning for the 
MCMC
samples are plotted against the actual values of their interactions
in the left and right figures, respectively.
The solid lines show the equal values between the ordinate and abscissa.
The overlapped points of $J_{ij}(a_k,a_l)$ in the units 0.001 are removed.
See \Table{\ref{tbl: PF00153_parameters}} for the regularization parameters employed.
\label{fig: PF00153MC_hJ}
}
\end{figure}
}

\TCBB{
\SUBSECTION{Reproducibility of the Evolutionary Energy Distribution of Natural Proteins 
by 
MCMC
Samples in the L2-GL1 Model}
}{
\SUBSECTION{Reproducibility of the evolutionary energy distribution of natural proteins 
by 
MCMC
samples in the L2-GL1 model}
}

\FigureInText{

\begin{figure}[h!]
\centerline{
\includegraphics[width=43mm,angle=0]{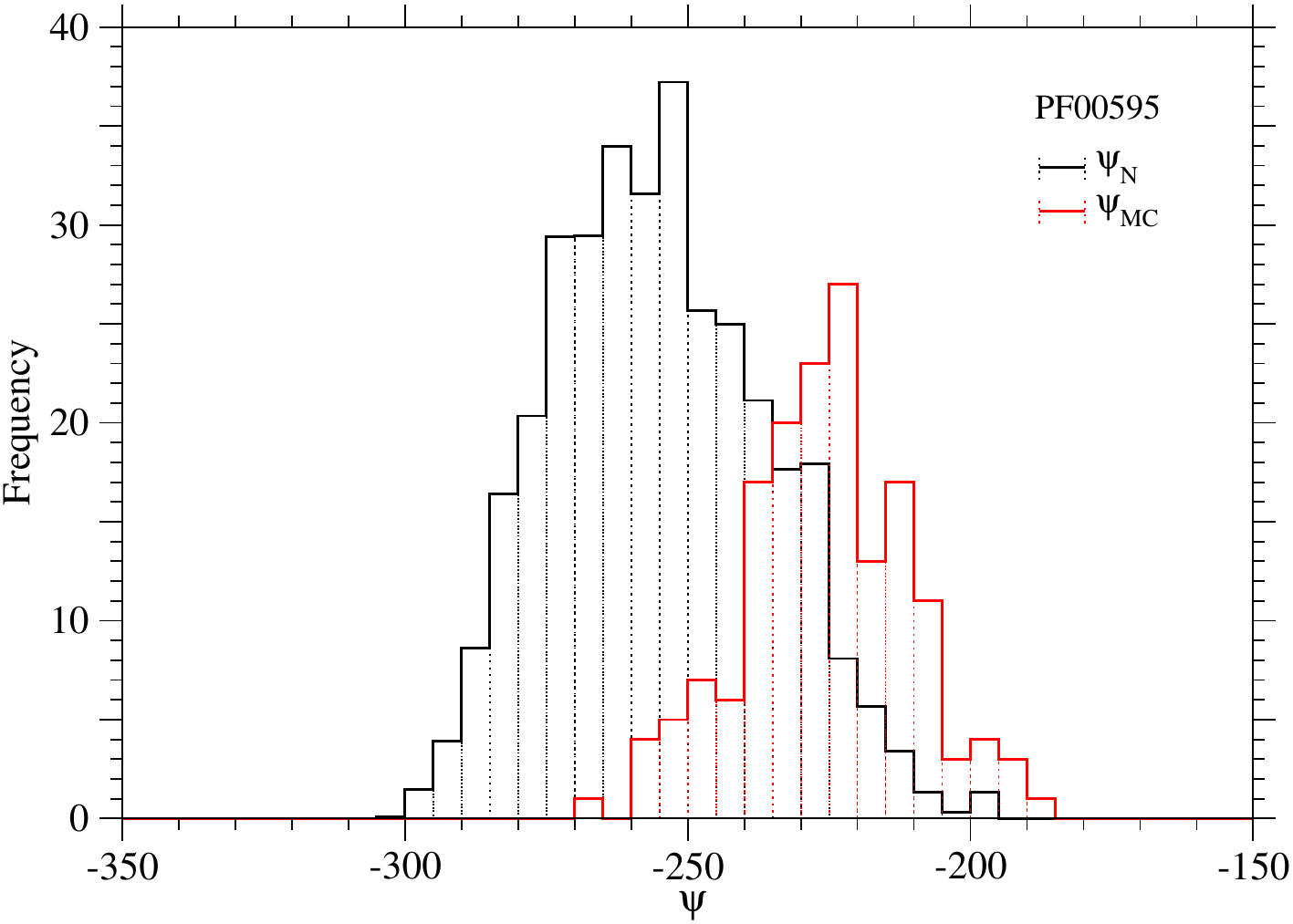}
\includegraphics[width=43mm,angle=0]{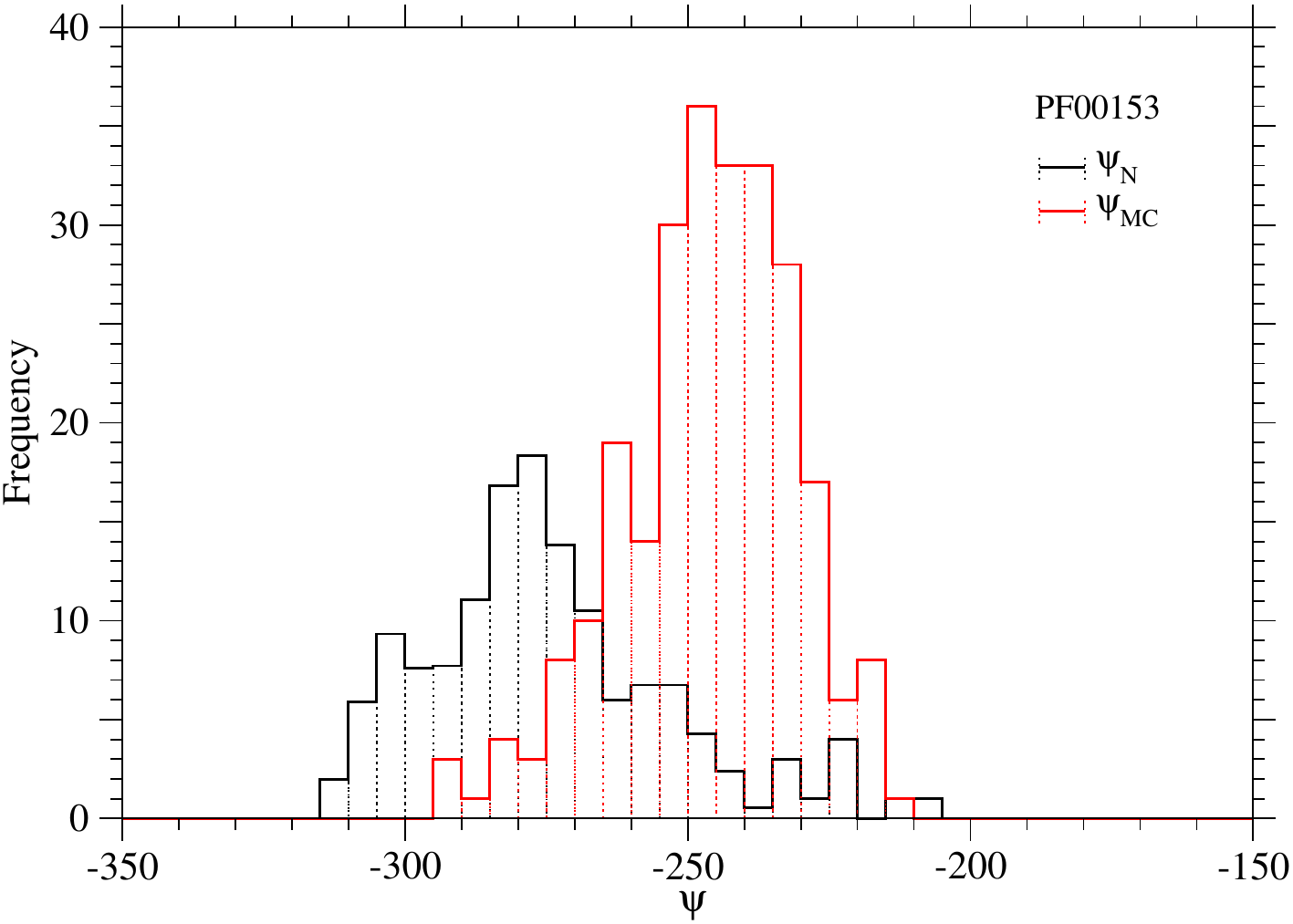}
}
\vspace*{1em}
\centerline{
\includegraphics[width=43mm,angle=0]{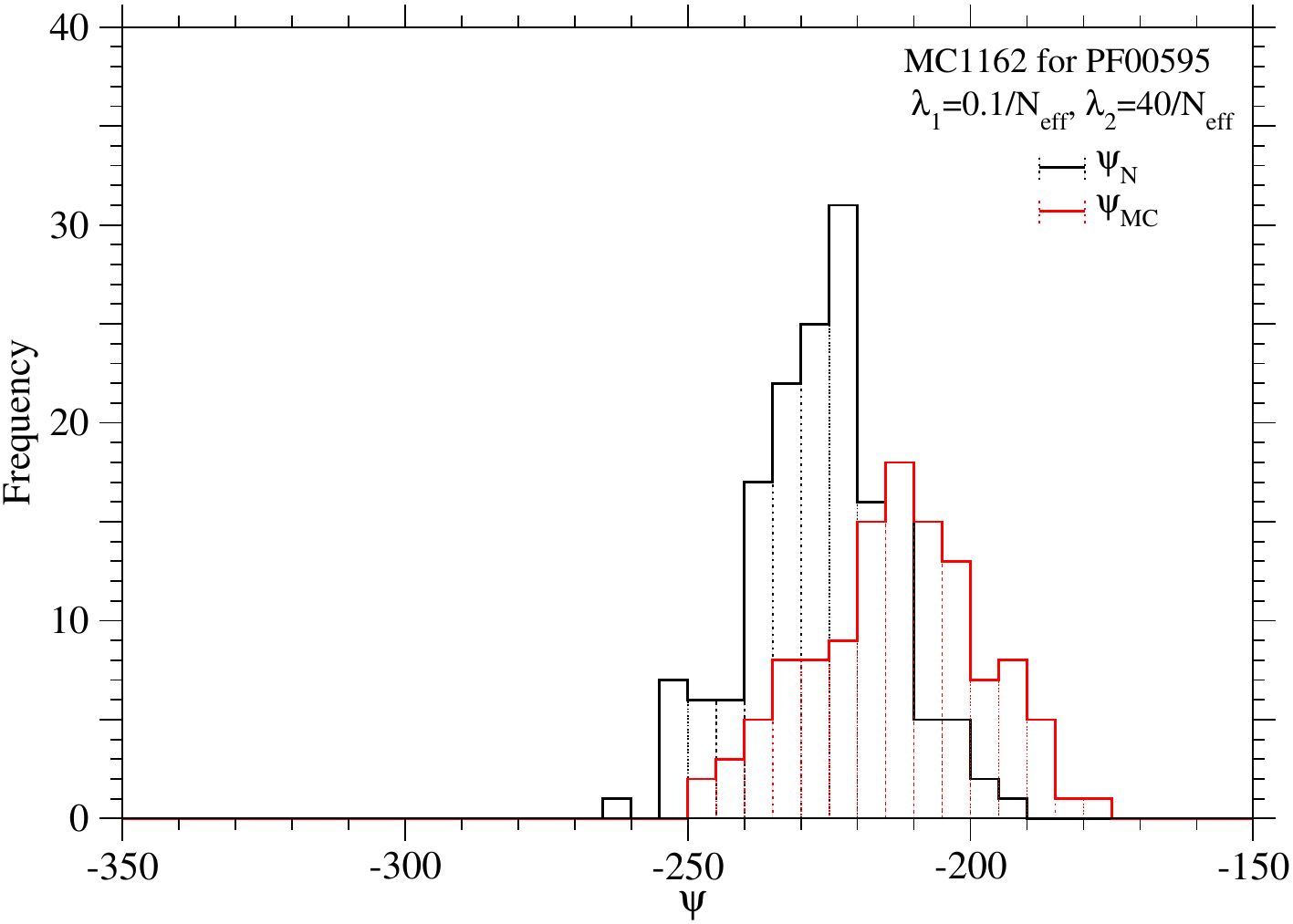}
\includegraphics[width=43mm,angle=0]{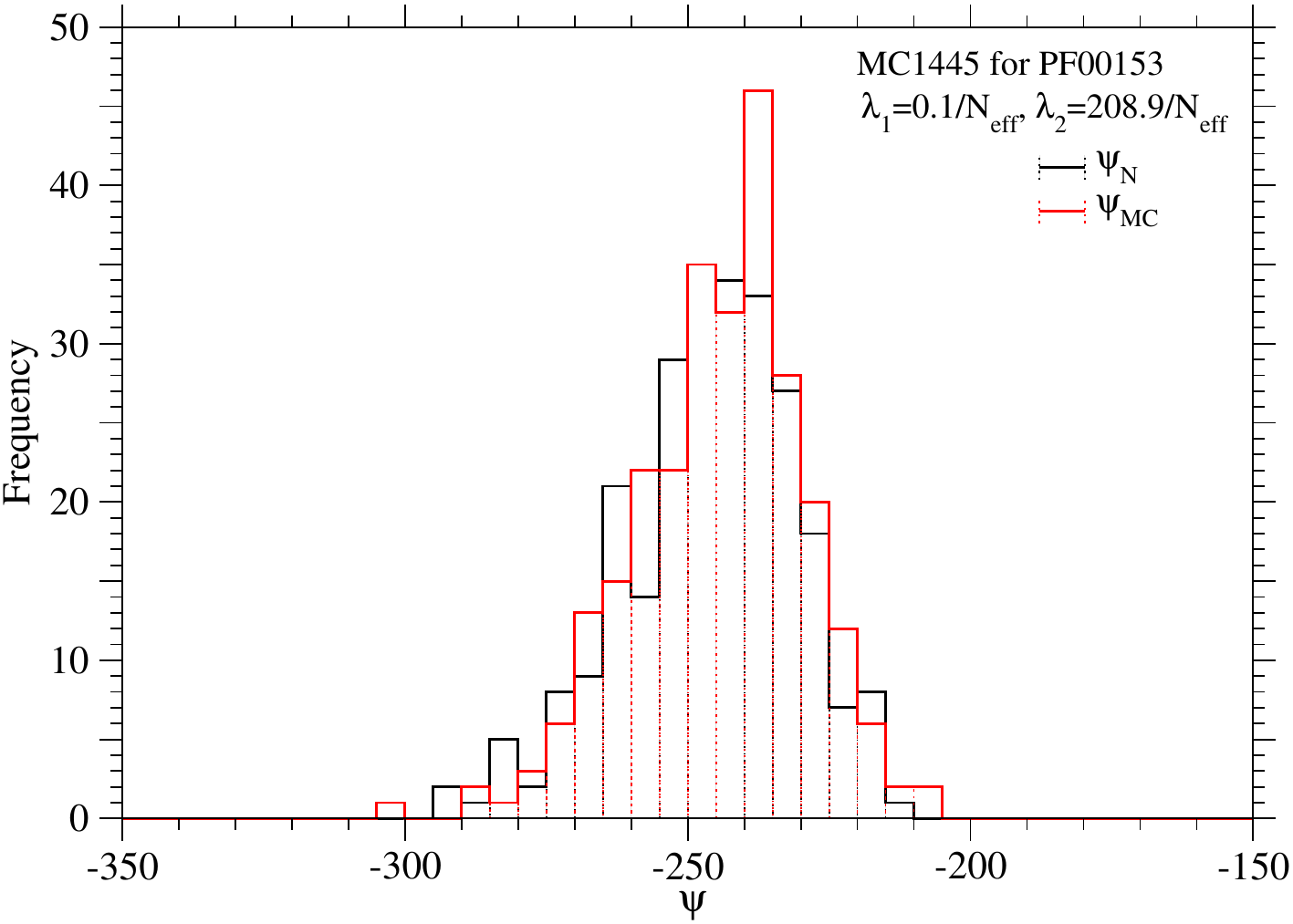}
}
\caption
{
\noindent
\TEXTBF{
Comparisons of the histograms of 
inferred evolutionary energies 
in the Ising gauge
between 
the target sequences ($\psi_N$) and the 
MCMC
samples ($\psi_{MC}$) obtained by the Boltzmann machine learnings.
}
In the upper left and right figures, 
the evolutionary energies ($\psi_N$) of the natural proteins
are compared with 
those ($\psi_{MC}$) of the 
MCMC
samples obtained by 
the Boltzmann machine learnings for PF00595 and PF00153,
respectively.
Sequences with no deletion
for PF00595 and with no more than 2 deletions for PF00153 
are employed;
the effective numbers $M_{\text{eff}}$ of 
sequences are
$340.0$ for PF00595, $139.8$ for PF00153.
The $162$ samples with no deletion in MC1162 and 
$254$ samples with no more than 3 deletions
in MC1445 are employed.
In the lower left and right figures,
the evolutionary energies ($\psi_N$) of 
the protein-like sequences, MC1162 and of MC1445,
are compared with those ($\psi_{MC}$) of the 
MCMC
samples obtained by the Boltzmann machine learnings 
for them. 
The same regularization parameters
as for the natural protein families are employed;
$\lambda_2 = 40.0 / N_{\text{eff}}$ for PF00595
and  
$\lambda_2 = 209/ N_{\text{eff}}$ for PF00153,		
and $\lambda_1 = 0.100 / N_{\text{eff}}$ for both.
The $118$ samples with no deletion in the 
MCMC
samples for MC1162 and 
$268$ samples with no more than 3 in the 
MCMC
samples for MC1445
are employed.
\label{fig: H_distr}
}
\end{figure}
}

It has been shown\CITE{FBW:18} that pairwise coevolutionary models capture the collective residue variability
across homologous proteins even for quantities which are not imposed by the inference procedure, 
like three-residue correlations and the sequence distances between homologs.
However, as listed in \Tables{\ref{tbl: PF00595_parameters} and \ref{tbl: PF00153_parameters}}
the average evolutionary energies of natural proteins in PF00595 and in PF00153
are significantly lower than those of 
MCMC
samples obtained 
by the Boltzmann machine learnings.
The same fact was also reported for the Pfam family PF00014\CITE{BLCC:16}.
In \Fig{\ref{fig: H_distr}}, the histograms of evolutionary energies 
are compared with between the natural protein families, PF00595 and PF00153, and 
the 
MCMC
samples obtained by the Boltzmann machine learnings. 
Here, only sequences with no deletion for PF00595 and with a few deletions
for PF00153 are employed.
It is clear that 
the 
MCMC
samples cannot well reproduce
the natural proteins with respect to the evolutionary energy distribution.
This discrepancy of the energy distributions may indicate 
the insufficient equilibration owing to frustrated interactions.
It is not improved, however, even by increasing the number of 
MCMC
samples 
from $N_{\text{MC}} = N_{\text{eff}}$ to 100000. 
Also, it should be noticed
that the discrepancy of the average evolutionary energy
between the target sequences and 
the MCMC
samples is improved 
when 
the MCMC
samples rather than the natural proteins are used as protein-like sequences;
see \Tables{\ref{tbl: PF00595_parameters} and \ref{tbl: PF00153_parameters}} and \Fig{\ref{fig: H_distr}}.
Thus, recovering the pairwise amino acid frequencies
in the resolution of a total energy may be harder for natural proteins than 
for protein-like sequences.

\TCBB{
\SUBSECTION{Selective Temperature, $T_s$}
}{
\SUBSECTION{Selective temperature, $T_s$}
}

\TableInText{
\begin{table*}[!h]
\begin{threeparttable}[b]
\caption{
\TCBB{
\TEXTBF{
Selective Temperatures ($T_s$) Estimated from the Fields and Couplings Inferred for PF00595 and PF00153.
}
}{
\TEXTBF{
Selective temperatures ($T_s$) estimated from the fields and couplings inferred for PF00595 and PF00153.
}
}
}
\label{tbl: all_single_mutations}
\ifdefined\ThreeDigitsInTable
\begin{tabular}{lllllllllllllll}
\hline
	& \multicolumn{1}{r}{$_a$}
	& \multicolumn{1}{r}{$_b$}
	& \multicolumn{1}{r}{$_b$}
	& \multicolumn{1}{r}{$_c$}
	& \multicolumn{1}{c}{$r_{\psi_N}$ $^d$} & $\alpha_{\psi_N}$ $^d$
	& \multicolumn{1}{c}{$r_{\psi_N}$ $^e$} & $\alpha_{\psi_N}$ $^e$
	& \multicolumn{1}{c}{$r$ $^f$}
	& slope $^f$ 
	& $\hat{T_s}$
	& $T_m^{\text{exp}}$ $^g$
	& $\hat{T_g}$ $^h$
	& $\hat{\omega}$ $^i$
	\\
\multicolumn{5}{l}{Pfam ID \hspace*{1em} $\overline{\psi_N}/L$ \hspace*{1em} $\overline{\overline{\Delta\psi_N}}$ $\overline{\text{Sd}(\Delta\psi_N)}$ $\text{Sd}(\text{Sd}(\Delta\psi_N))$}
	& \multicolumn{2}{c}{for $\overline{\Delta\psi_N}$ }
	& \multicolumn{2}{c}{for $\text{Sd}(\Delta\psi_N)$ }
	&
	& (kcal/mol)
	& ($^\circ$K)
	& ($^\circ$K)
	& ($^\circ$K)
	& ($k_B$) 
	\\
\hline
PF00595	& $-3.15$	
	& 3.94	
	& 2.64	
	& 0.113	
	& $-0.980$	
	& $-1.90$	
	& $-0.237$	
	& $-0.113$	
	& 0.920		
	& 0.400		
	& 201		
	& 313 $^k$	
	& 215	 	
	& 1.20		
	\\
PF00153	& $-2.84$	
	& 3.36	
	& 2.71	
	& 0.141		
	& $-0.981$	
	& $-1.91$	
	& $-0.537$	
	& $-0.338$	
	& 
	& 
        & \RED{196}           
	\\
\hline
\end{tabular}
\else
\fi
\begin{tablenotes}

\item [$^a$] The average of evolutionary energies per residue over representatives of homologous sequences;
the Ising gauge is employed.
The representatives of unique sequences with no deletions for PF00595 and with no more than 2 deletions for PF00153, 
which are at least 20\% different from each other, are used;
their numbers are 361 for PF00595 and 144 for PF00153.

\item [$^b$] The averages of $\overline{\Delta \psi_N}$ and $\text{Sd}(\Delta \psi_N)$,
which are the mean and the standard deviation of $\Delta \psi_N$ due to single nucleotide nonsynonymous substitutions in a sequence, 
over the representatives of homologous sequences.

\item [$^c$] The standard deviation of $\text{Sd}(\Delta \psi_N)$ over the representatives of homologous sequences.

\item [$^d$] The correlation ($r_{\psi_N}$) and regression coefficients ($\alpha_{\psi_N}$) of $\overline{\Delta \psi_N}$ on $\psi_N/L$;
$\overline{\Delta \psi_N} \sim \alpha_{\psi_N} \psi_N/L + \beta_{\psi_N}$.

\item [$^e$] The correlation ($r_{\psi_N}$) and regression coefficients ($\alpha_{\psi_N}$) of $\text{Sd}(\Delta \psi_N)$ on $\psi_N/L$;
$\text{Sd}(\Delta \psi_N) \sim \alpha_{\psi_N} \psi_N/L + \beta_{\psi_N}$.

\item [$^f$] The reflective correlation ($r$) and regression coefficient (slope) of the experimental values\CITE{GGCJVTVB:07} of 
folding free energy changes ($\Delta\Delta G_{ND}$) due to single amino acid substitutions on
$\Delta \psi_N (\simeq \Delta\Delta \psi_{ND})$ for the same types of substitutions;
the slope is equal to $k_B T_s$.

\item [$^g$] An experimental value of melting temperature; $T_m^{\text{exp}}$ for PF00595 is taken from \CITE{TES:12}.

\item [$^h$] Glass transition temperature; $(T_s/(2T_m))(1 + (T_m^2/T_g^2)) = 1$ . 	
Folding free energy, $\langle \RED{\Delta} G_{ND}(\VECS{\sigma}, T) \rangle_{\VECS{\sigma}} = k_B T_s \delta \psi^2 [(T_s/(2T))(1 + (T^2/T_g^2)) - 1]$, is
inferred to be \RED{$-1.4$} kcal/mol at $T=298^\circ$K for PF00595 , while its experimental value is equal to \RED{$-2.9 \pm 0.2$} kcal/mol\CITE{GCAVBT:05}.

\item [$^i$] Conformational entropy per residue; $\omega = (T_s/T_g)^2 \delta\psi^2/(2L)$ .

\end{tablenotes}
\end{threeparttable}
\end{table*}

}

\FigureInText{

\begin{figure}[hbt]
\centerline{
\includegraphics[width=43mm,angle=0]{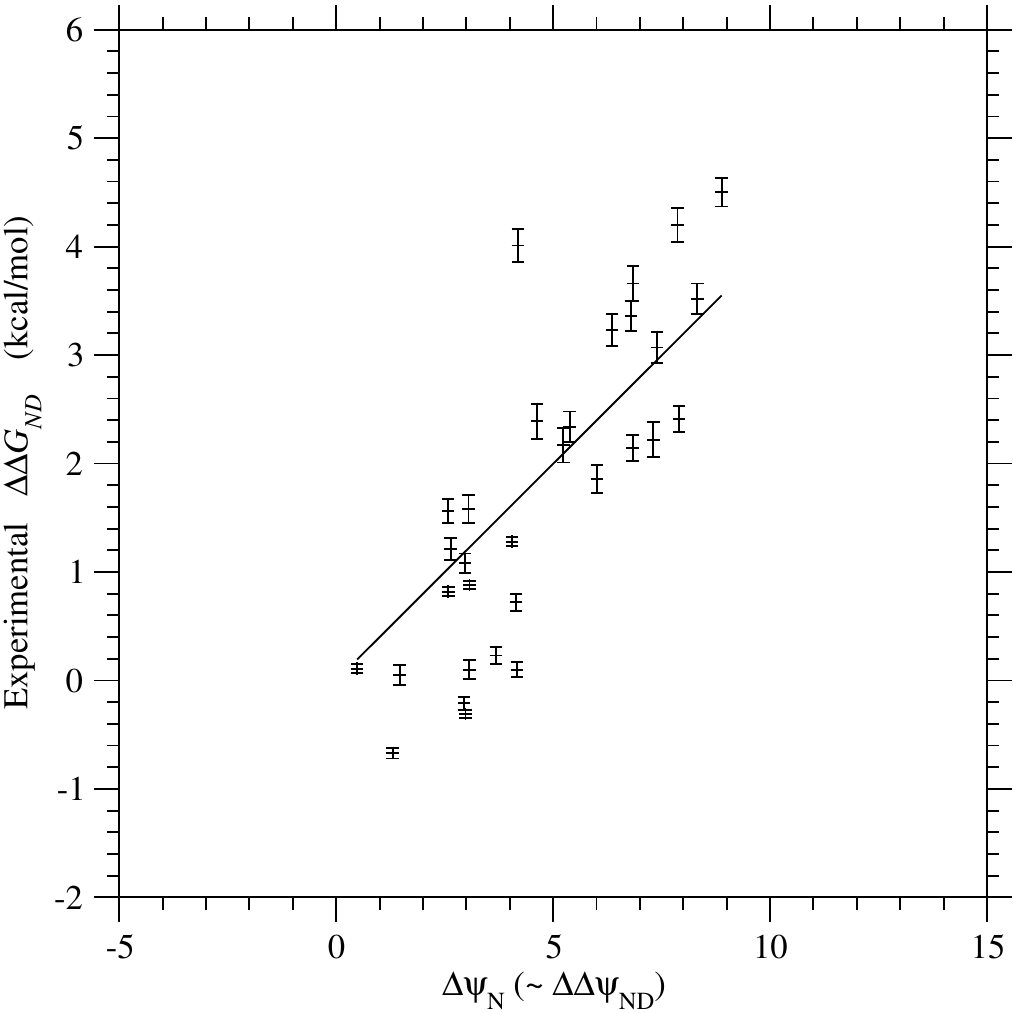}
}
\caption
{
\noindent
\TEXTBF{
Regression of the experimental values\CITE{GGCJVTVB:07} of folding free energy changes ($\Delta\Delta G_{ND}$)
due to single amino acid substitutions on
$\Delta \psi_N (\simeq \Delta\Delta \psi_{ND})$ for the same types of substitutions in PF00595.	
}
The solid line shows the least-squares regression line through the origin with the slope, $0.400$ kcal/mol, which is the estimates of $k_B T_s$.
The reflective correlation coefficient is equal to $0.92$.
The free energies are in kcal/mol units.
\label{fig: PF00595_dH_vs_ddG}
}
\end{figure}
}

\FigureInText{

\begin{figure}[hbt]
\centerline{
\includegraphics[width=43mm,angle=0]{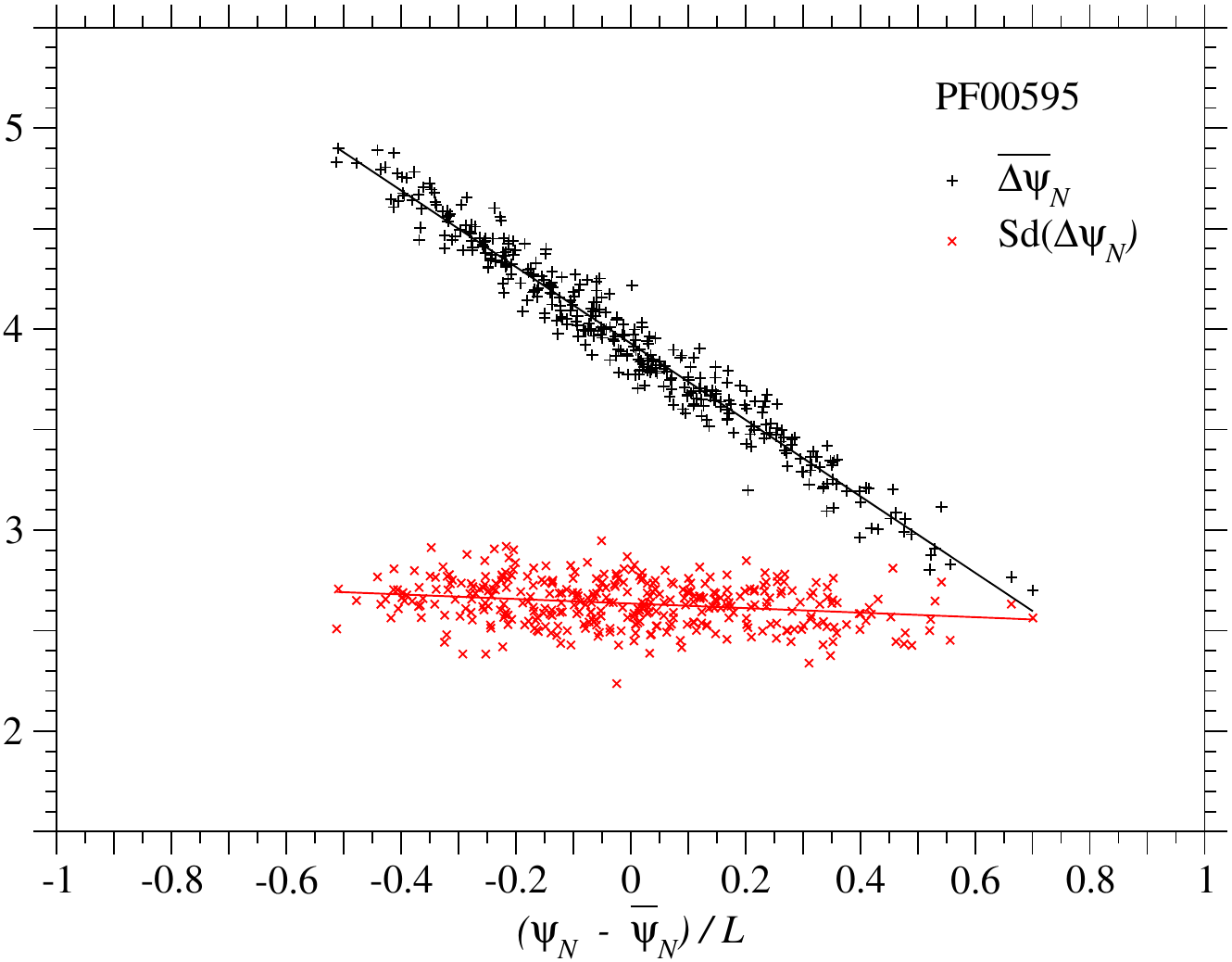}
\includegraphics[width=43mm,angle=0]{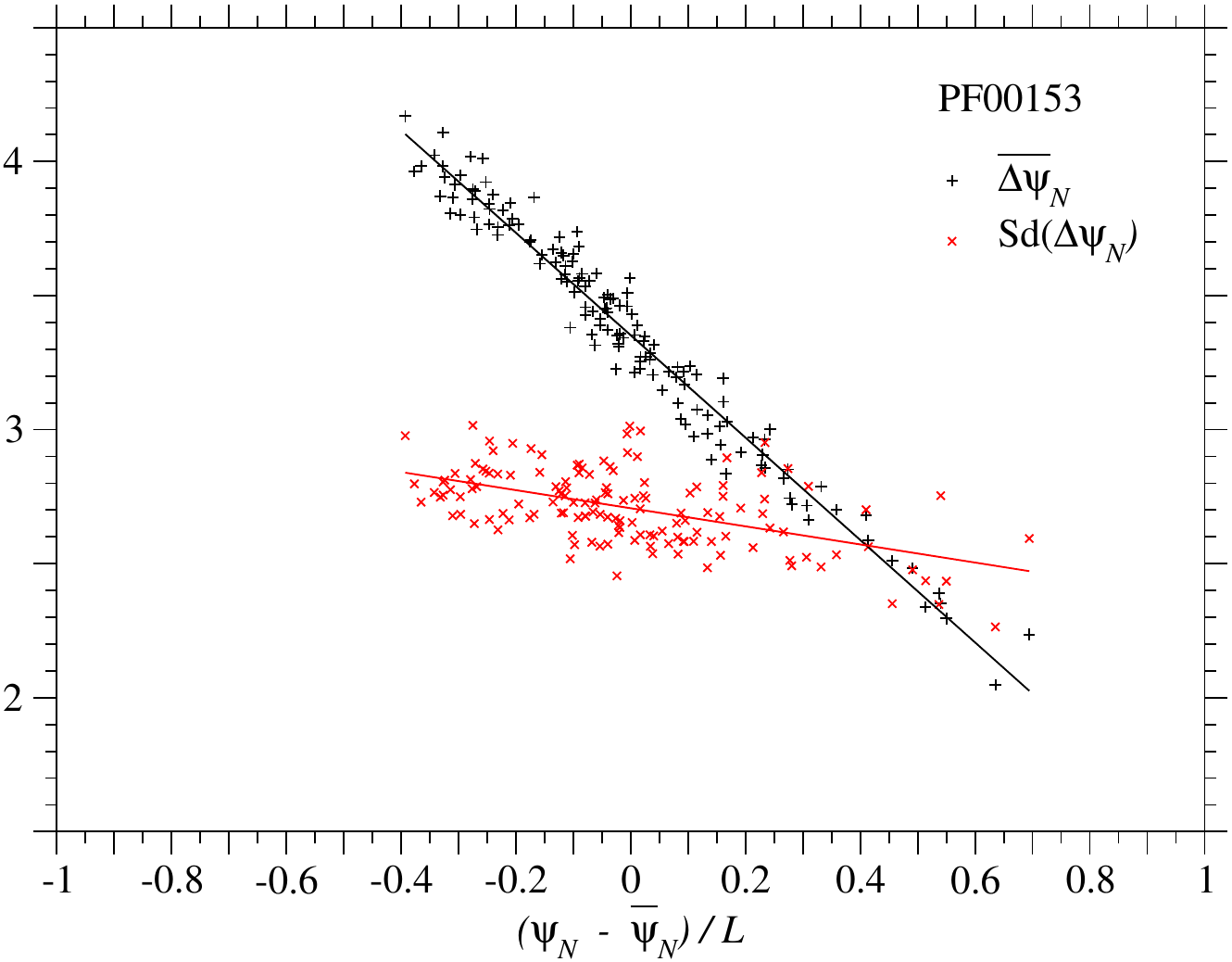}
}
\caption
{
\noindent
\TEXTBF{
Relationships between
$\Delta \psi_N$ due to single nucleotide nonsynonymous substitutions
and $\psi_N$ of the homologous sequences in PF00595 and in PF00153.
}
Each of the black plus and red cross marks corresponds to the mean and the standard deviation
of $\Delta \psi_N$ due to
all types of single nucleotide nonsynonymous substitutions
over all sites in each of the homologous sequences, respectively;
the left and right figures are for PF00595 and PF00153, respectively.
Representatives of unique sequences
with no deletion for PF00595 and with no more than 2 deletions for PF00153,
which are at least 20\% different from each other, are employed;
their numbers are 361 for PF00595 and 144 for PF00153.
The solid lines show the regression lines for the mean and the standard deviation of $\Delta \psi_N$.
The correlation and regression coefficients are listed in \Table{\ref{tbl: all_single_mutations}}.
\label{fig: all_singles}
}
\end{figure}
}

Selective temperature $T_s$ that is defined by \Eq{\EQTs}, 
which quantifies how strong 
the folding/structural constraints are in the evolution of a protein family,
has been estimated by various methods\CITE{DS:01,MSCOW:14,M:17}.
In principle, if folding free energy changes ($\Delta \Delta G_{ND}$) due to single amino acid substitutions
are known, the selective temperature $T_s$ can be 
estimated by comparing $\Delta \Delta G_{ND}$ with $\Delta \Delta \psi_{ND}$, that is, 
with the equation 
$\Delta \Delta G_{ND} / (k_B T_s) \simeq \Delta \Delta \psi_{ND} \simeq \Delta \psi_N $\CITE{M:17};
see \SecEvolution.
In \Fig{\ref{fig: PF00595_dH_vs_ddG}}, 
folding energy changes $\Delta \Delta G_{ND}$\CITE{GGCJVTVB:07}
due to single amino acid substitutions in PF00595
are plotted against their evolutionary energy changes $\Delta \psi_N$.
The slope of the least-squares regression line through the origin
gives the value of $k_B T_s$, 0.400 kcal/mol 
(201$^\circ$ K) for PF00595.	

For protein families for which folding energy changes are unknown, 
PF00595 may be employed as a reference protein to estimate $T_s$.
The standard deviation of the free energy changes 
due to single amino acid substitutions, $\text{Sd}(\Delta G_N)$,
must not explicitly depend on $k_B T_s$ but the free energy $G_N$.
On the other hand, 
the standard deviation of evolutionary energy changes due to single nucleotide nonsynonymous substitutions, 
$\text{Sd}(\Delta \psi_N )$, is independent of $\psi_N$ and
almost constant 
across homologous sequences in every protein family
as shown in \Fig{\ref{fig: all_singles}}.
In other words, $\text{Sd}(\Delta \psi_N )$ is not a function of
$G_N$ but only $T_s$.
Therefore, $\text{Sd}(\Delta G_N )$ 
is expected to be
nearly constant across protein families\CITE{M:17};
$\text{Sd}(\Delta G_N) = k_B T_s \text{Sd}(\Delta \psi_N ) \approx \text{constant}$ .
As a result, $T_s$ can be estimated for any protein family 
from the ratio of $\RED{\overline{\BLACK{\text{Sd}(\Delta \psi_N )}}}$ 
to that of a reference protein.
The $k_B T_S$ of PF00153 is estimated to be  
\RED{0.389} kcal/mol ($\RED{196}^\circ$K).	
In \Table{\ref{tbl: all_single_mutations}},
various quantities\CITE{M:17} derived from $\Delta \psi_N$ are listed.


\SECTION{Discussion}

In the Boltzmann machine learning,
the objective function, which is the cross entropy,
fluctuates in the minimization process,
because 
the partial derivatives are calculated from
the pairwise marginal distributions estimated by 
MCMC
sampling, and the first-order gradient-descent method is also used.
The cross entropy, however,
cannot be monitored in the minimization process, because it can be hardly evaluated. 
Here, the average ($D^{KL}_2$) of Kullback-Leibler divergences for pairwise marginal distributions
over all residue pairs is monitored instead
as a rough measure of fitting to the reference distribution.
As shown in \Fig{\ref{fig: PF00595_learning_process_KL}},
$D^{KL}_2$ significantly fluctuates in the learning process.
On the other hand, 
the sample and ensemble averages of evolutionary energy
smoothly change and slowly converge in the learning process; see \Fig{\ref{fig: average_energy}}.
Also, the profile of the sample and ensemble averages of evolutionary energy
along the learning process is well reproducible by another run of learning.
Here we stop learning 
and pick up the parameter estimates corresponding to $\min D^{KL}_2$,
if $\min D^{KL}_2$ does not change during a certain number of iterations
after a sufficient number of iterations are executed 
for the sample and ensemble averages of evolutionary energy to converge;
see \SecIterations\ for details.

Machine leaning methods such as Boltzmann machine
employ a gradient-descent method to minimize a loss function
or cross entropy. There are not a few gradient-descent methods invented.
In most machine learning problems,  a gradient-descent method 
is chosen from a view point of the speed and the degree of minimization.
Even if there is a unique minimum in an objective function,
there can be multiple sets of parameter values 
that take similarly approximate values for the minimum of the objective function.
Some of them may have reasonable characteristics for the solution but the others may not.
Thus, we must choose a gradient-descent method that yields
reasonable values for parameters.
For the present problem, 
\ifdefined\NAG
four 
\else
three
\fi
gradient-descent methods are examined;
the Adam method that is commonly employed in machine learning,
the RPROP-LR method that is the RPROP modified\CITE{FBW:18} for 
Boltzmann machine learnings of protein sequences,
and 
\RED{
the ModAdam method that is
}
the Adam method modified
\ifdefined\NAG
for sparse couplings\RED{,}
and the NAG method that is
not a per-parameter learning rate method like the ModAdam 
but not an adaptive method unlike the ModAdam.
\else
 for sparse couplings.
\fi

Surprisingly, it is shown in
the Boltzmann machine with the L2-L2 regularization for PF00595
that each method gives a very different solution.
The Adam method normalizes each partial derivative by a parameter
and therefore tend to yield similarly dense interactions
even for a sparse-interaction system like protein sequences.
\ifdefined\NAG
On the other hand, the NAG and the ModAdam, 
in which the stepsize of parameter updates 
is proportional to the partial gradient
and the proportional constant is the same for all parameters,
yield the reasonable parameters in which
the couplings 
well correlate with residue-residue distance. 
\else
On the other hand, the ModAdam, which is the Adam modified
in such a way that the stepsize of parameter updates 
is proportional to the partial gradient
and the proportional constant is adaptive but the same for all parameters,
yields the reasonable parameters in which
the couplings 
well correlate with residue-residue distance. 
\fi
This fact strongly indicates that
the stepsize must be proportional to the partial gradient
to yield reasonable solutions at least for sparse-interaction systems.
In other words,
the per-parameter methods such as AdaGrad, AdaDelta,
RMSprop, and Adam, in which each partial gradient of parameter is normalized,
and RPROP, in which the stepsize 
does not depend on the partial derivative but only on
the temporal behavior of its sign,
may be appropriate to similarly-dense-interaction systems as well as the $L_2$ regularization
but inappropriate to sparse-interaction systems for which the $L_1$ regularization
is often employed. 

On the other hand,
the RPROP-LR, 
in which the stepsize is proportional to the partial derivative,
yield similar solutions to the ModAdam, 
but their profiles of the average evolutionary energies
are very different, indicating that in the RPROP-LR convergence is very fast but
parameters converge to less favorable interactions to the natural proteins than
those inferred by the ModAdam.
Also, the RPROP-LR is inferior to the ModAdam in the recoverability of pairwise frequencies. 

The results indicate that the profile of the sample and ensemble averages
of evolutionary energy is very useful
not only for tuning hyper-parameters but also
to judge whether iterations of learning
converge or not, and which optimization method is better than the others.
\ifdefined\NAG
Although the fields and couplings inferred by the NAG in the L2-L2
almost coincide with those by the ModAdam, the energy profiles
indicate that those inferred by the ModAdam are slightly more optimized
than those by the NAG.
\else
\fi

In the inverse Potts problem, 
in which the evolutionary interactions, fields and coupling, are
inferred from a multiple sequence alignment of protein homologous sequences, 
regularization must be applied to objective functions
in order to prevent over-fitting or to solve ill-posed problems.
Commonly used regularizations are $L_2$, $L_1$ and their variants. 
Regularization, however, introduces hyper-parameters to be determined.
In almost all models for protein sequences,
regularization parameters have been arbitrarily determined or 
adjusted to increase the precision for residue-residue contact prediction.
In order to correctly infer fields and couplings,
however, 
the other requirement for them should be also taken into account;
the sample average of evolutionary energies over target sequences
must be equal to the ensemble average.
The estimates of the sample and ensemble averages of 
evolutionary energy
significantly
depend on regularization parameters as shown in \Fig{\ref{fig: tuning}}.
Here the regularization parameter $\lambda_1$ for fields
is first tuned as small as possible as long as 
the sample average
of evolutionary energies over target sequences 
is lower than the ensemble average and
both the averages of evolutionary energy decrease.
Then, the regularization parameter $\lambda_2$ for couplings 
is tuned as large as possible as long as
the sample average 
of evolutionary energies over target sequences is
lower than the ensemble average,
where the ensemble average is evaluated by 
approximating 
the distribution of the evolutionary energies of random sequences
as a Gaussian distribution;
the Gaussian approximation\CITE{PGT:97} is known to be appropriate
for native proteins.

The group $L_1$ regularization was employed for a pseudo-likelihood
minimization in a graphical model (GREMLIN\CITE{BKCLL:11}) 
for protein sequences. 
However, in the inverse Potts problem on protein sequences,
most analyses have employed the $L_2$ regularization because 
the $L_2$ function is differentiable, although 
the $L_2$ regularization is appropriate to similarly-dense-interaction systems
rather than sparse-interaction systems such as protein sequences.
The present analyses clearly show that the group $L_1$ regularization
for couplings makes their estimates more sparse and therefore
more reasonable than $L_1$ and $L_2$.
The present results strongly indicate that regularization models
and learning methods must be carefully chosen for particular
interaction systems.

The most important question is how precisely the evolutionary interactions
can be inferred.  
Recoverabilities of single-site frequencies and pairwise correlations, and
even higher correlations by the inverse Potts model have been examined 
\CITE{BLCC:16,CFFMW:17,FBW:18}.
On the other hand, reproducibilities of fields and couplings
have been examined mostly for artificial systems such as 
artificial data on Erd\"{o}s-Renyi models and 
a lattice protein\CITE{BLCC:16}.
Because the reproducibilities of fields and couplings 
depend on the mode of pairwise interactions,
we need to examine how well they can be reproduced
for protein sequences.
Here the 
MCMC
samples that are generated with the interactions inferred
by the Boltzmann machine learning for the natural protein families
are employed as protein-like sequences.
In other words,
fields and couplings are first inferred from protein sequences 
and compared with those inferred again by Boltzmann machine learning 
from 
MCMC
samples that are generated with the interactions first inferred.  
Both the fields and couplings are well recovered for PF00595 and PF00153
except for weak couplings.

However, the distribution of evolutionary energies over the natural proteins
significantly shifts towards lower energies in comparison with those
over the
MCMC
samples; see \Fig{\ref{fig: H_distr}}.
The same feature was also reported for
trypsin inhibitor\CITE{BLCC:16}.
This discrepancy is improved if 
the MCMC
samples are
employed as protein-like sequences for Boltzmann machine learning;
see \Fig{\ref{fig: H_distr}}.
Recovering the pairwise amino acid frequencies
in the resolution of a total energy is harder for the natural proteins than
for the protein-like sequences.

A computational load is very high for
the Boltzmann machine method to infer fields and couplings.
Recently, restricted Boltzmann machines that are equivalent to the present Boltzmann machine
have been studied\CITE{TCM:19,SW:19}. In these models,
the coupling interactions $J_{ij}(a_k, a_l)$ are estimated in the decoupled form, 
$\sum_{\mu=1}^{L_q} \xi_i^{\mu}(a_k) \xi_j^{\mu}(a_l)$, 
and approximated with the small numbers for $L_{q}$, reducing a computational load;
the number of parameters for coupling interactions 
may be reduced from $L(L-1)q^2/2$ to $LL_{q}q$;
$q = 21$.
Thus, using the restricted Boltzmann machines 
certainly has a merit, although $L_{q}$
seems to be large to well approximate sparse coupling interactions.

Accurate estimation of coupling interactions is useful 
in analyses of protein evolution\CITE{MSCOW:14,M:17} 
and protein foldings\CITE{BCCJM:16,JGSCM:16}.
On the basis of the constancy of 
the standard deviation of evolutionary energy changes
due to single nucleotide nonsynonymous substitutions ($\text{Sd}(\Delta \psi_N)$) over protein families,
selective temperature, $T_s$ in \Eq{\EQequilDistrOfSeqOfdG},
was estimated\CITE{M:17} for several proteins 
from fields and couplings inferred by the mean field DCA method.
The estimates of fields and couplings by the mean field DCA method\CITE{MPLBMSZOHW:11,MCSHPZS:11} 
are known to be rough\CITE{BLCC:16,CFFMW:17}. 
Therefore, the constancy of $\text{Sd}(\Delta \psi_N)$ over protein families
has been confirmed here with the present estimates of fields and couplings for PF00595 and PF00153.
Although $\text{Sd}(\Delta \psi_N)$ are estimated significantly smaller
for PF00595 by the Boltzmann machine than that by the mean fields DCA,
the methods and discussion\CITE{M:17} on protein evolution are still valid.


\vspace*{1em}
\noindent
\textit{The program written in Scala and the MSAs employed are available from
https://gitlab.com/sanzo.miyazawa/BM/.}

\TCBB{

}{

}

}



\ifCLASSOPTIONcaptionsoff
  \newpage
\fi



\bibliographystyle{IEEEtran}
\bibliography{jnames_with_dots,SM,Protein,MolEvol,Bioinfo}
%

%
\begin{IEEEbiography}[{\includegraphics[height=1in,clip,keepaspectratio,angle=90]{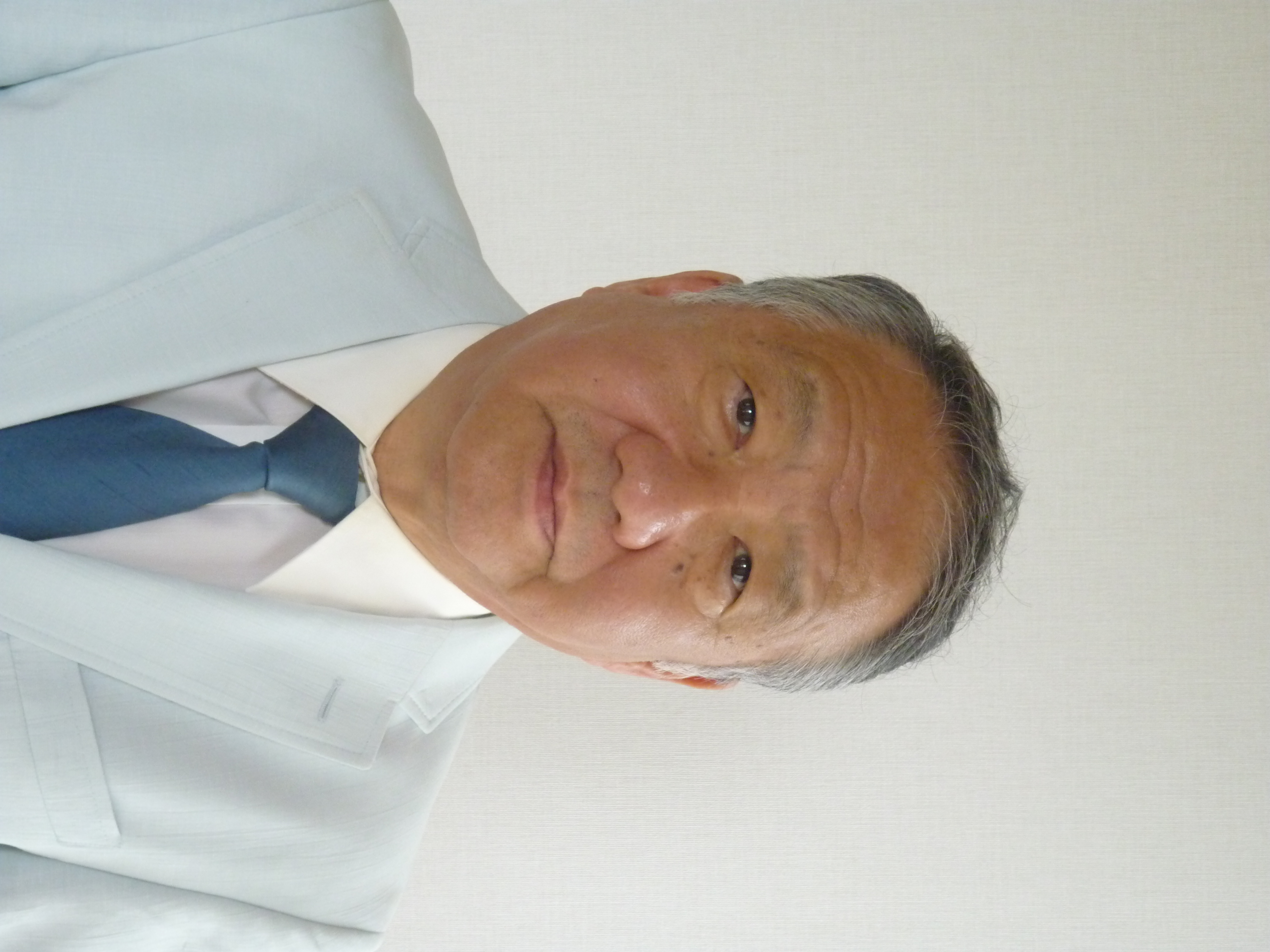}}]{Sanzo Miyazawa}
had worked for the Graduate School of Engineering in Gunma University, Japan until retired at age 65 in 2013.
His research interests include protein structure and evolution.
\end{IEEEbiography}




\TCBB{
}{
}

\renewcommand{\TEXT}[1]{}
\renewcommand{\SUPPLEMENT}[1]{#1}
\FiguresWithoutCaption{
\renewcommand{\SUPPLEMENT}[1]{}
}

\SUPPLEMENT{
\renewcommand{\SkipSupplToMerge}[1]{#1}
\SkipSupplToMerge{

\SUPPLEMENT{
\SkipSupplToMerge{

\renewcommand{\TEXT}[1]{}
\renewcommand{\SUPPLEMENT}[1]{#1}
\renewcommand{\EQ}{seq}
\renewcommand{\TBL}{stbl}
\renewcommand{\FIG}{sfig}

\clearpage
\newpage

\arXiv{

\setcounter{page}{1}
\renewcommand{\thepage}{S-\arabic{page}}

}{
\setcounter{page}{1}
\renewcommand{\thepage}{S-\arabic{page}}
}

\setcounter{section}{0}
\renewcommand{\thesection}{S.\arabic{section}}

\setcounter{equation}{0}
\renewcommand{\theequation}{S\arabic{equation}}

\setcounter{table}{0}
\renewcommand{\thetable}{S\arabic{table}}

\setcounter{figure}{0}
\renewcommand{\thefigure}{S\arabic{figure}}

\TCBB{

\begin{figure*}

\Large

\begin{center}
\textbf{Supplementary Material} \\
for
\\
Boltzmann Machine Learning and Regularization Methods for Inferring
\\
Evolutionary Fields and Couplings from a Multiple Sequence Alignment
\\
( Article DOI: 10.1109/TCBB.2020.2993232 )
\\
in
\\
IEEE/ACM Transactions on Computational Biology and Bioinformatics, 2020
\\
\end{center}

\vspace*{1em}
\begin{center}
Sanzo Miyazawa  \\
sanzo.miyazawa@gmail.com
\end{center}
\begin{center}
2021-06-07
\end{center}

\normalsize

\end{figure*}

\clearpage

\newpage

\SECTION{Methods}

\SUBSECTION{The Inverse Potts model for protein homologous sequences}
\label{section: Potts}

Let us consider probability distributions $P(\VECS{\sigma})$ of
amino acid sequences 
$\VECS{\sigma} (\equiv (\sigma_1, \ldots, \sigma_L) )$,
which satisfy the following constraints that 
single-site and two-site marginal probabilities must be equal to
a given frequency $P_i(a_k)$
of amino acid $a_k$ at each site $i$ and a given frequency $P_{ij}(a_k,a_l)$ of
amino acid pair $(a_k,a_l)$ for site pair $(i,j)$, respectively.
\begin{eqnarray}
P(\sigma_i = a_k) &\equiv&
\sum_{\VECS{\sigma}} P(\VECS{\sigma}) \delta_{\sigma_i a_k} = P_i(a_k)
        \label{\EQ: constraints_single_site_marginals}
        \\
        P(\sigma_i = a_k, \sigma_j = a_l) &\equiv&
        \sum_{\VECS{\sigma}} P(\VECS{\sigma}) \delta_{\sigma_i a_k} \delta_{\sigma_j a_l} = P_{ij}(a_k, a_l)
	\hspace*{2em}
        \label{\EQ: constraints_two_site_marginals}
\end{eqnarray}
where $\sigma_i, a_k \in \{ \text{amino acids, deletion} \}$
$k = 1, \ldots, q$,
$q \equiv | \{ \text{amino acids, deletion} \} | = 21$,
$i, j = 1, \ldots, L$,
and $\delta_{\sigma_i a_k}$ is the Kronecker delta. 
The sequence distribution 
$P(\VECS{\sigma}|h, J)$	
with the maximum entropy 
can be represented as
\begin{eqnarray}
   P(\VECS{\sigma}|h, J)
	= \frac{1}{Z_{\VECS{\sigma}}} e^{- \psi_{N}(\VECS{\sigma}|h,J) }
	\label{\EQ: max_entropy_distr}
  \hspace*{1em} , \hspace*{1em} Z_{\VECS{\sigma}} =
        \sum_{\VECS{\sigma}} e^{-\psi_{N}(\VECS{\sigma}|h,J) }
        \label{\EQ: Z_for_Potts}
	\\
  \psi_{N}(\VECS{\sigma}|h,J)
        = - \, [ \,
		\sum_i \{ \, h_i(\sigma_i) + \sum_{j(>i)} J_{ij}(\sigma_i, \sigma_j) \, \}
		\, ]
        \label{\EQ: H_for_Potts}
        \label{\EQ: psi_for_Potts}
\end{eqnarray}
where Lagrange multipliers $h_i(a_k)$ and $J_{ij}(a_k,a_l)$ are 
interaction potentials called fields and couplings,
and $\psi_{N}(\VECS{\sigma}|h,J)$ is referred to here as evolutionary energy.

Fields $h_i(a_k)$ and couplings $J_{ij}(a_k,a_l)$ provide 
useful information to understand protein evolution\CITE{M:17}
and also to predict residue-residue contacts in protein structures 
on the basis of coevolutional residue substitutions \CITE{LGJ:02,MPLBMSZOHW:11,MCSHPZS:11,M:13}.

For given
single-site $P_{i}(a_k)$ and two-site frequencies $P_{ij}(a_k,a_l)$,
which are evaluated from a multiple sequence alignment,
inferring $h_i(a_k)$ and $J_{ij}(a_k,a_l)$ have
been attempted as the Inverse Potts problem
by the Boltzmann machine learning\CITE{WWSHH:09,FBW:18},
by the mean field approximation\CITE{LGJ:02,MPLBMSZOHW:11,MCSHPZS:11}, 
by the Gaussian approximation\CITE{BZFPZWP:14},
by maximizing a pseudo-likelihood\CITE{ELLWA:13,EHA:14,BKCLL:11,KOB:13},
and
by minimizing a cross entropy in the adaptive cluster expansion\CITE{BLCC:16}.

\SUBSECTION{The sample average of evolutionary energy}
\label{section: sample_ave}

According to the Potts model, 
the sample average of
$\psi_N(\VECS{\sigma}_N)$ over 
natural sequences, $\VECS{\sigma}_N$, fixed in protein evolution is equal to
the ensemble average of $\psi_N(\VECS{\sigma})$ over sequences, $\VECS{\sigma}$.
Sample averages are calculated with a sample weight 
$w_{\VECS{\sigma}_N}$ for each homologous sequence,
which is used to reduce phylogenetic biases in the set of homologous sequences;
for example, the sample average of evolutionary energy is calculated as follows.
\begin{eqnarray}
	\overline{ \psi_N(\VECS{\sigma}_N) } 
	&\equiv &
		\frac{\sum_{\VECS{\sigma}_N} w_{\VECS{\sigma}_N} \psi_N(\VECS{\sigma}_N) }{\sum_{\VECS{\sigma}_N} w_{\VECS{\sigma}_N} } 
	\label{\EQ: def_sample_ave_of_psi}
		\\
	&=&
	\langle \psi_N(\VECS{\sigma}) \rangle_{\VECS{\sigma}} 	
	\label{\EQ: sample_ave_of_psi}
\end{eqnarray}
where $\overline{ \psi_N(\VECS{\sigma}_N) }$ denotes a sample average 
of $\psi_N(\VECS{\sigma}_N)$ 
with a sample weight 
$w_{\VECS{\sigma}_N}$ for each homologous sequence $\VECS{\sigma}_N$,
and $\langle \psi_N(\VECS{\sigma}) \rangle_{\VECS{\sigma}}$ is
the ensemble average of $\psi_N(\VECS{\sigma})$ that obeys a Boltzmann distribution.

\SUBSECTION{Ensemble average by a Gaussian Approximation for the distribution of the evolutionary energies of random sequences}
\label{section: ensemble_ave}

The ensemble average over sequences, for example, of $\psi_N(\VECS{\sigma})$
is estimated by the Gaussian approximation\CITE{PGT:97,M:17}, in which 
the distribution of the evolutionary energies of random sequences is approximated
as a Gaussian distribution, $\mathcal{N}(\bar{\psi}, {\delta \psi}^2)$.
The mean $\bar{\psi}$ and variance ${\delta \psi}^2$
are evaluated as those of evolutionary energies 
of random sequences whose amino acid composition is equal to
the average amino acid composition of sequences in a protein family.
\begin{eqnarray}
\lefteqn{
\langle \psi_N(\VECS{\sigma}) \rangle_{\VECS{\sigma}} 	
	\equiv 
	\, [ \, 
	\sum_{\VECS{\sigma}} \psi_{N}(\VECS{\sigma}) 
		\exp( - \psi_{N}(\VECS{\sigma}) )
	\, ] \,
	/ 
	\, Z_{\VECS{\sigma}}
}
	\\
	&\approx& \frac{ \int \psi_N \exp (- \psi_N)) \, \mathcal{N}(\bar{\psi}, {\delta \psi}^2) \, d\psi_N }  
		{\int \exp (- \psi_N) \, \mathcal{N}(\bar{\psi}, {\delta \psi}^2) \, d\psi_N }
	\\
	&=& \bar{\psi}( \overline{\VEC{f}(\VECS{\sigma}_N)} ) - {\delta \psi}^2( \overline{\VEC{f}(\VECS{\sigma}_N)} )
	\label{\EQ: ensemble_ave_of_psi}
\end{eqnarray}
where $\overline{\VEC{f}(\VECS{\sigma}_N)}$ is the sample-average amino acid composition of natural sequences in a protein family.


\SUBSECTION{Relationships between evolutionary energy $\psi_N(\VEC{\sigma})$,
fitness $m(\VEC{\sigma})$, and
folding free energy $\Delta G_{ND}(\VEC{\sigma})$ of protein $\VEC{\sigma}$ \CITE{M:17}}
\label{section: evolution}

In \CITE{M:17},
it was proved by assuming the detailed balance principle 
that the equilibrium distribution of protein sequences
must be the Boltzmann distribution of their
Malthusian fitness $m$ as well as that of $\Delta \psi_{ND}$.
On the other hand, a protein folding theory
\CITE{SG:93a,SG:93b,RS:94,PGT:97} 
based on a random energy model (REM)
indicates that it can be approximated to the Boltzmann distribution of
the folding free energy divided by selective temperature, $\Delta G_{ND}/(k_B T_s)$.

\begin{eqnarray}
        P^{\script{eq}}(\VEC{\mu}) &=& \frac{ P^{\script{mut}}(\VEC{\mu}) \exp( 4N_e m( \VEC{\mu} ) ( 1 - q_m )) }
                { \sum_\nu P^{\script{mut}}(\VEC{\nu})
                        \exp( 4N_e m(\VEC{\nu}) ( 1 - q_m )) )}
                \label{\EQ: equilibrium_distr_of_seq_of_m}
                \\
        &=& \frac{ P^{\script{mut}}(\overline{\VEC{\mu}}) \exp ( - (\psi_{N}(\VEC{\mu}) - \psi_{D}(\overline{\VEC{f}(\VEC{\mu})}, T) )) }
                { \sum_{\VEC{\nu}} P^{\script{mut}}(\overline{\VEC{\nu}})
                        \exp ( - (\psi_{N}(\VEC{\nu}) - \psi_{D}(\overline{\VEC{f}(\VEC{\nu})}, T) ) ) }
		\hspace*{2em}
                \label{\EQ: equilibrium_distr_of_seq}
                \label{\EQ: equilibrium_distr_of_seq_of_dPsi}
                \\
        &\simeq& 
                \frac{ P^{\script{mut}}(\VEC{\mu}) \exp( - \Delta G_{ND}(\VEC{\mu}, T ) / (k_B T_s) ) }
                { \sum_\nu P^{\script{mut}}(\VEC{\nu})
                        \exp( - \Delta G_{ND}(\VEC{\nu}, T) / (k_B T_s) )}
                \label{\EQ: equilibrium_distr_of_seq_of_dG}
\end{eqnarray}
where
$p^{\text{mut}}(\VEC{\sigma})$ is the probability of a sequence (\VEC{\sigma})
randomly occurring in a mutational process
and depends only on the amino acid composition of the sequence $\VEC{f}(\VEC{\sigma})$, 
$q_m$ is the frequency of a single mutant gene in a population,
$k_B$ is the Boltzmann constant, $T$ is growth temperature,
$T_s$ is selective temperature that quantifies how strong 
the folding constraints are in protein evolution,
$\overline{\VEC{f}(\VEC{\sigma}) } \equiv \sum_{\VEC{\sigma}} \VEC{f}(\VEC{\sigma}) P(\VEC{\sigma})$ and
$\log P^{\script{mut}}(\overline{\VEC{\sigma}}) \equiv \sum_{\VEC{\sigma}} P(\VEC{\sigma}) \log (\prod_i P^{\script{mut}}(\sigma_i))$.
Then, the following relationships are derived for sequences for which 
$f(\VEC{\mu}) = \overline{f(\VEC{\mu})}$.
\begin{eqnarray}
        4N_e m(\VEC{\mu}) (1 - q_m)
        &=& - \Delta \psi_{ND}(\VEC{\mu},T) + \mathrm{constant}
        \label{\EQ: relationship_between_m_and_dPsi}
        \\
        &\simeq& 
        \frac{- \Delta G_{ND}(\VEC{\mu},T)}{k_B T_s} + \mathrm{constant}
        \label{\EQ: relationship_between_m_and_dG}
\end{eqnarray}
The selective advantage of $\VEC{\nu}$ to $\VEC{\mu}$ is represented as follows for
$f(\VEC{\mu}) = f(\VEC{\nu}) = \overline{f(\VEC{\sigma})}$.
\begin{eqnarray}
\lefteqn{
        4N_e s(\VEC{\mu} \rightarrow \VEC{\nu}) ( 1 - q_m )
}
        \nonumber
        \\
        &=&
        (4N_e m(\VEC{\nu})
        - 4N_e m( \VEC{\mu} ) ) ( 1 - q_m )
        \\
        &=& - (\Delta \psi_{ND}(\VEC{\nu}, T) - \Delta \psi_{ND}(\VEC{\mu}, T) )
        =
        - ( \psi_N(\VEC{\nu}) - \psi_N(\VEC{\mu}) )
	\hspace*{2em}
        \label{\EQ: s_vs_dPsi}
        \label{\EQ: selective_advantage_and_dPsi_N}
        \\
        &\simeq& 
        - (\Delta G_{ND}(\VEC{\nu}, T) - \Delta G_{ND}(\VEC{\mu}, T) ) / (k_B T_s)
	\nonumber \\
	& & \hspace*{7em}
        =
        - ( G_N(\VEC{\nu}) - G_N(\VEC{\mu}) ) / (k_B T_s)
        \label{\EQ: s_vs_dG}
	\\
\lefteqn{
	\psi_N(\VEC{\mu}) \simeq G_N(\VEC{\mu}) / (k_B T_s)  \hspace*{2em} 
		\psi_D(\VEC{\mu}) \simeq G_D(\VEC{\mu}) / (k_B T_s)
}
        \label{\EQ: Ts}
\end{eqnarray}
where $G_N(\VEC{\sigma})$ and $G_D(\VEC{\sigma})$ are the free energies of the native and the denatured states
of sequence $\VEC{\sigma}$.
It should be noted here that
only sequences for which $f(\VEC{\sigma}) = \overline{f(\VEC{\sigma})}$
contribute significantly to the partition functions in
\Eq{\ref{\EQ: equilibrium_distr_of_seq_of_dPsi}}, and other sequences may be ignored.


\SUBSECTION{Relationships among selective temperature ($T_s$), 
glass transition temperature ($T_g$), and melting temperature ($T_m$) of protein}
\label{section: Ts_Tg_Tm}

The distribution of conformational energies
in the denatured state (molten globule state), which
consists of conformations as compact as the native conformation,
is approximated in 
the random energy model (REM), particularly the independent 
interaction model (IIM)\CITE{PGT:97}, to be equal to
the energy distribution of the randomized sequences,
which is approximated by 
the energy distribution of the random sequences with the same amino acid composition
and then by a Gaussian distribution,
in the native conformation.
That is, the partition function $Z$ for the denatured state is written as follows with 
the number density per energy $n(E)$ of conformations that is approximated by a product   
of a Gaussian probability density and the total number of conformations 
whose logarithm is proportional to the chain length.
\begin{eqnarray}
	Z &=& \int \exp (\frac{- E}{k_B T}) \, n(E) dE
		\\
	n(E) &\approx& \exp ( \omega L ) \mathcal{N}(\bar{E}(\VEC{f}(\VECS{\sigma}_N)), \delta E^2(\VEC{f}(\VECS{\sigma}_N)) )
\end{eqnarray}
where $\omega$ is the conformational entropy per residue 
in the compact denatured state, 
and $\mathcal{N}(\bar{E}(\VEC{f}(\VECS{\sigma}_N)), \delta E^2(\VEC{f}(\VECS{\sigma}_N)) )$ is
the Gaussian probability density with mean $\bar{E}$ and variance $\delta E^2$, 
which depend only on the amino acid composition, $\VEC{f}(\VECS{\sigma}_N)$, 
of the protein sequence, $\VECS{\sigma}_N$.
The free energy of the denatured state is approximated as follows.
\begin{eqnarray}
	G_D(\VECS{\sigma}_N,T)	
		&\approx&
		\bar{E}(\VEC{f}(\VECS{\sigma}_N))
	- \frac{\delta E^2(\VEC{f}(\VECS{\sigma}_N))}{2 k_B T} 
	- k_B T \omega L
	\hspace*{2em}
	\\
	&=& \bar{E}(\VEC{f}(\VECS{\sigma}_N))
	- \delta E^2(\VEC{f}(\VECS{\sigma}_N))
		\frac{\vartheta(T/T_g)}{k_B T}
	\\
	\psi_D(\VECS{\sigma}_N,T) 
		&\approx&
	\bar{\psi}(\VEC{f}(\VECS{\sigma}))
	- \delta \psi^2(\VEC{f}(\VECS{\sigma}))
		\vartheta(T/T_g)\frac{T_s}{T}
	\\
 \vartheta(\frac{T}{T_g}) &\equiv&
		\begin{array}{ll}
			(1 + T^2 / T_g^2) / 2 & \textrm{ for } T \geq T_g \\
		\end{array} 
	\label{\EQ: free_energy_of_denatured_state}
\end{eqnarray}
where $\bar{E}$ ($\bar{\psi}$) and $\delta E^2$ (${\delta \psi}^2$)
are estimated as the mean and variance of
interaction energies $E$ ($\psi_N$) of 
the randomized sequences, which are approximated by
random sequences,
in the native conformation;
$\bar{E} \simeq k_B T_s \bar{\psi}$ and ${\delta E}^2 \simeq (k_B T_s)^2 {\delta \psi}^2$.
$T_g$ is the glass transition temperature of the protein
\TEXT{
at which entropy becomes zero\CITE{SG:93a,SG:93b,RS:94,PGT:97}; 
$ - \partial G_D / \partial T |_{T=T_g} = 0$.
}
\SUPPLEMENT{
at which entropy becomes zero\CITE{SG:93a,SG:93b,RS:94,PGT:97}. 
\begin{eqnarray}
	- \frac{\partial G_D}{\partial T} |_{T=T_g} &=& 0
\end{eqnarray}
}
The conformational entropy per residue $\omega$ in the compact denatured state
can be represented with 
\TEXT{
$T_g$; $\omega L = \delta E^2 / (2 (k_B T_g)^2) $.
}
\SUPPLEMENT{
$T_g$.
\begin{eqnarray}
	\omega L &=& \frac{ \delta E^2}{ 2 (k_B T_g)^2 }
\end{eqnarray}
}
Thus, unless $T_g < T_m$, a protein will be trapped at local minima 
on a rugged free energy landscape before 
it can fold into a unique native structure.

The ensemble average of $\Delta G_{ND}(\VECS{\sigma}, T)$ over sequences, which 
is observable as the sample averages of $\Delta G_{ND}(\VECS{\sigma_N}, T)$ over 
homologous sequences fixed in protein evolution,
is estimated as follows\CITE{M:17}.
\begin{eqnarray}
\lefteqn{
\langle \Delta G_{ND}(\VECS{\sigma}, T) \rangle_{\VECS{\sigma}}
	\equiv \sum_{\VECS{\sigma}} \Delta G_{ND}(\VECS{\sigma}, T) P^{\text{eq}}(\VECS{\sigma})
} \nonumber \\
	&\approx& \sum_{\{ \VECS{\sigma} | \VEC{f}(\VECS{\sigma}) = \overline{\VEC{f}(\VECS{\sigma}_N)} \} }
		 \Delta G_{ND}(\VECS{\sigma}, T) P^{\text{eq}}(\VECS{\sigma})
	\\
	&=& \langle G_{N}(\VECS{\sigma}) \rangle_{\VECS{\sigma}} - G_{D}(\overline{\VEC{f}(\VECS{\sigma_N})} , T)
\end{eqnarray}
where the ensemble averages of $G_N(\VECS{\sigma})$ over sequences is also estimated
in the Gaussian approximation\CITE{PGT:97}.
\begin{eqnarray}
\lefteqn{
	\langle G_N(\VECS{\sigma}) \rangle_{\VECS{\sigma}} 	
} \nonumber \\
	&\approx& \int E \exp (- \frac{E }{ k_B T_s}) \, 
		\mathcal{N}(\bar{E}(\overline{\VEC{f}(\VECS{\sigma}_N)}), \delta E^2(\overline{\VEC{f}(\VECS{\sigma}_N)}) )
		    \, dE 
	\hspace*{2em}
	\\
	&=& \bar{E}( \overline{\VEC{f}(\VECS{\sigma}_N)} ) - 
		\frac{ {\delta E}^2( \overline{\VEC{f}(\VECS{\sigma}_N)} ) } { k_B T_s }
	\label{\EQ: ensemble_ave_of_G}
\end{eqnarray}

The sample averages of $\Delta G_{ND}(\VECS{\sigma_N}, T)$ and $\psi_N(\VECS{\sigma_N})$ over 
homologous sequences fixed in protein evolution are equal to
their ensemble averages over sequences\CITE{M:17}.
\begin{eqnarray}
\lefteqn{
\overline{\Delta G_{ND}(\VECS{\sigma_N}, T) } / (k_B T_s)
} \nonumber \\
	&=& 
	\langle \Delta G_{ND}(\VECS{\sigma}, T) \rangle_{\VECS{\sigma}} / (k_B T_s) 
	\\
\SUPPLEMENT{
	&\approx& \, [ \, {\delta E}^2( \overline{\VEC{f}(\VECS{\sigma_N})} ) 
	\, [ \, \vartheta(T/T_g) T_s / T - 1 \, ] / ( k_B T_s )^2
	\hspace*{3em}
	\\
	&=&
}
\TEXT{
	&\approx& 
}
	{\delta \psi}^2( \overline{\VEC{f}(\VECS{\sigma_N})} ) \, [ \,
	\vartheta(T/T_g) T_s / T - 1 \, ]
	\label{\EQ: ensemble_ave_of_ddG}
	\\
\SUPPLEMENT{
	&=& \overline{ \Delta G_{ND}(\VECS{\sigma_N}, T_g) } \, / \, ( k_B T^{\prime}_s )
	\\
	T^{\prime}_s &=& T_s (T_s/T_g - 1) / ( \vartheta(T/T_g) T_s / T - 1 )
	\\
}
\TEXT{
}
	\overline{ \psi_N(\VECS{\sigma_N}) } 
	&\equiv &
		\frac{\sum_{\VECS{\sigma}_N} w_{\VECS{\sigma}_N} \psi_N(\VECS{\sigma}_N) }{\sum_{\VECS{\sigma}_N} w_{\VECS{\sigma}_N} } 
		\\
	&=&
	\langle \psi_N(\VECS{\sigma}) \rangle_{\VECS{\sigma}} 	
	\approx \bar{\psi}( \overline{\VEC{f}(\VECS{\sigma_N})} ) - {\delta \psi}^2( \overline{\VEC{f}(\VECS{\sigma_N})} )
\end{eqnarray}
where the sample averages are calculated with a sample weight 
$w_{\VECS{\sigma}_N}$ for each homologous sequence,
which is used to reduce phylogenetic biases in the set of homologous sequences.
\SUPPLEMENT{
$\Delta G_{ND}(\VECS{\sigma_N}, T_g)$ corresponds to the energy gap\CITE{SG:93a} 
between the native and the glass states, and
$T^{\prime}_s$ will be the selective temperature
if $\Delta G_{ND}(\VECS{\sigma_N}, T_g)$ is used for selection instead of $\Delta G_{ND}(\VECS{\sigma_N}, T)$.
}

The folding free energy becomes equal to zero at the melting temperature $T_m$; 
$\langle \Delta G_{ND}(\VECS{\sigma_N}, T_m) \rangle_{\VECS{\sigma}} = 0$.  Thus, the following relationship must be 
satisfied\CITE{SG:93a,SG:93b,RS:94,PGT:97}.
\begin{equation}
	\vartheta(\frac{T_m}{T_g}) \frac{T_s}{T_m} = \frac{T_s}{2T_m}(1 + \frac{T_m^2}{T_g^2}) = 1
	\hspace*{1em} \textrm{ with } T_s \leq T_g \leq T_m
	\label{\EQ: relationship_among_characteristic_T}
\end{equation}


\newcommand{\norm}[1]{\left\lVert#1\right\rVert}
\newcommand{\argmin}{\text{argmin}}
\newcommand{\smalltext}[1]{\text{\script{#1}}}

\SUBSECTION{Boltzmann machine learning}
\label{section: BML}

The cross entropy with a regularization term, $S$, which corresponds to a negative log-posterior-probability per instance, is minimized.
\begin{eqnarray}
	S &\equiv& \frac{- 1}{\sum_{\tau} 1} \sum_{\tau} \log P(\VECS{\sigma}^{\tau}) + R
\end{eqnarray}
where $R$ is a regularization term, and $\tau$ denotes an instance. 
According to \CITE{FBW:18}, instead of $h_i$ and $J_{ij}$, 
we use the new parameters $\phi_{i}$ and $\phi_{ij}$ for minimization, 
which are Lagrange multipliers in the maximum entropy model corresponding
to $[ \sum_{\VECS{\sigma}}P(\VECS{\sigma})\delta_{\sigma_i a_k} - P_i(a_k) ]$ and
$[ \sum_{\VECS{\sigma}} P(\VECS{\sigma}) \delta_{\sigma_i a_k} \delta_{\sigma_j a_l} 
- P_{ij}(a_k,a_l) 
- \sum_{\VECS{\sigma}} P(\VECS{\sigma}) \delta_{\sigma_i a_k} P_j(a_l)
- P_i(a_k) \sum_{\VECS{\sigma}} P(\VECS{\sigma}) \delta_{\sigma_j a_l}
+ 2 P_i(a_k) P_j(a_l) ] $
in the maximum entropy model, respectively.
The partial derivatives of the cross entropy can be easily calculated:
\begin{eqnarray}
	\frac{\partial S}{\partial \phi_i(a_k)} &=& \sum_{\VECS{\sigma}} P(\VECS{\sigma}) \delta_{\sigma_i a_k}  - P_i(a_k)
					+ \frac{\partial R}{\partial \phi_i(a_k)} 
	\label{\EQ: gradients-h}
		\\
	\frac{\partial S}{\partial \phi_{ij}(a_k, a_l)} &=& 
	[
	\sum_{\VECS{\sigma}} P(\VECS{\sigma}) \delta_{\sigma_i a_k} \delta_{\sigma_j a_l} - P_{ij}(a_k,a_l) 
	\nonumber
	\\
 	& &
	- \sum_{\VECS{\sigma}} P(\VECS{\sigma}) \delta_{\sigma_i a_k} P_j(a_l)
	- P_i(a_k) \sum_{\VECS{\sigma}} P(\VECS{\sigma}) \delta_{\sigma_j a_l}
	\hspace*{1em}
	\nonumber
	\\
	& &
	+ 2 P_i(a_k) P_j(a_l) ]
	]
					+ \frac{\partial R}{\partial \phi_{ij}(a_k, a_l)} 
	\label{\EQ: gradients-J}
\end{eqnarray}
The relationships between $(h_i, J_{ij})$ and $(\phi_i, \phi_{ij})$
are as follows.
\begin{eqnarray}
	h_i(a_k) &=& \phi_i(a_k) - \sum_{j (\neq i) } \sum_l \phi_{ij}(a_k, a_l)) P_j(a_l)
	\\
	J_{ij}(a_k, a_l) &=& \phi_{ij}(a_k, a_l)
\end{eqnarray}

The single-site and two-site frequencies, $P_i(a_k)$ and $P_{ij}(a_k, a_l)$, are evaluated from
homologous sequences, each of which has a sample weight $w_{\VECS{\sigma}_{N}}$,
in a multiple sequence alignment.
\begin{eqnarray}
	P_i(a_k) &=& \sum_{\VECS{\sigma}_N} w_{\VECS{\sigma}_N} \delta_{\sigma_{N,i} a_k} / \sum_{\VECS{\sigma}_N} w_{\VECS{\sigma}_N}
		\\
	P_{ij}(a_k,a_l) &=& \sum_{\VECS{\sigma}_N} w_{\VECS{\sigma}_N} \delta_{\sigma_{N,i} a_k} \delta_{\sigma_{N,j} a_l} / \sum_{\VECS{\sigma}_N} w_{\VECS{\sigma}_N}
\end{eqnarray}
where $\VECS{\sigma}_N$ denotes natural sequences.

$\sum_{\VECS{\sigma}} P(\VECS{\sigma})  \delta_{\sigma_i a_k} $
and
$\sum_{\VECS{\sigma}} P(\VECS{\sigma})  \delta_{\sigma_i a_k} \delta_{\sigma_j a_l}$
are estimated by a Markov chain Monte Carlo method
with the Metropolis-Hastings algorithm\CITE{MRRTT:53,H:70},
and then a gradient-descent algorithm is used to minimize the cross entropy $S$;
the Metropolis-Hastings algorithm was employed
rather than the Gibbs sampler\CITE{GG:84}, because
calculating full conditionals require more computation time.

\SUBSECTION{Regularization}
\label{section: Regularization}

Couplings $\phi_{ij}(a_k, a_l)$ are expected to be 
significant between residues that are closely located 
in a 3D protein structure and complex.
Thus, they are expected to be sparse, because
the number of residue-residue contacts in a protein 3D structure 
is between 2 and 4 per residue depending on a criterion,
in comparison with the number of residue pairs, $L(L-1)/2$, 
where $L$ is a protein length\CITE{MJ:82}.
Here, to take account of the sparsity of the couplings, 
the elastic net \CITE{ZhHt:05,DDR:09,MRMVV:10} and group $L_1$ regularizations
are employed to see the effects of different regularizations.
The  elastic net regularization \CITE{ZhHt:05,DDR:09,MRMVV:10} is used instead of pure $L_1$ regularization,
which is not strictly convex and can occasionally
produce non-unique solutions\CITE{ZhHt:05}.
Group $L_1$ is employed to deal with pairwise couplings, $\phi_{ij}(a_k, a_l)$, 
between residues $i$ and $j$ as a group. 

\SUBSUBSECTION{An elastic net regularization}

An elastic net regularization\CITE{ZhHt:05,DDR:09,MRMVV:10}
is a mixture of $L_1$ and $L_2$.
\begin{eqnarray}
R \equiv
	\lambda_1 \sum_i \sum_k \{ \theta_1 | \phi_i(a_k) | + \frac{(1 - \theta_1)}{2} \phi_i(a_k)^2 \}
	+ \hspace*{5em} 
	\nonumber \\
	\lambda_2 \sum_i \sum_k \sum_{j (> i) }\sum_l \{ \theta_2 | \phi_{ij}(a_k, a_l) | + \frac{(1 - \theta_2)}{2} \phi_{ij}(a_k, a_l)^2 \}
	\label{\EQ: elastic_net}
\end{eqnarray}
where $0 \leq \theta _1 \leq 1$ and $0 \leq \theta _2 \leq 1$.
If $\theta = 0(1)$, the regularization will be $L_2(L_1)$.
In the present work,  $L1$ regularization means the elastic net with $\theta = 0.9$ rather than $1.0$.

\PARAGRAPH{The soft-thresholding function for $L_1$ regularization}

Let us assume that the learning of fields and couplings ($\phi_i$, $\phi_{ij}$) is iteratively updated as
follows.
\begin{eqnarray}
	\phi_{\mu} (t+1) &=& \phi_{\mu}(t) - [ \alpha_{\mu}(t+1) + \beta_{\mu}(t+1) (\frac{\partial S}{\partial \phi_{\mu}})_{\VEC{\phi}(t)} ]
	\hspace*{2em}
		\label{\EQ: parameter_updates}
		\\
	&=&  \phi_{\mu}(t+1 \smalltext{ without } L_1) - \gamma_{\mu}(t+1) (\frac{\partial | \phi_{\mu} | }{\partial \phi_{\mu}})_{\VEC{\phi}(t)}
	\\
	&=&  \text{prox} (\gamma_{\mu}(t+1) | \phi_{\mu} | \, , \,
	\phi_{\mu}(t+1 \smalltext{ without } L_1)
	)
\end{eqnarray}
where 
the suffix $\mu$ denotes $_i(a_k)$ or $_{ij}(a_k, a_l)$,
$\phi_{\mu}(t+1 \smalltext{ without } L_1)$ is $\phi_{\mu}(t+1)$ which does not include the $L_1$ regularization term, and
the second term is one corresponding to the $L_1$ terms of the regularization in \Eq{\ref{\EQ: elastic_net}},
and the derivative in the second term may be evaluated as a subderivative at a singular point.
Here the proximal operator\CITE{Sm:17.540L5} defined as follows is used for faster convergence.
\begin{eqnarray}
	\text{prox} (h(u), x )
	&\equiv&
	\argmin_{u} ( h(u) + \frac{1}{2} \norm{ u - x }_2^2 )
\end{eqnarray}

The proximal operator for $L_1$ regression is equal to:
\begin{eqnarray}
\lefteqn{
	\text{prox} ( \gamma_{\mu}(t+1)  | \phi_{\mu} | \, , \,
	\phi_{\mu}(t+1 \smalltext{ without } L_1)
	)
}
	\nonumber
	\\
	&=& 
        \RED{
	 \argmin_{\phi_{\mu}} (  \gamma_{\mu}(t+1) | \phi_{\mu} | +
        }
	\nonumber
	\\
	& &
        \RED{
	 \hspace*{2em} \frac{1}{2} ( \phi_{\mu} - \phi_{\mu}(t+1 \smalltext{ without } L_1) )^2 )
        }
	\\
	&=& 
        \RED{
	 \frac{ \phi_{\mu}(t+1 \smalltext{ without } L_1) }{ | \phi_{\mu}(t+1 \smalltext{ without } L_1) | }
        }
	\nonumber
	\\
	& &
        \RED{
	 \hspace*{2em} \text{max} \{ 0, | \phi_{\mu}(t+1 \smalltext{ without } L_1) | -  \gamma_{\mu}(t+1) \}
        }
	\\
\lefteqn{
 \gamma_{\mu}(t+1) \equiv \left\{ \begin{array}{ll} 
			\beta_{\mu}(t+1) \lambda_1 \theta_1	& \text{ for } \mu = _i(a_k)
			\\
			\beta_{\mu}(t+1) \lambda_2 \theta_2	& \text{ for } \mu = _{ij}(a_k,a_l)
			\end{array}
			\right.
}
\end{eqnarray}

\SUBSUBSECTION{$L_2$ regularization for $\phi_{i}(a_k)$ and group $L_1$ for $\phi_{ij}(a_k,a_l)$}

The regularization terms of the $L_2$ for $\phi_{i}(a_k)$ and the group $L_1$ for $\phi_{ij}(a_k,a_l)$
are as follows.
\begin{equation}
R \equiv
	\lambda_1 \sum_i \sum_k \frac{1}{2} \{ \phi_i(a_k)^2 \}
	+
	\lambda_2 \sum_i \sum_{j (> i) } \sqrt{\sum_k \sum_l \{ \phi_{ij}(a_k, a_l)^2 \} }
	\hspace*{1em}
	\label{\EQ: group_L1}
\end{equation}

\PARAGRAPH{The soft-thresholding function for group $L_1$ regularization}

\begin{eqnarray}
\lefteqn{
\VEC{\phi_{ij}}(t+1)
	= 
	\text{prox}( \VEC{\gamma_{ij}}(t+1) \norm{ \VEC{\phi_{ij}} }_2 \, , \, \VEC{\phi_{ij}}(t + 1 \smalltext{ without group } L_1 ) )
} 
	\nonumber
	\\
	&=&
        \RED{
	 \argmin_{\VEC{\phi_{ij}}} (  \VEC{\gamma_{ij}}(t+1) \norm{\VEC{\phi_{ij}} }_2  +
        }
	\hspace*{11em}
	\nonumber
	\\
	& &
        \RED{
	 \hspace*{2em} \frac{1}{2} \norm{ \VEC{\phi_{ij}} - \VEC{\phi_{ij}}(t+1 \smalltext{ without } L_1) }_2^2 )
        }
	\\
	&=& 
        \RED{
	 \frac{ \VEC{\phi_{ij}}(t+1 \smalltext{ without } L_1) }{ \norm{ \VEC{\phi_{ij}}(t+1 \smalltext{ without } L_1) }_2 }
        }
	\nonumber
	\\
	& &
        \RED{
	 \hspace*{2em} \text{max} \{ 0, \norm{ \VEC{\phi_{ij}}(t+1 \smalltext{ without } L_1) }_2 -  \VEC{\gamma_{ij}}(t+1) \}
        }
	\\
\lefteqn{
 \VEC{\gamma_{ij}}(t+1) 
	\equiv \VEC{\beta_{ij}}(t+1) \lambda_2
		\label{\EQ: def_gamma_ij}
}
\end{eqnarray}

\SUBSECTION{Parameter updates}
\label{section: Parameter_updates}

Given the convexity of the cross entropy function, its minimum can be found by
the gradient descent.

\SUBSUBSECTION{Modified Adam method (ModAdam)}

The modified version of the adaptive learning rate method\CITE{KB:14}, which is named ModAdam here, has been used.

\begin{eqnarray}
	m_{\mu}(t+1) &=& \rho_m m_{\mu}(t) + ( 1 - \rho_m ) [ (\frac{\partial S}{\partial \phi_{\mu}})_{\VEC{\phi}(t)} ]
	\hspace*{5em}
	\\
	v_{\mu}(t+1) &=& \rho_v v_{\mu}(t) + ( 1 - \rho_v ) [ (\frac{\partial S}{\partial \phi_{\mu}})_{\VEC{\phi}(t)} ]^2
	\\
	\kappa(t+1) &=& \kappa_0 \frac{(1-\rho_v^{t+1})^{1/2}}{1-\rho_m^{t+1}} \frac{1}{ \max_{\mu}(v_{\mu}(t+1)^{1/2}) + \epsilon  }
	\label{\EQ: ModAdam}
	\\
	\phi_{\mu}(t+1) &=& \phi_{\mu}(t) - \kappa(t+1) m_{\mu}(t+1)
\end{eqnarray}
where $\kappa_0$ is an initial learning rate, and 
$\rho_m$, $\rho_v$, and $\epsilon/(1-\rho_v^{t+1})^{1/2}$ 
have been set to 0.9, 0.999, and $10^{-8}$ according to the Adam method\CITE{KB:14}.
It should be noted here that unlike the Adam method $\kappa(t+1)$ 
takes the same value for all parameters, because $v_{\mu}(t+1)^{1/2}$ is replaced by its maximum
in \Eq{\ref{\EQ: ModAdam}}\RED{.}

An important property of Adam's update rule is its careful choice of stepsizes.
The effective stepsize is upper bounded by 
$
|\Delta \phi(t + 1) | \leq \kappa_0 \max((1- \rho_m) / \sqrt{(1 - \rho_v )}, 1)
$ \CITE{KB:14}
but essentially all elements of the increment vector $\Delta \phi(t + 1)$ are the same order.
However, unlike the original Adam, in which 
$\Delta \phi_{\mu}(t+1) = - \kappa_0 (\sqrt{(1-\rho_v^{t+1})}/ (1-\rho_m^{t+1}) ) ( m_{\mu}(t+1) / (v_{\mu}(t+1)^{1/2} + \epsilon) )$,
in this modified version the increment $\Delta \VEC{\phi}(t+1)$ is proportional to $- \VEC{m}(t+1)$.

Thus, $\alpha_{\mu}$ and $\beta_{\mu}$ in \Eq{\ref{\EQ: parameter_updates}} are defined as follows.
\begin{eqnarray}
	\alpha_{\mu}(t + 1) &=& \kappa(t+1) \rho_m m_{\mu}(t)
	\\
	\beta_{\mu}(t + 1) &=& \kappa(t+1) (1 - \rho_m)
\end{eqnarray}

\ifdefined\NAG

\SUBSUBSECTION{Nesterov's Accelerated Momentum/Gradient method (NAG)}

The algorithm of Nesterov's Accelerated Momentum/Gradient method (NAG) \CITE{N:04}
employed here is a simple version with the constant friction for velocity as follows.
This version includes a correction, which is employed in the Adam method, 
for the bias that the estimate of the first moment of the gradients 
will be biased towards zero if it is initialized as zero.
\begin{eqnarray}
m_{\mu}(t+1) &=& \rho_m m_{\mu}(t) + (1 - \rho_m) [ ( \frac{\partial S}{\partial \phi_\mu} )_{\VEC{\phi}(t)} ] 
	\\
\phi_{\mu}(t+1) 
	&=& \phi_{\mu}(t) - \kappa_0 \, [ (1 + \rho_m) m_{\mu}(t+1) - \rho_m  m_{\mu}(t)  ]
	\hspace*{4em}
	\\
	&=& \phi_{\mu}(t) - \kappa_0 \, [ \rho_m^2 m_{\mu}(t) +
		( 1 - \rho_m^2) ( \frac{\partial S}{\partial \phi_\mu} )_{\VEC{\phi}(t)} ]
	\\
	&=& \phi_{\mu}(t) - \kappa_0 \, [ \rho_m m_{\mu}(t + 1) +
		( 1 - \rho_m) ( \frac{\partial S}{\partial \phi_\mu} )_{\VEC{\phi}(t)} ]
	\label{eq: NAG_parameter_updates}
\end{eqnarray}
where $\kappa_0$ is an initial learning rate, and $\rho_m$ has been set
to $0.95$.
The $\alpha_{\mu}$ and $\beta_{\mu}$ in \Eq{\ref{\EQ: parameter_updates}} 
for the NAG method are defined as follows
by replacing $m_{\mu}(t + 1)$ in \Eq{\ref{eq: NAG_parameter_updates}} by its estimate, 
$m_{\mu}(t + 1)/(1-\rho_m^{t+1})$, for the initial condition, $m(0) \equiv 0$.
\begin{eqnarray}
	\alpha_{\mu}(t+1) 
		&=&
		\kappa_0 \, \rho_m^2 \, [ \, m_{\mu}(t) / ( 1 - \rho_m^{t+1} ) \, ]
		\\
	\beta_{\mu}(t+1) &=&
		\kappa_0 \, [ \, \rho_m / (1 - \rho_m^{t+1}) + 1 \, ] \, (1 - \rho_m) 
\end{eqnarray}

\fi

\SUBSUBSECTION{The number of iterations for learning}
\label{section: Iterations}

The objective function is expected to significantly fluctuate 
in the minimization process, when the first-order methods based on 
gradients are employed. In addition,
the partial derivatives of \Eqs{\ref{\EQ: gradients-h} and \ref{\EQ: gradients-J}},
which are calculated from the pairwise marginal distributions
estimated by Markov Chain Monte Carlo samplings,
include statistical errors.
Thus, even though the learning rate $\kappa$ is sufficiently small, 
the cross entropy/log-likelihood are not monotonically improved.
However, the cross entropy/log-likelihood can hardly be evaluated for the Boltzmann machine, 
although its partial-derivatives can be easily calculated and then it can be minimized/maximized. 
Thus, it is not obvious to judge which set of interactions is the best in the learning process.

Here we monitor the average, $D_{2}^{KL}$, of Kullback-Leibler divergences for pairwise marginal distributions 
over all residue pairs as a rough measure of fitting to the reference distribution.
\begin{equation}
D_{2}^{KL} 
	\equiv
	 \frac{2}{L(L-1)} \sum_{i} \sum_{j > i} 
		\sum_{k} \sum_{l} P_{ij}(a_k, a_l) \log \frac{ (P_{ij}(a_k, a_l) + \epsilon ) }
			{ (\sum_{\VECS{\sigma}} P(\VECS{\sigma}) \delta_{\sigma_i a_k} \delta_{\sigma_j a_l} + \epsilon) }
	\label{\EQ: KL2}
\end{equation}
\begin{eqnarray}
	D_{1}^{KL} \equiv \frac{1}{L} \sum_{i}
		\sum_{k} P_{i}(a_k) \log \frac{ ( P_{i}(a_k) + \epsilon) }
		{ ( \sum_{\VECS{\sigma}} P(\VECS{\sigma}) \delta_{\sigma_i a_k} + \epsilon ) }
	\hspace*{5em}
	\label{\EQ: KL1}
\end{eqnarray}
where $\epsilon = 10^{-5}$ is employed to prevent the logarithm of zero.
The iteration of parameter updates has been stopped when $\min D_{2}^{KL}$  
over the iteration numbers larger than 1000 does not improve during a certain number, 100, of iterations, 
and the number of iterations passes over a certain threshold, 1200 iterations.
Then the fields and couplings and Monte Carlo samples corresponding to the $\min D_{2}^{KL}$
over the iteration numbers larger than 1000 are selected.


\SUBSECTION{A gauge employed to compare $h_i(a_k)$ and $J_{ij}(a_k, a_l)$ between various models}
\label{sec: Ising_gauge_for_comparison}

The $\psi_N$ of \Eq{\ref{\EQ: psi_for_Potts}} is invariant
under a certain transformation of fields and couplings,
$J_{ij}(a_k,a_l) \rightarrow J_{ij}(a_k,a_l) - J^1_{ij}(a_k) - J^1_{ji}(a_l) + J^0_{ij}$,
$h_i(a_k) \rightarrow h_i(a_k) - h^0_i + \sum_{j\neq i} J^1_{ij}(a_k)$ for any
$J^1_{ij}(a_k)$, $J^0_{ij}$ and $h^0_i$.
Therefore, in order to compare $h$ and $J$ between various models, 
a certain gauge must be used. Here we use the following gauge that 
we call the Ising gauge.
\begin{eqnarray}
        h_i(\cdot) &=& \sum_q J_{ij}(a_k, \cdot) = \sum_q J_{ij}(a_q, \cdot) = 0
        \label{\EQ: Ising gauge}
\end{eqnarray}
where ``$\cdot$'' denotes the reference state, which is the average over all states 
for the Ising gauge. Any gauge can be transformed to this gauge by the following transformation.
\begin{eqnarray}
J^{\script{I}}_{ij}(a_k,a_l) &\equiv& J_{ij}(a_k,a_l) - J_{ij}(\cdot,a_l)
			- J_{ij}(a_k, \cdot) +  J_{ij}(\cdot,\cdot)
		\hspace*{3em}
		\\
h^{\script{I}}_{i}(a_k) &\equiv& h_{i}(a_k) - h_{i}(\cdot) + 
	\sum_{j \neq i} (J_{ij}(a_k, \cdot) - J_{ij}(\cdot, \cdot) )
\end{eqnarray}


\TEXT{

\TCBB{

}{

}

\TCBB{
}{

}

\TCBB{
}{

}

}

\SkipSupplToMerge{
\SUPPLEMENT{

\SECTION{Figures}

\vspace*{4em}

\begin{figure*}[!hb]
\centerline{
\includegraphics[width=63mm,angle=0]{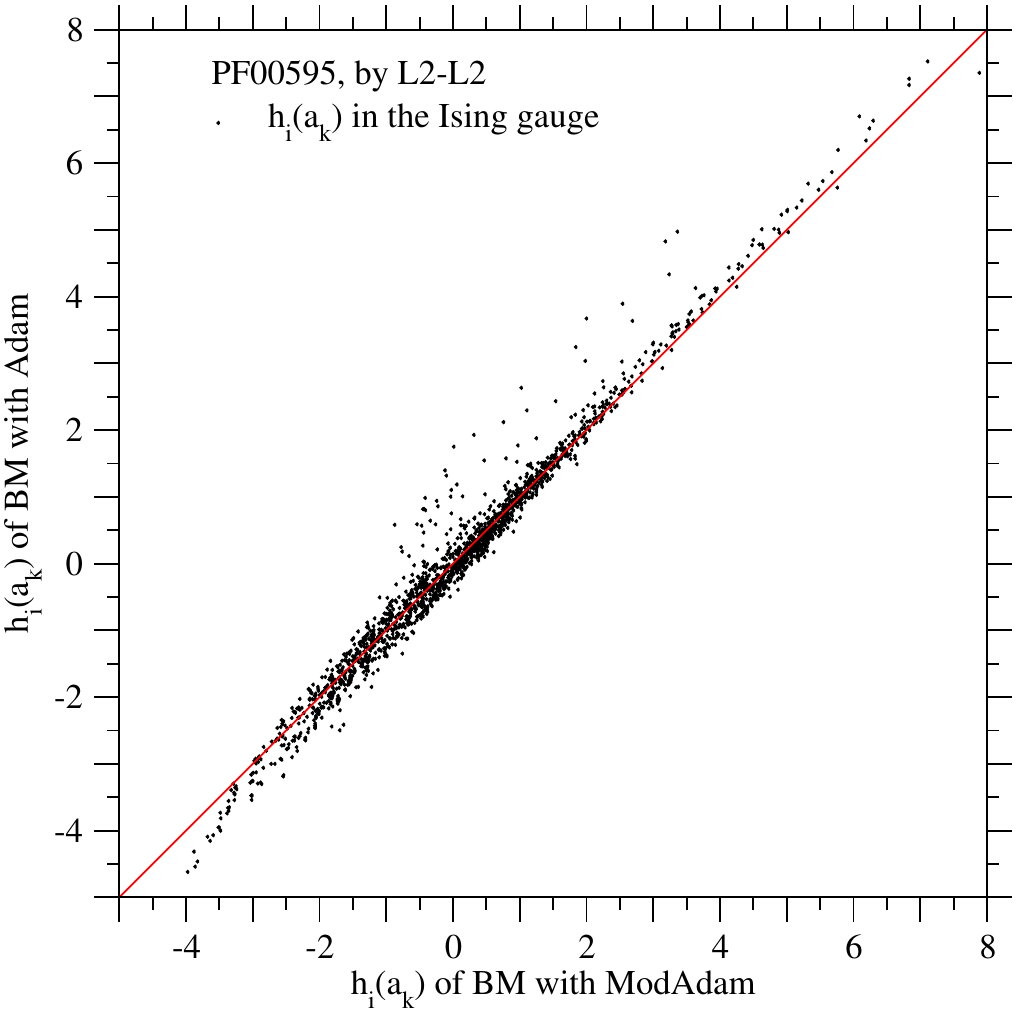}
\includegraphics[width=63mm,angle=0]{Figs_1/L2L2_PF00595uniq_25_1_0_1_Adam_again_hJ_1012_in_Ising_vs_ModAdam_1119_J}
}
\centerline{
\includegraphics[width=63mm,angle=0]{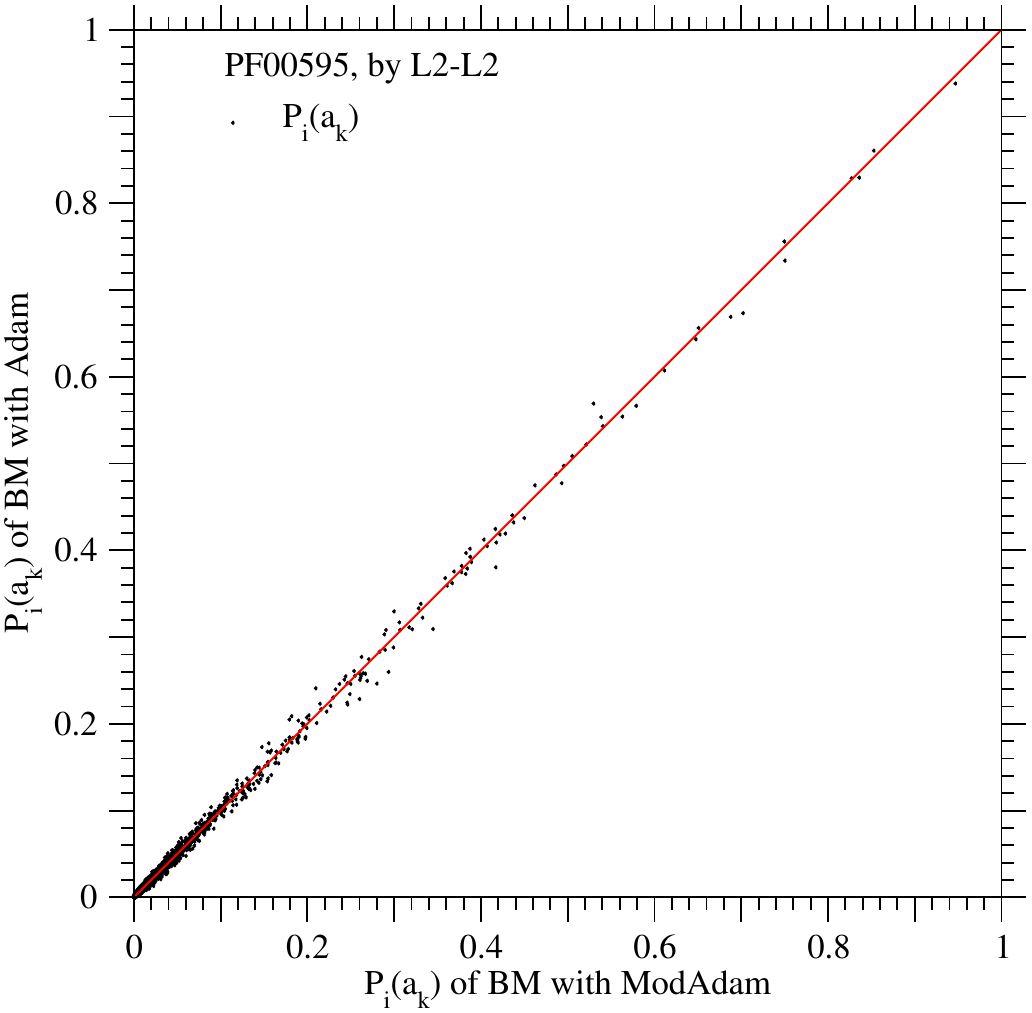}
\includegraphics[width=63mm,angle=0]{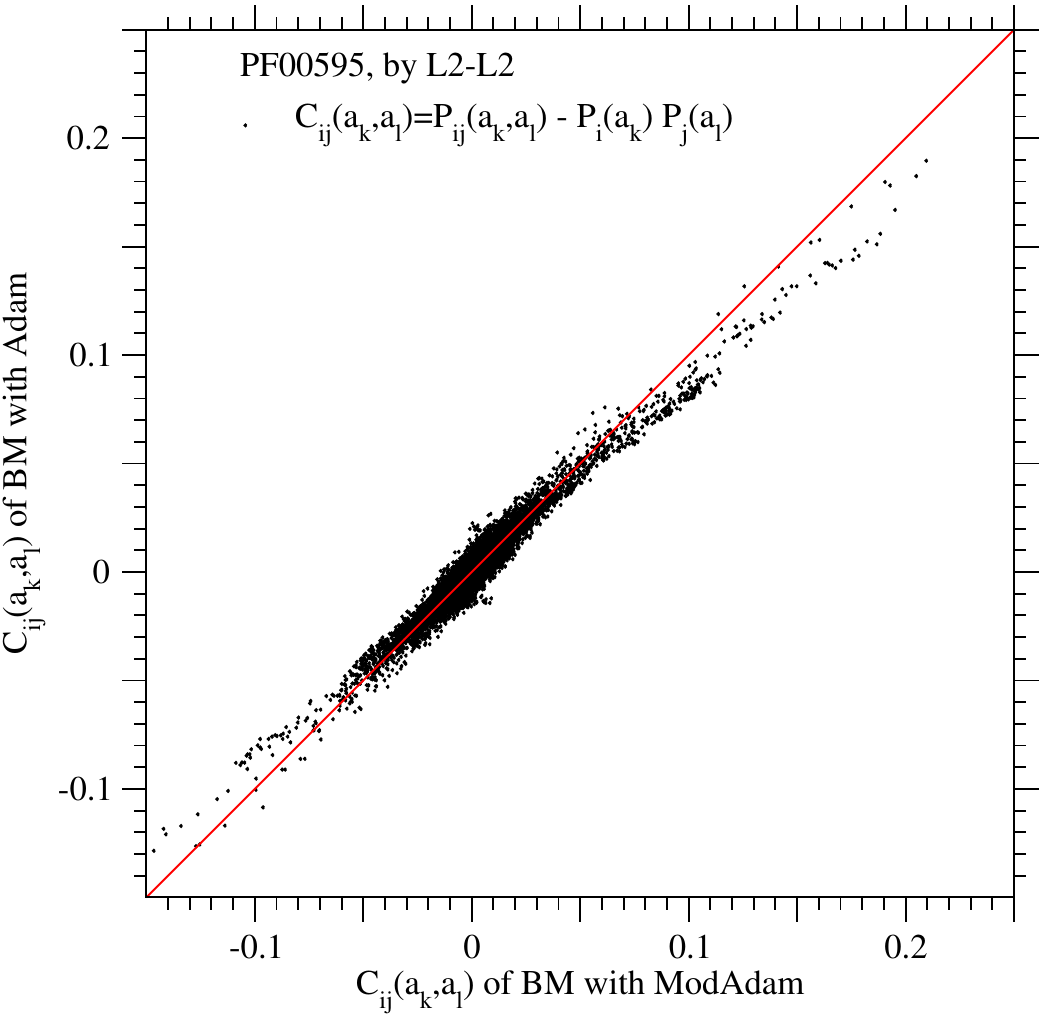}
}
\caption
{
\label{sfig: PF00595_ModAdam_vs_Adam_hJ}
\label{sfig: PF00595_ModAdam_vs_Adam_hJ_PiaCijab}
\noindent
\TEXTBF{Comparison of the Adam with the ModAdam gradient-descent method
in each of the inferred fields and couplings and the recovered single-site marginals and pairwise correlations for PF00595.}
The upper left and upper right figures are the comparisons of the inferred fields and couplings in the Ising gauge, respectively,
and 
the lower left and lower right figures are the comparisons of the recovered single-site frequencies and pairwise correlations, respectively.
The abscissas and ordinates correspond to 
the quantities estimated by
the modified Adam and Adam methods for gradient descent, respectively.
The regularization model L2-L2 is employed for both methods.
The solid lines show the equal values between the ordinate and abscissa.
The values of hyper-parameters are listed in \Table{\ref{tbl: PF00595_parameters}}.
The overlapped points of $J_{ij}(a_k,a_l)$ 
in the units 0.001
and of $C_{ij}(a_k,a_l)$ in the units 0.0001 are removed.
}
\end{figure*}

\begin{figure*}[hbt]
\centerline{
\includegraphics[width=63mm,angle=0]{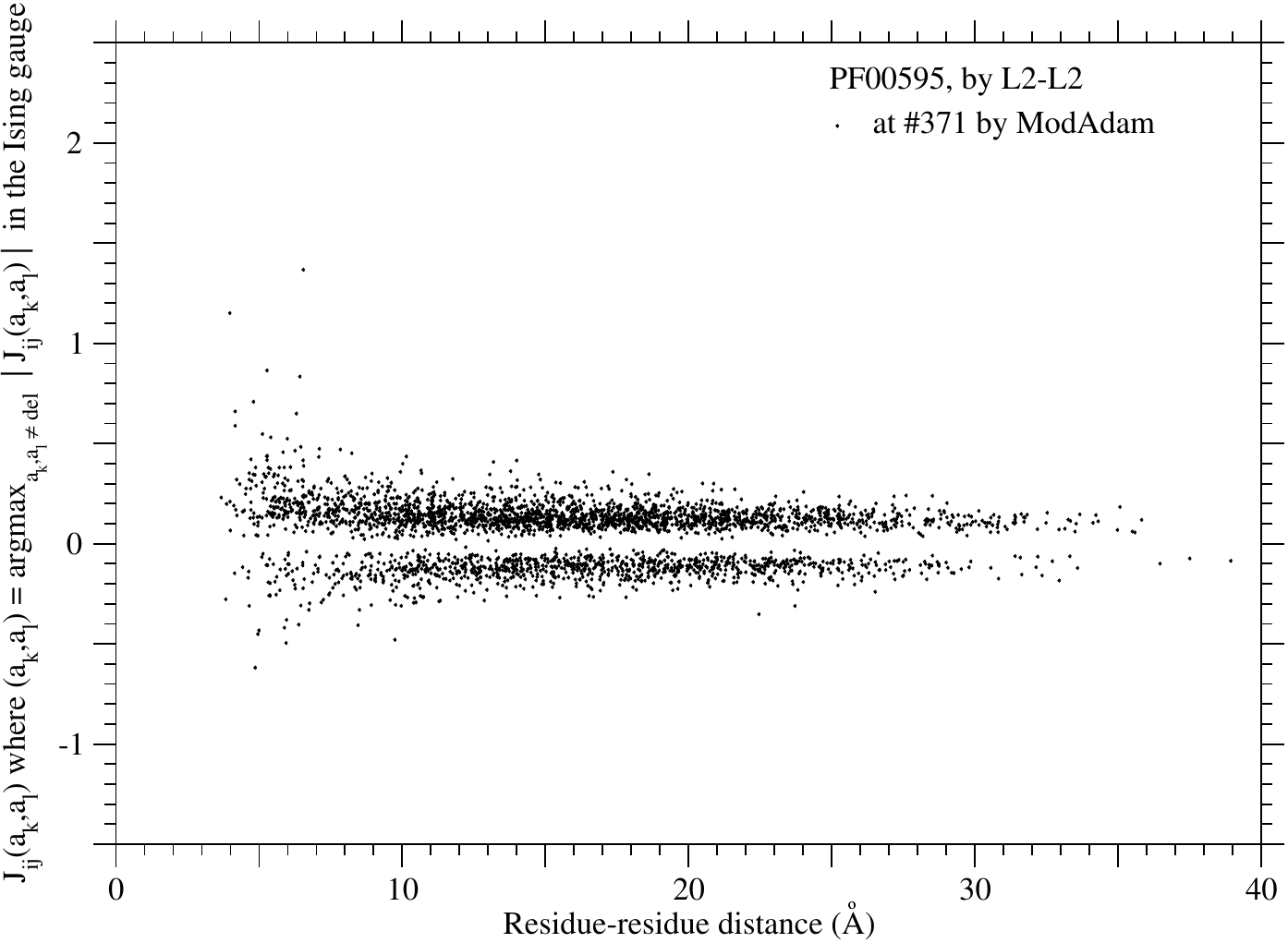}
\includegraphics[width=63mm,angle=0]{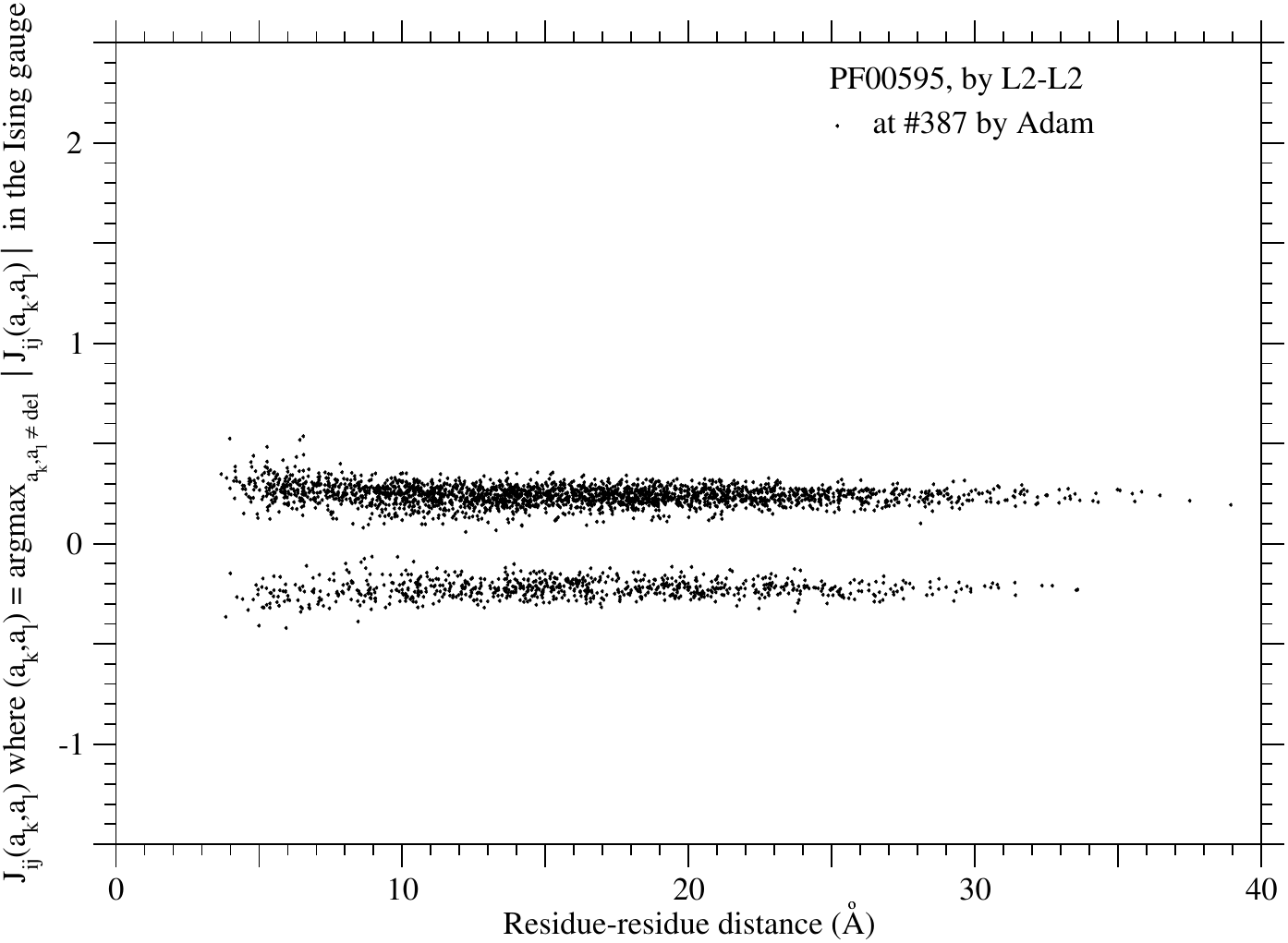}
}
\vspace*{1em}
\centerline{
\includegraphics[width=63mm,angle=0]{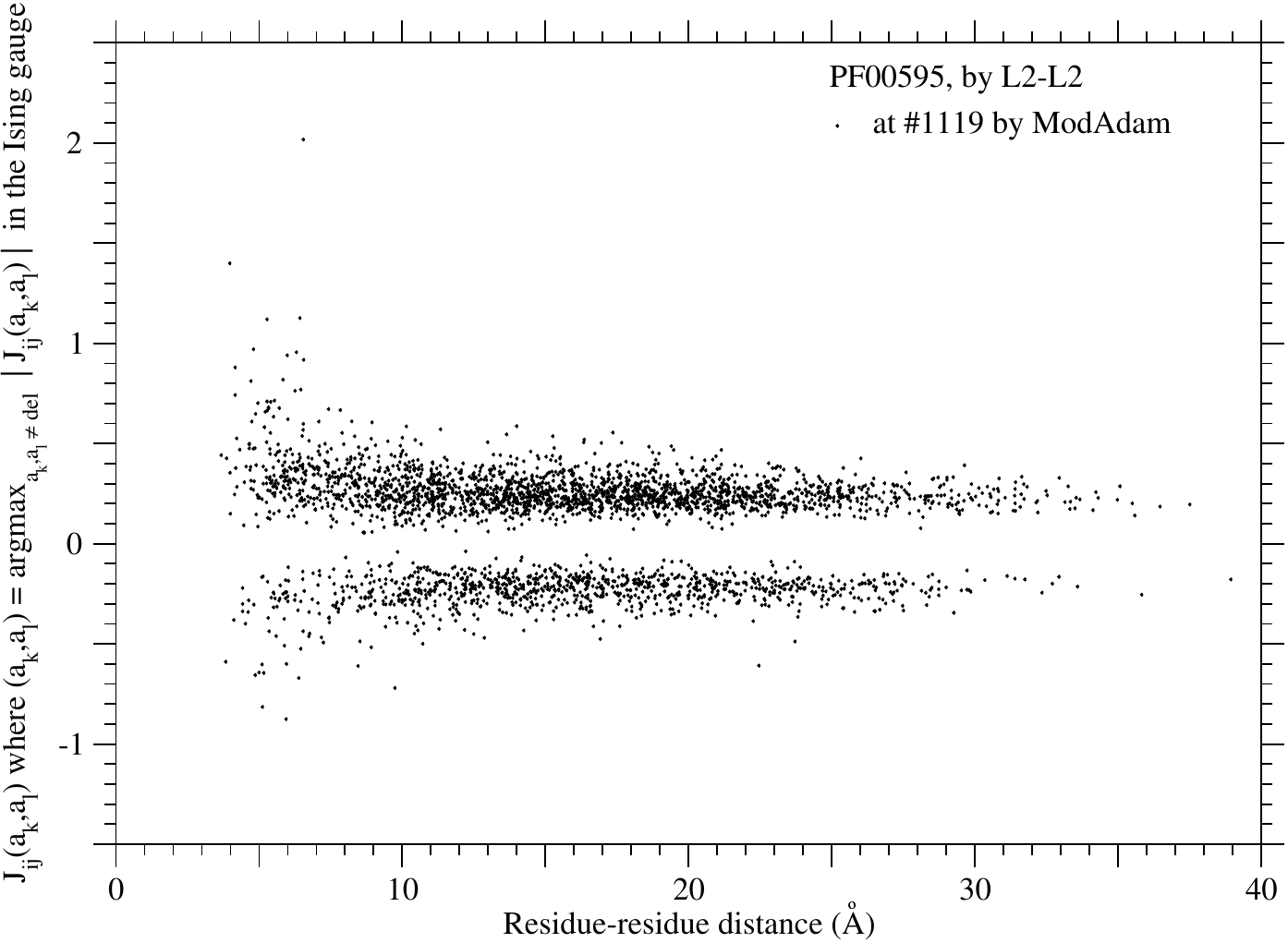}
\includegraphics[width=63mm,angle=0]{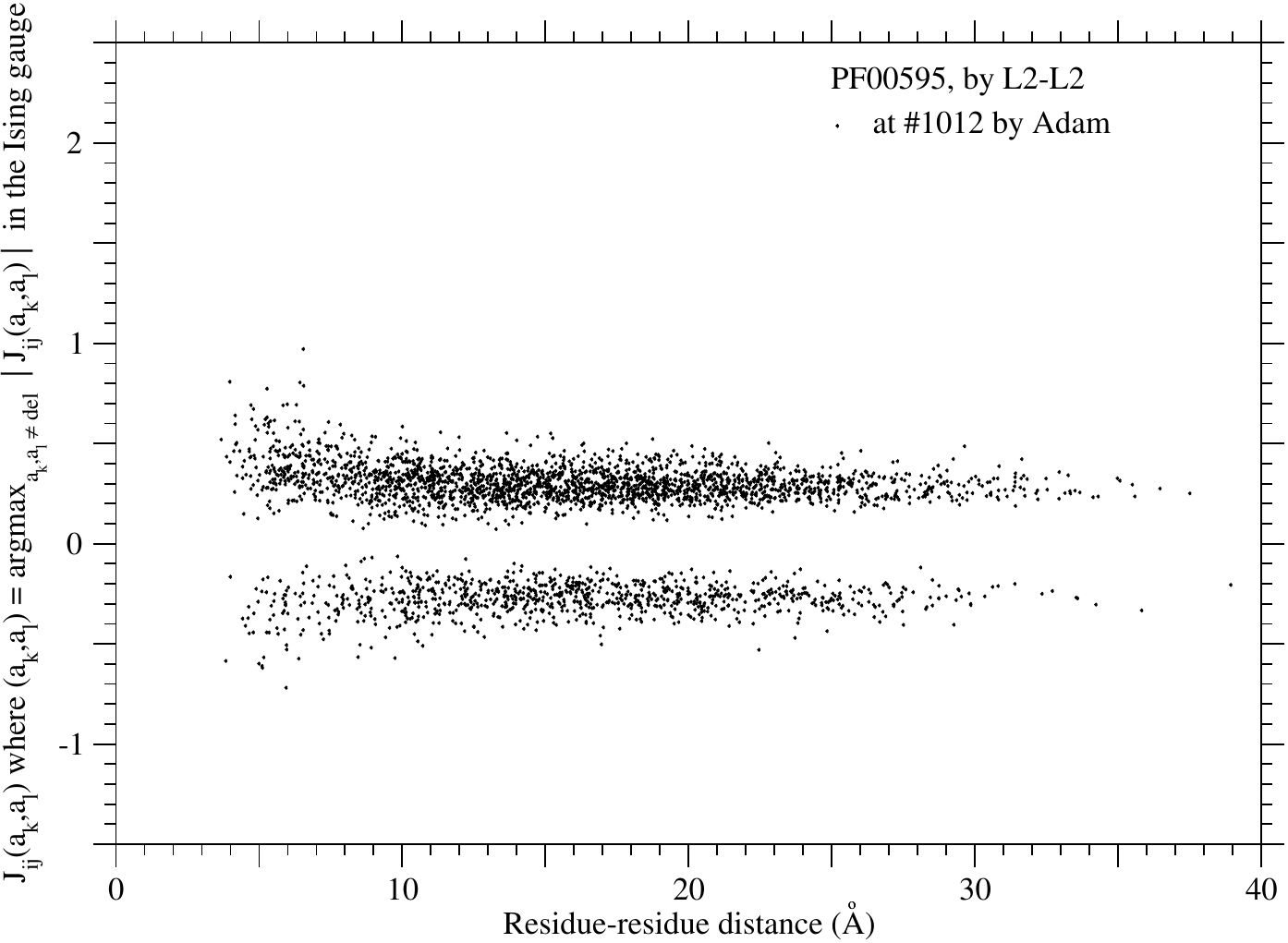}
}
\caption
{
\label{sfig: PF00595_maxJij_comparison}
\noindent
\TEXTBF{Differences in the learning of coupling parameters, $J_{ij}(a_k,a_l)$,
between the ModAdam and Adam gradient-descent methods for PF00595.}
All $J_{ij}(a_k,a_l)$ where $(a_k,a_l) = \text{argmax}_{a_k,a_l \neq \text{deletion}} | J_{ij}(a_k,a_l) |$ 
in the Ising gauge
are plotted
against the distance between $i$th and $j$th residues.
The upper left and lower left figures are for the iteration numbers
371 and 1119
in a learning process by the modified Adam method, respectively.
The upper right and lower right figures are for the iteration numbers 387 and 1012
in a learning process by the Adam method, respectively.
These iteration numbers correspond to $\min D^{\text{KL}}_2$ over the iteration numbers smaller than 400
and those over the iteration numbers larger than 1000.
The regularization model L2-L2 is employed for both methods.
The learning processes by both methods are shown in
\Figs{\ref{fig: PF00595_learning_process_KL_by_ModAdam_and_Adam}
and \ref{fig: average_energy_comparison}}.
Please notice that 
more strong couplings tend to be inferred for closely located residues pairs by 
the modified Adam method than by the Adam method.
The values of hyper-parameters are listed in \Table{\ref{tbl: PF00595_parameters}}.
}
\end{figure*}
\begin{figure*}[!hb]
\centerline{
\includegraphics[width=63mm,angle=0]{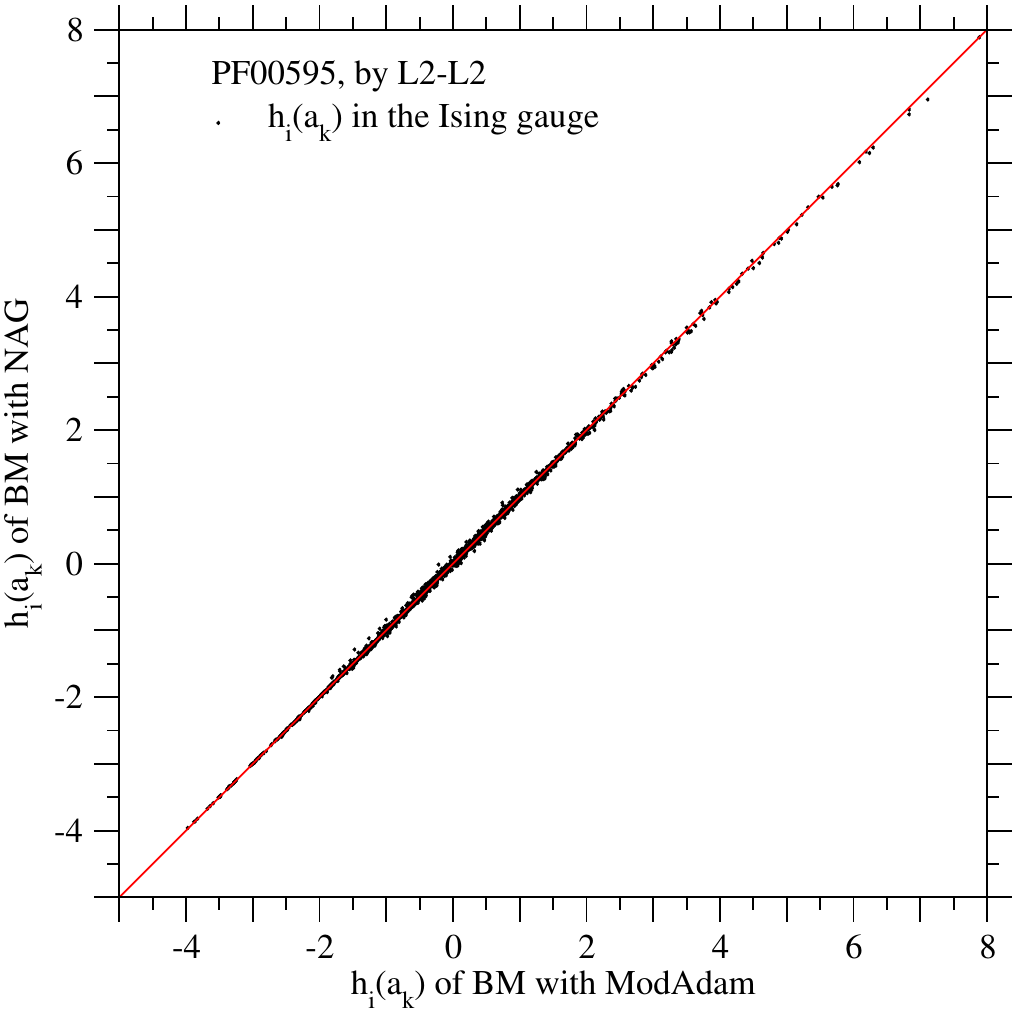}
\includegraphics[width=63mm,angle=0]{Figs_1/L2L2_PF00595uniq_25_1_0_1_NAG0_95_0_1_2000_hJ_1110_in_Ising_vs_ModAdam_1119_J}
}
\centerline{
\includegraphics[width=63mm,angle=0]{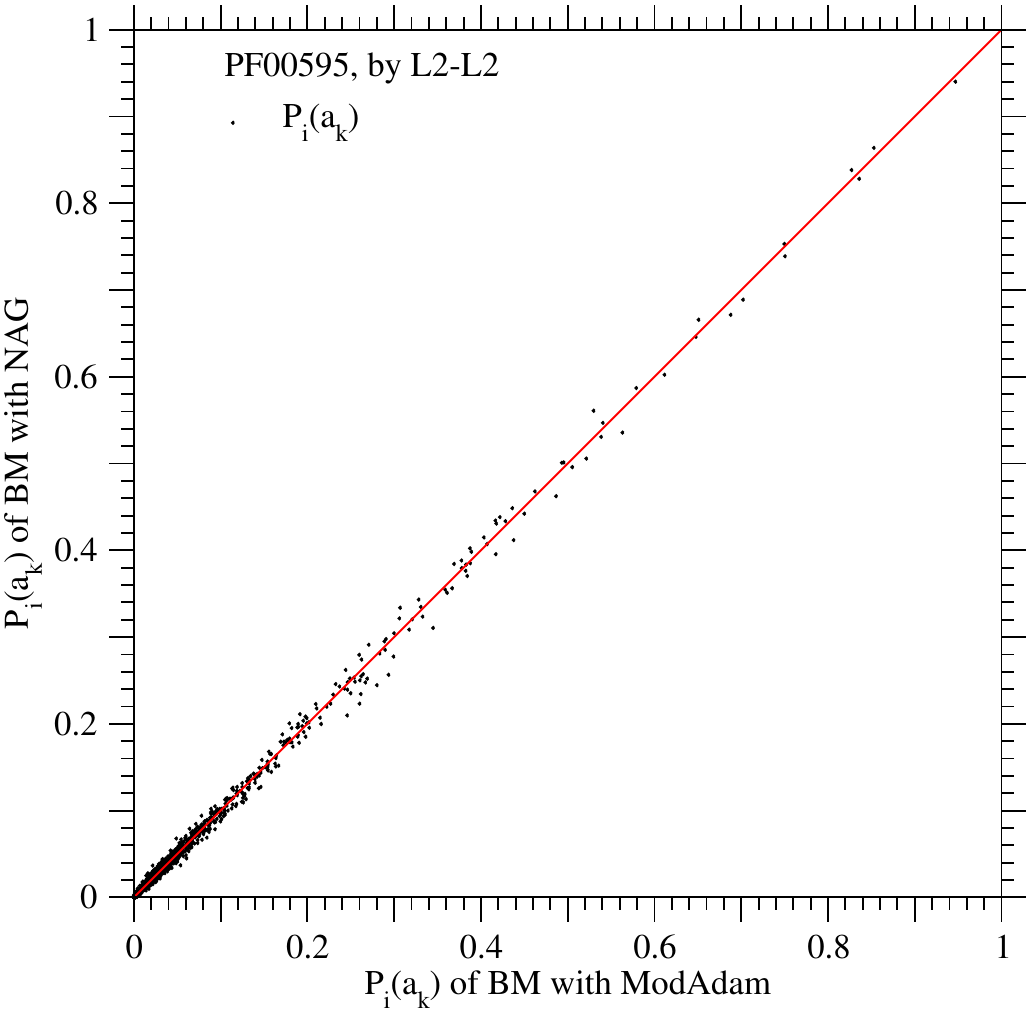}
\includegraphics[width=63mm,angle=0]{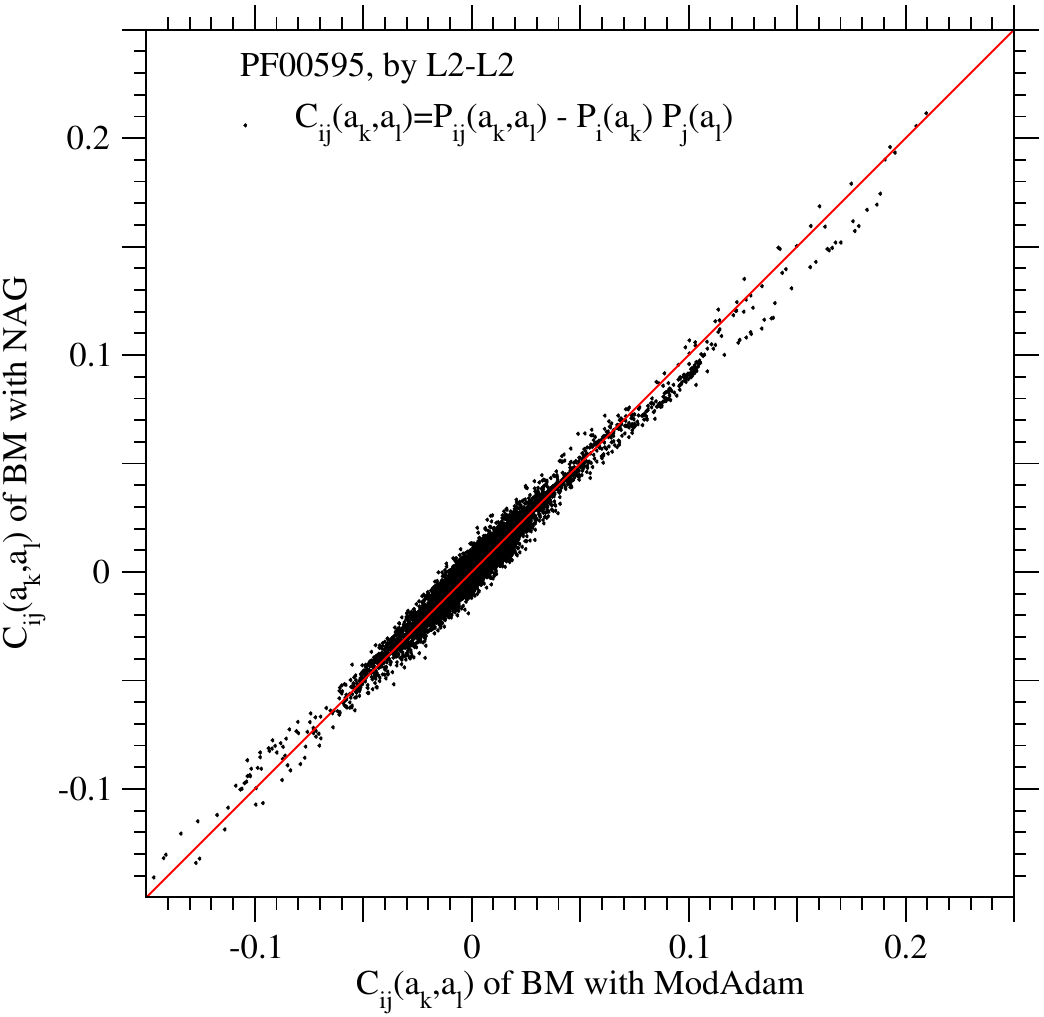}
}
\caption
{
\label{sfig: PF00595_ModAdam_vs_NAG_hJ}
\label{sfig: PF00595_ModAdam_vs_NAG_hJ_PiaCijab}
\noindent
\TEXTBF{Comparison of the NAG with the ModAdam gradient-descent method
in each of the inferred fields and couplings and the recovered single-site marginals and pairwise correlations for PF00595.}
The upper left and upper right figures are the comparisons of the inferred fields and couplings in the Ising gauge, respectively,
and
the lower left and lower right figures are the comparisons of the recovered single-site frequencies and pairwise correlations, respectively.
The abscissas and ordinates correspond to 
the quantities estimated by
the modified Adam and NAG methods for gradient descent, respectively.
The regularization model L2-L2 is employed for both methods.
The solid lines show the equal values between the ordinate and abscissa.
The values of hyper-parameters are listed in \Table{\ref{tbl: PF00595_parameters}}.
The overlapped points of $J_{ij}(a_k,a_l)$ 
in the units 0.001
and of $C_{ij}(a_k,a_l)$ in the units 0.0001 are removed.
}
\end{figure*}

\begin{figure*}[hbt]
\centerline{
\includegraphics[width=63mm,angle=0]{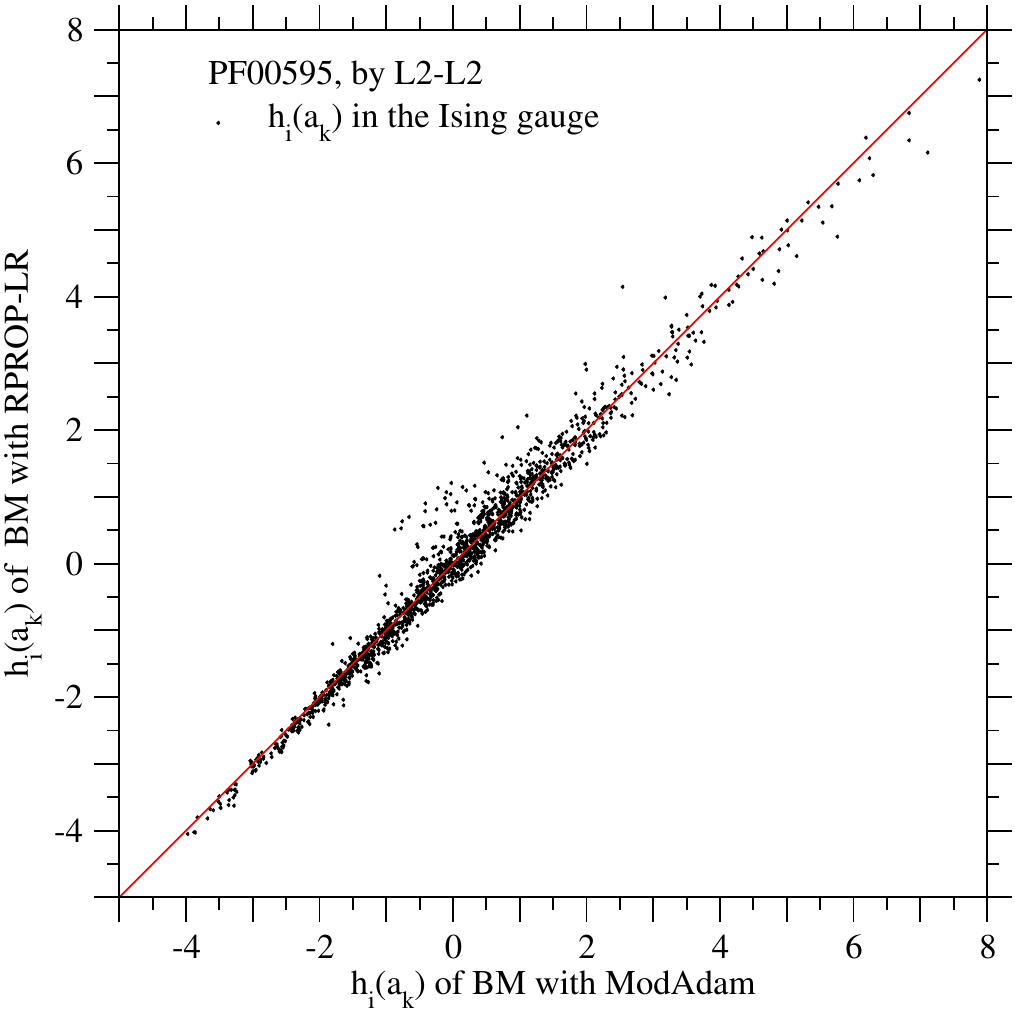}
\includegraphics[width=63mm,angle=0]{Figs_1/L2L2_PF00595uniq_25_1_0_1_MF0_00001-0_01-10_0_again_hJ_1052_in_Ising_vs_ModAdam_1119_J}
}
\centerline{
\includegraphics[width=63mm,angle=0]{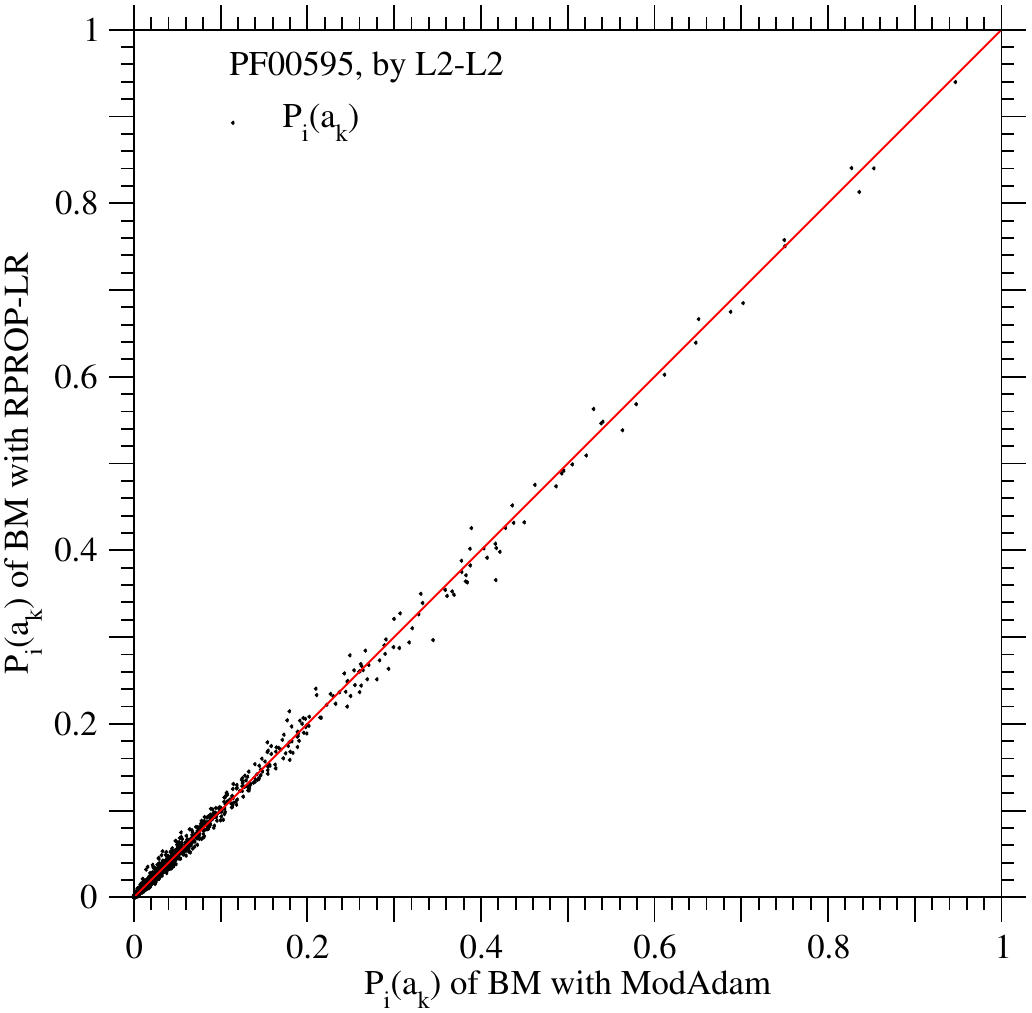}
\includegraphics[width=63mm,angle=0]{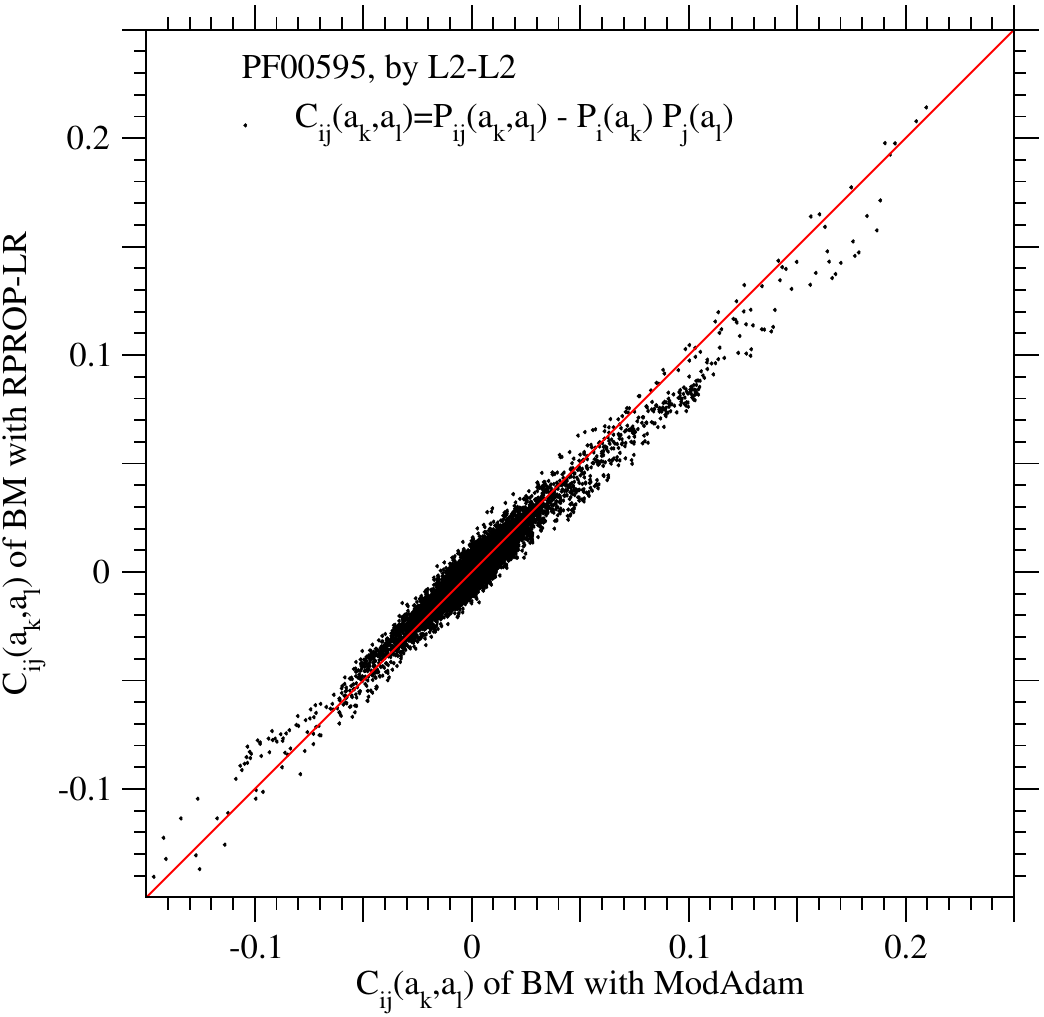}
}
\caption
{
\label{sfig: PF00595_ModAdam_vs_RPROP-LR_hJ}
\label{sfig: PF00595_ModAdam_vs_RPROP-LR_hJ_PiaCijab}
\noindent
\TEXTBF{Comparison of the RPROP-LR with the ModAdam gradient-descent method
in each of the inferred fields and couplings and the recovered single-site marginals and pairwise correlations
for PF00595.}
The upper left and upper right figures are the comparisons of the inferred fields and couplings in the Ising gauge, respectively,
and
the lower left and lower right figures are the comparisons of the recovered single-site frequencies and pairwise correlations, respectively.
The abscissas and ordinates correspond to
the quantities estimated by
the modified Adam and RPROP-LR method for gradient descent, respectively.
The regularization model L2-L2 is employed for both methods.
The solid lines show the equal values between the ordinate and abscissa.
The values of hyper-parameters are listed in \Table{\ref{tbl: PF00595_parameters}}.
The overlapped points of $J_{ij}(a_k,a_l)$ in the units 0.001 
and of $C_{ij}(a_k,a_l)$ in the units 0.0001 are removed.
}
\end{figure*}

\begin{figure*}[hbt]
\centerline{
}
\vspace*{1em}
\centerline{
\includegraphics[width=60mm,angle=0]{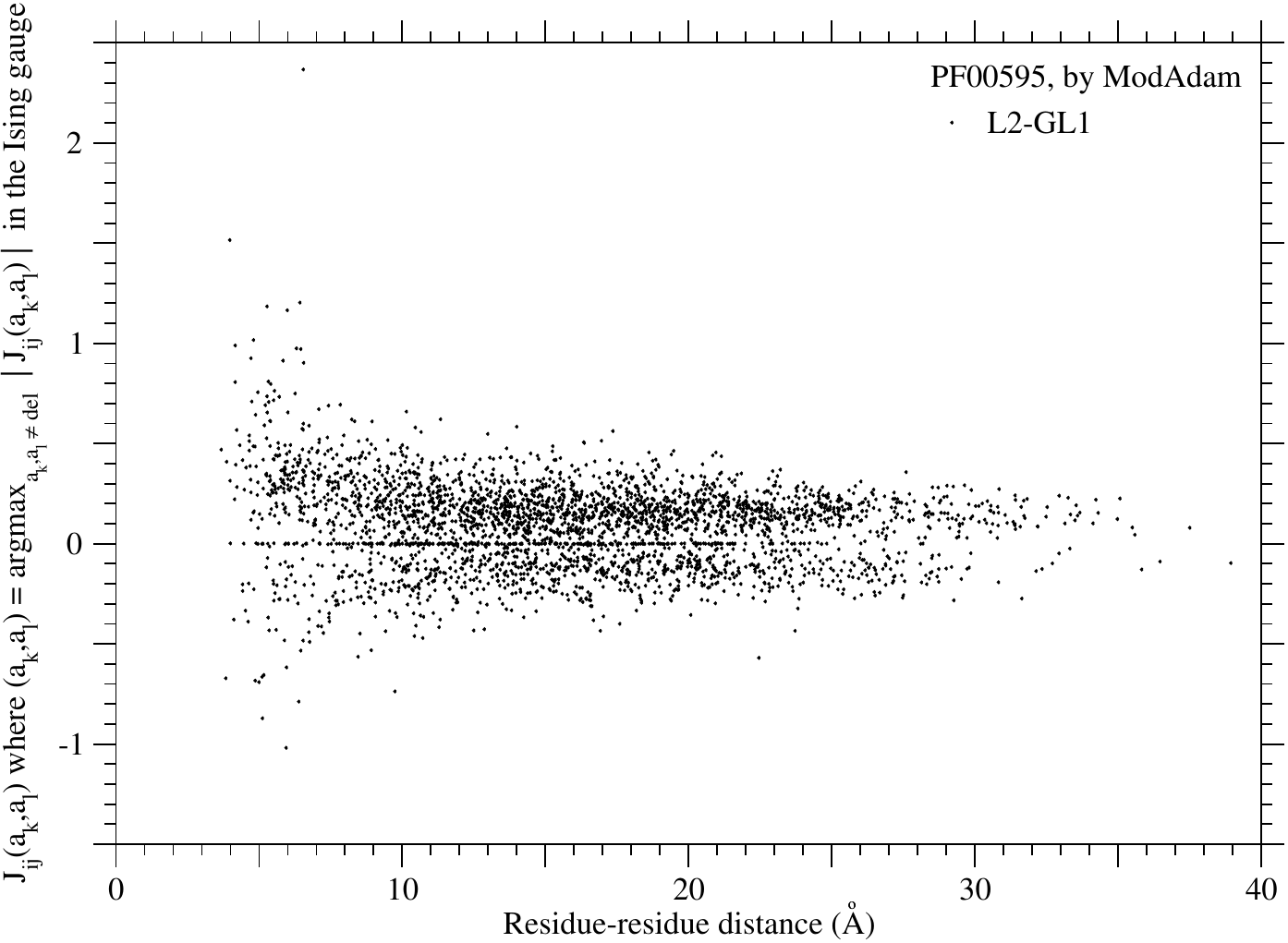}
\includegraphics[width=60mm,angle=0]{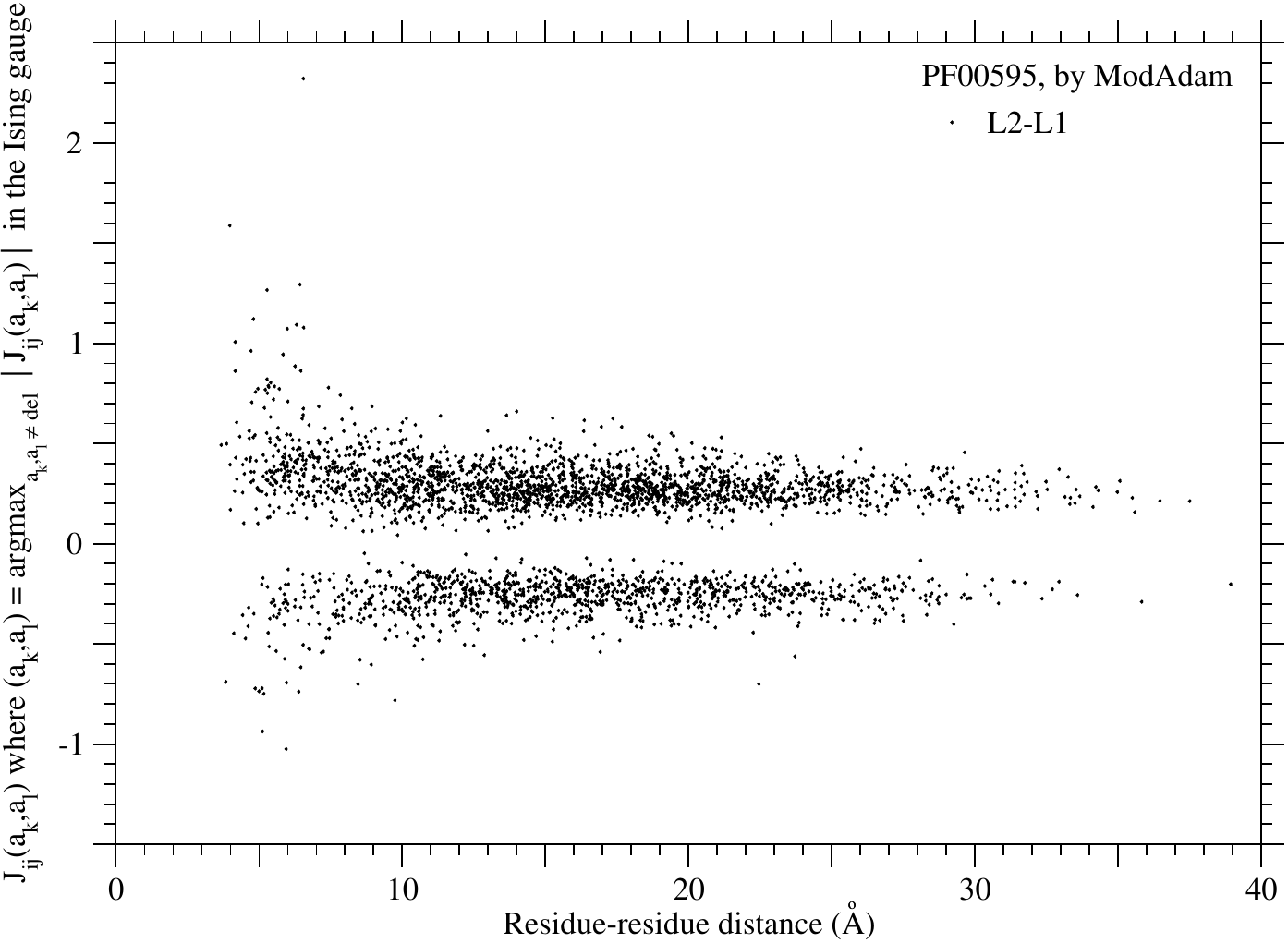}
\includegraphics[width=60mm,angle=0]{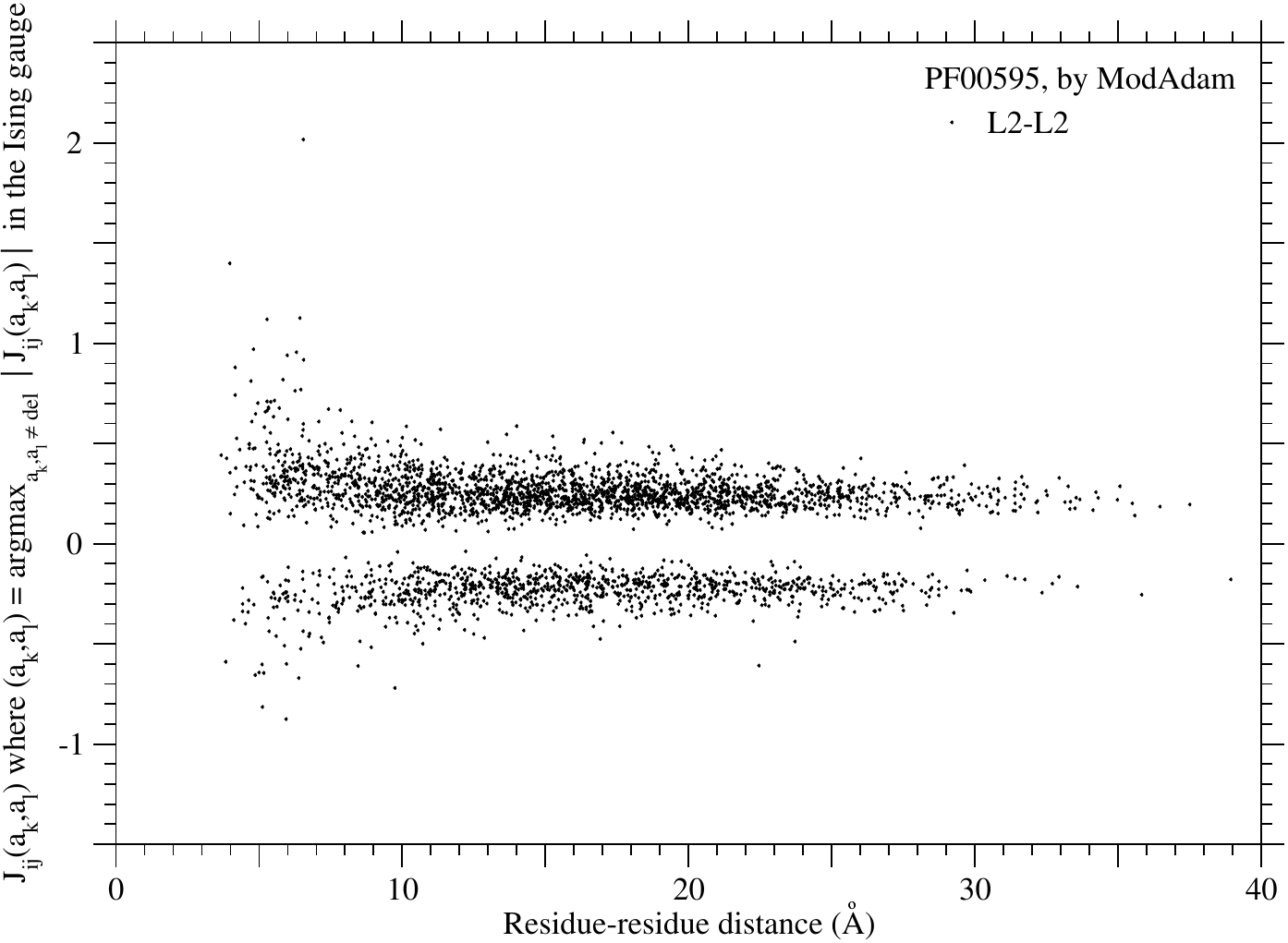}
}
\caption
{
\label{sfig: PF00595_maxJij_vs_rpd}
\noindent
\TEXTBF{Differences of inferred couplings $J_{ij}$ among the regularization models for PF00595.}
All $J_{ij}(a_k,b_l)$ where $(a_k,a_l) = \text{argmax}_{a_k,a_l \neq \text{deletion}} | J_{ij}(a_k,a_l) |$
in the Ising gauge are plotted 
against the distance between $i$th and $j$th residues.
The protein family PF00595 is employed.
The regularization models L2-GL1, L2-L1, and L2-L2 are
employed for the left, middle, and right figures, respectively.
The values of regularization parameters are listed in \Table{\ref{tbl: PF00595_parameters}}.
}
\end{figure*}

\begin{figure*}[hbt]
\centerline{
\includegraphics[width=63mm,angle=0]{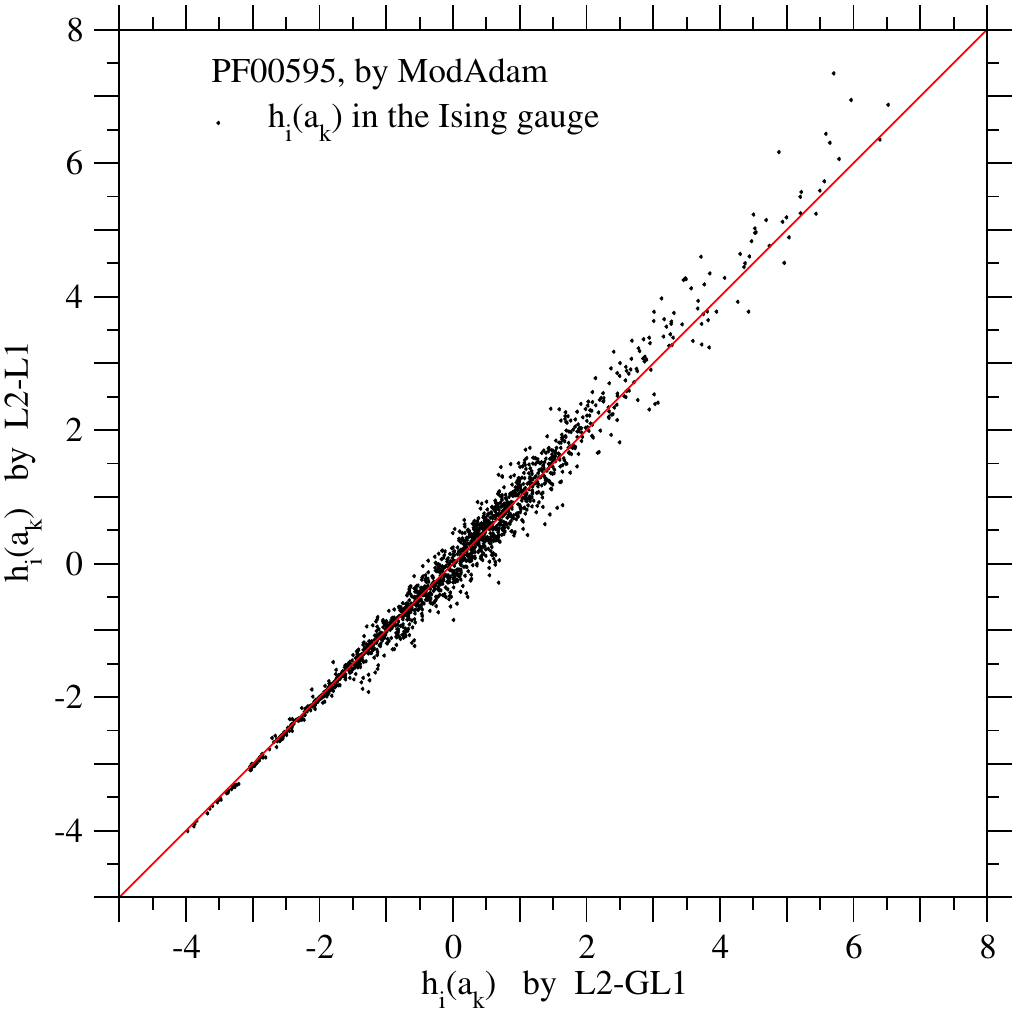}
\includegraphics[width=63mm,angle=0]{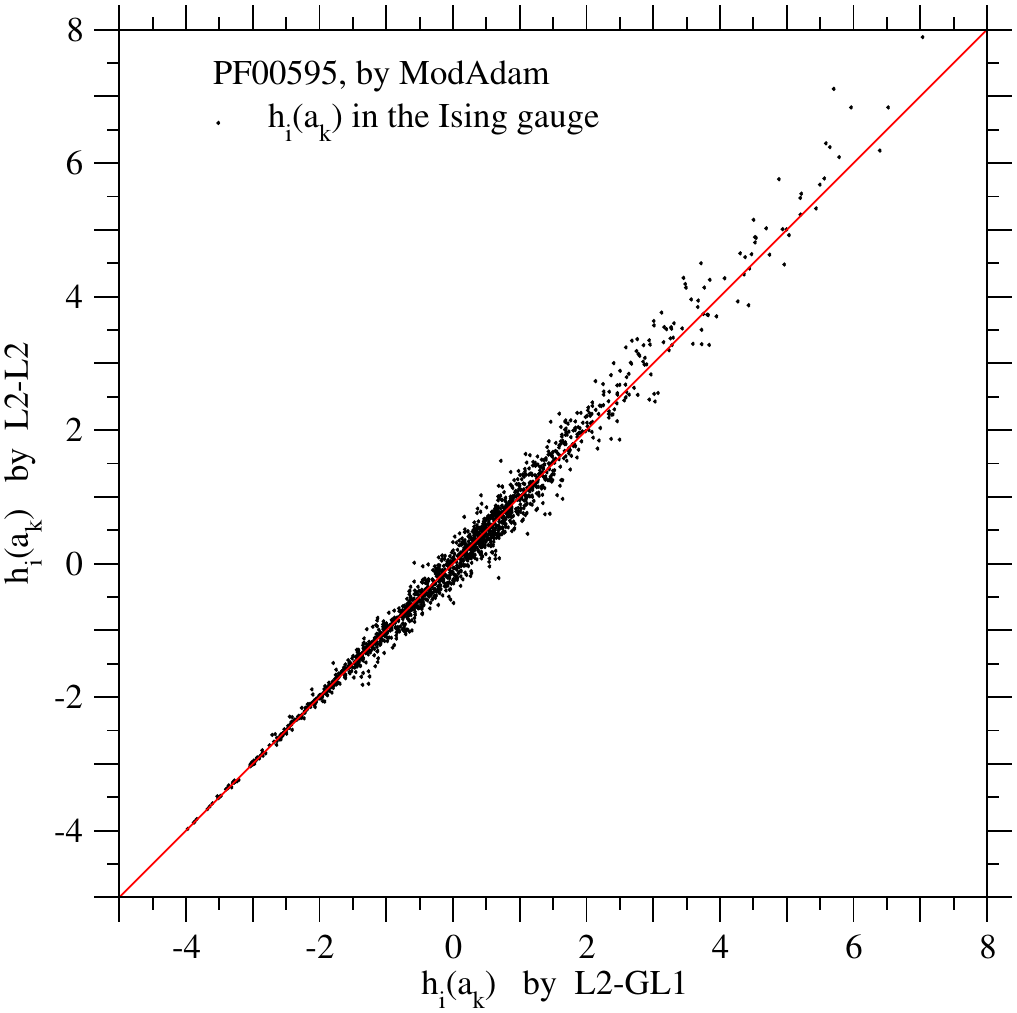}
}
\vspace*{1em}
\centerline{
\includegraphics[width=63mm,angle=0]{Figs_1/L2GL1_PF00595uniq_40_0_0_1_hJ_1162_in_Ising_vs_L2L1_J}
\includegraphics[width=63mm,angle=0]{Figs_1/L2GL1_PF00595uniq_40_0_0_1_hJ_1162_in_Ising_vs_L2L2_J}
}
\caption
{
\label{sfig: PF00595_hJ_comparison}
\noindent
\TEXTBF{Comparisons of inferred fields $h_i(a_k)$ and couplings $J_{ij}(a_k,a_l)$ 
in the Ising gauge
between the regularization models for PF00595.}
The upper and lower figures show the comparisons of fields and couplings in the Ising gauge,
respectively.
All abscissa correspond to the fields or couplings
inferred by the L2-GL1.
The ordinates in the left and right figures 
correspond to the fields or couplings
inferred by the L2-L1 and L2-L2 models,
respectively.
The values of regularization parameters are listed in \Table{\ref{tbl: PF00595_parameters}}.
The solid lines show the equal values between the ordinate and abscissa.
The overlapped points of $J_{ij}(a_k,a_l)$ in the units 0.001 are removed.
}
\end{figure*}

\begin{figure*}[hbt]
\centerline{
\includegraphics[width=63mm,angle=0]{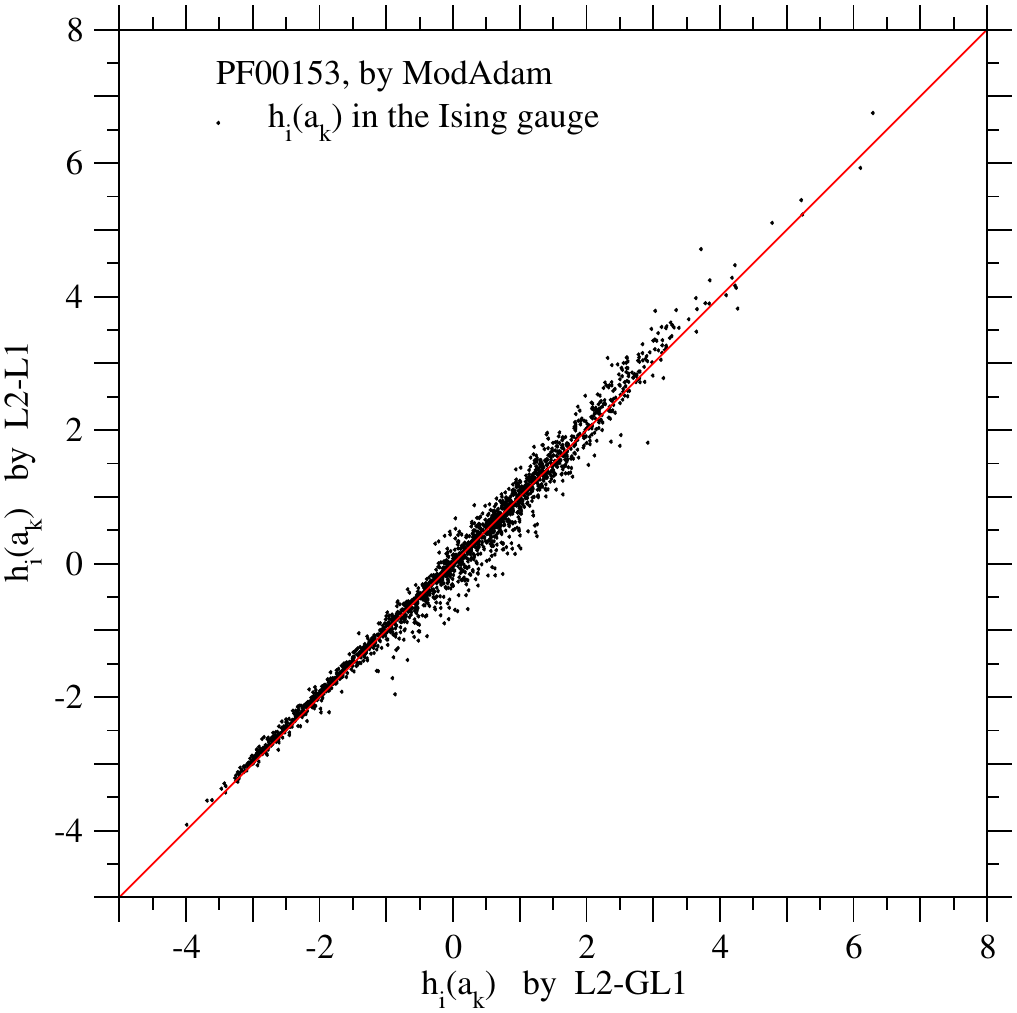}
\includegraphics[width=63mm,angle=0]{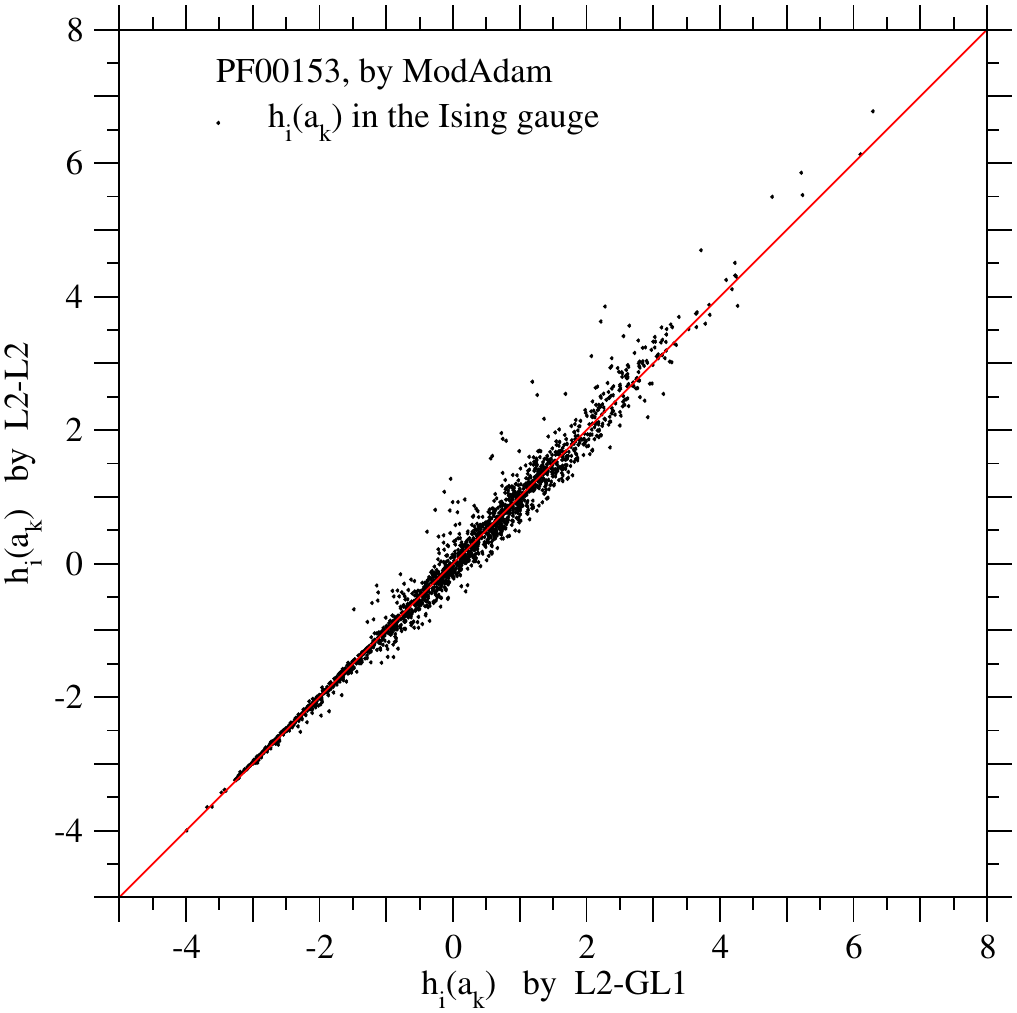}
}
\vspace*{1em}
\centerline{
\includegraphics[width=63mm,angle=0]{Figs_1/L2GL1_PF00153_208_9_0_1_hJ_1445_in_Ising_vs_L2L1_J}
\includegraphics[width=63mm,angle=0]{Figs_1/L2GL1_PF00153_208_9_0_1_hJ_1445_in_Ising_vs_L2L2_J}
}
\caption
{
\label{sfig: PF00153_hJ_comparison}
\noindent
\TEXTBF{Comparisons of inferred fields $h_i(a)$ and couplings $J_{ij}(a,b)$ 
in the Ising gauge
between the regularization models for PF00153.}
The upper and lower figures show the comparisons of fields and couplings
in the Ising gauge,
respectively.
All abscissa correspond to the fields or couplings inferred by the L2-GL1.
The ordinates in the left and right figures 
correspond to the fields or couplings inferred by the L2-L1 and L2-L2 models,
respectively.
The values of regularization parameters are listed in \Table{\ref{tbl: PF00153_parameters}}.
The solid lines show the equal values between the ordinate and abscissa.
The overlapped points of $J_{ij}(a_k,a_l)$ in the units 0.001 are removed.
}
\end{figure*}

\begin{figure*}[hbt]
\centerline{
\includegraphics[width=63mm,angle=0]{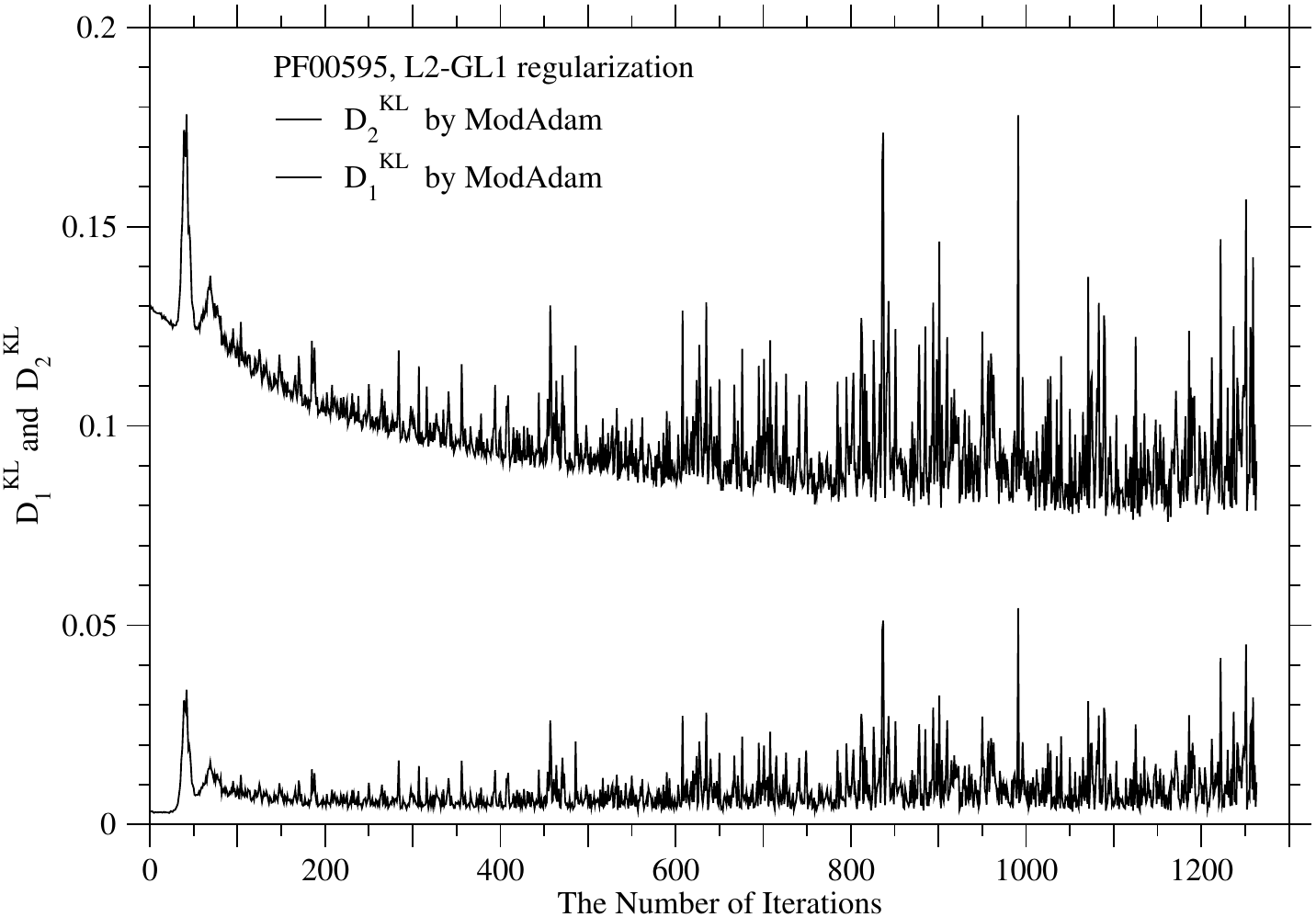}
\includegraphics[width=63mm,angle=0]{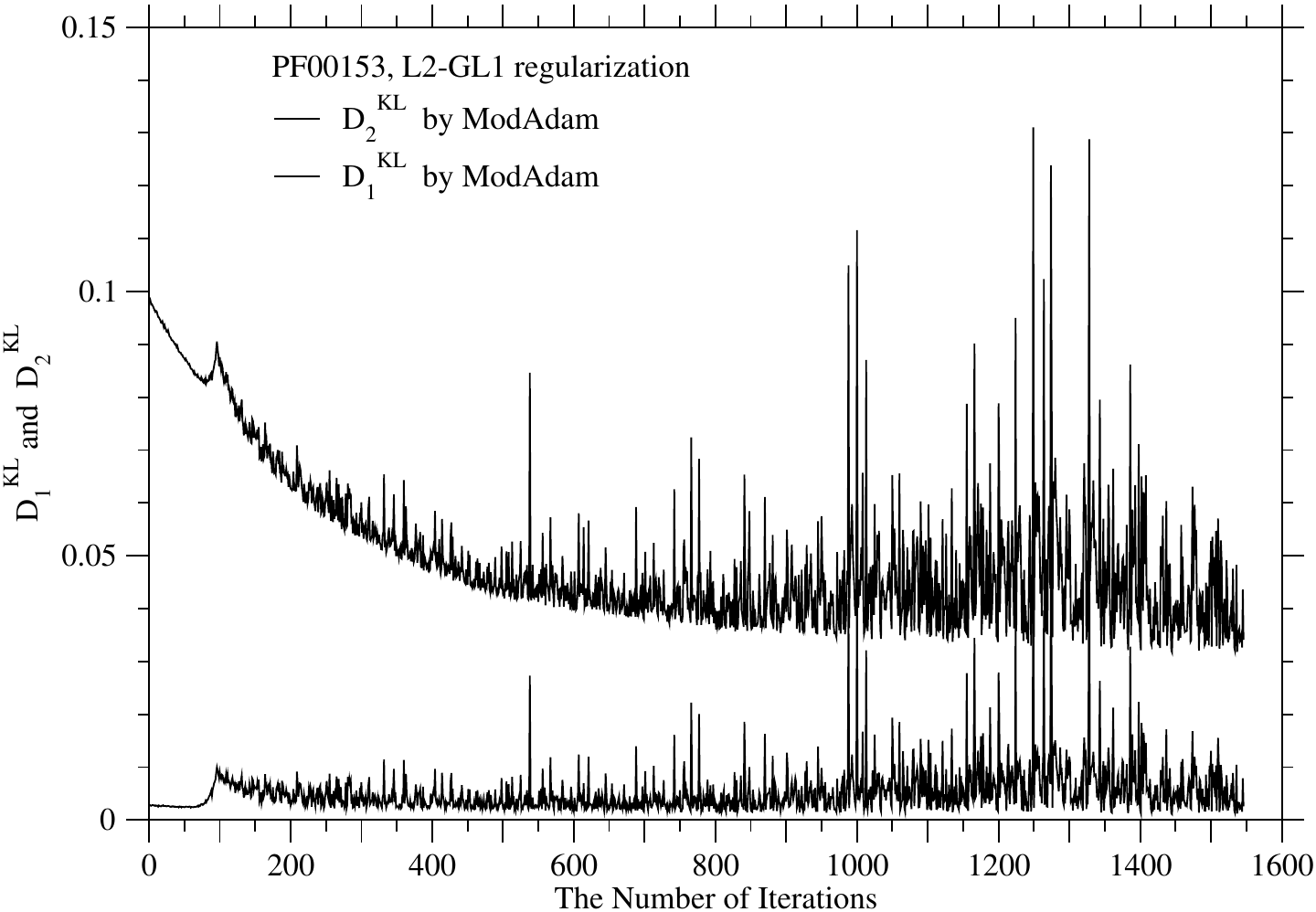}
}
\caption
{
\noindent
\TEXTBF{Learning processes by the L2-GL1 model and the ModAdam method for PF00595 and PF00153.}
The averages of Kullback-Leibler divergences,  
$D^{2}_{\text{KL}}$ for pairwise marginal distributions
and $D^{1}_{\text{KL}}$ for single-site marginal distributions, are
drawn
against iteration number in the learning processes
with the L2-GL1 model and the ModAdam method
for PF00595 and PF00153 in the left and right figures, respectively. 
The values of hyper-parameters
are listed in \Tables{\ref{tbl: PF00595_parameters} and \ref{tbl: PF00153_parameters}} as well as others.
}
\label{sfig: learning_process_KL}
\end{figure*}
\TCBB{

\begin{figure*}[hbt]
\centerline{
\includegraphics[width=63mm,angle=0]{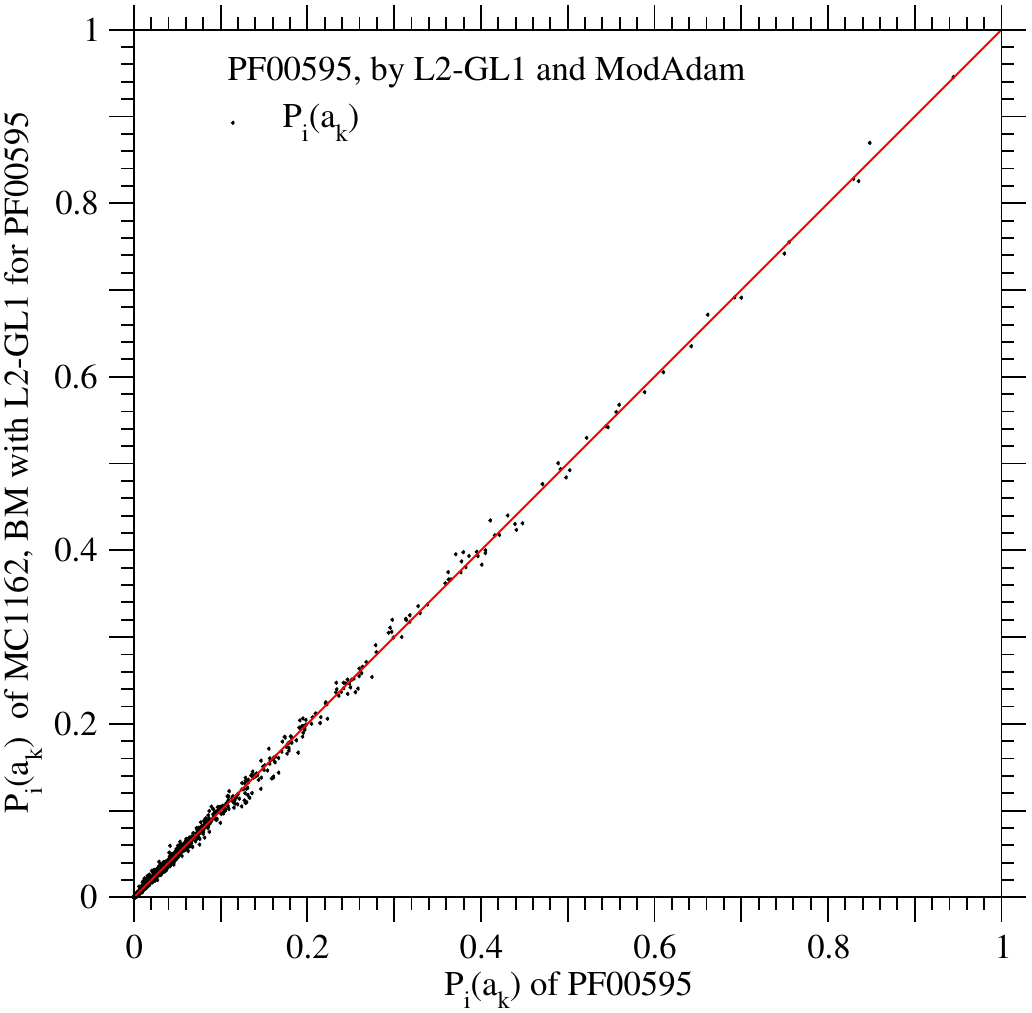}
\includegraphics[width=63mm,angle=0]{Figs_1/L2GL1_PF00595uniq_40_0_0_1_MC_1162_cijab_vs_PF00595uniq}
}
\caption
{
\noindent
\TEXTBF{Recoverabilities of the single-site frequencies and pairwise correlations
of PF00595
by the Boltzmann machine learning 
with the L2-GL1 model and the ModAdam method.}
The left and right figures are for single-site frequencies and pairwise correlations,
respectively;
$D_1^{KL} = 0.003695$ and $D_2^{KL} = 0.07594$.
The solid lines show the equal values between the ordinate and abscissa.
The overlapped points of $C_{ij}(a_k,a_l)$ in the units 0.0001 are removed.
See \Table{\ref{tbl: PF00595_parameters}} for the regularization parameters employed.
}
\label{fig: PF00595_PiaCijab}
\label{fig: PF00595_PiaPijab}
\end{figure*}

\begin{figure*}[hbt]
\centerline{
\includegraphics[width=63mm,angle=0]{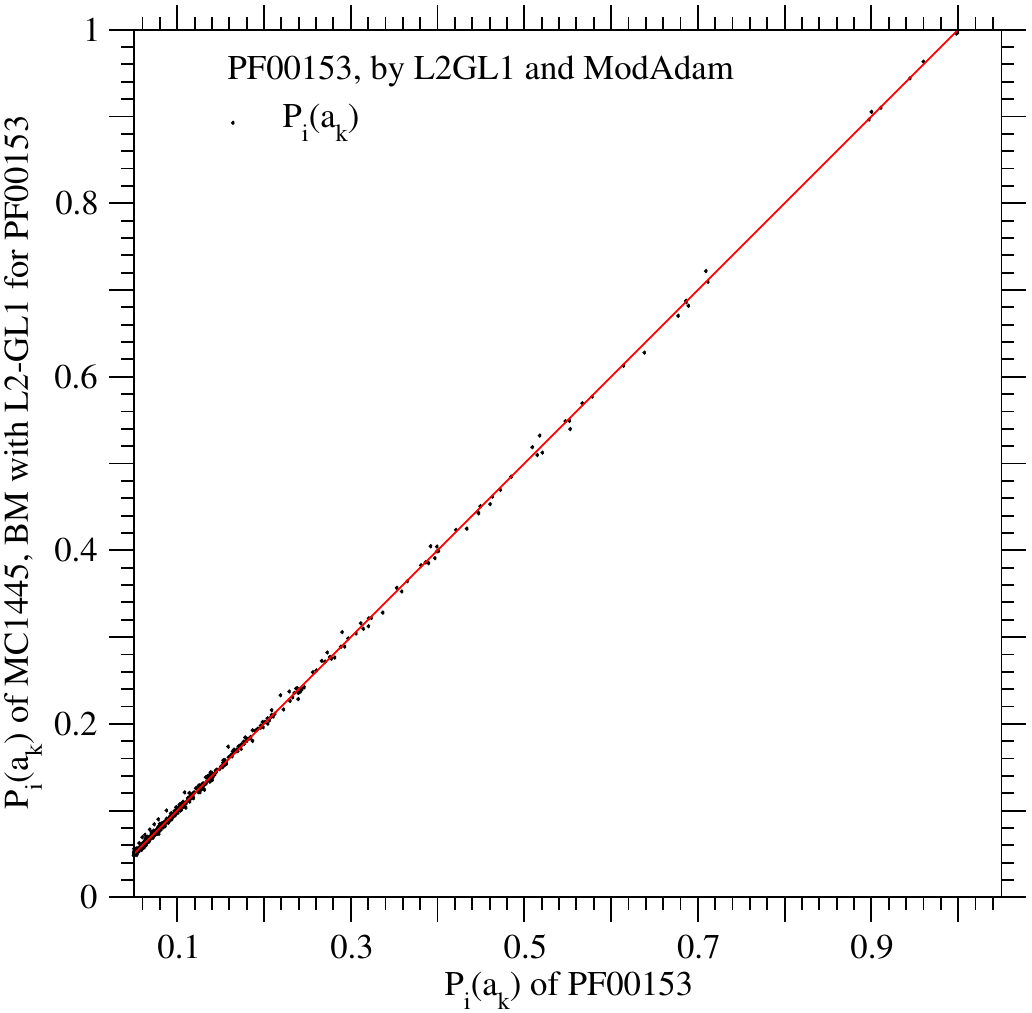}
\includegraphics[width=63mm,angle=0]{Figs_1/L2GL1_PF00153_208_9_0_1_MC_1445_cijab_vs_PF00153}
}
\caption
{
\noindent
\TEXTBF{Recoverabilities of the single-site frequencies and pairwise correlations
of PF00153
by the Boltzmann machine learning
with the L2-GL1 model and the ModAdam method.}
The left and right figures are for single-site frequencies and pairwise correlations,
respectively;
$D_1^{KL} = 0.001120$ and $D_2^{KL} = 0.03176$.
The solid lines show the equal values between the ordinate and abscissa.
The overlapped points
of $C_{ij}(a_k,a_l)$ in the units 0.0001 are removed.
See \Table{\ref{tbl: PF00153_parameters}} for the regularization parameters employed.
}
\label{fig: PF00153_PiaCijab}
\label{fig: PF00153_PiaPijab}
\end{figure*}
}{
}

\begin{figure*}[hbt]
\centerline{
\includegraphics[width=63mm,angle=0]{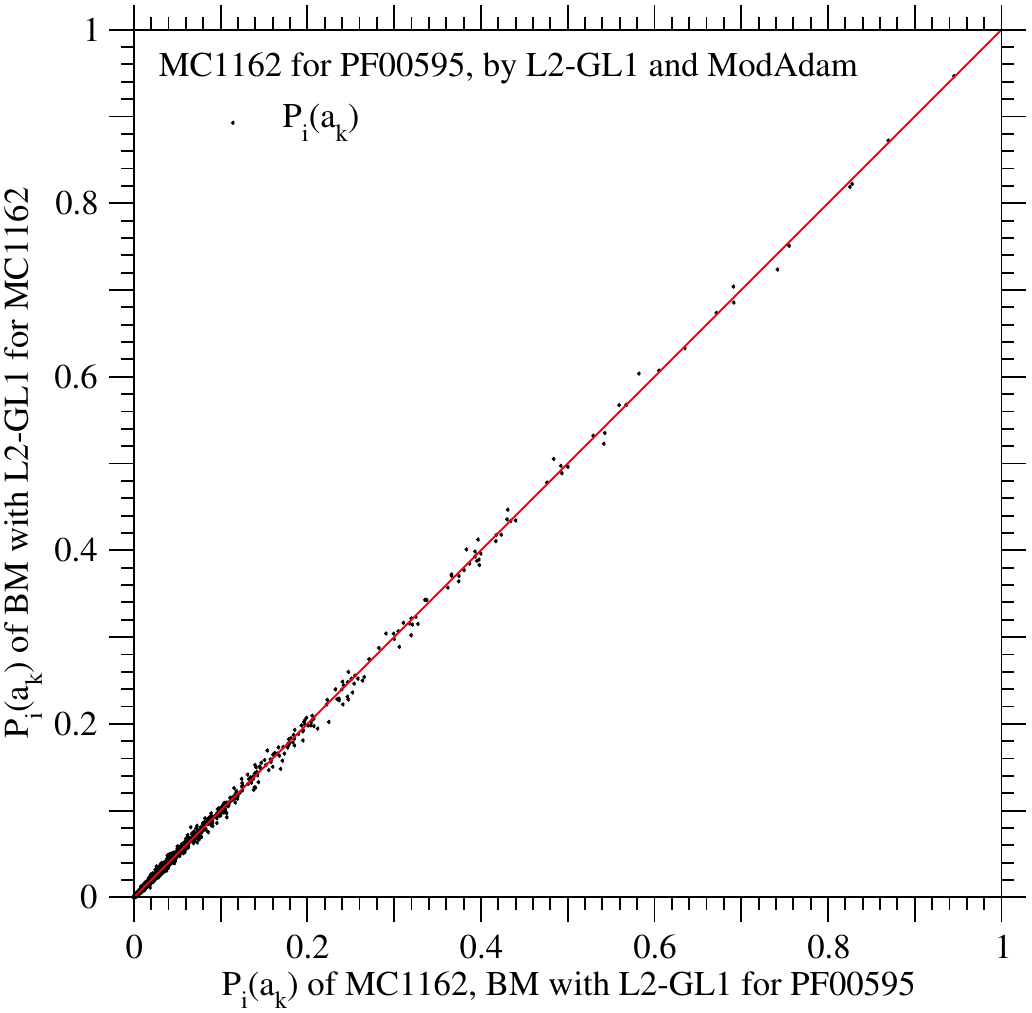}
\includegraphics[width=63mm,angle=0]{Figs_1/L2GL1_PF00595uniqMC1162_12_6_0_891_MC_1183_cijab_vs_input}
}
\vspace*{1em}
\centerline{
\includegraphics[width=63mm,angle=0]{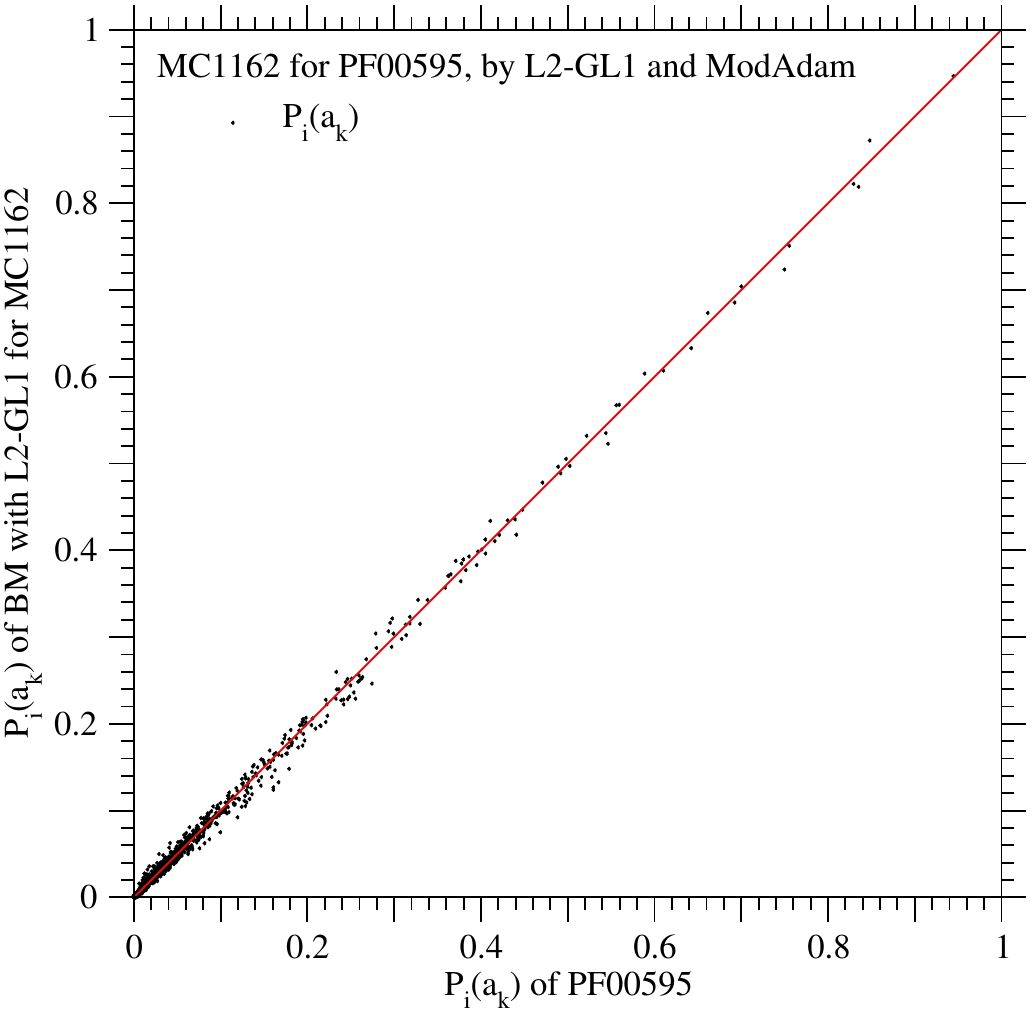}
\includegraphics[width=63mm,angle=0]{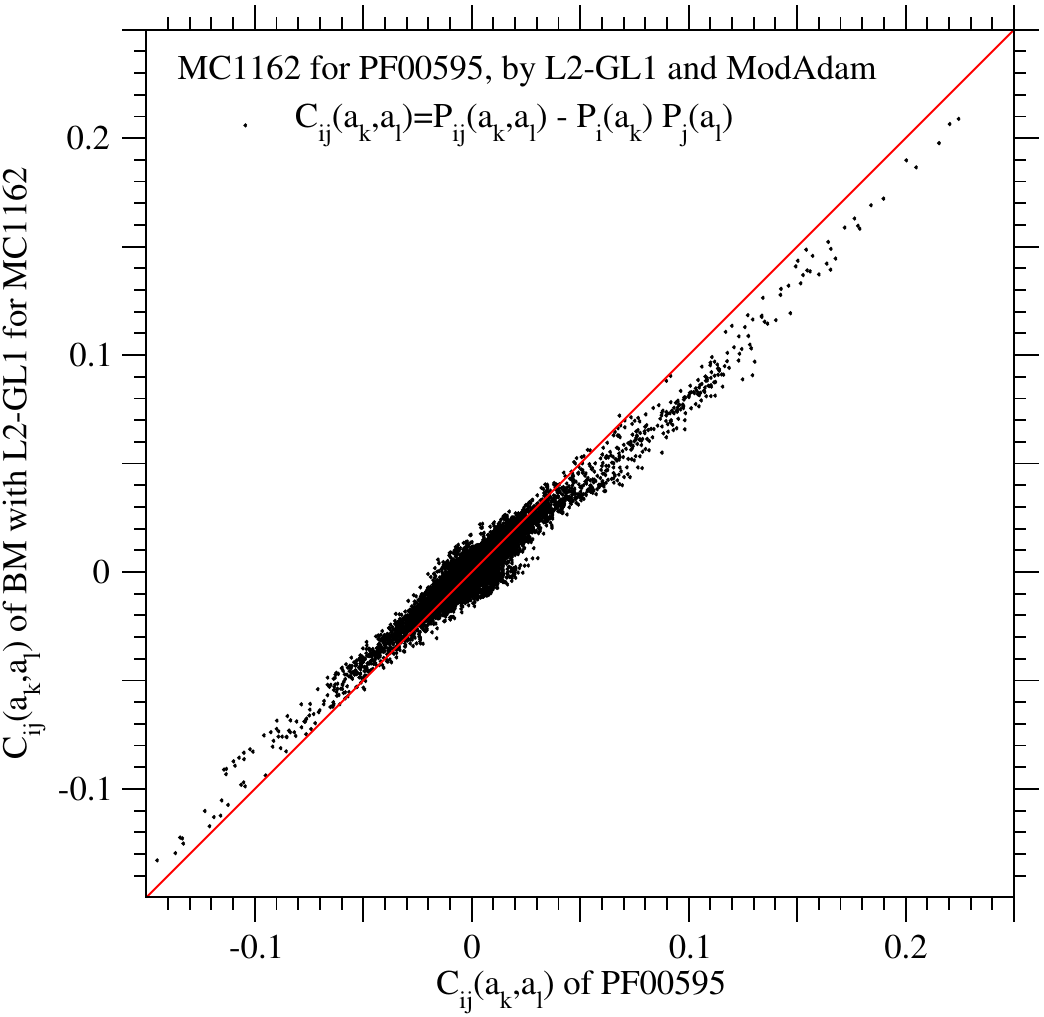}
}
\caption
{
\label{sfig: PF00595MC_PiaCijab}
\noindent
\TEXTBF{Recoverabilities of the single-site frequencies and pairwise correlations
by the Boltzmann machine learning
with the L2-GL1 model and the ModAdam method
for the protein-like sequences,
the MCMC
samples that are obtained by
the same Boltzmann machine for PF00595.}
The 
MCMC
samples obtained by the Boltzmann machine learning
with the L2-GL1 model and the ModAdam method for PF00595
are employed as protein-like sequences for which the Boltzmann machine learning 
with the same model and method is executed again
in order to examine
how precisely the marginals of the protein-like sequences
can be recovered.
The marginals recovered by the Boltzmann machine learning for the 
MCMC
samples
are compared to those of the 
MCMC
samples in the upper figures, and to those of PF00595 in the lower figures.
The left and right figures are for the single-site probabilities and pairwise correlations,
respectively.
The solid lines show the equal values between the ordinate and abscissa.
The overlapped points of $C_{ij}(a_k,a_l)$ in the units 0.0001 are removed.
See \Table{\ref{tbl: PF00595_parameters}} for the regularization parameters employed.
}
\end{figure*}

\begin{figure*}[hbt]
\centerline{
\includegraphics[width=63mm,angle=0]{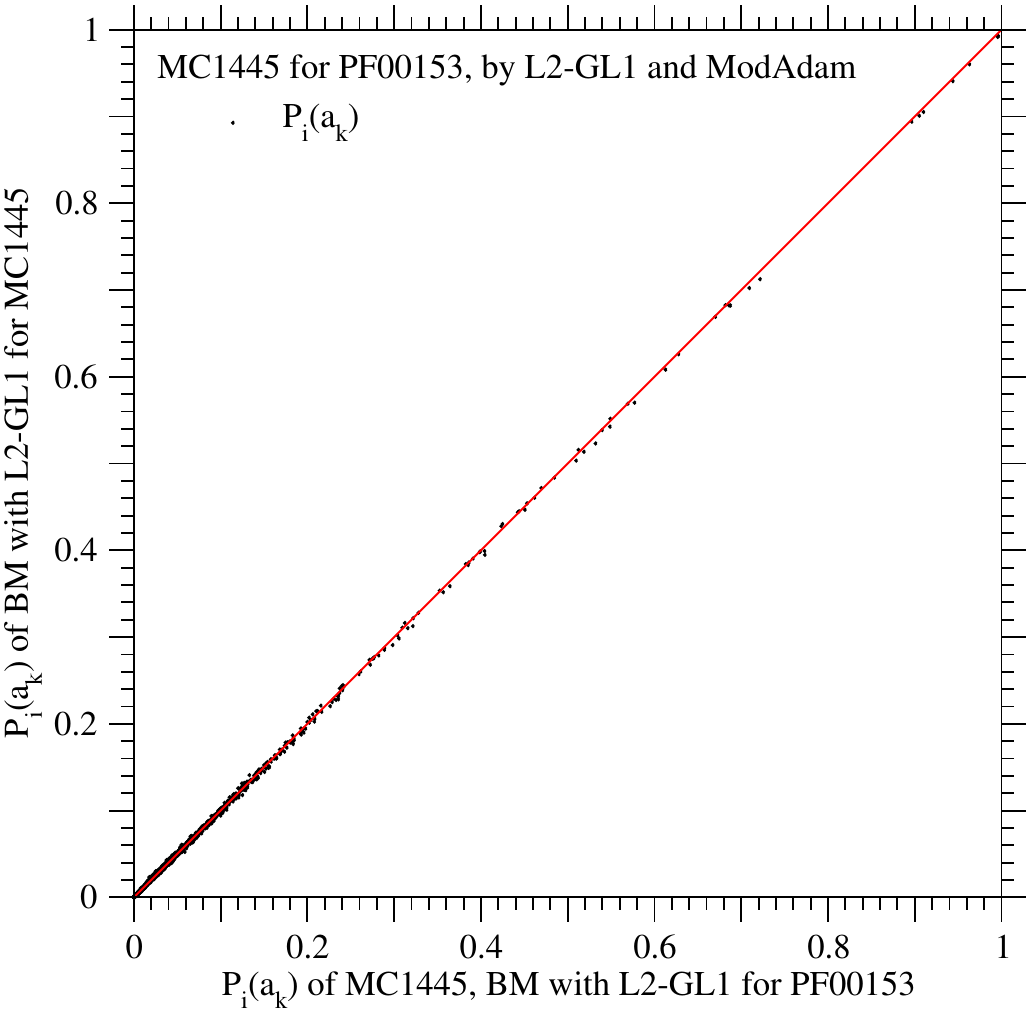}
\includegraphics[width=63mm,angle=0]{Figs_1/L2GL1_PF00153MC1445_19_95_7_94_MC_1197_cijab_vs_MC1445}
}
\vspace*{1em}
\centerline{
\includegraphics[width=63mm,angle=0]{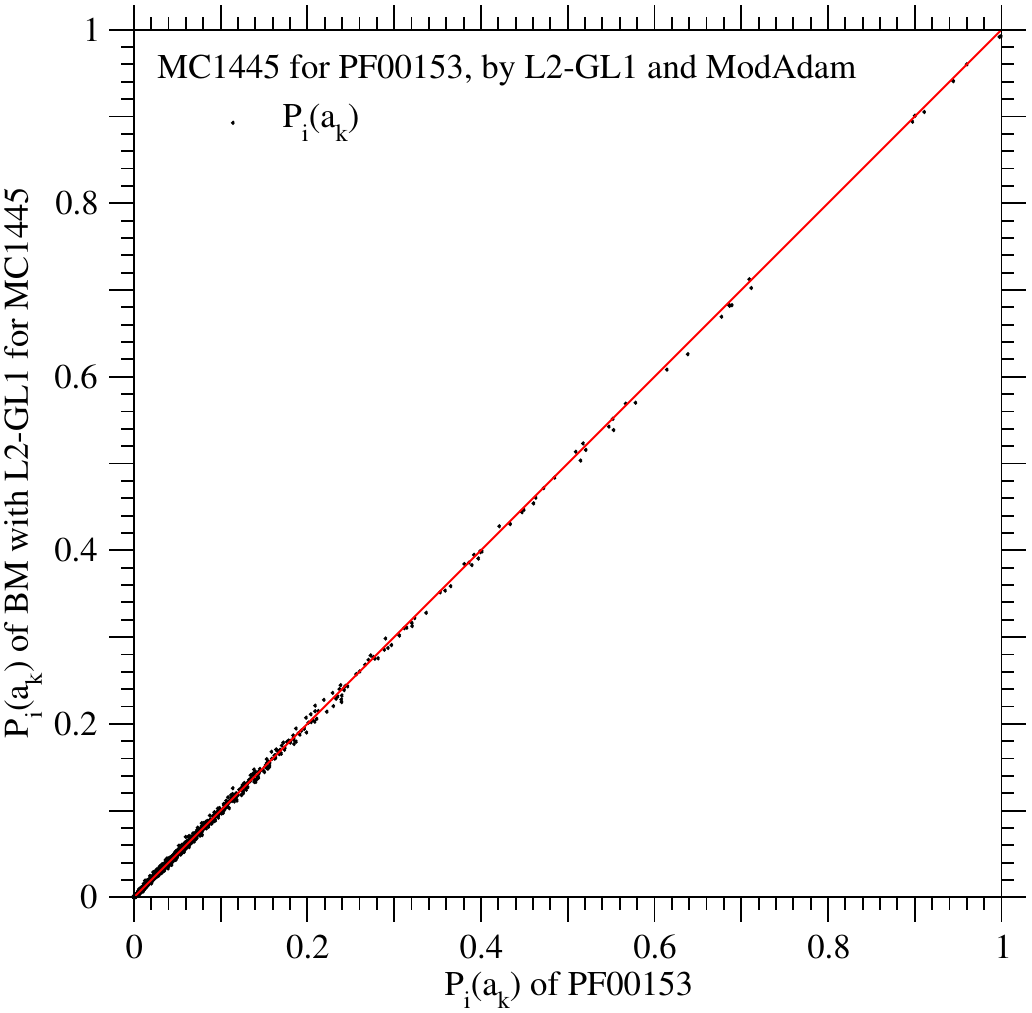}
\includegraphics[width=63mm,angle=0]{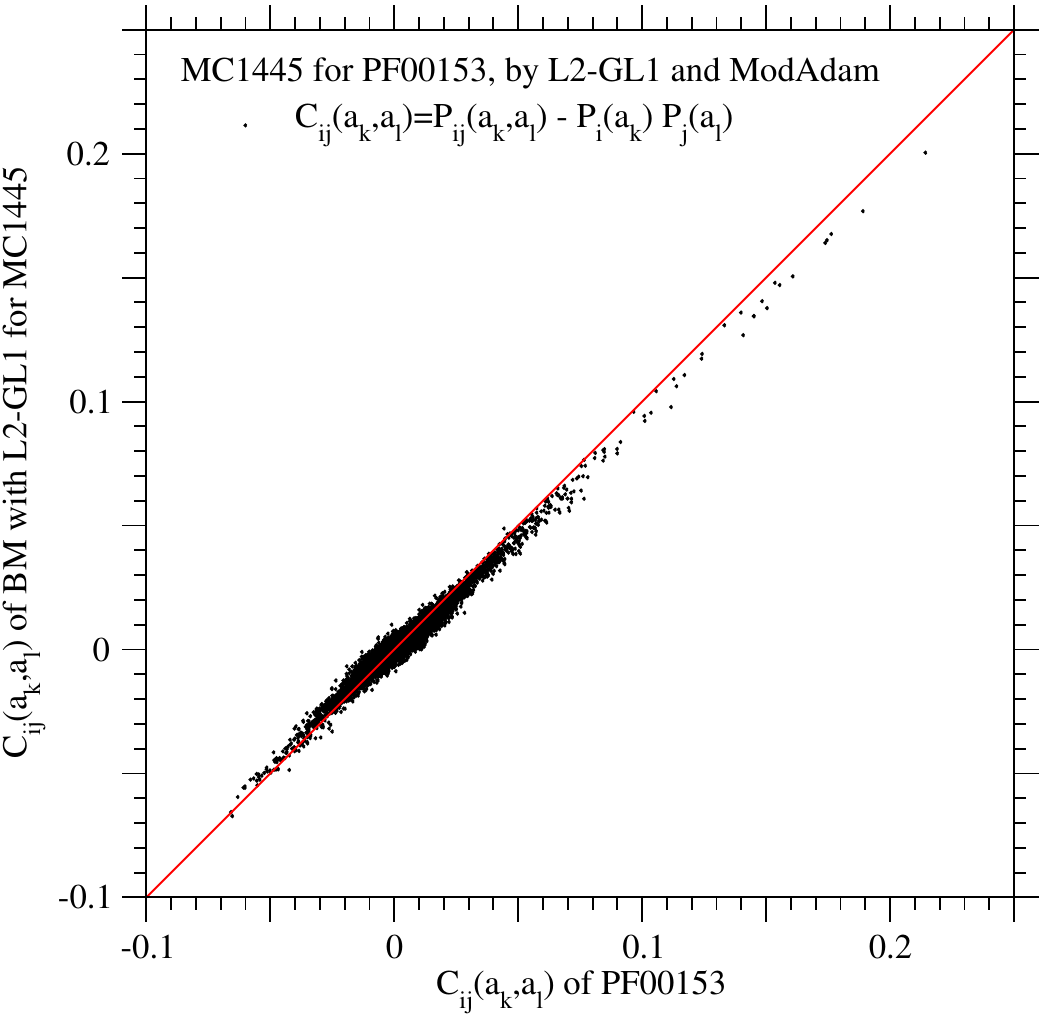}
}
\caption
{
\label{sfig: PF00153MC_PiaCijab}
\noindent
\TEXTBF{Recoverabilities of the single-site frequencies and pairwise correlations
by the Boltzmann machine learning
with the L2-GL1 model and the ModAdam method
for the protein-like sequences,
the MCMC
samples that are obtained by
the same Boltzmann machine for PF00153.
}
The 
MCMC
samples obtained by the Boltzmann machine learning
with the L2-GL1 model and the ModAdam method for PF00153
are employed as protein-like sequences for which the Boltzmann machine learning
with the same model and method is executed again
in order to examine
how precisely the marginals of the protein-like sequences
can be recovered.
The marginals recovered by the Boltzmann machine learning for the 
MCMC
samples
are compared to those of the 
MCMC
samples in the upper figures, and to those of PF00153 in the lower figures.
The left and right figures are for the single-site probabilities and pairwise correlations,
respectively.
The solid lines show the equal values between the ordinate and abscissa.
The overlapped points of $C_{ij}(a_k,a_l)$ in the units 0.0001 are removed.
See \Table{\ref{tbl: PF00153_parameters}} for the regularization parameters employed.
}
\end{figure*}
}
}

}{

}

}
}

}
}

\end{document}